\documentclass[twoside,a4paper,11pt]{book}
\usepackage{lmodern}

\usepackage{etex}

\usepackage{amssymb,amsfonts,amsthm}
\usepackage{extarrows}
\usepackage{lmodern}
\usepackage[british,spanish,es-nosectiondot]{babel} 
\usepackage[T1]{fontenc}
\usepackage[utf8]{inputenc} 

\usepackage{fancyhdr} 
\usepackage[nottoc]{tocbibind} 
\usepackage{graphicx}
\usepackage{graphbox}
\usepackage{epsfig}\parskip 5pt plus 1pt
\usepackage{multirow}
\usepackage{booktabs}  
\usepackage{epsfig}
\usepackage{epstopdf}
\usepackage{tikz-cd}
\usepackage{commath}

\usepackage{enumerate}
\usepackage{amsmath}
\usepackage{amssymb}
\usepackage{amsfonts}
\usepackage{bbm}
\usepackage{mathrsfs}
\usepackage{ytableau}
\usepackage{subfigure} 
\usepackage[font=small,labelfont=bf]{caption} 
\usepackage{comment}

\usepackage{amsbsy}
\usepackage{color}

\usepackage{braket}
\usepackage{cancel}

\def\IC{\mathbb{C}}

\def\IZ{\mathbb{Z}}
\def\IR{\mathbb{R}}
\def\IP{\mathbb{P}}

\newcommand{\cO}{\mathcal{O}}

\newcommand{\cN}{\mathcal{N}}

\newcommand{\cA}{\mathcal{A}}

\newcommand{\cB}{\mathcal{B}}

\newcommand{\cR}{\mathcal{R}}

\newcommand{\cV}{\mathcal{V}}

\usepackage{slashed}

\usepackage[hmargin=3cm,vmargin=3.4cm]{geometry}

\usepackage{tabularx} 

\newcolumntype{M}[1]{>{\centering\arraybackslash}m{#1}}
\newcolumntype{N}{@{}m{0pt}@{}}

\definecolor{MyGrey}{rgb}{0,0,0} 
\definecolor{MyDarkBlue}{rgb}{0.23,0.21,0.69} 
\definecolor{MyLightBlue}{rgb}{0.22,0.51,0.86}
\definecolor{MyDarkRed}{rgb}{0.69,0.21,0.23}

\usepackage[colorlinks=true,linkcolor=MyDarkRed,citecolor=MyDarkRed,urlcolor=MyLightBlue,bookmarksnumbered=true,bookmarksopen]{hyperref}

\usepackage{cite} 
\usepackage{cleveref}
\usepackage{float}

\usetikzlibrary{mindmap,trees}

\usepackage[times]{quotchap}

\def\cy {{\text{CY}}}
\newcommand{\pr}{\mathbbm{R}}

\def\beq{\begin{equation}}
\def\eeq{\end{equation}}
\def\beqa{\begin{eqnarray}}
\def\eeqa{\end{eqnarray}}

\def\Om{\Omega}
\def\om{\omega}
\def\th{\theta}

\def\-{\hphantom{-}}
\def\ov{\overline}
\def\s2{\frac{1}{\sqrt2}}

\def\oh{\frac{1}{2}}

\def\IC{\mathbb{C}}
\def\sig{{\sigma}}

\def\raw{\rightarrow}

\def\CK {{\cal K}}
\def\CM {{\cal M}}
\def\mM {{\cal M}}
\def\CR {{\cal R}}
\def\CN {{\cal N}}

\def\cK {{\cal K}}

\def\CO {{\cal O}}

\def\mm {{\mathcal M}}
\def\IF{\relax{\rm I\kern-.18em F}}
\def\II{\relax{\rm I\kern-.18em I}}
\def\IP{\relax{\rm I\kern-.18em P}}
\def\IC{\relax\hbox{\kern.25em$\inbar\kern-.3em{\rm C}$}}
\def\IR{\relax{\rm I\kern-.18em R}}

\def\cA {{\cal A}}
\def\cB {{\cal B}}
\usepackage{multicol}
\def\be{\begin{equation}}
\def\ee{\end{equation}}
\def\bea{\begin{eqnarray}}
\def\eea{\end{eqnarray}}
\def\bes{\begin{subequations}}
\def\ees{\end{subequations}}
\newcommand{\cM}{\mathcal{M}}
\def\vol  {{\text{vol}}}
\def\Vol  {{\text{Vol}}}
\def\mm {{\mathcal M}}
\def\re{{\text{Re}}}
\usepackage[font={small}]{caption}

\def\bZ{\mathbb{Z}}
\def\Z{\mathbb{Z}}
\def\IM{\text{Im}\,}
\def\im{\text{Im}\,}
\def\RE{\text{Re}\,}
\def\ov{\overline}
\usepackage[skins,theorems,most]{tcolorbox}
\usepackage{dsfont}
\def\bfG {{\textbf{G}}}
\def\bfC {{\textbf{C}}}
\def\mk {{\mathcal{K}}}

\def\oh{\frac{1}{2}}
\def\a{{\alpha}}
\def\b{{\beta}}
\def\d{{\delta}}
\def\eps{{\epsilon}}
\def\th{{\theta}}

\def\Om{{\Omega}}
\def\om{{\omega}}
\def\sig{{\sigma}}

\def\g{{\gamma}}

\def\p{{\partial}}

\def\r{{\rangle}}

\usepackage{dsfont}

\def\mk {{\mathcal K}}

\def\mm {{\mathcal M}}

\def\r {{\rho}}
\def\hr {{\hat{\rho}}}
\def\tr {{\tilde{\rho}}}
\def\p {{\partial}}
\def\e {{\epsilon}}
\def\g {{\gamma}}
\def\hg {{\hat{\gamma}}}
\def\te {{\tilde{\epsilon}}}
\def\he {{\hat{\epsilon}}}
\def\tg {{\tilde{\gamma}}}

\def\s {{\sigma}}

\def\th {{\theta}}
\def\hth {{\hat{\theta}}}
\def\ph   {{\phi}}
\def\re   {{\text{Re}}}
\def\vol  {{\text{vol}}}
\def\Vol  {{\text{Vol}}}

\def\tr{\operatorname{tr\:}}     \def\Tr{\operatorname{Tr\,}}

   \renewcommand{\Im}{\mathop{\rm Im}}
\newtheorem{Theorem}{Theorem}[section]
\newtheorem{Lemma}{Lemma}[section]
\newtheorem{Corrolary}{Corrolary}[section]

\newcommand{\bt}{\begin{Theorem}}   \newcommand{\et}{\end{Theorem}}
\newcommand{\bl}{\begin{Lemma}}     \newcommand{\el}{\end{Lemma}}
\newcommand{\bc}{\begin{Corrolary}} \newcommand{\ec}{\end{Corrolary}}

\graphicspath{{./Introduction/}{./1-Axion/}{./2-Instanton/}{./3-BFT/}{./4-AdS/}}

\usepackage{dsfont}

\renewcommand{\thesection}{\arabic{chapter}.\arabic{section}{}}

\def\R{\mathbb{R}}
\newcommand{\cF}{\mathcal{F}}

\usepackage[normalem]{ulem}

\def\mk {{\mathcal K}}

\def\r {{\rho}}
\def\hr {{\hat{\rho}}}
\def\tr {{\tilde{\rho}}}
\def\p {{\partial}}
\def\e {{\epsilon}}
\def\g {{\gamma}}
\def\hg {{\hat{\gamma}}}
\def\te {{\tilde{\epsilon}}}
\def\he {{\hat{\epsilon}}}
\def\tg {{\tilde{\gamma}}}

\def\s {{\sigma}}

\def\th {{\theta}}
\def\hth {{\hat{\theta}}}
\def\ph   {{\phi}}
\def\CN {{\cal N}}
\def\trh {{\tilde{\rho}}}
\def\sig{{\sigma}}
\def\d{{\delta}}
\def\raw{\rightarrow}

\def\eps{{\epsilon}}
\def\oh{\frac{1}{2}}
\def\CG {{\cal G}}

\usepackage[utf8]{inputenc}
\def\R{\mathbb{R}}
\def\C{\mathbb{C}}

\title{Swampland quien lo lea}
\author{Joan Quirant Pell\'in}
\makeindex

\begin{document}

\pagenumbering{roman} 
\pagestyle{headings}
\pagestyle{empty}

\newcommand{\HRule}{\rule{\linewidth}{1mm}}
\setlength{\parindent}{1cm}
\setlength{\parskip}{1mm}
\noindent




\pagestyle{fancy}
\fancyhead{}
\fancyfoot{}
\fancyhead[LE] {\itshape\nouppercase\leftmark}
\fancyhead[RO] {\itshape\nouppercase\rightmark}
\fancyfoot[C]{\thepage}
\renewcommand{\headrulewidth}{0.3pt}
\thispagestyle{empty}

\fancypagestyle{simple}{
  \fancyhead{}
  \fancyfoot[C]{\thepage}
 \renewcommand{\headrulewidth}{0.0pt} 
}

\noindent
\HRule
\begin{center}
\huge{\textbf{Aspects of type IIA AdS$_4$ orientifold vacua}}
 \vspace{0.2cm}
\end{center}
\HRule

\vspace{1.0cm}

\begin{center}

\large{	   
Memoria de tesis doctoral realizada por \\[3mm]
\textbf{\large{Joan Quirant Pell\'in}} \\[3mm]
presentada ante el Departamento de F\'isica te\'orica \\[1mm]                  
de la Universidad Aut\'onoma de Madrid \\[1mm]
para optar al t\'itulo de doctor en f\'isica te\'orica. \\[1mm]
}

\vspace{1.3cm}
Tesis doctoral dirigida por \textbf{\large{Fernando G. Marchesano Buznego}}, \\[2mm]
cient\'ifico titular del CSIC. \\

\end{center}

\vspace{0.7cm}

\begin{center}
{\Large {Departamento de F\'isica Te\'orica\\[1mm] Universidad Aut\'onoma de Madrid }}\\
\vspace{0.5cm}
{\Large {Instituto de F\'isica Te\'orica UAM-CSIC}}\\
\end{center}
\vspace{1.2cm}

\begin{figure}[ht]
\centering
\begin{tabular}{cc}
\includegraphics[height=2.2cm]{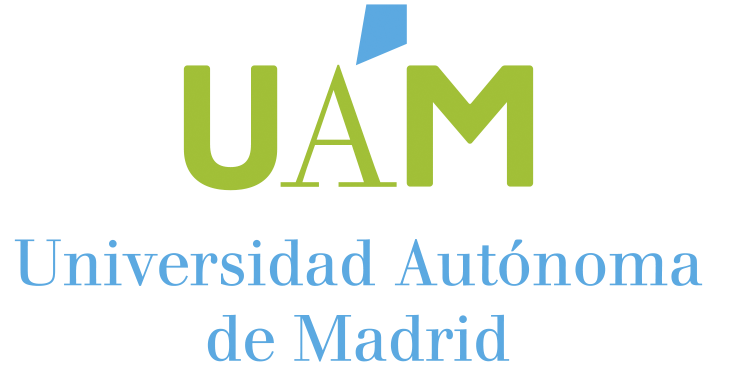}\qquad\qquad
\includegraphics[height=2.2cm]{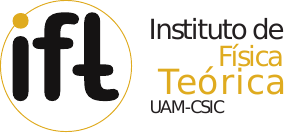}\\~\\~\\~\\
\includegraphics[height=1.3cm]{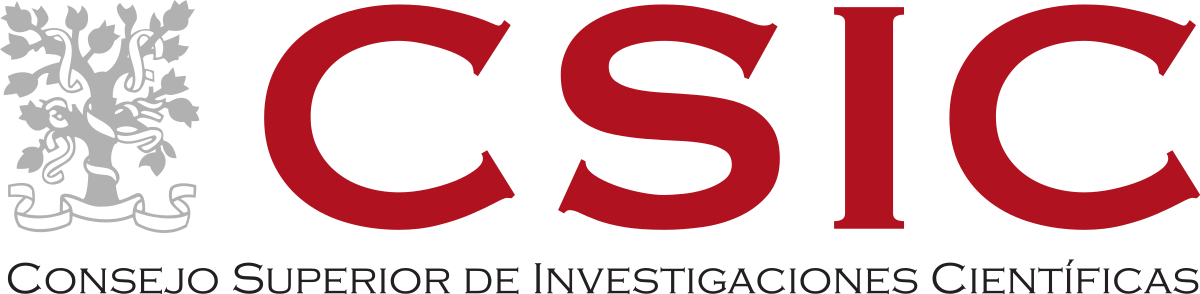}
%
%
%
\end{tabular}
\end{figure}

\vspace{0.5cm}

\begin{center}
{\large Junio de 2022}
\end{center}

\thispagestyle{empty}
\clearpage

\begin{flushright}
    \thispagestyle{empty}
    \vspace*{30mm}
	A tots aquells docents que van fer que volguera seguir aprenent.\\~\\~\\~\\  ``\textit{Chambea, ¡jala!}''\\ Porque el doctorado tambi\'en era eso.
    \vspace*{\fill}
\end{flushright}
\clearpage
\thispagestyle{empty}



\newpage
\vspace*{120pt}
\thispagestyle{empty}
\newpage
\thispagestyle{empty}

Esta tesis doctoral está basada en los siguientes artículos:
\begin{enumerate}
\item[{\hypersetup{hidelinks}\cite{Marchesano:2019hfb}}]{\emph{A Landscape of AdS Flux Vacua},\\
F. Marchesano, \textbf{J. Quirant},\\
\href{https://link.springer.com/article/10.1007/JHEP12(2019)110}{\emph{JHEP} \textbf{12} (2019) 110}
[\href{https://arxiv.org/abs/1908.11386}{arXiv:1908.11386}]};

\item[{\hypersetup{hidelinks}\cite{Marchesano:2020qvg}}]{\emph{On supersymmetric AdS$_4$ orientifold vacua}, \\
F. Marchesano, E. Palti, \textbf{J. Quirant}, A. Tomasiello,  \\
\href{https://link.springer.com/article/10.1007/JHEP08(2020)087}{\emph{JHEP} \textbf{08} (2020) 087} 
[\href{https://arxiv.org/abs/2003.13578}{arXiv:2003.13578}]};

\item[{\hypersetup{hidelinks}\cite{Marchesano:2020uqz}}]{\emph{Systematics of Type IIA moduli stabilisation}, \\
F. Marchesano, D. Prieto,  \textbf{J. Quirant}, P. Shukla,\\
\href{https://link.springer.com/article/10.1007/JHEP11(2020)113} {\emph{JHEP} \textbf{11} (2020) 113} 
[\href{https://arxiv.org/abs/2007.00672}{arXiv:2007.00672}]};

\item[{\hypersetup{hidelinks}\cite{Marchesano:2021ycx}}]{\emph{BIonic membranes and AdS instabilities}. \\
F. Marchesano, D. Prieto,  \textbf{J. Quirant},\\ 
\href{https://link.springer.com/article/10.1007/JHEP07(2022)118} {\emph{JHEP} \textbf{07} (2022) 118} 
[\href{https://arxiv.org/abs/2110.11370}{arXiv:2110.11370}]}.
\end{enumerate}

Otros art\'iculos escritos durante la realizaci\'on del doctorado y que no se incluyen en esta tesis son
\begin{enumerate}
\item[{\hypersetup{hidelinks}\cite{Quirant:2022fpn}}]{\emph{Noninteger conformal dimensions for type IIA flux vacua},\\
\textbf{J. Quirant},\\
\href{https://journals.aps.org/prd/abstract/10.1103/PhysRevD.106.066017} {\emph{Phys. Rev. D}  \textbf{106} (2022) 066017 
}[\href{https://arxiv.org/abs/2204.00014}{arXiv:2204.00014}]};
\item[{\hypersetup{hidelinks}\cite{nuevo}}]{\emph{New instabilities for non-supersymmetric AdS$_4$ orientifold vacua},\\
Fernando Marchesano, \textbf{Joan Quirant}, Matteo Zatti,\\
\href{https://link.springer.com/article/10.1007/JHEP10(2022)026} {\emph{JHEP} \textbf{10} (2022) 026} [\href{https://arxiv.org/abs/2207.14285}{arXiv:2207.14285}]}.
\end{enumerate}

\selectlanguage{spanish}
\thispagestyle{empty}
\chapter*{Agradecimientos}
No puedo sino empezar dándole las gracias a Fernando por, sin él saberlo, hacer que abandonara el \textit{caloret faller} de València y me viniera a Madrid a vivir su \textit{libertad}  y, sobre todo, a aprender qué era eso de la teoría de cuerdas. Gracias por iniciarme en esto de la investigación, por tener (literalmente) siempre la puerta abierta para todo lo que necesitara, por tener paciencia con todas mis preguntas y por enseñarme tanta física.  Aun con una pandemia y el teletrabajo de por medio (y un montón de meses sin poder discutir nada en persona), ha sido un privilegio haber podido aprender tanto. ¡Espero que durante estos años se me haya pegado algo de esa intuición que tienes para abordar un problema y saber qué está pasando!

Gracias también a Ángel y a Luis por enseñarme  tanta física en los SPLEs con vuestros comentarios y vuestras preguntas. ¡Espero que mis memes os gustaran! (o al menos no os disgutaran). Gracias Ángel por el curso en teoría de cuerdas (que me ayudó a orientarme cuando todavía iba un poco perdido) y por enseñarnos que el conocimiento no solo hay que buscarlo sino también divulgarlo. Thank also Eran and Alessandro for your illuminating discussions, and Eran for giving me the opportunity to continue learning. Gracias Mariana por acogerme esos tres meses en París, ¡fueron muy enriquecedores!.

Hablando de gente de la que he podido aprender, no puedo olvidarme de mi hermanito Max, I hope to see you soon! De Florent -sorry for skipping always all the mandatory beers, but you know me-, Wieland, que aguantó todas mis preguntas de novatillo, Álvaro, que hizo que en París me sintiera muy a gusto y Edu (senior) que, a ver cuándo hace queso por fin. Thank you also Pramod for your motivation and your work.

De estos años también me llevo unos cuantos amigos (de los que también aprendo cosas, claro, pero tenía que ir agrupando). Thought it was short, it was a pleasure meeting you Pierre! I will ask you for recommendations in Israel! Ginevra y Alessandro, que empezamos el doctorado juntos, ¡espero volver a veros! José, \textit{qué pasa crack, figura}, éramos  unos prigandillos y míranos ahora. Lo seguimos siendo, solo que en distintos países.  Y ojo con la nueva hornada del grupo de cuerdas del IFT, que viene pisando fuerte: abdominales Alberto, David -con el que sufrí todo ese camino de encontrar unos convenios consistentes, ¡ha sido un verdadero placer trabajar contigo!-, Jesús, Luca, Ignacio, Matteo, Matilda y Roberta. ¡Habéis hecho que los congresos hayan sido mucho más divertdos (cod cachopo, no olvidamos)!

Más allá de la gente del grupo de cuerdas, el ambiente en el IFT ha sido siempre súper bueno y no quiero dejar de mencionarlo. Fran, Jorge y Llorenç, habéis hecho que las comidas siempre fueran muy divertidas. Víctor, a ver si vuelves ya, que casi no hemos tenido tiempo de compartir despacho. A la gente de secretaría y de gestión del IFT, gracias por vuestra infinita paciencia con todos los papeleos y por ayudarnos con todo.

Edu y Martín, a vosotros os dejo en un párrafo aparte. Suena a tópico, pero estos 4 años hubieran sido muy diferentes sin vosotros. Martín, tío, si alguna vez volvemos a vivir juntos, ¡espero verte más de 4 días por el piso! Gracias Edu por nuestras charlas, por aguantarme siempre con mis historias y por tus consejos (aunque al final nunca te hiciera caso). ¡Sois unos grandes!

A la gent que vaig deixar en Val\`encia quan em vaig vindre a Madrid i amb la qual vaig donar els meus primers pasos en la f\'isica, ¡un sopar a l'any no \'es suficient! Kevin i Pepe: ens hem de vore mes, que vos trobe a faltar (que tornen l'amor i els ridicles). Esos tres meses en París, Kevin, fueron la bomba. Vos estime!

A toda la gente que ha hecho que en Madrid me haya sentido como en casa y que estos años hayan sido muy felices, muchas gracias. Ángel y Rafa, mira que nos lo pasamos siempre bien y qué poco quedamos. Tenemos que vernos más, ahora que todavía compartimos ciudad. ¿Cuándo volvemos al Nuria? Andrea, ha sido un placer conocerte y compartir cenas siempre súper ricas. ¡El yoga que no te lo quiten! Blanca, gracias por esos meses y por enseñarme tanto casi sin darte cuenta. Gràcies Miquel per tots els bons moments que hem compartit junts, per ser un amic i el millor company de cordada que algú podria tindre (de cara a la galeria direm que som uns fanàtics, però els dos sabem que som uns frikazos de l'escalada). Eixos dijousos patoneros s'han de repetir, ara que encara podem. Gracias también a Dani por motivarme (carnet de local de la cueva incluido)  y en general a toda la gente que he conocido en la roca y con la que he compartido muy buenos momentos.

No me quiero olvidar en esta tesis de Juan, Javi, Sergio, Machote, Pere y Jordi. Sé que últimamente ya nos vemos poco (menos de lo que me gustaría, si os soy sincero) pero gracias por haberme acompañado en este romance mío con los libros. Hacerse mayor con vostros está siendo (o ha sido, porque ya vamos teniendo una edad) muy divertido.

La Clau, tú, que ya estaba leyendo esto, viendo que se acababan los agradecimientos y pensando que no le iba a decir nada. Gracias por contribuir a que me sienta así de feliz (el queso en un plato de pasta súper rico). Gracias por quererme así de bien y por todas las experiencias tan chulas que estamos viviendo. Por ser la primera amiga que tuve en Madrid. ¿Quién te lo iba a decir a ti, ese día en los montaditos, que acabarías apareciendo así en esta tesis? \textit{En la discoteca, se ponen loco'...}

Esta tesis no podria existir -literalment- si no fora per la meua família. Gràcies papa i mama per recolzar-me, per deixar que estudiara el que jo volia, animant-me i sense posar-me pegues. Gràcies per escoltar-me, per donar-me sempre els millors consells i per fer que haja pogut arribar fins ací (ara podreu dir que teniu un fill doctor!). Gràcies Anna per comprendre  les meues inquietuds vitals i compartir junts tots els drames de jóvens formats i precaris. Vos estime molt als tres.

A todos los que he mencionado y a los que no (porque la tesis ya es larga y no es cuestión de añadir más páginas), gracias. Nos vemos en el camino.

\thispagestyle{empty}
\chapter*{Abstract}
\thispagestyle{empty}

In this thesis we explore the vacua structure of type IIA orientifold (CY) compactifications with fluxes, both from the 4d and the 10d point of view.

We start by reviewing type IIA Calabi-Yau orientifold compactifications with fluxes. First we consider only RR and NSNS fluxes, and then add (non)-geometric fluxes. We recall how the potential created by the fluxes can be rewritten as a bilinear expression, which is very useful to do a systematic search of vacua. We also review the main swampland conjectures that can be applied to these scenarios.

Once the basics have been setted, we perform a systematic search of vacua, using directly the 4d effective action and the potential generated.  We focus first on the case with only RR and NSNS fluxes. This generalises previous results in the literature, computed in toroidal examples, to any Calabi-Yau orientifold. We obtain several families of SUSY and non-SUSY AdS$_4$ vacua. We study their stability. These vacua have the property that scale separation between the internal radius and the AdS$_4$ radius can be obtained parametrically, tuning the $G_4$ flux.  Scale separation is in tension with the strong version of the AdS distance conjecture, we comment on this. We then repeat the same game by including geometric fluxes. We take an ansatz for the geometric fluxes in the vacuum -motivated by stability arguments- and find again several families of SUSY and non-SUSY AdS$_4$ vacua. We check which of them are stable. In this case, we are not able to find a regime with scale separation.

We then put on the 10d glasses to analyse from the 10d perspective the SUSY vacua derived in the 4d effective theory.  We comment on how an uplift to $SU(3)$ structure manifolds does not exist unless the O6 planes are smeared along the internal dimensions. This is the so-called smeared approximation. Though these results are consistent with  the previous literature, it is important to point out that in this case we are not describing the true physical situation, in which the sources must be localised in the internal manifold. We then go beyond this approximation and look for an uplift in $SU(3)\times SU(3)$ structure manifolds, considering only the case with RR and NSNS fluxes. To do so, we expand the equations of motion in terms of $g_s$. At zeroth order we recover the smeared approximation. We solve the expansion at first order, where the localised nature of the O6-planes is taken into account -but not the intersecting terms between the different O6-planes, which would appear at next order-.

After this, we use the same machinery to study the non-perturbative stability of the family of non-SUSY vacua with only RR and NSNS fluxes characterised by having $G_4^{\text{non-SUSY}}=-G_4^{\text{SUSY}}$ (they are perturbatively stable). According to a refined version of the weak gravity conjecture, there should be membranes in the spectrum with $Q>T$, triggering its decay. We see that D8 branes wrapping the internal manifold with D6 branes ending on them can satisfy this requirement, making these vacua unstable. 

We finish the thesis by recapping all the results and making some comments about which questions are still open, as well as possible future lines of research.

\newpage
\thispagestyle{empty}

\chapter*{Resumen}
\selectlanguage{spanish}\thispagestyle{empty}
En esta tesis exploramos la estructura de vacíos de la teoría de cuerdas tipo IIA cuando se compactifica en un \textit{orientifold} (de una variedad de Calabi-Yau) con flujos, tanto desde el punto de vista 4-dimensional, como desde el punto de vista  10-dimensional.

Lo primero que hacemos en repasar la acción efectiva que se obtiene al compactificar la teoría de de cuerdas tipo IIA en un \textit{orientifold} de un Calabi-Yau cuando se incluyen flujos.  Primero consideramos solo flujos de tipo RR y NSNS y luego añadimos también flujos (no)-geométricos. Recordamos cómo el potencial creado por los flujos se puede reescribir en una formulación bilineal, lo que resultará desupués muy útil para realizar una búsqueda sistemática de vacíos. También repasamos las principales conjeturas de la ciénaga (\textit{swampland conjectures}) que se pueden aplicar a estos escenarios.

Una vez hemos establecido los coneptos básicos, realizamos una búsqueda sistemátia de vacíos, usando directamente la acción efectiva en 4d y el potencial generado. Nos centramos primero en el caso en el que solo incluímos flujos del tipo RR y NSNS. Esto generaliza los resultados previamente obtenidos en la literatura, donde los cálculos se habían hecho usando modelos toroidales, a compactificaciones en \textit{orientifolds} de cualquier variedad de Calabi-Yau. Obtenemos varias familias de vacíos AdS$_4$ tanto supersimétricos como no-supersimétricos. Estudiamos su estabilidad. Todos estos vacíos tienen la característica de que uno puede ir a un régimen en el que el radio de la variedad interna es paramétircamente mucho más pequeño que el radio de AdS$_4$, simplemente ajustando el flujo $G_4$. Esta separación de escalas está en coflicto con la \textit{AdS distance conjecture}, como discutimos posteriormente. Hecho el análisis con flujos RR y NSNS, repetimos la jugada pero incluyendo también flujos geométricos. Para ello asumimos un \textit{ansatz} sobre la forma que los flujos geométricos deben tener en el vacío, guiados por argumentos de estabilidad. Obtenemos de nuevo varias familias de vacios AdS$_4$ tanto supersimétricas como no-supersimétricas. Comprobamos cuáles de ellas son estables. En este caso no somos capaces de encontrar un régimen con separación de escalas entre la variedad interna y la externa.

A continuación nos ponemos las gafas 10-dimensionales para analizar desde esta perspectiva los vacíos supersimétricos obtenidos con la acción efectiva en 4d. Discutimos por qué un \textit{uplift}\footnote{Esto es, una solución a las ecuaciones 10 dimensionales que reproduzca lo que vemos en 4d} de estos vacíos en variedades con una estructura $SU(3)$ no es posible a menos que los O6-planos estén deslocalizados en las dimensiones internas. Es la llamada aproximación \textit{smeared} (que podríamos traducir como aproximación \textit{deslocalizada}). Estos resultados son consistentes con  lo obtenido previamente en la literatura. Hay que tener en cuenta que no describen la verdadera situación física, en la que los O6-planos deben estar localizadas en las dimensiones internas.  El siguiente paso es ir más allá de esta aproximación y buscar un \textit{uplift} en variedades con estructura $SU(3)\times SU(3)$, para lo que nos centramos en el caso solo con flujos  de tipo RR y NSNS. Para ello expandimos las ecuaciones de movimiento en términos de $g_s$. A orden zero recuperamos la aproximación deslocalziada. Resolvemos entonces la expansión a primer orden, en la que la naturaleza localizada de los O6-planos se hace presente -aunque no los términos de intersección entre varios O6-planos, que aparecer\'ian a siguiente orden en la expansión-.

En la última parte de la tesis utilizamos este mismo formalismo para estudiar la estabilidad no perturbativa de la familia de vacíos no-supersimétricos obtenidos con flujos RR y NSNS y caracterizada por tener $G_4^{\text{no-SUSY}}=-G_4^{\text{SUSY}}$ -a nivel perturbativo sí son estables-. Según una versión refinada de la conjetura de la gravedad débil (la \textit{WGC} por sus siglas en inglés, \textit{Weak Gravity Conjecture}), estos vacíos deberían contener en su espectro al menos una membrana con $Q>T$, provocando su decaimiento. Comprobamos que D8-branas enrollando la variedad interna y con D6-branas acabando en ellas parecen satisfacer este requerimiento, haciendo estos vacios inestables.

Finalizamos la tesis recapitulando todos los resultados obtenidos y haciendo algunos comentarios sobre qué preguntas quedan aun por responder, así como posibles futuras líneas de investigación. 

\newpage

\selectlanguage{british}

\newpage

\selectlanguage{british}
\thispagestyle{empty}
{\hypersetup{hidelinks}
\tableofcontents}
\clearpage
\thispagestyle{empty}
\cleardoublepage
\pagenumbering{arabic}


\chapter{Introduction}
\label{ch:introducion}
\thispagestyle{simple}

``Someone in the crowd could be the one you need to know'' they sing in the first minutes of La La Land. And I just can not resist starting my thesis by paraphrasing this song \footnote{La La Land is one the favourite films of the author.} and writing ``String theory could be the one you need to know''. But let us start by the \textit{beginning}. How the world is understood changed drastically during the 20th century. General relativity and quantum mechanics were discovered, and a lot of new measurements and experiments were done. Not only this, but also the most precise description of nature as of today was formulated: the standard model, a theory of  \emph{almost} all the fundamental interactions and particles, accurate in some cases to 12 decimal places \cite{Parker:2018vye}. If this was the end of the story, probably this thesis would be different. But it is not. We have two frameworks, general relativity and quantum mechanics, that work astonishing well. The first one allows us to understand gravity (and so basically the universe) at large distances, whereas the second one explains the processes we can detect at small distances with a formidable accuracy. But these frameworks do not talk to each other.

Unfortunately -or fortunately for the young generation of scientists-, we know that there must be some deeper theory, some framework that does not have the word \emph{almost} in its description. And this is one of the hardest and most beautiful problems we are trying to solve, the description of gravity at the quantum scale. Or, in other words, the seek of a theory of quantum gravity.

String theory (ST) is nowadays the most promising candidate we have for it.  Besides describing gravity at small distances, it also aims to describe \textit{all} other interactions, so it can be seen as the theory of everything we were looking for. And though we do not have yet any proof regarding its validity as a description of our universe, we do know that string theory \emph{is} a theory of quantum gravity, and this makes interesting its study per se.

In the era of the big experiments, with the discovery of the Higgs Boson in 2012 \cite{HiggsDiscovery1}, the observation of gravitational waves in  2016 \cite{LIGO} and the first picture of a black hole in 2019 \cite{EventHorizonTelescope:2019dse}, one could ask when we will be able to proof or disproof string theory. To do this, we need to connect it with our universe, so we need to understand what kind of theories can arise from strings. This is precisely the core of the swampland program \cite{Vafa:2005ui}, initiated a few years ago. In a sense, the swampland program provides tools for a down-top approach, cataloguing the properties that effective field theories must satisfy to be embeddable in string theory. 

At the same time, it is also important to continue studying pure top-down constructions: both to learn if string theory is useful to describe our reality and, more generally, to list the properties that string compactifications have. In fact, in the era of the swampland program, this field is more active than ever. It is in this way that one may able to \textit{classify} string compactifications and provide, at some point, a recipe with the ingredients that theories coming from strings have. 

In all this game of connecting string theory with a low energy model that is falsifiable, \emph{flux compactifications} play a key role. On the one hand, they are the main tool we have to build 4d vacua configurations with realistic properties (scale separation, moduli stabilisation...). On the other hand, they supply us with a plethora of examples from which one can chart the distinct vacua of string theory -the string landscape- and extract patterns to formulate and test the so-called swampland conjectures. They are the heart of string phenomenology. The path that could lead us to the most detailed comprehension of our universe.

Let me finish this section, before moving to a more detailed discussion of all the concepts introduced here, in the same spirit as we started, paraphrasing a movie. This time we could turn to Kubrick and say `How I Learned to Stop Worrying and Love the Landscape''. 

\textbf{String theory: a quick start guide}

The basic idea behind string theory is quite simple: considering that the fundamental constituents of nature are not punctual objects but strings.
\begin{figure}[h]
\centering
\includegraphics[width=8cm]{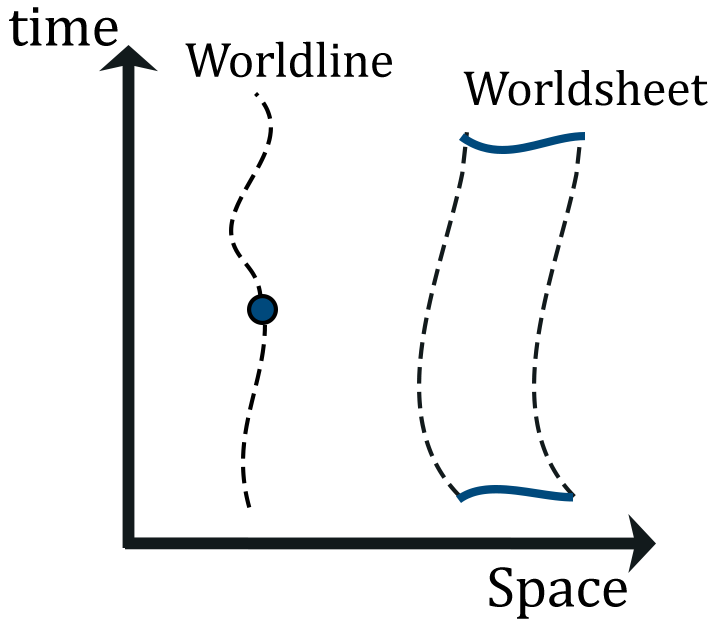}
\caption{Propagation of a point particle and an open string through spacetime. The first one sweeps a one-dimensional trajectory, whereas for the second one the trajectory is 2d. The picture can be easily generalised for closed strings and branes
}
\end{figure}
This seemingly inoffensive change of paradigm modifies completely how we understand the world at its smallest scales. There would not be different particles anymore, but just one only component, the string, whose vibrational modes would correspond to the different particles we observe. Among these modes, there appears a  massless spin 2 field corresponding to the graviton \cite{Scherk:1974ca}. One has then a quantum description of gravity since \textit{quantum} gravity questions in this framework yield always finite answers, unlike the point particle approach. Besides strings, the theory also contains extended objects, the $p$-branes -where $p$ refers to its spatial dimensions-, that appear when one studies the theory beyond its perturbative description.

It is important to remind that string theory has only one free parameter: the string length $l_s$, which can only be measured experimentally and determines the energy scale $M_s\sim1/l_s$ at which the extended nature of the string becomes relevant. Even the dimensions of the space-time are fixed by the internal consistency of the theory and cannot be put by hand. For superstring theories, the ones in which we will be interested, the space-time must have $d=10$. Four of these ten dimensions are identified with the ones we detect. The other 6 extra dimensions are assumed to be so tiny that cannot be seen in our current experiments. We say that they are compactified.

\begin{figure}[h]
\centering
\includegraphics[width=8cm]{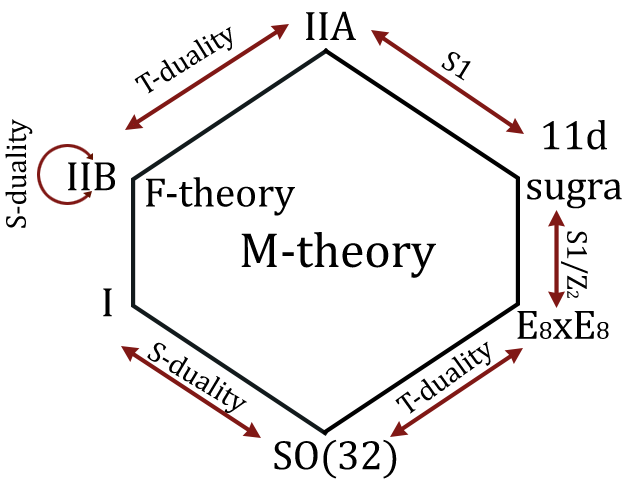}
\caption{Graphical representation of the dualities of string theory. }
\label{dualities}
\end{figure}

And how are the geometry and the topology of these compact dimensions? Since current experiments cannot say much about it, one needs to guess what properties are needed to reproduce the world we observe in four dimensions. But the problem arises in the freedom one has in doing so. The vast amount of theories that can be obtained from string theory is the so-called string landscape -see for instance \cite{Douglas:2003um, Acharya:2006zw, Taylor:2015xtz} for estimations in the number of string vacua-. Among the different choices one has to do when constructing  4d vacua of string theory, there are the vev of the field strengths in the compact dimensions, the \textit{fluxes}. A compactification where some of them are non-vanishing is called a flux compactification. Fluxes cannot be arbitrarily big and are subjected to the tadpole cancellation conditions, the fact that in a compact space the total charge must be zero. At the same time, they restrict the kind of geometry of the internal dimensions, since not any manifold admits any flux. We will also see below that fluxes create a potential in the effective theory that fixes the moduli -the plethora of massless scalar fields that appear when compactifying- of the compactification. The search for vacua of this potential and the study of their properties from the 4d and the 10d perspective will be the main topic of this thesis. Among the different superstring theories, all of them related by dualities -see picture \ref{dualities}-, we will focus on flux compactifications of type IIA.
We will not spend much more time talking about string theory basics and refer the reader to the classical books \cite{Green:1987sp, Green:1987mn, Polchinski:1998rq, Polchinski:1998rr, Ibanez:2012zz, Blumenhagen:2013fgp} for a more paused and detailed discussion.

\textbf{A swampland in the room}

One can not talk nowadays about string theory without referring to the swampland program. Initiated a few years ago in \cite{Vafa:2005ui}, its spirit is against the picture that ``any theory can arise in string theory''. 
\begin{figure}[h]
\centering
\includegraphics[width=8cm]{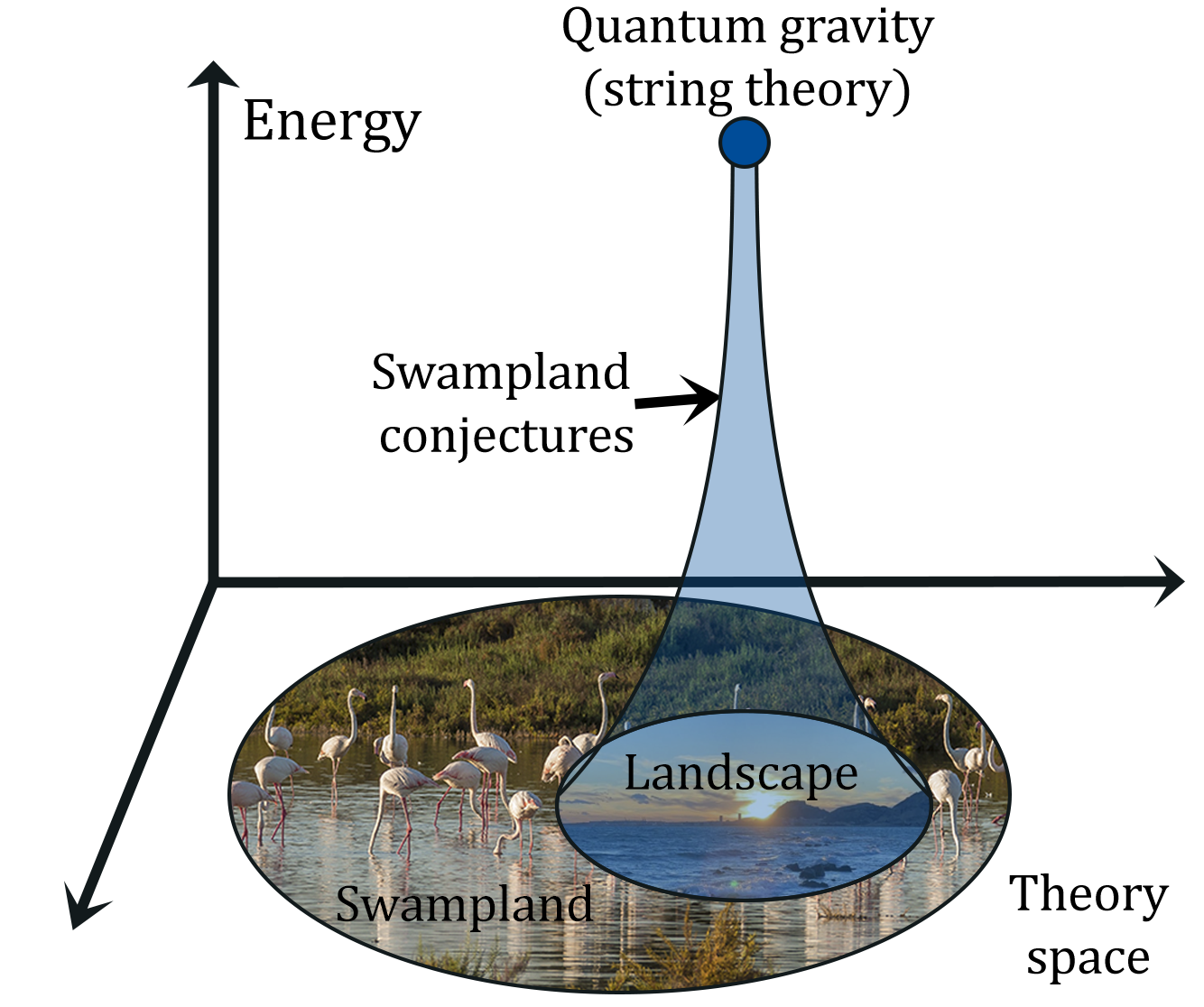}
\caption{Adapted from  \cite{vanBeest:2021lhn}. The swampland (quantum field theories that cannot be consistently coupled to quantum gravity) and the landscape.}
\label{aliswamp}
\end{figure}Contrary to what was widely believed in the early 2000s, the idea is that though the string landscape is big, it would be of measure zero compared to the number of theories that are semiclassically consistent but inconsistent when coupled to quantum gravity -see figure \ref{aliswamp}-. Such theories are said to lay in the \emph{swampland}. The community is now trying to list which are precisely the properties that effective theories must satisfy to be in the landscape and not in the swampland. For the moment they are formulated in terms of conjectures, which are the so-called \emph{swampland conjectures}.

In this thesis, we will always keep an eye on the swampland program and the swampland conjectures, especially in the ones that can be applied to our constructions. 

\textbf{Plan of the thesis}

The goal of this thesis is to study the vacua structure of type IIA orientifold compactifications. We will do so by following a logical path. The thesis then is structured as follows:
\begin{itemize}
\item We will start in chapter \ref{ch:review} by reviewing type IIA CY orientifold compactifications from the 4d perspective. We will recall how the 4d action is derived and the role of the RR, NSNS and (non)-geometric fluxes. We will explain how the potential generated by these fluxes admits a bilinear formulation. This formulation will be very useful when we look for vacua in the following sections. We will end this chapter by summarising the main swampland conjectures that can be applied to these scenarios.
\item Having the 4d action and the potential generated by the fluxes, we will initiate our exploration of the type IIA vacua landscape in chapter \ref{ch:rrnsnsvacua} by considering the case with only RR and NSNS fluxes. We will look for the extrema of the potential trying to be as systematic as possible. We will obtain SUSY and non-SUSY vacua. We will close the chapter by commenting on the properties of the vacua and analysing its consistency and (possible) instabilities.
\item We will then add (non)-geometric fluxes in chapter \ref{ch:geometricflux} and will repeat the same strategy, looking for the extrema of the potential and trying to be as general as possible. We will take an ansatz that can be motivated by stability arguments and that switches off non-geometric fluxes. Once we obtain the extrema, we will analyse their properties to determine which of them are true (stable) vacua.
\item Having done a 4d systematic search of vacua, we will move to study the full 10d background in chapter \ref{ch:review10d}. This chapter will serve as a bridge between the 4d and the 10d pictures. We will review the 10d equations of motion and recall how SUSY equations can be expressed in a very elegant way using the language of polyforms. We will initiate the discussion of the uplift of the vacua derived in the previous chapters by checking if they can be embedded in $SU(3)$ structure manifolds. We will see that this is not enough unless one \textit{smears} the sources, which is not describing the physical true situation. 
\item In chapter  \ref{ch:uplift10d} we will continue the study of the 10d uplift of the SUSY 4d vacua derived in chapter \ref{ch:rrnsnsvacua}. We will analyse ten-dimensional solutions to type IIA string theory of the form AdS$_4 \times X_6$ which contain orientifold planes and preserve ${\cal N}=1$ supersymmetry. They would correspond to the uplift of CY orientifold compactifications. To do so we will expand the equations in powers of $g_s$. We will see that at zero-order we recover the smeared solution. We will give the explicit solution of the 1st order correction, which sees the localised nature of the orientifold sources.
\item The results obtained in chapter \ref{ch:uplift10d} will be useful in chapter \ref{ch:nonsusy}, which will be dedicated to study the non-perturbative stability of the non-SUSY solutions of chapter \ref{ch:rrnsnsvacua}. This requires having the 10d picture since we will be interested in the membrane spectrum of the theory. According to a refined version of the WGC, there should be a membrane in the spectrum of these non-SUSY vacua triggering its decay. We will see that indeed this is the case.
\item We will finish the thesis in chapter \ref{ch:conclusions} with some general remarks and conclusions (translated to Spanish in chapter \ref{ch:conclusionss}). Some long computations, technicalities and conventions are left to appendices \ref{app:relations}, \ref{ap:Hessiannsrr}, \ref{ap:convl}, \ref{ap:10dconv}, \ref{ap:SU33}, \ref{ap:end}.

\end{itemize}

\newpage


\chapter{Type IIA orientifold compactifications: a 4d perspective}\label{ch:review}

In this chapter, we will present and review the main ingredients regarding flux compactifications on Calabi-Yau (CY) orientifolds. We do not aim to give a detailed introduction to the topic -for that one could go to the original references \cite{Grimm:2004uq, Grimm:2004ua, Grimm:2005fa} or to the excellent reviews \cite{Grana:2005jc, Douglas:2006es, Blumenhagen:2006ci, Denef:2007pq, Ibanez:2012zz, Tomasiello:2022dwe}- but just presenting the framework and fixing notation. We will be working mainly using the 4d effective action, and not worrying about the 10d equations -which will be the subject of study of chapter \ref{ch:review10d}-. We will start by considering compactifications with NSNS and RR fluxes, and then adding geometric and non-geometric fluxes. We well afterwards discuss the bilinear formulation of \cite{Herraez:2018vae} and present the main swampland conjectures involving these scenarios.

As argued in the introduction, flux compactifications play a central role in string phenomenology, being the most promising way of constructing effective theories which can describe our universe. In string compactifications with no fluxes, the spectrum contains a large number of massless fields, the moduli, including the dilation and the fields parametrising the geometry of the internal manifold. In the 4d effective theory these massless scalars couple to matter and create long-range interactions. Because these couplings are not necessarily universal, for a 4d observer they would be seen as ``fifth forces'', leading to violations in the principle of equivalence that have not been detected experimentally so far. Adding fluxes (non-vanishing vevs for the field strengths) to the compactification, a potential for the moduli is generated in the 4d action. The fields are then stabilised at their value in the minima of the potential and the moduli acquire a mass, overcoming the previous problem. This process is called \emph{moduli stabilisation}.

We will focus on flux compactifications of (massive) type  IIA and will study moduli stabilisation. In this scenario  \cite{DGKT,Camara:2005dc}, NSNS and RR fluxes alone can stabilise all moduli in a regime -weak coupling large volume- in which the supergravity description can be trusted, which is very suggestive from the phenomenological point of view. The price to pay is that because of the backreaction of  fluxes and  sources, the internal manifold can not be Calabi-Yau \cite{Lust:2004ig} and the construction is not totally well understood. One usually ignores this problem and works in the smearing approximation \cite{Acharya:2006ne}, where the sources needed to cancel the tadpole equations for the fluxes are \textit{smeared} along the internal manifold, which is Calabi-Yau in this approximation. We will address this problem in chapters   \ref{ch:review10d} and \ref{ch:uplift10d}.  For completeness, despite the fact we will not study them in this thesis, we can say that the same problem happens in Type I/heterotic flux compactifications, where moreover one needs to take into account non-perturbative corrections to cancel the tadpoles -see \cite{Grana:2005jc, Blumenhagen:2006ci} for more references-. On the other side, for type IIB the backreaction of the fluxes makes the compact manifold be a warped Calabi-Yau, which is better understood, but in this case NSNS and RR fluxes alone are not enough to stabilise all moduli \cite{Giddings:2001yu}.

We have been already talking a while about the \emph{tadpole equations} but we have not yet defined them properly. The point is that fluxes cannot be put in any arbitrary manifold at will, but there are some ``rules'' that must be satisfied. It is just Gauss law applied to compact spaces and field strengths of p-forms. Consider for instance type IIA and the Bianchi identity for the 2-form $G_2$. Without sources, it reads 
\begin{align}
dG_2=F_0H\, .
\end{align}
Integrating this equation over a 3-cycle, we see that the LHS always vanishes whereas the RHS does not if there is some flux of $H$ on it. This inconsistency can be cured by adding sources $\delta_3$ that compensate the fluxes and make that the RHS also vanishes when integrated over any 3-cycle,
\begin{align}
dG_2=F_0H+\delta_3 \quad \longrightarrow \quad \int \left( F_0H +\delta_3\right)=0.
\end{align}
These sources will be in general (anti)D-branes and orientifold planes. Since we are interested in preserving some amount of supersymmetry in the effective theory, we will use D-branes and orientifold planes to cancel the fluxes. These cancellation conditions are the so-called tadpole equations and we will talk about flux compactifications on Calabi-Yau orientifolds. They are described by a $N=1$ supergravity theory in 4d. We will follow mostly \cite{Grimm:2005fa,Ibanez:2012zz} (with the conventions of our latest works) in this section.

\section{Massive type IIA on CY orientifolds}
\label{sec:4deffaction}

The bosonic part of the low energy description of massive IIA string theory in the Einstein frame  at leading order in $\alpha'$ and $g_s$  and in the democratic formulation is \cite{Bergshoeff:2001pv}
\begin{align}
\label{10daction}
S_{\mathrm{IIA}}^{10d}=& \frac{1}{2 \kappa_{10}^{2}} \int d^{10} x \sqrt{|g|} e^{-2 \phi}\left(R+4(\partial \phi)^{2}\right)-\frac{1}{4 \kappa_{10}^{2}} \int e^{-2 \phi} \star_{10} H \wedge H-\frac{1}{8 \kappa_{10}^{2}}\int\star_{10}\textbf{G}\wedge\textbf{G},
\end{align}
where $\kappa_{10}^{2}=\ell_s^8/4\pi$, $\ell_s=2\pi\sqrt{\alpha'}$. To obtain the complete action one should add possible sources and fermions.  In this formulation we double the degrees of freedom and define the polyform $\bfG$ as the formal sum $\textbf{G}=G_0+G_2+G_4+G_6+G_8+G_{10}$. Then we impose 
\begin{align}
\bfG=\star_{10}\lambda\left(\bfG\right)\, ,
\end{align}
where $\lambda$ is the operator reversing the indices. The field strengths $G_n$ are related with the RR $p$-form potentials $C_p$ as
\begin{align}
\label{gdef}
\bfG=d_H\bfC+e^B\wedge\bar{\bfG}\, ,
\end{align}
with $\bfC=C_1+C_3+C_5+C_7+C_9$, $H=dB+\bar H$ is the three-form NS flux, $d_H\equiv d-H\wedge$ is he $H$-twisted differential and $\{\bar{H}, \bar{\bfG}\}$ will be the background values (the fluxes) for the NSNS and the RR field strengths respectively. 

We can now put this theory on an orientifold of $\mathds{R}^{1,3}\times\mathcal{M}_6$ being $\mM_6$ a compact Calabi-Yau three-fold.

\begin{tcolorbox}[breakable, enhanced]
\begin{small}
\textbf{\emph{Intermezzo}: CY manifolds and its relevance in ST compactifications}
\\

CY manifolds play a key role in string theory compactifications. For a detailed discussion of their properties and how they appear in this context one can go for instance to \cite{Candelas:1985en,Green:1987mn, Strominger:1990pd, Candelas:1990pi, Hubsch:1992nu, Greene:1996cy, Joyce:2001xt}. We will just say a few words to be as self-contained as possible. The basic idea is that if we choose a background $\mM_{10}=\mathds{R}^{1,3}\times\mathcal{M}_6$ and demand that some supersymmetry\footnote{One could forget about it and just look for generic solutions to the 10d equations of motion. But, as a first step to understanding ST, having some supersymmetry makes life easier since we have more control on the compactification.} is preserved in the 4d action, $\mM_6$ is constrained severely. In short, supersymmetry demands $\mathcal{M}_6$ to admit one non-trivial globally defined spinor that is covariantly constant. This is satisfied by manifolds with $SU(3)$ holonomy. Calabi-Yau manifolds are precisely $d$-dimensional complex, K\"ahler manifolds with  $SU(d)$ holonomy. They are  characterised by having a globally defined and closed $(1,1)$-form $J$  -the K\"ahler form- and a globally defined and closed $(3,0)$-form $\Omega$, the holomorphic form, satisfying  
\begin{align}
-\frac{1}{6} J\wedge J\wedge J=-\frac{i}{8}\Omega\wedge\bar\Omega=d\text{vol}_{\text{CY}}\, .
\end{align}
If we put Heterotic or Type I string theory on these backgrounds, we obtain a $\CN=1$ supergravity theory in 4d, and the same is true for type IIA/B but with $\CN=2$. In this last case, modding out the internal manifold by the appropriate symmetries (so compactifying not on a CY but on a CY orientifold)  supersymmetry is reduced to $\CN=1$, as it will be in our case.
\\ 

In the situation in which the radius of $\mathds{R}^{1,3}$ is much bigger than the one of $\mM_6$, the low energy description of the theory is obtained by performing a Kaluza-Klain reduction of the fields and considering only the massless modes. They are given by the eigenfunctions with null eigenvalue of the internal Laplacian or,  in other words, by the harmonic forms of $\mM_6$. In turn, the harmonic forms of a manifold are counted by the dimension of the different cohomology groups, the Hodge numbers. For CY manifolds, the only non-vanishing cohomology groups are \cite{Grimm:2005fa}
\begin{align}
H^{\text{even}}\quad=\quad H^{(0,0)}\oplus H^{(1,1)}\oplus H^{(2,2)}\oplus H^{(3,3)}\, , \nonumber\\
H^{\text{odd}}\quad=\quad H^{(3,0)}\oplus H^{(2,1)}\oplus H^{(1,2)}\oplus H^{(0,3)}\, ,
\end{align}
where the indices in $H^{(m,n)}$ refer to $p$-forms with $m$ holomorphic and $n$ antiholomorphic indices. The dimensions of each group $\text{dim} H^{(m,n)}=h^{(m,n)}$ can be summarized in the Hodge diamond
\begin{align}
\begin{array}{r r r r r r r r r r r r r r r }
 & & &h_{0,0}& && & & & & &1& && \\
& &h_{1,0}& &h_{0,1}& & & & & &0& &0& &\\
& h_{2,0}& &h_{1,1} & &h_{0,2} & & & &0& &h_{1,1}& &0& \\
h_{3,0} & &h_{2,1}& &h_{1,2}& & h_{0,3}&=&1& &h_{2,1}& &h_{2,1}& &1\\
  & h_{3,1}& &h_{2,2} & &h_{1,3} & & & &0& &h_{1,1}& &0& \\
    & &h_{3,2}& &h_{2,3}& & & & & &0& &0& &\\
    & & &h_{3,3}& && & & & & &1& && 
\end{array}
\end{align}
\end{small}
\end{tcolorbox}
Let us consider massive type IIA string theory compactified on an orientifold of $\mathbb{R}^{1,3} \times \CM_6$ with $\CM_6$ a compact Calabi-Yau three-fold. More precisely, we take the standard orientifold quotient by $\mathcal{O}=\Omega_p (-)^{F_L} {\cal R}$ \cite{Blumenhagen:2005mu,Blumenhagen:2006ci,Marchesano:2007de,Ibanez:2012zz},  with $\Omega_p$ the whorldsheet parity  and $F_L$  the left-moving spacetime fermion number. ${\cal R}$ is an anti-holomorphic Calabi-Yau involution acting on the K\"ahler 2-form $J$ and the holomorphic 3-form $\Omega$ as ${\cal R}(J) = - J$ and ${\cal R} (\Omega) = -\ov \Omega$. The $p$ dimensional hyper-surfaces left invariant by $\cal R$ are the orientifold planes, the O$p$-planes. In type IIa there are generically only O$6$-planes\footnote{In type IIB there can be either $O3/O5$ or $O5/O9$, depending on the orientifold projection.}. These $O6$-planes are  source of RR charge and consequently appear in the Bianchi identities.  By construction, they span  $\mathbb{R}^{1,3}$ and wrap special Lagrangian 3-cycles $\Pi_3$ characerised by
\begin{align}
J|_{\Pi_3}&=0\, ,	&	\im\left(\Omega\right)|_{\Pi_3}=0\, .
\end{align}
On the other hand, under the action of $\mathcal{R}$, the harmonic forms split into even and odd  $H^p\left(\mathcal{M}_6\right)=H^p_+\oplus H^p_-$. We can introduce a basis for each of these subspaces:
\begin{table}[h]
\begin{center}
\begin{tabular}{| c || c | c| c | c | c | c |} \hline
   \rule[-0.3cm]{0cm}{0.9cm} cohomology group &  $\ H^{(1,1)}_+\ $ & 
   $\ H^{(1,1)}_-\ $ & $\ H^{(2,2)}_+\ $ & $\ H^{(2,2)}_-\ $ & $\ H^{3}_+\ $ & $\ H^{3}_-\ $
   \\ \hline
   \rule[-0.3cm]{0cm}{0.8cm} dimension &  $h^{(1,1)}_+$  & $h^{(1,1)}_- $  
                                       &  $h^{(1,1)}_-$  & $h^{(1,1)}_+$ 
                                       &  $h^{(2,1)}+1$  &  $h^{(2,1)}+1$ 
   \\ \hline
   \rule[-0.3cm]{0cm}{0.8cm} basis     & $\ell_s^{-2}\omega_\alpha$ & $\ell_s^{-2}\omega_a$
                                       & $\ell_s^{-4}\tilde \omega^a$ & $\ell_s^{-4}\tilde \omega^\alpha$
   & $\ell_s^{-3} \alpha_\mu$ & $\ell_s^{-3}\beta^\mu$ \\ \hline
\end{tabular}
\caption{Extracted from \cite{Grimm:2005fa} and adapted to our conventions. Basis forms for the different cohomology groups. The correspondent factors of the string length $\ell_s = 2 \pi \sqrt{\alpha'}$ are introduced to made the $p$-forms dimensionless.}
\label{base}
\end{center}
\end{table}

\vspace*{-5mm}
\noindent where Hodge duality imposes $h^{(1,1)}_+=h^{(2,2)}_-$,  $h^{(1,1)}_-=h^{(2,2)}_+$. The elements of the basis are normalised such that 
\begin{align}
\frac{1}{\ell_s^6}&\int_{\mathcal{M}_6}\omega_a\wedge\tilde{\omega}^b=\delta^b_a\, ,	&	\frac{1}{\ell_s^6}&\int_{\mathcal{M}_6}\omega_\alpha\wedge\tilde{\omega}^\beta=\delta^\beta_\alpha\, ,	&	\frac{1}{\ell_s^6}&\int_{\mathcal{M}_6}\alpha_\mu\wedge\beta^\nu=\delta^\beta_\alpha\, .
\end{align}
For completeness, we also list  how the 10d fields have to transform under  $\mathcal{R}$ to be invariant under the action of $\mathcal{O}$ 
\begin{table}[h!]
\begin{center}
\begin{tabular}{c |c | c | c | c | c }
&$g$ &		 $\phi$&	$B_2$	& $C_1$	&  $C_3$\\
\hline
$\mathcal{R}$	& $+$&	$+$	&	$-$	& $-$	&  $+$\\
\end{tabular}
\caption{Action of $\cal R$ on the 10d massless spectrum of IIA.}
\label{transform}
\end{center}
\end{table}

\vspace*{-5mm}
\noindent since this is illustrative to understand how part of the spectrum is projected out -compared to a compactification on a CY manifold- and supersymmetry is reduced from $\mathcal{N}=2$ to $\mathcal{N}=1$ in the 4d effective theory. 

Once the presentations have been made, we are ready to compute the 4d effective theory. We will divide this calculation into three pieces:   massless spectrum (moduli) in \ref{modulisection},  NSNS and RR fluxes in \ref{rrnnsection} and  geometric and non-geometric fluxes in \ref{generalsection} .

\subsection{4d effective action: massless fields (moduli)}\label{modulisection}

Neglecting worldsheet and D-brane instanton effects, dimensional reduction of massive type IIA to 4d yields several massless chiral and vectors fields, whose components can be described as follows \cite{Grimm:2004ua}. 
\begin{itemize}
\item There are the massless modes coming from the free parameters of the 10d metric $g$. We will omit here the discussion -for which one can go for instance to \cite{Candelas:1990pi}- and limit ourselves to present the results. On the one hand, the are the deformations of the K\"ahler form $J$, which give rise to $h^{(1,1)}_-$ real scalars $t^a$ -recall that $J$ is odd under $\mathcal{R}$-. They control the volume of the 2-cycles of the CY manifold. We can pair them with the real scalars $b^a$ arising from the $B=b^a\omega_a$ field to define the complexified K\"ahler moduli $T^a = b^a + it^a$,
\begin{equation}
J_c \equiv B + i\, e^{\frac{\phi}{2}} J = \left( b^a + i t^a\right) \omega_a= T^a\omega_a \, , \qquad \quad a \in \{1, \ldots, h^{1,1}_- \},
\end{equation} 
where $J$ is expressed in the Einstein frame,  $\phi$ represents the ten-dimensional dilaton and, recalling table \ref{base}, $\ell_s^{-2}\omega_a$ correspond to harmonic representatives of the classes in $H_-^{1,1}\left(\mm_6\right)$. The kinetic terms for these moduli is encoded in their K\"ahler potential
\begin{equation}
K_K \,  = \, -{\rm log} \left(\frac{i}{6} \CK_{abc} (T^a - \bar{T}^a)(T^b - \bar{T}^b)(T^c - \bar{T}^c) \right) \, = \,  -{\rm log} \left(\frac{4}{3} \cK\right) \, ,
\label{KK}
\end{equation}
where ${\cal K}_{abc} = - \ell_s^{-6} \int_{{\cal M}_6} \omega_a \wedge \omega_b \wedge \omega_c$ are the Calabi-Yau triple intersection numbers and $\cK = \cK_{abc} t^at^bt^c = 6 {\rm Vol}_{\CM_6} = \frac{3}{4} {\cal G}_T$ is a homogeneous function of degree three on the $t^a$. 
\item On the other hand, there are the deformations of the complex 3-form $\Omega$, parametrising the volumes of the internal 3-cycles. We can pair them with the axio-dilaton and the internal $3$-form coming from the RR potential $C_3$, giving rise to the so-called complex structure moduli. Introducing $\Omega_c\equiv C_3+i\re\left(\mathcal{C}\Omega\right)$ where $\mathcal{C}=e^{-\phi}e^{\frac{1}{2}\left(K_{cs}-K_K\right)}$ is a compensator, with $K_{cs}=-\log\left(-i\ell_s^{-6}\int\Omega\wedge\bar\Omega\right)$ and $\phi$ the 10d dilaton, the complex structure moduli are defined as
\begin{align}
\label{cpxmoduli}
U^\mu=\xi^\mu+i u^\mu=\ell_s^{-3}\int\Omega_c\wedge \beta^\mu\, ,	\quad	\mu\in\left\{0,\dots, h^{2,1}\right\}\, ,
\end{align}
where we are using the basis introduced in table \ref{base}, $\beta^\mu\in H_-^3\left(\mm_6\right)$.
Their kinetic terms are given in terms of the following piece of the K\"ahler potential:
\begin{align}
\label{KQ}
K_Q=4\log{\left(\frac{e^\phi}{\sqrt{\Vol_{\mm_6}}}\right)}\equiv -\log\left(e^{-4D}\right)\, ,
\end{align}
where $D$ is the four-dimensional dilaton defined through $e^{D} \equiv \frac{e^{\phi}}{\sqrt{{\rm Vol}_{\CM_6}}}$. The function ${\cal G}_Q= e^{-K_Q/2}$ is a homogeneous function of degree two in $u^ \mu$. The complex structure moduli \eqref{cpxmoduli} are redefined in the presence of D6-brane moduli, and so is the K\"ahler potential \eqref{KQ}. For simplicity, we will not consider this case for now, leaving its discussion to section \ref{s:D6branes}.
\item There is the $C_1$, which is projected out, since it is odd under $\mathcal{R}$,  $\Omega_p\left(-1\right)^{F_L}$ act trivially on it and there are no harmonic 1-forms on a CY.
\item Finally\footnote{We will not discuss $\{C_5,C_7,C_9\}$ since they do no introduce new degrees of freedom, and are just dual to the $p$-forms already considered.}, there are the pieces of $C_3$ with external legs. $C_3$ can be expanded as
\begin{equation}
\label{eq:c3}
C_3=c_3(x)+A^\alpha(x)\wedge\omega_\alpha+C_3\, ,
\end{equation}
We are abusing a bit of notation and are calling in the same way the RR 3-form potential $C_3$ and its internal 3-form component, which we have already discussed two paragraphs ago when $\Omega_c$ was introduced. $c_3(x)$ and $A^\alpha(x)$ give rise to a 3-form and a 1-form in four dimensions respectively. $c_3$ does not propagate any degree of freedom and corresponds to a flux parameter. The gauge kinetic functions for the field strengths $F^\alpha=d A^\alpha$ are
\begin{align}
\label{eq:fg}
2f_{\alpha\beta}=i\hat{\cal K}_{a\alpha\beta}T^ a\, ,
\end{align}
with $\hat{\cal K}_{a\alpha\beta}= - \ell_s^{-6} \int_{{\cal M}_6} \omega_a \wedge \omega_\alpha \wedge \omega_\beta$.
\end{itemize}
Collecting all the results, the spectrum of massive type IIA compactified on a CY orientifold boils down to a $\CN=1$ 4d supergravity theory whose bosonic part of the spectrum is summarised in table \ref{spectrum4d}.\begin{table}[h!]
\begin{center}
\begin{tabular}{|c |c | c | }
 Multiplet &Multiplicity			&	Bosonic field Content\\ [0.25cm]\hline
gravity multiplet&1	&$g_{\mu\nu}$\\[0.25cm]
vector multiplet	&$h_+^{1,1}$	& $A^\alpha$\\[0.25cm]
Chiral multiplets	& $h_-^{(1,1)}$	& $t^a$, $b^a$\\[0.25cm]
	Chiral multiplets		& $h^{2,1}+1$	& $u^\mu$, $\xi^\mu$\\[0.25cm]
\end{tabular}
\caption{4d $\mathcal{N}=1$ bosonic massless spectrum of IIA compactified on a CY orientifold.}
\label{spectrum4d}
\end{center}
\end{table}
The metric for the kinetic terms of the chiral multiplets is derived from the K\"ahler potentials $\{K_K, K_Q\}$, introduced in \eqref{KK} and \eqref{KQ} for the K\"ahler and complex moduli respectively, see appendix \ref{ap:relations}. They satisfy some nice properties, which will be of utility later on, that we list in that same appendix. Explicitly, the bosonic part of the action for this theory is
\begin{equation}
\begin{aligned}
S_{\mathrm{IIA}}^{4 d}=& \int-\frac{1}{2} R \star 1-K_{a b} d T^{a} \wedge \star d \bar{T}^{b}-K_{\mu\nu} dU^\mu\wedge\star d\bar{U}^\nu\\
&-\frac{1}{2} \operatorname{Im} f_{\alpha \beta} F^{\alpha} \wedge F^{\beta}-\frac{1}{2} \operatorname{Re} f_{\alpha \beta} F^{\alpha} \wedge \star F^{\beta}
\end{aligned}
\end{equation}
with the gauge kinetic functions defined in \eqref{eq:fg}.

\subsection{4d effective action: NSNS and RR fluxes}\label{rrnnsection}

On top of the above orientifold background one may add RR and NSNS  fluxes. Using the democractic formulation introduced previously, the Bianchi identities for the field strengths read
\begin{equation}\label{IIABI}
\ell_s^{2} \,  d (e^{-B} \wedge {\bf G} ) = - \sum_\a \lambda \left[\delta (\Pi_\alpha)\right] \wedge e^{\frac{\ell_s^2}{2\pi} F_\alpha} \, ,  \qquad d H = 0 \, ,
\end{equation} 
with $\ell_s  =  2\pi \sqrt{\a'}$ the string length. Here $\Pi_\alpha$ hosts a D-brane source with a quantised worldvolume flux $F_\alpha$, and $\delta(\Pi_\alpha)$ is the bump $\delta$-function form with support on $\Pi_\alpha$ and indices transverse to it, such that $\ell_s^{p-9} \d(\Pi_\a)$ lies in the Poincar\'e dual class to $[\Pi_\a]$. O6-planes contribute as D6-branes but with minus four times their charge and $F_\alpha \equiv 0$. Finally, $\lambda$ is the operator that reverses the order of the indices of a $p$-form. As introduced in \eqref{gdef},the solution to \eqref{IIABI} can then be decomposed as
\begin{equation}\label{IIABIsol}
{\bf G} = e^{B}\wedge (d {\bf A} + \ov{\bf G})\,, \qquad H = dB + \ov H\, ,
\end{equation}
where ${\bf A} = {\bf C}\wedge e^{-B}$ and $\ov{\bf G}$ is a sum of closed $p$-forms to be thought as the background values for the internal RR fluxes.  One may now impose Page charge quantisation \cite{Marolf:2000cb},
\begin{equation}
\frac{1}{\ell_s^{2p-1}} \int_{\pi_{2p}} d A_{2p-1} + \ov G_{2p} \in \Z, \qquad  \frac{1}{\ell_s^{2}}\int_{\pi_3} dB + \ov H \in \Z,
\end{equation} 
where $\pi_{2p}$ with $p=1,2,3$ and $\pi_3$ are internal cycles of $\cM_6$. In the absence of localised sources such as D-branes, the gauge potentials~${\bf A}$ are well-defined everywhere and the cohomology class of $\ov{G}_{2p}$, $\ov H$ along $\cM_6$ capture  the internal flux quanta. For orientifold compactifications the internal $p$-cycles have to comply with the orientifold projection, such that    
the background flux can be characterised by virtue of flux quanta $(m,m^a,e_a,e_0)$. These are defined as
\begin{equation}
m \, = \,  \ell_s G_0\, ,  \quad  m^a\, =\, \frac{1}{\ell_s^5} \int_{X_6} \bar{G}_2 \wedge \tilde \omega^a\, , \quad  e_a\, =\, - \frac{1}{\ell_s^5} \int_{X_6} \bar{G}_4 \wedge \omega_a \, , \quad e_0 \, =\, - \frac{1}{\ell_s^5} \int_{X_6} \bar{G}_6 \, .
\label{RRfluxes}
\end{equation}
The internal RR-fluxes ${\ov{\bf G}}$ are known to generate a perturbative superpotential for the K\"ahler moduli \cite{Gukov:1999ya,Taylor:1999ii}:
\begin{equation} \label{WT}
\ell_s W_K = - \frac{1}{\ell_s^5} \int_{X_6} {\ov{\bf G}} \wedge e^{J_c} = e_0 + e_a T^a + \frac{1}{2} {\cal K}_{abc} m^a T^b T^c + \frac{m}{6} {\cal K}_{abc} T^a T^b T^c\,,
\end{equation}
where ${\cal K}_{abc} = - \ell_s^{-6} \int_{X_6} \omega_a \wedge \omega_b \wedge \omega_c$ are the Calabi--Yau triple intersection numbers, $\cK = \cK_{abc} t^at^bt^c = 6 {\rm Vol}_{X_6}$ and we have made used of the flux quanta defined in \eqref{RRfluxes}.
The NS 3-form flux ${\ov H}_3$ on the other hand threads the ${\cal R}$-odd three-cycles $(B^\mu) \in H_3^-({\cal M}_6)$, which are the de Rham duals to the ${\cal R}$-odd three-forms $(\beta^\mu)$ introduced earlier. Similar to the RR-fluxes, the quantised Page charge for the NS-flux background can be expressed in terms of the integer flux quanta $(h_\mu)$:
\begin{equation}\label{Hflux}
 \frac{1}{\ell_s^2} \int_{B^\mu} {\ov H} = h_\mu\,,
\end{equation}
and generate a linear superpotential for the complex structure moduli
\begin{equation} \label{WQ}
\ell_s W_{Q} = -\frac{1}{\ell_s^5} \int_{{\cal M}_6} \Omega_c \wedge {\ov H}_3 =h_\mu U^\mu\, . 
\end{equation}
The combination of RR and NS-fluxes suffices to generate a four-dimensional F-term scalar potential for the geometric moduli $(t^a, u^\mu)$ and closed string axions $(b^a,\xi^\mu)$, given by
\begin{equation}
\label{eq:vf}
V_{F}=e^{K}\left(K^{\alpha \bar{\beta}} D_{\alpha} W D_{\bar{\beta}} \bar{W}-3|W|^{2}\right)\, ,
\end{equation}
where $D_\alpha W=\partial_\alpha W+W\partial_\alpha K$, $K=K_K+K_Q$ -computed in \eqref{KK}-\eqref{KQ}-, $W=W_K+W_Q$ -see \eqref{WT}-\eqref{WQ}- and the indices $\alpha,\beta=\{a,\mu\}$ go over the complex and K\"ahler moduli. We will see in the next section \ref{sec:4forms}  that this potential exhibits a remarkable factorisation into a geometric moduli piece, an axion piece and a flux piece~\cite{Bielleman:2015ina,Herraez:2018vae} which is very useful to do a systematic search of vacua.

\subsection{4d effective action: general fluxes}\label{generalsection}

As we have already commented, type IIA compactifications with NSNS and RR fluxes represent a particularly interesting corner of the string landscape, as already the classical potential generated by $p$-form fluxes suffices to stabilise all moduli \cite{DeWolfe:2005uu, Camara:2005dc}. Even so, as pointed out in \cite{Shelton:2005cf} one may consider a larger set of NS fluxes for this class of compactifications, related to each other by T-duality. Taking them into account results into a richer scalar potential, as analysed in \cite{Aldazabal:2006up,Shelton:2006fd,Micu:2007rd,Ihl:2007ah,Wecht:2007wu,Robbins:2007yv}. In this last part of the derivation of the 4d effective action, we will consider the potential generated when we turn on geometric and non-geometric fluxes.

Again, we do not aim to give a detailed review of the topic -for that one can go to \cite{Wecht:2007wu, Plauschinn:2018wbo}- but just a minimalist introduction to motivate the scenario and fix the notation. The basic idea is that geometric and non-geometric fluxes appear when we consider T-dualities in flux compactifications. As it is known, T-duality relates a theory compactified on a circle of radius $R$ with another theory compactified on a circle of radius $1/R$. An example of this symmetry is the duality between type IIA and IIB string theories, as pictorially presented in figure \ref{dualities}. Subtleties appear when we play this same game in circles threaded by fluxes. Let us review the example studied in \cite{Shelton:2005cf}. Take IIB compactified on a $T^6$ with NSNS flux $H_{abc}$ on some 3-cycle, with the indices $a,b,\dots \in\{1,2,\dots,6\}$ denoting the internal directions. Performing a first T-duality, for instance in the direction $a$, the effect of the flux H is that in the IIA description the compactification is not on a torus but on a twisted torus. That is, in the dual picture, the flux is seen in the metric, which now has a component $\left(dx^a-f^a_{bc}x^ cdx^b\right)^2$, where $f^a_{bc}\in \mathds{Z}$  is the \emph{geometric flux}. One can then perform a second T-duality, say in the direction $b$.  The dual theory still admits a local description in terms of geometry (so we can define a metric locally) but globally the Riemannian description fails\footnote{Basically when one goes around the circle $x^b\rightarrow x^b+2\pi R^b $ the metric and the B-field mix in a non-trivial way and the manifold is not globally well defined.} and the space is called \emph{non-geometric}. In this exotic background, the presence of $H$ is encoded through the flux $Q^{ab}_c$. One can perform a third T-duality, which yields a dual description that does not even admit a local description of the internal space in terms of geometry, and so is also non-geometric. We denote by $R^{abc}$ the  T-dual lux arising in this case in the 4d effective theory. 
\begin{figure}[h]
\centering
\includegraphics[width=14cm]{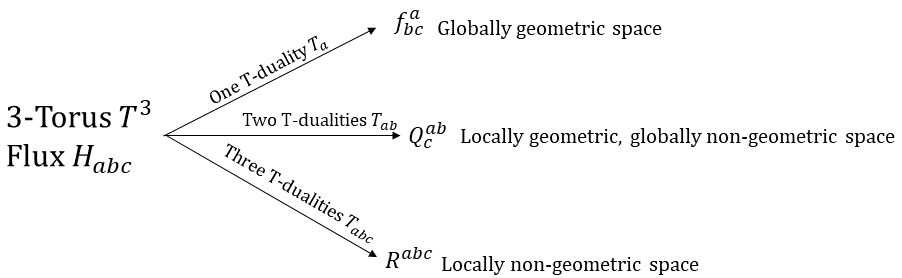}
\caption{(Non)-geometric fluxes and T-dualities. }
\label{dualitiess}
\end{figure}
It is important to point out that compactifications with non-geometric fluxes ($Q, R$) are not well understood. Though we will take them into account when constructing the 4d potential, when we look for vacua we will limit ourselves to just considering RR, NSNS and geometric fluxes.

The flux superpotential including geometric and non-geometric NS fluxes is described in terms of a twisted differential operator \cite{Shelton:2006fd}
\be
\label{eq:twistedD}
{\cal D} = d - H \wedge  +\  f \triangleleft  +\ Q \triangleright  +\  R\, \bullet \, ,
\ee
where $H$ is the NS three-form flux, $f$ encodes the geometric fluxes, $Q$ that of globally-non-geometric fluxes and $R$ is the locally-non-geometric fluxes, see e.g. \cite{Wecht:2007wu} for more details. The action of various fluxes appearing in ${\cal D}$ is such that for an arbitrary $p$-form $A_p$, the pieces $H\wedge A_p$, $f \triangleleft A_p$, $Q \triangleright A_p$ and $R \bullet A_p$ denote a $(p+3)$, $(p+1)$, $(p-1)$ and $(p-3)$-form respectively. Given these definitions, the NSNS flux-generated superpotential $W_Q$ is now \cite{Shelton:2006fd,Aldazabal:2006up}
\be
\label{eq:Wrrnsns}
   W_{\rm Q} =  - \frac{1}{\ell_s^6}  \int_{X_6} \Omega_c \wedge {\cal D}\left( e^{J_c} \right)\,.
\ee
Expanding in the basis introduced in table \ref{base} and using the action of the NS fluxes on such a basis as given in \eqref{eq:fluxActions0}, one obtains the following expression for the superpotential
\bea
\label{eq:Wgen}
\ell_s W_{\rm Q} &= & U^\mu \Bigl[ h_\mu + f_{a\mu} T^a + \frac{1}{2} {\cal K}_{abc} \, T^b \, T^c \, Q^a{}_\mu + \frac{1}{6}\, {\cal K}_{abc} T^a T^b T^c \, R_\mu \Bigr] \, ,
\label{eq:WgenNS}
\eea
which, compared to \eqref{WQ}, has some extra terms.  $W_K$ remains the same -recall equation \eqref{WT}- and is not altered by the presence of (non)-geometric fluxes.

\subsubsection{The F-term flux potential}

Under the assumption that background fluxes do not affect the K\"ahler potential pieces \eqref{KK} and \eqref{KQ},\footnote{The validity of this assumption should not be taken for granted and will depend on the particular class of vacua. The results in \cite{Junghans:2020acz,Buratti:2020kda,Marchesano:2020qvg} suggest that it is valid in the presence of only $p$-form fluxes $F_{\rm RR}$, $H$. However,  \cite{Font:2019uva} gives an example of compactification with metric fluxes in which the naive KK scale is heavily corrected by fluxes, and so should be the K\"ahler potential.} one can easily (re)compute the F-term flux potential for closed string moduli via the standard supergravity expression \eqref{eq:vf}. As in \cite{Bielleman:2015ina,Herraez:2018vae}, one can show that this F-term potential also displays a bilinear structure that will be exploited in chapter \ref{ch:geometricflux}

\subsubsection{The D-term flux potential}

In the presence of a non-trivial even cohomology group $H_+^{1,1}$,  geometric and non-geometric fluxes will generate a D-term contribution to the scalar potential. This can be computed as
\be
\label{eq:VDgen}
V_D = \frac{1}{2}  \left( {\rm Re} f\right)^{-1\: \alpha\beta}\, D_\alpha \, D_\beta\, ,
\ee
where $D_\alpha$ is the $D$-term for the $U(1)$ gauge group corresponding to a 1-form potential $A^\alpha$ arising from $C_3$ -recall decomposition \eqref{eq:c3}-
\be
D_\alpha = i \partial_{{\cal A}} K \, \delta_\alpha \varphi^{\cal A} + \zeta_\alpha \,,
\ee
where $\delta_\alpha \varphi^{\cal A}$ is the variation of the scalar field $\varphi^{\cal A}$ under a gauge transformation, and $\zeta_\alpha$ is the corresponding Fayet-Iliopoulos term. In order to find the explicit expression of the D-term potential we perform a gauge transformation on the gauge bosons in $C_3$\footnote{We are focusing on $C_3$ to maintain the discussion simple but, in the democratic formulation, similar operations have to be done for the  1-form $A_\alpha$ arising from $C_5$.}
\be
A^\alpha\, \quad \longrightarrow  \quad A^\alpha + d \lambda^\alpha\, .
\ee
The transformation of the RR $p$-form potential $C_{RR}  \equiv  C_1 + C_3  + \dots $ can then be given in terms of the twisted differential ${\cal D}$ given in  \eqref{eq:twistedD}
\bea
\label{eq:C3change}
C_{RR}   & \longrightarrow &  C_{RR} + {\cal D}\left(\lambda^\alpha \, \omega_\alpha + \lambda_\alpha \, \tilde{\omega}^{\alpha} \right) \\
& = & \left(\xi^\mu + \lambda^\alpha \,f_{\alpha}{}^\mu + \lambda_\alpha\, Q^{\alpha \mu}\right) \, \alpha_\mu  + \dots \nonumber \, ,
\eea
where we have used the flux actions given in (\ref{eq:fluxActions0}), with $f_{\alpha}{}^\mu$, $Q^{\alpha \mu}$, $f_{\alpha\mu}$, $Q^{\alpha}{}_\mu$ integers. This transformation shows that the scalar fields $\xi^\mu$ are not invariant under the gauge transformation, leading to the following shift in the ${\cal N} = 1$ coordinates $U^\mu$,
\be 
\delta U^\mu =  \lambda^\alpha\,f_{\alpha}{}^\mu +  \lambda_\alpha \, Q^{\alpha \mu}\, .
\label{transU}
\ee
Note that due to the Bianchi identities \eqref{eq:bianchids2} only the combinations of fields  $U^\mu$ invariant under \eqref{transU} appear in the superpotential and, as a result, the Fayet-Iliopoulos terms vanish.  Interpreting \eqref{transU} as gaugings of the U(1) gauge fields and their magnetic duals one obtains the D-terms
\be
D_\alpha = \frac{1}{2} \partial_\mu K \, \left(f_\alpha{}^\mu +\hat{\cal K}_{a\alpha\beta} b^a Q^{\beta \mu}  \right) \, , \qquad   D^\alpha =\frac{1}{2}  \partial_\mu K \, Q^{\alpha \mu} \, .
\ee
Taking into account the kinetic couplings \eqref{eq:fg} we end up with the following D-term scalar potential 
\be
\label{eq:DtermGen}
V_D =-\frac{1}{4}  \partial_\mu K \partial_\nu K \biggl({\rm Im}\, \hat{\cal K}^{-1\: \alpha\beta} \left(f_\alpha{}^\mu +\hat{\cal K}_{a\alpha\gamma} b^a Q^{\gamma \mu}  \right) \left(f_\beta{}^\nu +\hat{\cal K}_{c\beta\delta} b^c Q^{\delta \nu}  \right)
 +  {\rm Im}\, \hat{\cal K}_{\alpha\beta}  Q^{\alpha \mu} \, Q^{\beta \nu} \biggr) \, ,
\ee
where $\hat{\cal K}_{\alpha\beta} = \hat{\cal K}_{a\alpha\beta} \, T^a$. Alternatively, one may obtain the same potential by following the tensor multiplet analysis of \cite{Grimm:2004uq,Louis:2004xi}.\footnote{This result is different from the type IIA D-term potential of \cite{Robbins:2007yv}, and recovers the expected discrete gauge symmetries related to $b$-field shifts. The same strategy can be applied to type IIB setups with non-geometric fluxes, recovering the full scalar  obtained by DFT dimensional reduction in   \cite{Blumenhagen:2015lta}.} The total potential has then two contributions
\begin{align}
V=V_F+V_D\, ,
\label{eq:vfvd}
\end{align}
though when we do a systematic search of facua in chapter  \ref{ch:geometricflux}, we will consider only the case with $V_D=0$.

\section{Flux potential, 4-forms and bilinear formalism}\label{sec:4forms}

The main goal of this thesis is to shed light on the landscape of (massive) type IIA orientifold compactifications. To do that, a crucial ingredient is to understand the vacua structure of the flux potential arising in the 4d theory, see \eqref{eq:vf} and \eqref{eq:vfvd}. In doing so, the bilinear formalism introduced in \cite{Bielleman:2015ina, Carta:2016ynn, Herraez:2018vae, Escobar:2018tiu, Escobar:2018rna} turns out to be a very powerful tool, since doing a systematic search of extrema in this context is much simpler. In this section we will review and motivate briefly how the bilinear formulation was derived -for a detailed discussion the reader interested can go to the original references-. We will use this form of writing the potential in chapters \ref{ch:rrnsnsvacua} and \ref{ch:geometricflux}.

The importance of the 4-forms in the vacua structure of string theory can be traced back to Bousso and Polchinski (BP)  \cite{Bousso:2000xa} (building on previous ideas of Brown and Teitelboim \cite{BT, BT2}). They noticed that the existence of a large set of non-propagating 3-forms $C_3^A$ in string theory could help to explain the smallness of the cosmological constant. The basic idea is that these 3-forms give rise to constant 4-forms (fluxes) $F_4^A=d C_3^A$ that contribute to the  vacuum energy as
\begin{align}
V_{\text{effective}}=\sum_{A,B} Z_{AB} F^AF^B+\Lambda_0\, ,
\end{align}
being $\Lambda_0$ the bare cosmological constant (c.c). They showed that for a large enough number of 4-forms, one can construct in a natural way an exponentially small cosmological constant $V_{\text{effective}}$, without strong requirements in the bare c.c. $\Lambda_0$. 

With this fact in mind, in \cite{Herraez:2018vae} they studied the role of the four-dimensional 4-forms in massive type IIA flux compactifications. They showed that the different contributions to the scalar potential $V_F$  can be written  -just playing with the equations- as
\begin{align}
16 S_{4d}\supset -Z_{AB}F_4^A\wedge \star_4 F_4^B+2F_4^A\rho_a-Z^{AB}\rho_a\rho_B\, ,
\end{align}
with the indices $A$ running over the fluxes of the compactifications -the four-dimensional four forms $F_4^A$-, $\rho_A$ depending on the axions and the topological data of the internal manifold and $Z_{AB}$ depending only on the saxions. On shell $\star F_4^A=Z^{AB}\rho_B$ and the scalar potential can be written as
\begin{align}
V=\frac{1}{8\kappa_4^2}Z^{AB}\rho_A\rho_B\, ,
\end{align}
which has a bilinear structure and is reminiscent of the  (BP) structure. This was first obtained in \cite{Herraez:2018vae} for massive type IIA compactifications with RR and NSNS fluxes, and then generalised  in the presence of mobile D6 branes \cite{Escobar:2018tiu} and including $\alpha'$ corrections \cite{Escobar:2018rna}. We will see in chapter \ref{ch:geometricflux} that even when we include (non)-geometric fluxes, the potential exhibits a bilinear factorisation.

It is important to stress that the bilinear expression for the potential is derived by just regrouping, defining, and rewriting some terms. There is no any hide assumption and the result is completely general. The point, as we will see, is that working with the $\rho$s one can perform quite easily a systematic search of vacua. We will give explicit expressions for them in the correspondent chapters.

\section{Swampland conjectures}
\label{sec:swamplandcon}
Finally, the last leg of this review of type IIA flux compactifications is the \emph{swampland program}\cite{Vafa:2005ui}. As we have mentioned in the introduction, the swampland  goes against the idea that any theory can be obtained from string theory. Contrary to this belief, quantum gravity is expected to constrain severely the kind of effective theories that are compatible with it. We say that effective theories that have a UV completion in quantum gravity are in the landscape, whereas those that do not have it are said to belong to the swampland. As of today, this field is actively in progress and we do not have many rigorous facts, but we work with \emph{conjectures}, the so-called swampland conjectures. They can be derived in two ways. On the one hand, by thinking about which properties a theory of quantum gravity should satisfy. This is the case for example of the no global symmetries conjecture  \cite{Banks:2010zn, Banks:1988yz}, formulated using black hole arguments. On the other hand, analysing systematically string compactifications and extracting general patterns. An example of a conjecture derived in this way is the swampland distance conjecture \cite{Ooguri:2006in}.

Under the swampland approach -for reviews see \cite{Brennan:2017rbf, Palti:2019pca,vanBeest:2021lhn, Grana:2021zvf}- (massive) type IIA compactifications on CY orientifolds have received a renewed interest and are now being scrutinised. Among the different swampland conjectures, three of them involve these scenarios directly, the Non-Supersymmetric AdS Instability Conjecture \cite{Ooguri:2016pdq}, the AdS distance conjecture \cite{Lust:2019zwm} and the dS swampland conjecture \cite{Obied:2018sgi, Ooguri:2018wrx}. Let us recall them briefly.

\subsection{Non-Supersymmetric AdS Instability Conjecture}
\label{sec:nonsusyconjecture}
As it was formulated originally, it states \cite{Ooguri:2016pdq}
\begin{tcolorbox}[breakable, enhanced]
\begin{small}
\textit{Any non-supersymmetric AdS geometry supported by flux is unstable in a consistent quantum theory of gravity with low energy description in terms of the Einstein gravity coupled to a finite number of matter fields.}
\end{small}
\end{tcolorbox}
The conjecture was not derived independently but obtained as a consequence of a refined version of the weak gravity conjecture (WGC). The vanilla version of the WGC \cite{ArkaniHamed:2006dz} says that in any gravitational theory coupled to a $U(1)$ with gauge coupling $g$, there must be a particle in the spectrum satisfying the inequality
\begin{align}
\label{eq:wgcc1}
m\leq  \sqrt{\frac{d-2}{d-3}}gq\left(M_p^d\right)^{\frac{d-2}{2}}\, .
\end{align}
To see arguments in its favour, other versions of it and references, we refer the reader to the mentioned swampland reviews and  \cite{Harlow:2022gzl} for a specific recent review of the WGC. The basic idea behind this conjecture combines the absence of global symmetries in quantum gravity, the evaporation of black holes and the violation of certain entropy bounds. These same arguments can be applied in theories with general $C_p$ gauge fields, from which the existence of $p-1$ dimensional objects follows, analogously to \eqref{eq:wgcc1}. In other words, the WGC states that in any theory of gravity coupled to a $p$-form Abelian gauge field, there must be a $p-1$ membrane whose tension and charge satisfy
\begin{align}
\label{eq:extendedwgc}
\sqrt{\frac{p\left(d-p-2\right)}{d-2}}T\leq QM_p^{\frac{d-2}{2}}\, .
\end{align}
In \cite{Ooguri:2016pdq} a stronger claim was made and a refined version of the WGC was proposed, stating that the WGC \eqref{eq:extendedwgc} is saturated if and only if the theory is supersymmetric and the state in question is BPS. From this, it follows that any non-SUSY AdS supported by fluxes is at least metastable, as we will argue in a moment.

Consider a non-SUSY AdS$_d$ vacuum supported by fluxes $f=\int_{p-\text{cycle}} F_p$. Using Hodge duality, the fluxes $f$ can be described in the $d$-dimensional theory in terms of top forms field strengths $F_d=dC_{d-1}$, where $d$ is the dimensions of the spacetime. Now, the refined version of \eqref{eq:extendedwgc} applied to the $C_{d-1}$ forms implies the existence of some $d-2$ brane satisfying $T<QM_p^2$. But precisely, a co-dimension one brane whose charge is smaller than its tensions implies an instability in the AdS \cite{Maldacena:1998uz}, and so the vacuum can be at best metastable.

Type IIA orientifold compactifications provide a big arena where this conjecture can be tested. In chapter \ref{ch:nonsusy} we will examine some of the obtained non-SUSY vacua under the lens of this conjecture and will see how is non-trivially satisfied.

\subsection{AdS distance conjecture}
\label{sec:adsdc}
Originally formulated in \cite{Lust:2019zwm} it states
\begin{tcolorbox}[breakable, enhanced]
\begin{small}
\textit{Consider quantum gravity  on d-dimensional AdS space with cosmological constant $\Lambda$. There existis an infinite tower of states with mass scale $m$ which, as $\Lambda\rightarrow 0$, behaves (in Planck units) as
\begin{align}
\label{eq:adsdc}
m\sim|\Lambda|^\alpha
\end{align}
where $\alpha$ is a positive order-one number.}
\end{small}
\end{tcolorbox}
In string theory, the AdS distance conjecture is usually satisfied through Kaluza Klain modes, that is, the mass scale $m$ appearing in the LHS of \eqref{eq:adsdc} can be identified with $M_{KK}$. 

In the same original paper, a stronger version was proposed, claiming that 
\begin{tcolorbox}[breakable, enhanced]
\begin{small}
for SUSY AdS vacua $\alpha=\frac{1}{2}$.
\end{small}
\end{tcolorbox}
Straightforwardly, from the strong version and the fact that $m\sim M_{KK}$ and $\Lambda\sim M_{AdS}^2$ it follows that $R_{KK}\sim R_{AdS}$ and so SUSY AdS vacua with scale separation would be in the swampland. Though there are many examples in string theory supporting this form of the conjecture, IIA compactified on a CY orientifold does not satisfy it. We will see that they yield $\alpha=7/18$ for both the SUSY and the non-SUSY AdS vacua, and so one can achieve $R_{\mathcal{M}_6}\ll R_{AdS}$ parametrically.

As we have explained in the introduction of this chapter, the problem of AdS$_4$ orientifold vacua is that the complete solution to the 10d equations of motion is not known, because of the presence of intersecting orientifold planes. One usually works in the smearing approximation, where the localised sources are \textit{smeared} along the internal dimensions -we will comment more on this approximation in the following chapters-. It is therefore an open question if the fully back-reacted solution still allows for separation of scales. In chapter \ref{ch:uplift10d} we will make some steps in this direction, showing that the smeared solution can be understood as the 0th order of an expansion whose 1st order we will construct. 

In the meanwhile, there have been refinements of the strong AdS distance conjecture that make it compatible with scale separation \cite{Buratti:2020kda}\footnote{Though this version does not seem to be satisfied in 3d \cite{Apers:2022zjx}.}, and the study of the would-be conformal duals of scale separated AdS vacua was initiated recently in \cite{Aharony:2008wz, Conlon:2021cjk, Apers:2022zjx, Apers:2022tfm, Quirant:2022fpn}. There are other proposals that can be applied to these scenarios with scale separation, as the one of \cite{Gautason:2018gln}, which states that SUSY AdS vacua whose mass of the lightest mode $m^2$ is very large compared with the AdS mass scale $L_{AdS}^{-1}$,
\begin{align}
m^2L^2_{AdS}\gg 1\, ,
\end{align}
are in the swampland. In any case, as of today, it is still unclear if string theory admits solutions with separation of scales, which is crucial to constructing 4d theories describing our universe.

\subsection{de Sitter conjecture}

The refined version of the dS conjecture\footnote{Just after the original version of the conjecture \cite{Obied:2018sgi} was released, several counterexamples were found -see for instance \cite{Andriot:2018wzk}- and so a refinement was proposed a few months later.}  states \cite{Ooguri:2018wrx}
\begin{tcolorbox}[breakable, enhanced]
\begin{small}
A A potential $V\left(\phi\right)$ for scalar fields in a low energy effective theory of any consistent quantum gravity must satisfy either,
\begin{align}
\label{eq:dsc}
|\nabla V| &\geq \frac{c}{M_p}\cdot V\,		&	 &\text{or}		&	\text{min}\left(\nabla_i\nabla_j V\right)&\leq -\frac{c'}{M_p^2}\cdot V\, ,
\end{align}
for some universal constants $c, c'\geq 0$ of order 1, where the left-hand side of \eqref{eq:dsc} is the minimum eigenvalue of the Hessian $\nabla_i\nabla_j V$ in an orthonormal frame.
\end{small}
\end{tcolorbox}
Though this is one of the most controversial conjectures,  it is widely believed to be true at least in asymptotic regions of the moduli space (near-infinite distance singularities), where it is satisfied in all the cases studied. In other regions of the moduli space, it is not clear that the conjecture holds. Notice that it has strong implications in the description of our universe since it forbids stable dS vacua in string theory. For reviews of the current status of this problem, one can go to \cite{Danielsson:2018ztv, Cicoli:2018kdo}. Related to this conjecture, we also need to mention the Transplanckian Censorship conjecture (TCC) \cite{Bedroya:2019snp}, which nevertheless we will not use in this work.

When looking for vacua, we will always have in mind the dS conjecture and will study what kind of set-ups could rise to violations of this conjecture. Anticipating the results, we will see first in equation \eqref{dss} that the conjecture always holds in the presence of only NSNS and RR fluxes. Then, adding (non)-geometric fluxes, in section \ref{subsec: no-go's}, we will see that the conjecture could in principle be violated, though not with the ansatz we will consider.

\clearpage


\chapter{Search of vacua: RR and NSNS fluxes}
\label{ch:rrnsnsvacua}

A fundamental question in the context of string compactifications is the characterisation of the string Landscape, that is the collection of isolated, metastable 4d vacua that are obtained from string theory. In this regard compactifications with background fluxes \cite{Grana:2005jc,Douglas:2006es,Becker:2007zj,Denef:2007pq,Ibanez:2012zz} have proven to be a remarkably fruitful framework. To great extent, this is because they provide a simple mechanism for moduli stabilisation that at the same time generates a discretum of vacua, which allows developing our intuition on how the full string Landscape may look like.

Within the flux landscape, a very interesting corner is given by (massive) type IIA compactifications with RR and NSNS fluxes, in the sense that one may achieve full moduli stabilisation using only classical ingredients, as already mentioned. Early results on this subject display a non-trivial set of classical IIA flux vacua to AdS$_4$ \cite{Behrndt:2004km,Behrndt:2004mj,Derendinger:2004jn,Lust:2004ig,Villadoro:2005cu,DeWolfe:2005uu,House:2005yc,Camara:2005dc,Koerber:2007jb,Aldazabal:2007sn,Tomasiello:2007eq,Koerber:2008rx,Caviezel:2008ik,Lust:2008zd,Cassani:2009ck,Lust:2009zb,Lust:2009mb,Koerber:2010rn,Narayan:2010em}. Some of these solutions are based on the results of \cite{Grimm:2004ua}, which combines the classical K\"ahler potential of Calabi-Yau (CY) orientifolds and the superpotential induced by RR and NS background $p$-form fluxes to obtain an effective F-term potential, as reviewed in chapter \ref{ch:review}. In particular, ref.\cite{DeWolfe:2005uu} obtains a discretum of $\CN=1$ AdS$_4$ vacua from such an effective 4d approach. The same strategy was implemented in \cite{Camara:2005dc} for the specific case in which the Calabi-Yau is a six-torus, finding different branches of supersymmetric and non-supersymmetric AdS$_4$ vacua. 

In this chapter we extend the general analysis of \cite{DeWolfe:2005uu} to find further vacua of the classical 4d potential of \cite{Grimm:2004ua}, which are not necessarily supersymmetric. The motivation to analyse this particular setup is two-fold: on the one hand, it has been recently shown in \cite{Herraez:2018vae} that the type IIA CY flux potential can be expressed as a bilinear on the flux quanta, in which the dependence of axions and saxions factorises -see section \ref{sec:4forms}-. As such, the extremisation conditions take a particularly simple form, already exploited in \cite{Escobar:2018tiu,Escobar:2018rna} in the search for new vacua. On the other hand, Calabi-Yau orientifolds with fluxes constitute an interesting arena to test the recent Swampland conjectures involving string compactifications to AdS \cite{Ooguri:2016pdq,Lust:2019zwm}, and in principle they could provide counterexamples to them -recall \ref{sec:swamplandcon}-. In order to properly address whether or not this is the case, it is important to determine the full set of vacua that corresponds to this construction. 

Needless to say, solving for general vacua of a potential is more involved than restricting the search to supersymmetric ones. In the last case, even when the K\"ahler metrics for moduli fields are not fully specified, the vanishing conditions for the F-terms allow rewriting the vacua conditions algebraically, significantly simplifying the analysis. Interestingly, the factorised form of the potential found in \cite{Herraez:2018vae}, which features a number of flux-axion polynomials invariant under discrete shift symmetries, allows implementing  a similar strategy in the search of more general vacua. Indeed, we find that by imposing a simple off-shell Ansatz for the derivatives of the potential in terms of the flux-axion polynomials, the extrema conditions can also be expressed algebraically. By solving them we find several branches of extrema, one of which corresponds to supersymmetric AdS vacua, other to the non-supersymmetric Minkowski vacua discussed in \cite{Palti:2008mg}, and the rest are different branches of non-supersymmetric AdS solutions. Compared to previous results in the literature, on the one hand we find a one-to-one correspondence between our branches of solutions and those found in \cite{Camara:2005dc} for isotropic toroidal compactifications. On the other hand, we find that some of the extrema found in \cite{Narayan:2010em} are incompatible with our results. Our approach also permits to analyse the perturbative stability of these new AdS solutions, solving for the spectrum of flux-induced masses for the simplest branches of extrema. In those cases we find some branches where tachyons are absent, and some others where they are present but satisfy the Breitenlohner-Freedman bound. Finally, our strategy can be easily generalised to include moduli and fluxes in the open string sector, providing an even richer landscape of AdS flux vacua. 

The rest of the chapter is organised as follows. In section \ref{s:IIAorientifoldnsrr} we rewrite  the classical F-term potential associate to IIA Calabi-Yau orientifolds with fluxes in a bilinear formulation. In section \ref{s:IIAvacua} we implement our Ansatz to solve for the extrema conditions, finding several branches of solutions which are summarised in table \ref{vacuresulnsrr}. In section \ref{s:stability} we analyse the perturbative stability of some of these branches and find that they can be considered perturbatively stable, see table \ref{tableensrr}. Section \ref{s:validity} discusses the validity of these solutions from a 4d viewpoint. Section \ref{s:D6branes} generalises the setup to include D6-brane with moduli and the corresponding worldvolume fluxes. We draw our conclusions in section \ref{s:conclu1}, and relegate some technical details to the appendices. Appendix \ref{ap:relations} contains some K\"ahler metric relations used in the main text, already introduced introduced in the previous chapter, while appendix \ref{ap:Hessiannsrr} performs a detailed analysis of the Hessian for several branches of solutions.


\section{Type IIA orientifolds with fluxes: bilinear formalism}
\label{s:IIAorientifoldnsrr}

In the following we will focus on (massive) type IIA flux vacua whose internal geometry can be approximated by a Calabi-Yau orientifold, as assumed in \cite{Grimm:2004ua} to derive the F-term potential used in \cite{DeWolfe:2005uu}.\footnote{Using such potential to search for vacua is justified a posteriori, by arguing that the flux-induced scale can be made parametrically smaller than the Kaluza-Klein scale, in the same region where corrections to the potential can be neglected, see \cite{DeWolfe:2005uu} and section \ref{s:validity4d}. Therefore, even if in the presence of fluxes the compactification metric is not Calabi-Yau, it is expected that the fluxless K\"ahler potential is a good approximation to capture the 4d dynamics. See also \cite{McOrist:2012yc,Gautason:2015tig} for some objections to this approach.} We will express the scalar potential directly  in the factorised bilinear form of \cite{Herraez:2018vae}. As pointed out in there, the bilinear form of the potential is independent on whether the background geometry is Calabi-Yau or not and, as it will be clear from the computations in the next section, so will be the strategy to extract the vacua from it. 

Let us consider massive type IIA string theory compactified on an orientifold of $\mathbb{R}^{1,3} \times \cM_6$ with $\cM_6$ a compact Calabi-Yau three-fold. As discussed in section \ref{ch:review}, the 4d effective theory is described by a $\mathcal{N}=1$ supergravity. The presence of RR and NS-fluxes suffices to generate a four-dimensional F-term scalar potential for the geometric moduli $(t^a, u^\mu)$ and closed string axions $(b^a,\xi^\mu)$, whose precise shape exhibits a remarkable factorisation into a geometric moduli piece, an axion piece and a flux piece~\cite{Bielleman:2015ina,Herraez:2018vae} -see section \ref{sec:4forms}-. Namely, we have a bilinear structure of the form
\begin{equation}\label{VF1}
V  = \frac{1}{\kappa_4^2} \vec{\rho}^{\ t}\, {\bf Z} \, \vec{\rho}\, , 
\end{equation}  
where the matrix ${\bf Z}$ only depends on the saxions $\{t,u\}$, while the vector $\vec{\rho}$ only on  the flux quanta and the axions $\{b, \xi\}$. More precisely, the dependence of the flux quanta is linear, and so one may write $\ell_s\vec{\rho}  = R' \cdot \vec{q}$, with $R'$ an axion-dependent rotation matrix and $\vec{q} = (e_0, e_a, m^a, m, h_\mu)^t$ the vector of flux quanta. In general the entries of $\vec{\rho}$ are axion polynomials with flux-quanta coefficients that are invariant under the discrete shift symmetries of the combined superpotential $W = W_T + W_Q$. In the case at hand they read
 \begin{equation}\label{rhos1}
\begin{array}{lcl}
\ell_s \rho_0 &=& e_0 + e_a b^a + \frac{1}{2} {\cal K}_{abc} m^a b^b b^c + \frac{m}{6} {\cal K}_{abc} b^a b^b b^c + h_\mu \xi^\mu, \\
\ell_s \rho_a &=& e_a + {\cal K}_{abc}  m^b b^c + \frac{m}{2} {\cal K}_{abc} b^b b^c, \\
\ell_s \tilde \rho^a &=& m^a + m b^a , \\
\ell_s \tilde \rho &=& m\,, \\
\ell_s \hat \rho_\mu & = & h_\mu\,.
\end{array}
\end{equation}
In this basis, the saxion-dependent matrix ${\bf Z}$ reads
\begin{equation}\label{ZAB}
{\bf Z} = e^{K} \left(\begin{array}{ccccc} 
4 & \\
& K^{a b} \\
&&\frac{4}{9} {\cal K}^2 K_{a b} \\
&&& \frac{1}{9} {\cal K}^2 & \frac{2}{3} {\cal K} u^\mu \\ 
&&&  \frac{2}{3} {\cal K} u^\nu & K^{\mu\nu}  
 \end{array} \right)\, ,
\end{equation} 
where $K = K_K + K_Q$, $K_{ab} = \frac{1}{4} \partial_{t^a} \partial_{t^b} K_K$, and $K_{\mu\nu} = \frac{1}{4} \partial_{u^\mu}\partial_{u^\nu} K_Q$, and with upper indices denote their inverses.


\section{Type IIA flux vacua}
\label{s:IIAvacua}

As already exploited in \cite{Escobar:2018tiu,Escobar:2018rna}, the bilinear structure of  F-term potential \eqref{VF1} can be used to look for vacua in type IIA flux compactifications. In this section we will generalise this approach and implement a quite general strategy for the search of extrema of $V$, that will lead to different branches of solutions for the case of CY orientifold flux backgrounds. These branches will mostly describe new non-supersymmetric AdS solutions, but they will also contain the supersymmetric AdS solutions of \cite{DeWolfe:2005uu} and the non-supersymmetric Minkowski solutions of \cite{Palti:2008mg}. As we will see, these vacua correspond to the branches of the toroidal type IIA flux vacua found in \cite{Camara:2005dc}, but now generalised to the much broader context of Calabi-Yau geometries. In the next section we will analyse the  spectrum of some of these extrema and see that they are, in fact, classically stable AdS vacua.

\subsection{Extrema conditions}

Let us start by writing explicitly the different extrema conditions, grouped into the first order derivatives of the F-term potential \eqref{VF1} with respect to the axions $\{\xi^\mu, b^a\}$ and saxions $\{u^\mu, t^a\}$ of the compactification. Using the explicit expressions for {\bf Z} and $\vec{\rho}$ we find:

\vspace*{.5cm}

\textbf{Axionic directions}

\bes
\label{paxions1}
\begin{equation}
\label{paxioncpx1}
\left.\frac{\partial V}{\partial \xi^\mu}\right|_{\rm vac}  = \left. 8 e^K\rho_0 \hat{\rho}_\mu\right|_{\rm vac} =0
\end{equation}
\begin{equation}
\label{paxionk1}
\left.\frac{\partial V}{\partial b^a}\right|_{\rm vac}  = e^K\left[ 8 \rho_0\rho_a+\frac{8}{9}\mathcal{K}^2\tilde{\rho}^cK_{ca}\tilde{\rho}+ 2\rho_c K^{cd}\mathcal{K}_{dla}\tilde{\rho}^l\right]_{\rm vac}=0 
\end{equation}
\ees

\vspace*{.5cm}

\textbf{Saxionic directions}

\bes
\label{psaxions1}
\begin{equation}
\label{psaxioncpx1}
\left.\frac{\partial V}{\partial u^\mu}\right|_{\rm vac} = e^K\left[e^{-K} V \partial_\mu K+\frac{4}{3}\mk\tr\hr_\mu +\partial_\mu K^{\kappa\sigma}\hr_\kappa\hr_\sigma\right]_{\rm vac}=0
\end{equation}
\begin{equation}
\label{psaxionk1}
\left.\frac{\partial V}{\partial t^a}\right|_{\rm vac} = e^K\left[e^{-K} V\partial_{a}K+\partial_{a}\left(\frac{4}{9}\mathcal{K}^2\tilde{\rho}^b\tilde{\rho}^c
 K_{bc}\right)+\partial_{a}K^{cd}\rho_c\rho_d+\mathcal{K}_a\tilde{\rho}\left(\frac{2}{3}\mathcal{K}\tilde{\rho}+4u^\mu\hat{\rho}_\mu \right)\right]_{\rm vac}=0
\end{equation}
\ees

Interestingly, manipulating these condition one may rederive the inequality found in \cite{Hertzberg:2007wc} that in turn prevents the existence of de Sitter vacua. Indeed, using the properties listed in appendix \ref{ap:relations} it is straightforward to see that, off-shell: 
\begin{align}
\label{dss}
 u^\mu\partial_{u^\mu} V+\frac{1}{3}t^a\partial_{t^a} V =-3V-\frac{8e^K}{27}\mathcal{K}^2\tilde{\rho}^a\tilde{\rho}^bK_{ab}-8e^K\rho_0^2-\frac{4e^K}{3}K^{ab}\rho_a\rho_b\, .
\end{align}
At each extremum, where $\p V=0$, this equation shows that $V|_{\text{extremum}}$ must be negative or vanishing. In particular at a vacuum $V|_{\text{vac}}\leq 0$, forbidding any dS vacuum at the classical level. It would be interesting to see if the above kind of relation is preserved or violated by the different corrections to the classical approximation, along the lines of \cite{Banlaki:2018ayh}.

\subsection{The Ansatz}
\label{ss:ansatz}

Rather than solving the extrema conditions \eqref{paxions1} and \eqref{psaxions1} by brute force, in the following we will use the algebraic properties of the axion polynomials $\rho_A$ to set up an Ansatz to look for vacua. To describe such Ansatz, we will first convert the vector $\vec{\rho}$ into a different vector $\vec{\gamma}$, of the form
\be
\vec{\rho} \ \raw \ \vec{\g} \, =\, 
\left(
\begin{array}{c}
\g_0  \\ \g_a \\ \tilde{\g}^a \\ \hat\g_\mu  \\ \tilde{\rho} 
\end{array}
\right)\, =\, 
\left(
\begin{array}{c}
\rho_0 -  \tilde{\rho} \epsilon_0  \\ \rho_a - \tilde{\rho} \eps_a \\ \tilde{\rho}^a - \tilde{\rho} \tilde{\eps}^a  \\ \hat\rho_\mu - \tilde{\rho} \hat\eps_\mu \\ \tilde{\rho}
\end{array}
\right) \, ,
\label{Hbasis}
\ee
which can be seen as a (field-dependent) change of basis. The moduli-dependent functions $\eps$ are such that $\vec{\gamma}$ has only one non-vanishing component at the vacuum. Namely we define them such that $\vec{\gamma}|_{\rm vac} = ( 0 \quad 0 \quad 0 \quad 0 \quad  \tilde{\rho})^t$. Of course, this does not really constrain what the $\epsilon$'s may be, because there is an infinite number of functions with the same value at a single point. However, we will impose an Ansatz that will significantly constrain this freedom. Indeed, in the following we will look at vacua such that, off-shell,
\begin{equation}
\boxed{\partial_\alpha V\, =\, \chi_\alpha^A \gamma_A}\, 
\label{Ansatz00djk0}
\end{equation}
where $\gamma_A = \{\gamma_0, \gamma_a, \tilde{\gamma}^a, \hat{\gamma}_\mu \}$ runs over all the components of $\vec{\gamma}$ except $\tilde{\rho}$, and $\chi_\alpha^A$ are some regular functions of the moduli, with the latter indexed by $\alpha$. Notice that this essentially implies that the $\epsilon_A$ are also regular functions of the moduli.

In order to implement this Ansatz, it proves useful to rewrite the potential and the extrema conditions in terms of the $\vec{\gamma}$ basis. We have that
\begin{equation}
    V  = \frac{1}{\kappa_4^2} \vec{\gamma}^{\ t}\, \hat{\bf Z} \, \vec{\gamma},
\end{equation}
where, unlike the $\rho_A$, the elements of $\vec{\g}$ are regular functions that depend on both the axions and the saxions. The bilinear product now reads
\be
\hat{\bf Z} = 
e^{K} \left(
\begin{array}{c  c  c  c c}
4 & & & & \eps_0  \\
 & K^{ab} & & & K^{ab}\eps_a \\
& & \frac{4}{9}\CK^2K_{ab} & & \frac{4}{9}\CK^2K_{ab}\tilde{\eps}^a \\
& & & K^{\mu\nu} &  \frac{2}{3} \CK u^\nu +  K^{\mu\nu} \hat\eps_\mu \\
\eps_0 & K^{ab}\eps_b &  \frac{4}{9}\CK^2K_{ab}\tilde{\eps}^b  &  \frac{2}{3} \CK u^\mu +  K^{\mu\nu} \hat\eps_\nu  & \frac{1}{9}\CK^2 + \a
\end{array}
\right)\, ,
\label{Zg}
\ee
where
\be
\a = \eps_0^2 + K^{ab}\eps_a\eps_b + \frac{4}{9}\CK^2K_{ab}\tilde{\eps}^a\tilde{\eps}^b + K^{\mu\nu}\hat\eps_\mu \hat\eps_\nu+ \frac{4}{3} \CK u^\mu \hat\eps_\mu\, .
\ee
The strategy will now be to extremise $V$ in this basis, in order to obtain the different expressions for the $\eps$'s or, in other words, the functional dependence of $\g_A$. To each class of solutions will correspond a different class of vacua. 

Notice that we can split the scalar potential as
\be\label{Vsplit}
V = V_1 + V_2 =  \vec{\g}^{\, t}\,  \hat{\bf Z}_{1} \, \vec{\g} +  \vec{\g}^{\, t}\,  \hat{\bf Z}_{2} \, \vec{\g}\, ,
\ee
where
\be
\hat{\bf Z}_{1} = 
e^{K} \left(
\begin{array}{c  c  c  c c}
4 & & &  \\
 & K^{ab} & & &  \\
& & \frac{4}{9}\CK^2K_{ab} & &  \\
& & & K^{\mu\nu} &  \\
& & & & 0
\end{array}
\right)\, ,
\label{Zg1}
\ee
and
\be
\hat{\bf Z}_{2} = 
e^{K} \left(
\begin{array}{c  c  c  c c}
0 & & & & \eps_0 \\
 & 0 & & & K^{ab}\eps_a \\
& & 0 & & \frac{4}{9}\CK^2K_{ab}\tilde{\eps}^a \\
& & & 0 &  \frac{2}{3} \CK u^\nu +  K^{\mu\nu} \hat\eps_\mu \\
\eps_0 & K^{ab}\eps_b &  \frac{4}{9}\CK^2K_{ab}\tilde{\eps}^b  &  \frac{2}{3} \CK u^\mu +  K^{\mu\nu} \hat\eps_\nu  & \frac{1}{9}\CK^2 + \a
\end{array}
\right)\, .
\label{Zg2}
\ee
Note also that $V_1$ is positive semidefinite, while $V_2$ is not. Because $V_1$ is quadratic on quantities that vanish at the vacuum, the extremisation conditions are equivalent to taking derivatives with respect to $V_2$ only
\be
\p V|_{\rm vac} = 0 \quad \iff \quad \p V_2|_{\rm vac} = 0\, .
\label{extrema1}
\ee
In this sense, our Ansatz \eqref{Ansatz00djk0} requires something stronger than \eqref{extrema1}. Namely that, off-shell, $\p V_2$ is a function which is at least linear in the $\g_A$. In the following we will classify the different classes of solutions that arise from this requirement. 

\subsection{Branches of vacua} 
\label{ss:branches}

Let us now turn to solve for the extrema conditions \eqref{paxions1} and \eqref{psaxions1}. As we will see, rewriting them in the form \eqref{extrema1} makes it easier to classify the different branches of solutions. Later on we will discuss how such branches reproduce and generalise previous vacua found in the literature. 


\subsubsection*{Axionic derivatives}

Already from the initial expression \eqref{VF1}, \eqref{rhos1}, \eqref{ZAB}, one can see that $V$ depends quadratically on $\rho_0$, which is the only quantity that depends on the complex structure axions $\xi^\mu$. Moreover, as it depends linearly we have that 
\begin{equation}
\label{rho0}
\partial_{\xi_\mu} V=  8 e^K\rho_0 \frac{\partial \rho_0}{\partial \xi_\mu}= 8e^K\rho_0 \hat{\rho}_\mu\, , \qquad \partial_{\xi_\mu} V|_{\rm vac} =0  \rightarrow \boxed{\rho_0|_{\rm vac}=0}\, .
\end{equation}
Therefore, in our Ansatz \eqref{Hbasis} one may take $\eps_0 \equiv 0$, as we will do in the following. 

Let us now look at the derivative with respect to the B-field axions:
\be
\p_{b^a} V_2 =  \tilde{\rho}^2 e^K \left[ \frac{8}{9}\CK^2K_{ab} \tilde{\eps}^b+2 K^{bd}\eps_d \CK_{abc}\tilde{\epsilon}^c \right]+ \dots
\label{bax}
\ee
where we have used that $\p_{b^a} \rho_b = \CK_{abc}\tilde{\rho}^c$ and $\p_{b^a} \tilde\rho^b = \tilde{\rho} \d_a^b$, and the dots stand for terms linear in the $\gamma_A$. The Ansatz \eqref{Ansatz00djk0} has then two possible solutions:
\begin{itemize}
\item \textbf{Branch A1:} 
\begin{align}
&\boxed{\tilde\epsilon^b=0}	&	\rightarrow&	& \tilde{\rho}^b\rvert_{\rm vac}&=0\, .
\end{align}

\item \textbf{Branch A2:} 

Let us assume that $\tilde\epsilon^b\neq 0$ and multiply \eqref{bax} by $t^a$. Using the relations in appendix \ref{ap:relations}, one sees that a necessary condition for the bracket in the rhs of \eqref{bax} to vanish off-shell is
\begin{align}
&\boxed{\epsilon_d=-\frac{1}{4}\mathcal{K}_d	}	&	&\rightarrow 	&	\rho_a\rvert_{\rm vac}=&-\frac{1}{4}\tilde{\rho}\mathcal{K}_a\, .
\end{align}
Replacing this result in \eqref{bax} one obtains a 2nd condition:
\begin{align}
&\boxed{\tilde \epsilon^a=Bt^a}	&	&\rightarrow	&	\tilde{\rho}^a\rvert_{\rm vac}&=Bt^a\, ,
\end{align}
with $B\neq0$ some regular function of the moduli.

\end{itemize}



\subsubsection*{Saxionic derivatives}

The saxionic derivatives conditions are, for the complex structure moduli:
\begin{align}
&\p_{u^\sig} V_2 =\trh^2e^{K}\left[\p_{u^\sig} K \left( \frac{\CK^2}{9} + \a\right) + \left( \p_{u^\sig} K^{\mu\nu}\right) \hat\eps_\mu \hat\eps_\nu + \frac{4}{3} \CK  \hat\eps_\sig \right]+\nonumber\\&+\p_{u^\sig}\left(\frac{4e^K}{3}\mathcal{K}\tilde{\rho}u^\mu+2e^K\tilde{\rho} K^{\mu \nu } \hat{\epsilon }_{\nu }\right)\hat{\gamma}_\mu+\p_{u^\sig}\left(\frac{8e^K}{9}\mathcal{K}^2\tilde{\rho}K_{ba}\tilde{\epsilon}^b\right)\tilde{\gamma}^a+\p_{u^\sig}\left(2e^K\tilde{\rho }K^{ab} \epsilon _a\right) \gamma_b\, .
\label{cpxsax}
\end{align}
Notice that if one contracts \eqref{cpxsax} with $u^\sig$ and uses that $u^\sig \p_{u^\sig} K^{\mu\nu} = 2 K^{\mu\nu}$ one obtains:
\be
-\frac{ e^{-K}}{4\trh^2}u^\sig\p_{u^\sig} V_2= \oh K^{\mu\nu} \hat\eps_\mu \hat\eps_\nu + \CK u^\mu  \hat\eps_\mu+ \left(\frac{1}{9}\CK^2 + K^{ab}\eps_a\eps_b + \frac{4}{9}\CK^2K_{ab}\tilde{\eps}^a\tilde{\eps}^b\right)+\dots \, 
\label{trcpxsax}
\ee
where the dots stand for terms linear in $\g_A$.

Finally, the K\"ahler saxionic derivative reads:
\begin{align}
&\p_{t^a} V_2 =e^K\trh^2 \left[\p_{t^a} K \left( \frac{\CK^2}{9} + \a\right) +\frac{1}{9} \p_{t^a} \CK^2 + \left( \p_{t^a} K^{bc}\right) \eps_b\eps_c + \frac{4}{9}\p_{t^a} \left( \CK^2K_{ab}\right) \tilde{\eps}^a\tilde{\eps}^b + {4} \CK_a  u^\mu \hat\eps_\mu\right] +\nonumber\\ &\p_{t^a}\left(\frac{4e^K}{3}\mathcal{K}\tilde{\rho}u^\mu+2e^K\tilde{\rho} K^{\mu \nu } \hat{\epsilon }_{\nu }\right)\hat{\gamma}_\mu+\p_{t^a}\left(\frac{8e^K}{9}\mathcal{K}^2\tilde{\rho}K_{bc}\tilde{\epsilon}^b\right)\tilde{\gamma}^c+\p_{t^a}\left(2e^K\tilde{\rho }K^{cb} \epsilon _c\right)\gamma_b\, .
 \label{kahsax}
\end{align}
Proceeding as before, one can contract  \eqref{kahsax} with $t^a$ to obtain:
\be
\frac{e^{-K}}{\trh^2} t^a\p_{t^a} V_2=\frac{1}{3} \CK^2  - K^{bc} \eps_b\eps_c + \frac{4}{9} \CK^2K_{ab} \tilde{\eps}^a\tilde{\eps}^b  - 3 K^{\mu\nu}\hat\eps_\mu \hat\eps_\nu+\dots\, 
\label{trkahsax}
\ee
where again the dots stand for terms linear in the $\g_A$ and we have used that $t^a  \p_{t^a} K^{bc} = 2  K^{bc}$ and  $t^a \p_{t^a} \left( \CK^2K_{ab}\right) = 4 \CK^2K_{ab}$. 
Notice that both the first line of \eqref{cpxsax} and of \eqref{kahsax} depend on $\tilde{\rho}$ but not on any other component of $\vec{\rho}$. As such, they cannot depend on the $\g_A$. Following our strategy, we will then demand them to vanish off-shell, ensuring our Ansatz \eqref{Ansatz00djk0} and therefore that $\p_{u^\sig} V|_{\rm vac} = \p_{t^a} V|_{\rm vac} =0$.

To proceed, let us consider the general Ansatz for $\hat{\eps}_\mu$:
\be\label{Ansatz00djk}
\hat{\eps}_\mu = A \CK \p_{u^\mu} K +\mk \hat{\eps}_\mu^{\rm p} \quad {\rm with}\quad    u^\mu \hat{\eps}_\mu^{\rm p} = 0, 
\ee
where $A$ is some function of the moduli, and the factor of $\CK$ has been introduced for later convenience. The term $\hat{\eps}_\mu^{\rm p}$ is a `primitive' component of $\hat{\eps}_\mu$. We will first consider the case where $\hat{\eps}_\mu^{\rm p} = 0$, which we dub:
\begin{itemize}
\item \textbf{Branch S1:} \boxed{\hat{\eps}_\mu^{\rm p} = 0}

On the one hand the vanishing of \eqref{trcpxsax} becomes 
\be
4A -8 A^2 = \frac{1}{9} + \CK^{-2} K^{ab}\eps_a\eps_b + \frac{4}{9} K_{ab}\tilde{\eps}^a\tilde{\eps}^b ,
\label{ptrcpxsax}
\ee
which we impose off-shell. On the other hand the vanishing of \eqref{trkahsax} reads
\be
48 A^2  = \frac{1}{3}   - \CK^{-2} K^{bc} \eps_b\eps_c + \frac{4}{9} K_{ab} \tilde{\eps}^a\tilde{\eps}^b ,
\label{ptrkahsax}
\ee
to be understood also off-shell. Combining these two equations we find
\bea
K_{ab}\tilde{\eps}^a\tilde{\eps}^b& =&  - \frac{1}{2} + \frac{9}{2}A  + 45 A^2  ,\\
 \frac{K^{ab}\eps_a\eps_b}{\CK^2} & = & \frac{1}{9}  +2A - 28A^2  .
 \label{quickcheck}
\eea

For the {\bf Branch A1} one finds the following solutions:
\bea\label{A1S1}
A = \frac{1}{15} & \raw & \boxed{\eps_a = \pm \frac{3}{10}\CK_a}\,  \boxed{\hat{\eps}_\mu = \frac{\CK}{15} \p_\mu K} \,\\
A = - \frac{1}{6} & \raw & \eps_a = \pm i \sqrt{\frac{3}{4}}\CK_a\, ,
\eea
the second one being unphysical. For the {\bf Branch A2} one finds
\bea\label{A2S1}
A = \frac{1}{12} & \raw & B^2 = \frac{1}{4} \raw \boxed{\tilde \epsilon^a=\pm \oh t^a} \,  \boxed{\hat{\eps}_\mu = \frac{\CK}{12} \p_\mu K}\, \\
A = - \frac{1}{84} & \raw & B^2 < 0 \, ,
\eea
again the second solution being unphysical. 

\item \textbf{Branch S2:} \boxed{\hat{\eps}_\mu^{\rm p} \neq 0}

Finding solutions in this branch is in general more involved, as one needs some more specific information on the K\"ahler potential for the dilaton and complex structure moduli. Things however simplify if one considers a K\"ahler potential of the form
\be
K_Q = - {\rm log} (2s) - 2 {\rm log}\left( \tilde{\CG} (u^i) \right)\, ,
\label{KS2}
\ee
where $\tilde{\CG}$ is a homogeneous function of degree $3/2$ on the geometric complex structure moduli. This kind of K\"ahler potential was used in \cite{Palti:2008mg,Escobar:2018tiu,Escobar:2018rna} to construct $\CN=0$ Minkowski flux vacua. Since in this case the metric for the dilaton and other complex structure moduli decouple, it is natural to make the following Ansatz
\be
\hat{\eps}_0 = E_0 \CK \p_s K = - E_0\frac{\CK}{s}\, , \quad \quad \hat{\eps}_i = E \CK \p_{u^i} K = - 2E \CK \frac{\p_i \tilde{\CG}}{\tilde{\CG}}\, ,
\ee
with $E$, $E_0$ functions of the moduli. Then we may easily derive two equations from \eqref{cpxsax}, namely
\bea
\p_s V_2 = 0 & \raw & 8  E_0^2\CK^2 - \frac{4}{3} E_0\CK^2 =  \left( \frac{1}{9}\CK^2 + \a\right) \, , \\
u^i \p_{u^i} V_2 = 0 & \raw & 8 E^2 \CK^2 - \frac{4}{3} E\CK^2 = \left( \frac{1}{9}\CK^2 + \a\right)\, .
\label{nptrcpxsax}
\eea
Notice that $E$, $E_0$ are solutions to the same quadratic equation, so if $E \neq E_0$ then necessarily
\be
E + E_0 = \frac{1}{6}\, .
\label{sumofAs}
\ee
Using this we can rewrite \eqref{nptrcpxsax} as
\be
- 8 E^2  + \frac{8}{3} E = \CK^{-2} K^{ab}\eps_a\eps_b + \frac{4}{9} K_{ab}\tilde{\eps}^a\tilde{\eps}^b .
\ee
Moreover, from  \eqref{trkahsax} and using \eqref{sumofAs} one obtains
\be
48 E^2 - 4 E  = - \CK^{-2} K^{bc} \eps_b\eps_c + \frac{4}{9} K_{ab} \tilde{\eps}^a\tilde{\eps}^b  .
\ee
To sum up, one finds the equations
\bea
K_{ab}\tilde{\eps}^a\tilde{\eps}^b& =& 9 \left(5 E^2 - \frac{1}{6} E\right)  \, ,\\
 \frac{K^{ab}\eps_a\eps_b}{\CK^2} & = & - 28E^2 + \frac{10}{3} E \, .
\eea
In the following we will analyse the possible solutions for the two axionic branches.

For the {\bf Branch A1} one finds the following solutions:
\bea
\label{solMink}
E = 0 & \raw & \boxed{\eps_a = 0}\, , \ \boxed{\hat{\eps}_0 = -\frac{\CK}{6s}}\, , \ \boxed{\hat{\eps}_i = 0}  \, \\
E = \frac{1}{30} & \raw &  \boxed{\eps_a = \pm \frac{ \sqrt{6}}{10}\CK_a} \,  \ \boxed{\hat{\eps}_0 = -\frac{2\CK}{15s}}\, \ \boxed{\hat{\eps}_i = \frac{\CK}{30} \p_{u^i}K} \, .
\label{A1S2}
\eea
One can check that \eqref{solMink} corresponds to the Minkowski vacua analysed in \cite{Palti:2008mg,Escobar:2018tiu}.

For the {\bf Branch A2} one finds
\bea
\label{solS2S1}
E = \frac{1}{12} & \raw & {\tilde\eps^a = \pm \oh t^a},\quad {\hat{\eps}_0 = \frac{\CK}{12}\p_{u^0}K},\quad {\hat{\eps}_i = \frac{\CK}{12} \p_{u^i}K}  \, \\
E = \frac{1}{28} & \raw &  \boxed{\tilde\eps^a = \pm \frac{ 1}{14} t^a} \,  \ \boxed{\hat{\eps}_0 = -\frac{11\CK}{84s}}\,  \ \boxed{\hat{\eps}_i = \frac{\CK}{28} \p_{u^i}K} \, .
\label{A2S2}
\eea
Note that \eqref{solS2S1} is in fact a special case of the Branch S1. For all the other solutions one can express things in terms of the Ansatz \eqref{Ansatz00djk} as
\be
\hat{\eps}_\mu = \left(\frac{E}{2} + \frac{1}{24}\right) \CK \p_{u^\mu} K + \mk\hat{\eps}_\mu^{\rm p} \, ,
\ee
with
\be
\hat{\eps}_0^{\rm p} = \left( \frac{1}{8} - \frac{3E}{2}\right)  \p_{s} K\, , \quad \quad \hat{\eps}_i^{\rm p} = \left(\frac{E}{2} - \frac{1}{24}  \right) \p_{u^i} K\, .
\ee

So in total we find two (double) classes of AdS solutions in the Branch {\bf S1} and two (double) classes of AdS solutions in the Branch {\bf S2}, where in the latter we have assumed the factorised metric Ansatz \eqref{KS2}. 

\end{itemize}

\textbf{Uniqueness of the solutions}

Some comments are in order regarding the uniqueness of these solutions. An implicit assumption of the above discussion is that the K\"ahler metric $K_K^{ab}$ is irreducible. If the metric display a block-diagonal structure, as for instance in toroidal orientifolds, then more solutions are recovered. Indeed, one can check that in that case the choice of sign for the $\epsilon_a$'s in \eqref{A1S1} and \eqref{A1S2} can be made independently on each block. Each choice corresponds in principle to a different solution, as it is related to different signs of the flux quanta. The election of the signs will not be reflected in the value of the $V_{\text{vacuum}}$ - which is invariant - but it will affect the F-terms and the spectrum of light modes. Unless stated differently, in the following we will consider a generic irreducible K\"ahler metric, for which the choice of sign must be equal for all $\eps_a$'s.

\subsection{Summary of the vacua and physical properties}

\label{ss:summary}

Let us recap the previous results and compute some of the properties of these extrema:

\textbf{General structure}

All the solutions found for the vacuum equations satisfy:
\begin{align}
\rho_0&=0\, ,		&	\hat{\rho}_\mu &= \trh \CK \left(A \p_{u^\mu} K + \hat{\eps}_\mu^{\rm p}\right) \, ,	& 	\tilde{\rho}^a&=B\tilde\rho t^a\, ,	&	\rho_a&=C\tilde\rho\CK_a\, ,
\label{solutions}
\end{align}
with $A,B,C \in \mathds{R}$. The \textbf{Branch A1} has $B=0$, whereas the \textbf{Branch A2} has $B\neq 0$, $C=-1/4$. The \textbf{Branch S1} has $ \hat{\eps}_\mu^{\rm p}=0$ whereas the \textbf{Branch S2} has $ \hat{\eps}_\mu^{\rm p}\neq0$. It is convenient to point out that as long as $A\neq0$, $C\neq 0$ - ignoring the complex structure axions for the moment - there are as many equations as moduli so in principle all the moduli can be fixed. Regarding the complex structure axions, only the linear combination that appears in the superpotential \eqref{WQ} is fixed. As pointed out in \cite{Camara:2005dc}, this allows the remaining axions to participate in the St\"uckelberg mechanism present in the presence of space-time-filling D6-branes, while guaranteeing the gauge invariance of the flux superpotential.

\bigskip

\textbf{K\"ahler moduli stabilisation}

The structure \eqref{solutions} provides several relations between the K\"ahler moduli and the axion polynomials of the compactification. In particular, the last two equations involving $\rho_a$ and $\tilde{\rho}^a$ provide $2 h^{1,1}_-$ relations between the quantised zero-, two- and four-form fluxes and the complexified K\"ahler moduli. By using \eqref{rhos1} one may derive an explicit relation between the geometric K\"ahler moduli and the quantised fluxes. Namely we have that
\begin{equation}
 \hat{e}_a \equiv e_a - \frac{1}{2} \frac{\CK_{abc}m^am^b}{m} = \ell_s \tr \CK_a \left(C - \frac{1}{2} B^2 \right)\, ,
 \label{Kstab}
\end{equation}
where we have defined a shifted four-form flux $\hat{e}_a$ analogous to the one in \cite{DeWolfe:2005uu}, invariant under discrete shifts involving K\"ahler axions and fluxes. It follows from this relation that whenever $B^2 = 2C$ one needs to impose $\hat{e}_a = 0$ in order to have a sensible solution for the extrema conditions, and that then the individual K\"ahler moduli are not stabilised. One can check that this is the case for the branch  \eqref{solMink},  corresponding to the non-supersymmetric Minkowski solutions analysed in \cite{Palti:2008mg}, see also \cite{Escobar:2018tiu,Escobar:2018rna}. As pointed out in there, for Minkowski vacua the constraint on the fluxes $\hat{e}_a = 0$ is lifted once that $\alpha'$ corrections for the K\"ahler sector are taken into account.

\bigskip

\textbf{Vacuum energy}

Using the expressions \eqref{dss} and \eqref{solutions} it is straightforward to see that the vacuum energy has the following general expression:
\begin{equation}
\Lambda= V\rvert_{\rm vac}=-\left(\frac{2}{27}B^2+\frac{16}{27}C^2\right)\frac{e^K}{\kappa_4^{2}}\CK^2\trh^2\, .
\end{equation}

\bigskip

\textbf{F-terms}

Using \eqref{solutions} and the expression for the F-terms derived in \cite{Escobar:2018tiu} one can directly compute them for each of the above extrema
\begin{equation}
F_{T^a}=\trh\CK_a\left(-\frac{C}{2}-\frac{1}{4}+6A\right)+i\CK_a\trh\frac{B}{4}\, ,
\end{equation}
\begin{equation}
F_{U^\mu}=\trh\CK\p_{u^\mu}K\left(\frac{C}{2}-\frac{1}{12}-A+i\frac{B}{4}\right)+\mk\trh\hat\epsilon^p_\mu\, .
\end{equation}

\bigskip
\textbf{Summary}

Finally, we gather all the above results in table \ref{vacuresulnsrr}:

\begin{table}[h]
\begin{center}
\scalebox{1}{%
    \begin{tabular}{| c || 	c | c | c | c |c |c |}
    \hline
  Branch & $A$  & $B$  & $C$  & $\kappa_4^{2}\Lambda$  &$F_{T^a}$ & $F_{U^\mu}$\\
  \hline \hline
  \textbf{A1-S1}  &$\frac{1}{15}$ &  $0$  & $\frac{3}{10}$  & $-\frac{4e^K}{75}\CK^2\trh^2$ &0	&0\\ \hline
    \textbf{A1-S1}  &$\frac{1}{15}$  & $0$  & $-\frac{3}{10}$  & $-\frac{4e^K}{75}\CK^2\trh^2$ &$\frac{3\trh}{10}\CK_a$	&$-\frac{3\CK\trh}{10}\p_{u^\mu}K$\\ 
\hline
  \textbf{A1-S2}   & $\frac{7}{120}$  & $0$  & $\pm\frac{\sqrt{6}}{10}$  & $-\frac{8e^K}{225}\CK^2\trh^2$ 	&	$\left(-\frac{C}{2}+\frac{1}{10}\right)\trh\CK_a$&
   {$\begin{aligned}
         F_S&=\left(-\tfrac{1}{15}+\tfrac{C}{2}\right)\CK\trh\p_{s}K\\
          F_{U^i}&=\left(-\tfrac{1}{6}+\tfrac{C}{2}\right)\CK\trh\p_{u^i}K
      \end{aligned}$}  \\ \hline
     \textbf{A1-S2}    & $\frac{1}{24}$ & $0$  & $0$  & $0$ 	&0	& $F_S=0$,\quad  $F_{U^i}=-\frac{\mathcal{K}\tilde{\rho}}{6}\p_{u^i}K$\\ 
\hline
     \textbf{A2-S1}    & $\frac{1}{12}$  & $\pm\frac{1}{2}$  & $-\frac{1}{4}$  & $-\frac{e^K}{18}\CK^2\trh^2$ 	&$\left(\frac{3}{8}+\frac{iB}{4}\right)\tilde{\rho}\mathcal{K}_a$	& $\left(-\frac{7}{24}+\frac{i}{4}B\right)\CK \trh\p_{u^\mu}K$ \\ 
 \hline
      \textbf{A2-S2}    & $\frac{5}{84}$ & $\pm\frac{1}{14}$  & $-\frac{1}{4}$  & $-\frac{11e^K}{294}\CK^2\trh^2$	&$\frac{13}{56}\trh\CK_a$	& 
      {$\begin{aligned}
         F_S&=\left(-\tfrac{11}{56}+\tfrac{iB}{4}\right)\CK\trh\p_{s}K\\
          F_{U^i}&=\left(-\tfrac{7}{24}+\tfrac{iB}{4}\right)\CK\trh\p_{u^i}K
      \end{aligned}$}  \\
      \hline
    \end{tabular}}      
\end{center}
\caption{Different branches of solutions with the corresponding vacuum energy and F-terms. The solutions in the branch {\bf S2} assume the K\"ahler potential \eqref{KS2}. \label{vacuresulnsrr}}
\end{table}
As already mentioned, when the structure of the metric in the K\"ahler sector is block diagonal, this allows to choose the sign of $C$ independently in each block and therefore the corresponding value of the F-term. In particular, in the \textbf{Branch A1-S1} one can then break SUSY independently in each of the block-diagonal sectors.

\subsection{Relation to previous results}
\label{sec:comp}
As a cross-check of formalism and the solutions discussed so far, let us compare them with some of the existing results in the literature. We will analyse three different papers, presenting their main results schematically. We refer the reader to the original papers for further details. 

\begin{enumerate}
\item \textbf{Comparison with DGKT} \cite{DeWolfe:2005uu}

This paper analyses the general conditions for $\cN=1$ Calabi-Yau orientifold vacua, which are then applied to the particular orbifold background  $\otimes^3_{j=1}T_j^2/\mathds{Z}_3^2$. At the general level, one can easily map our conditions for the {\bf A1-S1} SUSY branch with the equations of section 4 of \cite{DeWolfe:2005uu}. For instance, the condition
\begin{align}\label{DGKT1}
\tr^a=0 \rightarrow b^a =-\frac{m^a}{m}\, ,
\end{align}
is equivalent\footnote{There are some signs differences which arise form the different conventions in the flux quanta definitions.} to (4.33) in  \cite{DeWolfe:2005uu}. This implies that
\begin{align}\label{DGKT2}
\rho_a&=\frac{3}{10}\tilde\rho\CK_a\longrightarrow \hat{e}_a=\frac{3}{10}m\mk_{abc}t^bt^c ,
\end{align}
which is equivalent to (4.36) in \cite{DeWolfe:2005uu}. Regarding the dilaton/complex structure sector, on the one hand one can see that the equations (4.24) and (4.25) in \cite{DeWolfe:2005uu} are equivalent to (3.37) of \cite{Escobar:2018tiu} and to the second condition in \eqref{A1S1}. On the other hand, one can check that the eq.(4.26) of \cite{DeWolfe:2005uu} that fixes one linear combination of axions $\xi$ is equivalent to $\rho_0=0$. 

The same statements hold when applying the above to the specific background $\otimes^3_{j=1}T_j^2/\mathds{Z}_3^2$. Before the inclusion of fluxes, the moduli space of this compactification consists of the axio-dilaton and 12 complexified K\"ahler moduli: 3 of them inherited form the toroidal geometry, and 9 associated with the blow-ups of the orbifold singular points. Since there are no complex structure moduli, the only necessary inputs to solve our equations are the intersection numbers, given by:
\begin{align}
\mk_{ijk}&=\kappa \iff i\neq j\neq k\, ,	&	\mk_{AAA}&=\beta\, ,
\end{align}
where $i,j...$ label the toroidal K\"ahler moduli and $A,B...$ the blow-up modes. Applying \eqref{DGKT1} and \eqref{DGKT2} to this model one finds 
\begin{align}
\tr_i=&\frac{3}{10}\tr\mk_i\rightarrow t_i=\sqrt{\frac{5\hat{e}_j\hat{e}_k}{3m\kappa\hat{e}_i}}\, , 	&	\tr_A&=\frac{3}{10}\tr\mk_A\rightarrow t_A=\sqrt{\frac{10\hat{e}_A}{3\beta m}} \, ,
\end{align}
with $\hat{e}_i=e_i-\kappa\frac{m_jm_k}{m}$, $\hat{e}_A=e_A-\beta\frac{e_A^2}{2m}$, which is equivalent to (5.5) and (5.8) in  \cite{DeWolfe:2005uu}. One can equally recover eqs.(5.10) and (5.12) from applying the conditions of the {\bf A1-S1} SUSY branch. Therefore our results reproduce the analysis in \cite{DeWolfe:2005uu}, as expected.

\item \textbf{Comparison with NT}\cite{Narayan:2010em}

This paper considers the same orbifold background as \cite{DeWolfe:2005uu}, but searches for non-supersymmetric vacua as well. By approximating the potential to its leading terms in certain flux quotients, more solutions to the extremisation equations are found, which are labelled as $\{\text{Case 1), \dots , Case 8)} \}$. Case 1) stands for the supersymmetric solutions already found in \cite{DeWolfe:2005uu}. Case 2) is related to Case 1) by an overall sign flip in all the RR four-form fluxes, that is by an overall sign flip in the $\rho_a$ or equivalently in the $\eps_a$. Therefore, Case 1) and 2) correspond to the two components of the branch \textbf{A1-S1} in table \ref{vacuresulnsrr}. Finally, Cases 3), \dots , 8) are obtained by partial sign flips in the four-form fluxes corresponding to the toroidal and blow-up two-cycles, and some of these cases are identified as classically stable vacua while others are not. 

However, one can check that once that the blow-up moduli are introduced the metric in the K\"ahler sector is irreducible. Therefore, from the viewpoint of our analysis, none of the cases 3), \dots , 8) would be actual extrema of the scalar potential. This can be seen for instance by means of the equation \eqref{quickcheck}: performing partial sign flips in the $\epsilon_a$'s will change the LHS for an irreducible K\"ahler metric, while the RHS remains invariant. The fact that the analysis in \cite{Narayan:2010em} identifies these cases as extrema is presumably due to the approximations made in the potential, which effectively removes the kinetic mixing between the different K\"ahler modes.

\item \textbf{Comparison with CFI}\cite{Camara:2005dc}

In this case the CY orientifold is  given by $\otimes^3_{j=1}T_j^2/\Omega_p\left(-1\right)^{F_L}\sigma$, so there are three complexified K\"ahler moduli,  three complex structure moduli and the axio-dilaton. To find different branches of vacua the simplification $T_1=T_2=T_3=T$ is imposed in the K\"ahler sector. The relevant data to apply our results are: 
\begin{align}
\mk_{ijk}&=1 \iff i\neq j\neq k\, ,		&			K_Q&\sim -\log\left(u_0u_1u_2u_3\right)\, ,
\end{align}
where we are using $i,j...$ to label the K\"ahler moduli and $\mu,\nu...$ to label the complex structure moduli ($U^i$) and the axio-dilaton ($U^0)$.
The two branches \textbf{A1} and \textbf{A2} become:
\begin{align}
\tilde{\rho}^b\rvert_{\rm vac}&=0\rightarrow b=-\frac{c_2}{\trh}\, ,	&	\rho_a\rvert_{\rm vac}=&-\frac{1}{4}\tilde{\rho}\mathcal{K}_a\rightarrow b=\frac{-c_2\pm\sqrt{\Gamma-\trh^2t^2/2}}{\trh}\, ,
\end{align}
respectively. Here, as in \cite{Camara:2005dc}, we have dropped the indices in the K\"ahler sector, renamed $m^a=c_2$, $e_a=c_1$ and defined $\Gamma= c_2^2-mc_1$. Notice that these are precisely the two branches found in eq.(4.23) of \cite{Camara:2005dc}, up to some sign due to different conventions in defining flux quanta. Inside each branch, we have distinguished between the subbranches  \textbf{S1} and \textbf{S2} that read:
\begin{align}
\hat{\eps}_\mu^{\rm p}& = 0\rightarrow \hat\rho_k u_k=\hat\rho_0 u_0\, ,	&	E+ E_0 &= \frac{1}{6}\rightarrow \hat\rho_k u_k=\trh t^3-\hat\rho_0u_0 \, ,
\end{align}
which are precisely  the two sub-branches in eq.(4.24) of \cite{Camara:2005dc}. Once that we have matched the branches, is direct to see that, in the vacuum:
\begin{itemize}
\item Branch \textbf{A1}-\textbf{S1}
\begin{align}
\hat\rho_\mu=\frac{\CK}{15}\p_{u^\mu}K\rightarrow \hat\rho_\mu u_\mu&=-\frac{2}{5}\trh t^3\,,	&	\rho_a&=\pm\frac{3\trh}{10}\CK_a\rightarrow t^2\trh^2=\mp\frac{5}{3}\Gamma\,,
\end{align}
equivalent to (4.25) in \cite{Camara:2005dc}.
\item Branch \textbf{A1}-\textbf{S2}
\begin{align}
\hat\rho_i=\frac{\CK}{30}\p_{u^i}K\rightarrow \hat\rho_0 u_0&=-\frac{4}{5}\trh t^3\, ,	&	\rho_a&=\pm\frac{\sqrt{6}\trh}{10}\CK_a\rightarrow t^2\trh^2=\mp\frac{5}{\sqrt{6}}\Gamma\,,
\end{align}
equivalent to (4.26) in \cite{Camara:2005dc}.
\item Branch \textbf{A2}-\textbf{S1}
\begin{align}
\hat\rho_\mu=\frac{\CK}{12}\p_{u^\mu}K\rightarrow \hat\rho_\mu u_\mu&=-\frac{1}{2}\trh t^3	\, ,&	\tilde\rho^a &=\pm\frac{\trh t^a}{2}\rightarrow t^2\trh^2=\frac{4}{3}\Gamma\,,
\end{align}
equivalent to (4.27)-(I) in \cite{Camara:2005dc}.
\item Branch \textbf{A2}-\textbf{S2}
\begin{align}
\hat\rho_i=\frac{\CK}{28}\p_{u^i}K\rightarrow \hat\rho_0 u_0&=-\frac{11}{14}\trh t^3\,,	&	\tilde\rho^a&=\pm\frac{\trh t^a}{14}\rightarrow t^2\trh^2=\frac{196}{99}\Gamma\,,
\end{align}
equivalent to (4.27)-(II) in \cite{Camara:2005dc}. 

\end{itemize}

\end{enumerate}

\section{Perturbative stability of the solutions} 
\label{s:stability}

Given the above families of extrema of the flux-induced potential, a natural question is which ones are actual vacua. In the following we would like to analyse this question at the classical level, by computing the spectrum of flux-induced masses on the former moduli fields. In particular, we will check whether the non-supersymmetric AdS extrema have any tachyonic direction with a mass below the BF found \cite{Breitenlohner:1982bm}. For simplicity, we will do this computation focusing only on the \textbf{A1-S1} and \textbf{A2-S1} branches, leaving the \textbf{S2} branch for further work. In chapter \ref{ch:nonsusy} we will go one step further and study the non-perturbative stability of branch  \textbf{A1-S1}.

\subsection{The Hessian}
\label{ss:hessian}

By construction, we have a potential whose first derivatives are of the form
\begin{equation}
\partial_\a V = \chi_\a^A \gamma_A \, ,
\end{equation}
with $\chi_a^A$ some regular functions of the saxions and the $\rho$'s. Therefore we have that
\be
\p_\a \p_\b V|_{\rm vac} = \chi_\a^A \p_\b \gamma_A \,,
\ee
where we have imposed our extremisation conditions $\gamma_A=0$. In fact, since $V_1$ is quadratic in the $\vec\gamma$, $\partial^2V_1|_{\rm vac}$ must be quadratic in $\partial \vec\gamma$. Indeed, one easily sees that
\begin{equation}
\boxed{ \p_\alpha\p_\beta V_1\rvert_{\rm vac}=2\left(\p_\alpha\vec{\g}^{\, t}\right) \hat{\bf Z}_{1} \left(\p_\beta\vec{\g}\right)}\
\label{HV1}
\end{equation}
where $\alpha=\{\xi^\mu,b^d,u^\delta,t^d\}$, $\hat{\bf Z}_{1}$ is defined as in \eqref{Zg1} and we have defined
\begin{align}
\vec{\g}^{\, t}=&\left(\begin{matrix}
\rho_0 &\g_a &\tg^a & \hg_\nu & \tr
\end{matrix}\right)\, , \\
\partial_{\xi^\mu}\vec{\g}^{\, t}=&\left(\begin{matrix}
h_\mu &0 &0 &0 & 0
\end{matrix}\right)\, ,\nonumber \\
\partial_{b^c}\vec{\g}^{\, t}=&\left(\begin{matrix}
\rho_c &\mk_{acd}\tr^d &\delta^a_c\tr &0 & 0
\end{matrix}\right)\, ,\nonumber \\
\partial_{u^\alpha}\vec{\g}^{\, t}=&\left(\begin{matrix}
0 &0 &0 &-\tr A\mk\p_\alpha\partial_\nu K-\trh\CK\p_\alpha\hat{\epsilon}^p_\nu & 0
\end{matrix}\right)\, ,\nonumber \\
\partial_{t^c}\vec{\g}^{\, t}=&\left(\begin{matrix}
0 & -2\tr C\mk_{ac} &-\tr B\delta^a_c &-3\tr A\mk_c\partial_\nu K & 0
\end{matrix}\right)\, . \nonumber
\end{align}
Notice that \eqref{HV1} is a product of two vectors with a positive definite metric. Therefore it corresponds to a positive definite Hessian, in agreement with the fact that $V_1$ is a sum of squares. 
The matrix of second derivatives of $V_2$ yields, by direct computation,
\begin{equation}
\boxed{\partial_\alpha \partial_\beta V_2\rvert_{\rm vac}=2\vec{\eta_\alpha}^t \hat{\bf Z}_{1} \p_\beta\vec{\gamma}= 2\p_\alpha\vec{\g}^{\, t} \hat{\bf Z}_{1} \vec{\eta_\beta}}\, 
\label{HV2}
\end{equation}
where we have defined
\begin{align}
\vec{\eta}_{\xi^\mu}^{\, t}=& \left(\begin{matrix}
0 &0  &0  &0 & 0
\end{matrix}   \right)\, ,
\\
\vec{\eta}_{b^d}^{\, t}=\tr&\left(\begin{matrix}
0 &0    &    \frac{3C}{\mk}K^{bc}\mk_{cd}    &    0 & 0
\end{matrix}\right)\, , \nonumber \\
\vec{\eta}_{u^\alpha}^{\, t}=&\tr\left(\begin{matrix}
0 & C\p_\alpha K\mk_a    &B\p_\alpha K t^a          &    \left(\frac{2}{3}-4A\right)\frac{\mk}{4} \left(\p_\alpha\p_\mu K-\p_\mu K \p_\alpha K\right)+e^{-K}\mk\tr K_{\beta\mu}\p_\alpha \left(e^KK^{\gamma\beta}\he^p_\gamma \right) & 0\end{matrix}\right)\, ,\nonumber \\
\vec{\eta}_{t^d}^{\, t}=&\tr\left(\begin{matrix}
0 & \frac{4C\mk}{3} K_{bd}    &    \frac{3B}{2\mk}K^{bc}\mk_{cd}     &    \mk\p_{t^d}\he^p_\mu & 0
\end{matrix}\right)\, .\nonumber
\end{align}
Unlike \eqref{HV1}, the term \eqref{HV2} is in general not definite, and may yield tachyonic directions. Putting both results together we find that the matrix of second derivatives is given by 
\begin{align}
\label{Hfin}
&\boxed{\p_\alpha\p_\beta V\rvert_{\rm vac}=2\left(\p_\alpha\vec{\g}^t\right) \hat{\bf Z}_{1} \left(\p_\beta\vec{\g}+\vec{\eta_\beta}\right)}\, 
\end{align}
which can also be written as:
\begin{align}
\p_\alpha\p_\beta V\rvert_{\rm vac}= \left(\p_\alpha\vec{\g_r}^t+\vec{\eta_\alpha}^t\right) \hat{\bf Z}_{1}\left(\p_\beta\vec{\gamma_r}+\vec{\eta_\beta}\right)+\left(\p_\alpha\vec{\g_r}^t\right) \hat{\bf Z}_{1} \left(\p_\beta\vec{\g_r}\right)-\vec{\eta_\alpha}^t \hat{\bf Z}_{1} \vec{\eta_\beta}\, .
\end{align}

\subsection{Flux-induced masses and perturbative stability}

The explicit form of the Hessian for the different branches {\bf A1-S1} and {\bf A2-S1} is given in Appendix \ref{ap:Hessian}, where the computation of its physical eigenvalues along  tachyonic directions is also performed. The relevant results for classical stability are summarised in table \ref{tableensrr}:

\begin{table}[h]
\begin{center}
\scalebox{1}{%
    \begin{tabular}{| c |     c | c | c | }
    \hline
 Branch & Tachyons & Physical eigenvalues & Massless modes \\
 \hline
  \textbf{A2-S1}& $0$    &    -& $2N$\\
 \hline
 \textbf{A1-S1, SUSY}&    $N$     &    $m^2_{tach}=\frac{8}{9}m_{BF}^2$&  $N$\\
  \hline
 \textbf{A1-S1, Non-SUSY}&     $N+1$    &    $m^2_{tach}=\frac{8}{9}m_{BF}^2$&  $N$\\
   \hline
    \end{tabular}}
    \caption{Massless and tachyonic modes for the extrema in the branch {\bf S1}. Here  $N$ stands for the number of complex structure moduli. The extra zero modes in the branch {\bf A2-S1} are discussed in appendix \ref{ap:complex}.}
    \label{tableensrr}
    \end{center}
\end{table}

Let us highlight some of the features resulting from this analysis:

\begin{itemize}

\item[-] Each vacuum has at least $N$ zero modes, which are the complex structure axions that do not appear in the superpotential \eqref{WQ}. As such, they do not appear in the F-term classical scalar potential, as one can check directly from eqs.\eqref{VF}-\eqref{ZAB}. Therefore they constitute $N$ flat directions of the classical potential. These unlifted axions may be eaten by D6-brane gauge bosons via the St\"uckelberg mechanism \cite{Camara:2005dc}.

\item[-] As expected from the analysis in \cite{Conlon:2006tq}, there are $N$ tachyons with mass $\frac{8}{9}|m_{BF}|^2$ in supersymmetric vacua. Such modes correspond to the saxionic directions that pair up with the flat axionic directions into complex fields. That is, they correspond to the saxions that do not appear in the superpotential \eqref{WQ}.

\item[-] As shown in appendix \ref{ap:ha1s1} the same tachyons are present in the non-supersymmetric vacua within the branch \textbf{A1-S1}, with the same mass in terms of the BF bound. Moreover, such non-SUSY vacua contain an extra tachyon which is a combination of complex and K\"ahler axionic directions, with exactly the same mass as the rest.  

\item[-] All these tachyons are absent in the \textbf{A2-S1} branch of solutions. Indeed, as shown in appendix \ref{ap:ha2s1}, all the solutions of this branch have a positive semidefinite Hessian. The tachyonic modes of the saxionic sector of the branch \textbf{A1-S1} are zero modes in this branch. They are however not flat directions and develop a positive quartic potential, see appendix \ref{ap:complex} for a detailed discussion. The rest of the spectrum does not arrange into mass-degenerate complex scalars. 

\item[-] A general analysis is more involved for the {\bf S2} branches. Following \cite{Camara:2005dc}, we have analysed them for the particular case of isotropic toroidal compactifications (i.e., where all three complex structure and three K\"ahler moduli are identified as $U_i = U$ and $T_i = T$, respectively). We have found that AdS solutions in this branch contain tachyons not satisfying the BF bound, and are therefore perturbatively unstable. It would be interesting to see if this feature is also present for more general solutions and compactifications within this branch.


\end{itemize}

\section{Validity of the solutions} 
\label{s:validity}

In the following we analyse the validity of our solutions from the 4d perspective. We relegate the analysis from the 10d perspective to chapters \ref{ch:review10d} and \ref{ch:uplift10d}.

\subsection{4d analysis and swampland conjectures}
\label{s:validity4d}

Since the different branches of solutions have been found via a classical potential $V$, one  should check that they fall in the compactification regime in which the corrections to $V$ are negligible. More precisely, a necessary condition to trust the above solutions is that the K\"ahler moduli are stabilised at sufficiently large values - so that $\alpha'$ corrections can be neglected - and the string coupling at small enough values - so that quantum corrections can also be neglected. In the following we will generalise the 4d validity analysis made in  \cite{DeWolfe:2005uu,Camara:2005dc} to our solutions, obtaining similar results. In short, the scaling of the volumes and couplings with the fluxes follows the same pattern as in these references, which allows to fall in the required regime for large values of the shifted four-form flux. Indeed, we have that
\begin{align}
&\ell_s^{-1} \hat{e}_a =  \tr \CK_a \left(C - \frac{1}{2} B^2 \right) \quad \rightarrow \quad t^2\left(C-\frac{B^2}{2}\right) \sim  \frac{\hat{e}}{m}\ ,
\\
&\hat{\rho}_\mu=\tr \mk A\p_\mu K \quad \rightarrow \quad u\sim \frac{t^3m A}{h}\, ,
\end{align}
where for simplicity we have assumed isotropic fluxes $h_\mu \sim h$, $\hat{e}_a \sim \hat{e}$. Since the $\hat{e}_a$ are unconstrained by tadpole equations, in principle we are free to scale them to be as large as needed. Assuming that $2C \neq B^2$, the moduli dependence on this scaling is given by
\begin{align}
&t\sim \hat{e}^{1/2}\, , &		 u&\sim \hat{e}^{3/2}\, .
\end{align}
In addition we have that
\begin{align}
&e^{-4D}\sim u^4 \sim \hat{e}^6 \rightarrow e^D\sim \hat{e}^{-3/2}\, , &	e^\phi&=\sqrt{{\rm Vol}_{\cM_6}}e^D\sim t^{-3/2}\sim \hat{e}^{-3/4}	\, .
\end{align}
These are the same scaling relations found in \cite{DeWolfe:2005uu} and so, for large $\hat{e}$, we are in a regime of large volume and weak coupling that prevents large corrections.  Finally, one can check that the four-form density scaling is similar to \cite{DeWolfe:2005uu} and therefore the corresponding higher derivative corrections are equally suppressed.

Additionally, one can check the scaling of the different mass scales, following for instance the relations given in \cite{Escobar:2018tiu}:
\begin{align}
\frac{M_{\rm KK}}{M_{\rm P}} & \sim \frac{g_s}{V^{2/3}}\sim t^{-7/2}\sim \hat{e}^{-7/4}\, , \nonumber\\
\frac{\Lambda}{M_{\rm P}^2} & \sim e^K\mk^2 m^2 \sim  \frac{t^3}{u^4}\sim t^{-3} \sim \hat{e}^{-9/2}\, , \\
R_{\rm AdS}M_{\rm P} & \sim \Lambda^{-1/2} M_{\rm P}\sim \hat{e}^{9/4}\, .\nonumber
\end{align}
We then recover the same scaling as found in \cite{DeWolfe:2005uu}, and in particular the same  parametric separation between the compactification scale and the AdS radius:
\begin{align}
\frac{R_{\rm AdS}}{R_{\rm KK}}&\sim \hat{e}^{1/2}\, ,	& \hat{R}_{\rm KK}\sim \hat{R}_{\rm AdS}^{7/9}\ \longrightarrow	\ \frac{{M}_{\rm KK}}{M_{\rm P}} \sim \hat{\Lambda}^{7/18}\, ,
\end{align}
where $\hat{R} = R M_{\rm P}$ and $\hat{\Lambda} = \Lambda/M_{\rm P}^2$.  Lastly, using the results derived in appendix \ref{ap:ha1s1}, for the \textbf{A1-S1} branch we have:
\begin{equation}
\label{amigo}
m^2_{\text{moduli}}\sim m_{BF}^2\sim \Lambda\sim\frac{1}{R_{AdS}^2}\, ,
\end{equation}
where $m^2_{\text{moduli}}$ refers to the canonically normalised mass of the moduli becoming massive. 

Regarding the swampland conjectures, the last relation \eqref{amigo} satisfies the criterium suggested in  \cite{Gautason:2018gln} for the lightest scalars -recall section \ref{sec:adsdc}-. It would be interesting to check if the spectrum of  St\"uckelberg masses associated with the zero modes as well as the spectrum of the other branches still satisfy this relation. In terms of the  AdS conjectures formulated in \cite{Lust:2019zwm}, all the AdS vacua found in our analysis satisfy the plain AdS Distance Conjecture, while the supersymmetric ones would fail to satisfy its strong version. It was suggested in \cite{Lust:2019zwm} that this failure could be related to the lack of knowledge of the full 10d supergravity background describing such vacua -which are only satisfied in the smearing approximation- and that the back-reaction of the sources could spoil the separation of scales. In fact, to date the absence of a solution to the 10d equations of motion holds for each of the AdS vacua found in our 4d analysis, and is to be expected that finding their 10d description is at the same level of difficulty. This problem will be addressed in detail in chapters \ref{ch:review10d}-\ref{ch:uplift10d}. Finally, there is also the non-Supersymmetric AdS Instability Conjecture, which predicts that the $\mathcal{N}=0$ vacua found should suffer from instabilities. We have seen in the previous section that these non-susy AdS vacua are stable at the classical level. We will examine branch $\textbf{A1-S1}$ in light of this conjecture in chapter \ref{ch:nonsusy}

\section{Including mobile D6-branes}
\label{s:D6branes}

As in \cite{Grimm:2011dx,Kerstan:2011dy,Carta:2016ynn} we may generalise the above setup by considering type IIA orientifold compactifications where D6-branes have deformation and Wilson line moduli. In order to preserve supersymmetry such D6-branes must wrap special Lagrangian three-cycles $\Pi_\alpha \subset {\cal M}_6$ with vanishing worldvolume flux \cite{Becker:1995kb,Martucci:2005ht}. Then the open string moduli space is characterised by $b_1(\Pi_\a)$ complex moduli \cite{Mclean96deformationsof,Hitchin:1997ti}. These are defined as \cite{Herraez:2018vae,Escobar:2018tiu}
\begin{equation}
\Phi_\alpha^i = T^a f_{\alpha\, a}^i - \theta^i_\alpha = \hat  \theta_\alpha^i + i\, \phi^i_\alpha\,  ,
\end{equation}
where $i$ runs over the integer harmonic one-forms $\zeta_i$ of $\Pi_\alpha$, $\theta^i_\alpha$ is the Wilson line corresponding to each of them and $f_{\alpha\, a}^i$ is a function of the corresponding geometric deformation of $\Pi_\alpha$ defined in terms of a chain integral. We refer the reader to \cite{Carta:2016ynn,Herraez:2018vae,Escobar:2018tiu} for further details on these definitions. 

For each harmonic one-form $\zeta_i \in H^1(\Pi_\a, \mathbb{Z})$ there is  a two-form $\eta^i \in H^2(\Pi_\a, \mathbb{Z})$ along with a worldvolume flux $F = n_{F\, i}^\alpha$ that can be turned on. This enters the D6-brane DBI action and therefore the scalar potential in the combination $n_{F\, i}^\a - \oh g_{i\, \a}^\mu h_\mu$, where $g_{i\, \a}^\mu$ is also defined in terms of a chain integral \cite{Carta:2016ynn,Herraez:2018vae,Escobar:2018tiu}. As such the presence of such fluxes generates a potential, captured by the superpotential
\be
\ell_s {W}_{\rm D6}  \, = \, \Phi_\a^i( n_{F\, i}^\a - n_{a\, i}^\a T^a ) + \ell_s W_0\, ,
\label{WD6}
\ee
where 
\be
n_{a\, i}^\a  =  \frac{1}{\ell_s^3} \int_{\Pi_\a} \omega_a \wedge \zeta_i \in \mathbb{Z}
\label{nai}
\ee
are non-vanishing whenever the two-cycles of $\Pi_\alpha$ are non-trivial in $H_2(\cM_6, \mathbb{Z})$. Indeed, as pointed out in \cite{Marchesano:2014iea} in this case the open string moduli develop a potential due to the D6-brane backreaction on a compact space.

An important effect to take into account is the field redefinition of the closed string moduli in the dilaton-complex structure sector in the presence of open string moduli. We have that the new variables read \cite{Herraez:2018vae,Escobar:2018tiu}
\be
U^\mu = U_\star^\mu  + \oh \sum_\a \left(g^\mu_{i\, \a} \th^i_\a -T^a  H_{a \, \a}^\mu \right)  \, ,
\label{redefN}
\ee
where $U_\star^\mu$ stand for the complex structure moduli in the absence of mobile D6-branes, namely \eqref{cpxmoduli}, and $U^\mu$ are the redefined 4d variables. Finally $H_{a \, \a}^\mu$ are functions of the saxions defined in terms of $f_{\alpha\, a}^i$ and $g_{i\, \a}^\mu$ \cite{Herraez:2018vae}. Notice that \eqref{KQ} is a function of $u^\mu_\star$, which is to be written in terms of the new 4d variables by means of \eqref{redefN}. 

A similar statement holds for the scalar potential, which still has the form \eqref{VF1} but now with
\begin{equation}
{\bf Z} = 
\left(
\begin{matrix}
 4 & 0 & 0 & 0 & 0 & 0 & 0 \\
 \\
 0 & K^{ab} & 0 & 0 & 0 & 0 & 0 \\
 \\
 0 & 0 & \frac{4}{9} \mk^2 K_{ab} & 0 & 0 & 0 & 0 \\
 \\
  0 & 0 & 0 & G^{ij} & 0 & 0 & 0\\
 \\
  0 & 0 & 0 & 0 & t^{a}t^b G^{ij} & 0 & 0\\
  \\
0 & 0 & 0 & 0 & 0 &  K^{\mu \nu }  & \frac{2}{3} \mk u^{\mu}_\star  \\
 \\
 0 & 0 & 0 & 0 & 0 & \frac{2}{3}\mk  u^{\nu}_\star  &\frac{\mk^2}{9} \\
\end{matrix}
\right)\, , \qquad 
\vec{\rho} \, =\, 
\left(
\begin{array}{c}
\rho_0' \\ \rho_a' \\ \tilde{\rho}^{\prime \, a} \\ \rho_i' \\ \rho_{a\, i} \\ \hat\rho_\mu  \\ \tilde{\rho}
\end{array}
\right)\, ,
\label{ZrhoD6}
\end{equation}
where
\begin{align}
 \r_0'&=\r_0+\hth^i\r_i\, ,\nonumber
\\
\r_a'&=\r_a-\hth^i\r_{ai}+f^i_a\r_i-\frac{1}{2}H^\mu_a\hr_\mu\, ,
\nonumber
\\
 \tr'^a&=\tr^a-\left(\mk^{ab}\ph^i+\mk^{ac}t^bf^i_c\right)\r_{bi} \label{newrhos}\, ,
\\
\r_i'&=\underbrace{\ell_s^{-1} \left(n_{i}-b^an_{ai}\right)}_{\r_i} -\frac{1}{2}g^\mu_{i}\hr_\mu=\r_i -\frac{1}{2}g^\mu_{i}\hr_\mu\, ,
\nonumber
\\
\r_{ai}&= \ell_s^{-1} n_{ai}\, . \nonumber
\end{align}
A few comments are in order. To simplify the notation we have absorbed the D6-brane index $\alpha$ into the open string moduli index $i$. Here $\rho_0$ is defined as in \eqref{rhos1} but now in terms of the redefined RR axion $\xi^\mu = \xi_\star^\mu - \frac{1}{2} b^a H_a^\mu + \frac{1}{2} g_i^\mu \theta^i$. Finally, notice that the $\rho$'s defined in \eqref{newrhos} not only depend on fluxes and axions, but also on some saxions,  differently from those defined in \cite{Herraez:2018vae}. This is just as well for the purpose of this analysis, as we are going to combine them right away in terms of saxion-dependent  polynomials $\gamma_A$. Indeed, applying the strategy of section \ref{ss:ansatz} we define
\begin{equation}
\label{gammaopen}
\vec{\gamma}'=
\left(\begin{matrix}
\g'_0\\
\g'_a\\
\tg'^a\\
\g'_i\\
t^a\rho_{ai}\\
\hg_\mu\\
\tr
\end{matrix}\right)=\left(\begin{matrix}
\r'_0\\
\r'_a-\tr\e_a\\
\tr'^a-\tr\te^a\\
\r'_i\\
t^a\rho_{ai}\\
\hr_\mu-\tr\he_\mu\\
\tr
\end{matrix}\right)\, ,
\end{equation}
where we are not relabelling $t^a\rho_{ai}$ in order to not to overload the notation and, as before, we assume that each of the terms of this vector vanishes in the vacuum, except $\tr$. The potential can again be split in two terms
\begin{align}
\label{openpotential}
V'&=\underbrace{e^K\left[\frac{4}{9}\mathcal{K}^2\tilde{\gamma}'^a\tilde{\gamma}'^b K_{ab}+4\g_0'^2+K^{ab}\gamma_a'\gamma_b'+K^{\mu\nu}\hat{\gamma}_\mu\hat{\gamma}_\nu+G^{ij}\left(\g'_i\g'_j+t^a\r_{ai}t^b\r_{bj}\right)\right]}_{V_1'}+\nonumber\\&+\underbrace{e^K \left[\frac{4}{3}\mathcal{K}\tilde{\rho}u_\star^\mu\hat{\gamma}_\mu+\frac{8}{9}\mathcal{K}^2\tilde{\rho}K_{ba}\tilde{\gamma}'^a\tilde{\epsilon}^b+2\tilde{\rho }K^{ab} \epsilon _a \gamma'_b+2\tilde{\rho} K^{\mu \nu }\gamma _{\mu } \hat{\epsilon }_{\nu }+\tilde{\rho}^2\left(\alpha+\frac{\mathcal{K}^2}{9}\right)\right]}_{V_2'}\, ,
\end{align}
where $\alpha=K^{ab}\epsilon_a\epsilon_b+\frac{4}{9}\mathcal{K}^2K_{ba}\tilde{\epsilon}^b\tilde{\epsilon}^a+K^{\mu \nu }\hat{\epsilon }_{\nu } \hat{\epsilon }_{\nu }+\frac{4}{3} \mathcal{K}u_\star^{\mu } \hat{\epsilon }_{\nu }$ and
\begin{align}
\te^a&=Bt^a\, ,	&	&\e_a=C\mk_a\, ,	&	\he_\mu=A\mk\p_\mu K+\mk\te^p_\nu\, .
\end{align}

Again, as $V_1$ is quadratic in the $\gamma'_A = \{\gamma'_0, \gamma'_a, \tilde{\gamma}'^a, \gamma'_i, t^a\rho_{ai}, \hat{\gamma}_\mu \}$, the extremisation conditions only depend on $V_2$. As before, one can take derivatives of $V_2$ along axionic and saxionic directions, and impose an Ansatz of the form \eqref{Ansatz00djk0}. The discussion parallels to a large extent the one in section \ref{ss:branches}, so we will provide fewer details of the derivation.

\subsubsection*{Axionic sector}
\begin{align}\nonumber
\p_{ \xi^\mu} V_2' & =8e^K\g_0'\hat\rho_\mu\,, \\
\p_{\hth^i} V_2' & =-\frac{8\mk e^K}{3}C\tr \left( t^b \rho_{b\, i}\right)\,, \\ \nonumber
\p_{b^a} V_2' & = e^K\frac{2}{3}\CK \tr \left[ \CK_{a} B \tilde{\rho} + 4 C t^b \left( \CK_{abc}\tilde{\rho}'^c + f_a^i \rho_{b\, i}\right)\right] 
 =  e^K\frac{8}{3}\CK \tr   \CK_{ac} \left( C \tilde{\gamma}'^c + B \left( C+\frac{1}{4}\right) t^c \tilde{\rho}\right) \,. 
\end{align}
The last expression is linear on $\tilde{\gamma}'^a$ for either $B=0$ (branch {\bf A1}) or $C=-\frac{1}{4}$ (branch {\bf A2}).

\subsubsection*{Saxionic sector}
\begin{align}
\partial_{u^\alpha}V_2'&=\left(\frac{1}{3}-2A\right)4\mk e^K\tr\left(u^\mu\p_\alpha K+\delta^\mu_\alpha\right)\hat{\gamma}_\mu+\p_\alpha K \left(\frac{2Be^K}{3}\mathcal{K}\tilde{\rho}\mk_b\right)\tilde{\gamma}'^b \\ &+\p_\alpha K\left(\frac{8C}{3} e^K\tilde{\rho }\mk t^b\right) \gamma'_b+2\mk\tr\p_\alpha\left(e^KK^{\mu\nu}\te^p_\nu\right)\hg_\mu\nonumber\\& +\trh^2e^{K}\left[\p_{u^\sig} K \left( \frac{\CK^2}{9} + \a\right) + \left( \p_{u^\sig} K^{\mu\nu}\right) \hat\eps_\mu \hat\eps_\nu + \frac{4}{3} \CK  \hat\eps_\sig \right]\, ,\nonumber\\
\partial_{t^a}V_2'&=\frac{\left(H^\alpha_a-f^i_ag^\alpha_i\right)\p_{u^\alpha} V_2}{2}+\left(\frac{4Be^K}{3}\mathcal{K}\tilde{\rho}\mk_{ab}\right)\tilde{\gamma}'^b+\left(\frac{8C}{3} e^K\tilde{\rho }\mk \right)\gamma'_a \\ &+\left(2e^K\tr\mk K^{\mu\nu}\p_{t^a}\te^p_\mu\right)\hat{\gamma}_\mu-\frac{4e^K}{3}\mk\tr f^i_a\left(2C\r'_i-Bt^d\r_{di}\right)\nonumber\\&+e^K\trh^2 \left[\p_{t^a} K \left( \frac{\CK^2}{9} + \a\right) +\frac{1}{9} \p_{t^a} \CK^2 + \left( \p_{t^a} K^{bc}\right) \eps_b\eps_c + \frac{4}{9}\p_{t^a} \left( \CK^2K_{ab}\right) \tilde{\eps}^a\tilde{\eps}^b + {4} \CK_a  u^\mu \hat\eps_\mu\right]\,, \nonumber \\
\partial_{\phi^i}V_2'& =\frac{g^\mu_i\p_{u^\mu} V_2}{2}+\frac{4e^K}{3}\mk\tr\left(2C\r_i'-Bt^b\r_{bi}\right)\,.
\end{align}
One can see that the conditions for $\p V$ to be linear on the $\gamma_A'$ are exactly the same as in the case without mobile D6-branes with the extra conditions $\{\gamma_i'=\r'_i=0, t^a\rho_{ai}=0\}$. Therefore, the same branches of vacua are recovered replacing the previous $\gamma$'s by the new ones. Notice that one of these branches corresponds to non-supersymmetric Minkowski vacua with D6-branes and that the conditions \eqref{Ansatz00djk0} precisely reproduce those of the vacua found in \cite{Escobar:2018tiu}. In general, we expect that the vacua of section \ref{s:stability} remain perturbatively stable in the presence of mobile D6-branes, generalising the results of \cite{Escobar:2018tiu} to AdS vacua. A detailed analysis of the Hessian, whose expression is given in appendix \ref{ap:openH}, is however left for future work.


\section{Summary}
\label{s:conclu1}

Let us close the chapter by recapping what we have done. In this chapter we have performed a general search for vacua of the classical type IIA flux potential in generic Calabi-Yau orientifold compactifications. Our analysis extends the one made in \cite{DeWolfe:2005uu} in the sense that we allow for non-supersymmetric vacua as well, the only requirement being the Ansatz of section \ref{ss:ansatz}. Implementing it we find several branches of vacua, including the supersymmetric AdS branch of \cite{DeWolfe:2005uu}, a Minkowski $\CN=0$ branch mirror to type IIB with three-form fluxes \cite{Palti:2008mg} and several new branches of non-supersymmetric AdS vacua. Remarkably, when restricted to the isotropic torus, these branches reduce to precisely the ones found in \cite{Camara:2005dc}. In this sense, our results can also be seen as an extension of the AdS type IIA flux landscape familiar from toroidal compactifications to the plethora of Calabi-Yau geometries. 

The technical ingredient behind this progress is essentially the bilinear form of the flux potential developed in \cite{Bielleman:2015ina,Carta:2016ynn,Herraez:2018vae}. This expression for $V$ conveniently factorises the saxionic and axionic degrees of freedom of the compactification, and arranges the latter in flux-axion polynomials $\rho_A$ invariant under discrete shift symmetries. This permits a more economic and organised description of the extrema conditions and their solutions, which arrange themselves into branches parametrised by real constants $A, B, C$ - see table \ref{vacuresulnsrr}. Moreover, it also allows incorporating into the analysis the light degrees of freedom of mobile D6-branes, together with their worldvolume fluxes. As a result one is able to extend the above landscape of solutions to the open string sector, in the spirit of \cite{Gomis:2005wc}.

Given these branches of critical points of the potential, the next step is to verify if they correspond to (possibly metastable) vacua. We have performed the analysis of the classical stability for the simplest branches of solutions, namely the {\bf Branch S1} of section \ref{ss:branches}, where the homogeneity properties of the K\"ahler potential allow to compute the mass spectrum of the would-be moduli. We have compared such masses with the Breitenlohner-Freedman bound, finding that {\it i)} the {\bf Branch A1-S1} develops tachyons satisfying the bound and {\it ii)} the {\bf Branch A2-S1} is absent of any tachyons. Therefore, this set of extrema already constitute a Landscape of AdS flux vacua. It would remain to analyse the non-perturbative stability of this collection of vacua, which could represent an interesting playground to test the recent Non-Supersymmetric AdS Instability Conjecture, reviewed in section \ref{sec:nonsusyconjecture}. We will examine branch  {\bf Branch A1-S1} from this point of view in chapter \ref{ch:nonsusy}.

The results of this chapter can be applied and generalised in different directions. For instance, they could be extended to include non-Calabi-Yau geometries, like SU(3) compactifications with metric fluxes or with non-geometric fluxes. Such compactifications can also be described by an effective scalar potential bilinear in the fluxes \cite{House:2005yc,Camara:2005dc,Caviezel:2008ik} and our strategy can be applied to them as well. Chapter \ref{ch:geometricflux} will be dedicated to do a systematic search of vacua in these set-ups. In fact, such bilinear structure arises as well in any supersymmetric effective field theory based on three-forms, like the ones recently developed in \cite{Farakos:2017jme,Bandos:2018gjp,Lanza:2019xxg}. One could combine our results with the said formalism to have a (partial) EFT description of the landscape of AdS flux vacua, together with membrane-mediated transitions between them. In this context, one may analyse the phenomenological properties of this landscape of vacua as an ensemble \cite{Douglas:2003um}.  For instance, given the F-terms for each of these vacua, one could extend the analysis of \cite{Escobar:2018tiu} to compute the spectrum of supersymmetry-breaking soft terms induced on the open string sector, and then analyse its statistical distribution. 

For each of these developments, a crucial step is to establish the perturbative stability of the extrema of the potential. In this sense, it would be interesting to extend the results of Appendix B to other branches not analysed in there, including solutions with mobile D6-branes. The same type of analysis could also be carried out for further examples of classical AdS vacua, like those involving metric fluxes. Some of these have the advantage that their 10d description is well understood, so analysing them with the formalism used here for Calabi-Yau orientifolds may help to better understand the 10d description of the latter. In this sense, the (smeared) 10d uplift of the vacua derived here will be studied in chaper \ref{ch:review10d}. In general, we expect that a global understanding of type IIA flux vacua from a 4d perspective will shed light on their microscopic description, helping to comprehend the ensemble of type IIA flux compactifications and eventually the string Landscape.
\clearpage


\chapter{Search of vacua: geometric fluxes}
\label{ch:geometricflux}

One of the major challenges in the field of string theory (and the subject of this thesis) is to determine the structure of four-dimensional meta-stable vacua, a.k.a. the string Landscape. In this sense, type IIA  flux compactifications with RR and NSNS fluxes have played a prominent role, as we have discussed in the previous chapters.  To some extent this is because, in appropriate regimes, type IIA moduli stabilisation can be purely addressed at the classical level  in these scenarios \cite{Derendinger:2004jn,Villadoro:2005cu,DeWolfe:2005uu,Camara:2005dc}, opening the door for a direct 10d microscopic description of such vacua. 

But, as we introduced in section \ref{generalsection}, this is not the end of the story since one can add more ingredients to the game: geometric and non-geometric fluxes. It is fair to say that the general structure of geometric type IIA flux compactifications is less understood that their type IIB counterpart \cite{Grana:2005jc,Douglas:2006es,Becker:2007zj,Denef:2007pq,Ibanez:2012zz}. Part of the problem is all the different kinds of fluxes that are present in the type IIA setup, which, on the other hand, is the peculiarity that permits to stabilise all moduli classically. Traditionally, each kind of flux is treated differently, and as soon as geometric fluxes are introduced the classification of vacua becomes quite involved. 

The purpose of this chapter is to improve this picture by providing a unifying treatment of moduli stabilisation in (massive) type IIA orientifold flux vacua. Our main tool will be again the bilinear form of the scalar potential $V = Z^{AB} \rho_A \rho_B$, introduced in section \ref{sec:4forms}. While this bilinear structure was originally found for the case of Calabi--Yau compactifications with $p$-form fluxes, we will show that it can be extended to include the presence of geometric and non-geometric fluxes, even when these fluxes generate both an F-term and a D-term potential. 
 
With this form of the flux potential, one may perform a systematic search for vacua, as we already did in chapter \ref{ch:rrnsnsvacua} for the case with only RR and NSNS fluxes. We will now generalize that analysis by also considering geometric fluxes, which are one of the main sources of classical AdS$_4$ and dS$_4$ backgrounds in string theory, and have already provided crucial information regarding swampland criteria. On the one hand, the microscopic 10d description of AdS$_4$ geometric flux vacua has been discussed in several instances  \cite{House:2005yc,Grana:2006kf,Aldazabal:2007sn,Koerber:2008rx,Caviezel:2008ik,Koerber:2010rn}. On the other hand, they have provided several no-go results on de Sitter solutions \cite{Hertzberg:2007wc,Haque:2008jz,Caviezel:2008tf,Flauger:2008ad,Danielsson:2009ff,Danielsson:2010bc,Danielsson:2011au}, as well as examples of unstable de Sitter extrema that have served to refine the original de Sitter conjecture \cite{Andriot:2018wzk}.  Therefore, it is expected that a global, more exhaustive description of this class of vacua  and a systematic understanding of their properties  leads to further tests, and perhaps even refinements, of the de Sitter and AdS distance conjectures.
  
To perform our search for vacua we consider a certain pattern of on-shell F-terms, that is then translated into an Ansatz. Even if this F-term pattern is motivated from general stability criteria for de Sitter vacua \cite{GomezReino:2006dk,GomezReino:2006wv,GomezReino:2007qi,Covi:2008ea,Covi:2008zu}, it turns out that in our setup de Sitter extrema are incompatible with such F-terms, obtaining a new kind of no-go result. Compactifications to AdS$_4$ are on the other hand allowed, and using our Ansatz we find both a supersymmetric  and a non-supersymmetric branch of vacua, intersecting at one point. In some cases we can check explicitly the perturbative stability of the non-SUSY AdS$_4$ branch, finding that the vacua are stable for a large region of the parameter space of our Ansatz, and even free of tachyons for a large subregion. 

The chapter is organised as follows. In section \ref{s:IIAorientifoldgeom} we consider the classical F-term and D-term potential of type IIA  compactifications with all kind of fluxes and express both potentials in a bilinear form. In section \ref{s:fluxpot} we propose an F-term pattern to avoid tachyons in de Sitter vacua, and build a general Ansatz from it. We also describe the flux invariants present in this class of compactifications. In section \ref{s:geovacua} we apply our results to configurations with $p$-form and geometric fluxes, in order to classify their different extrema. We find two different branches, that contain several previous results in the literature. In section \ref{s:stabalidity} we discuss which of these extrema are perturbatively stable. We draw our conclusions in section \ref{s:concluu}. 

Some technical details have been relegated to the Appendices. Appendix \ref{ap:conv} and  \ref{ap:convm} contain several aspects regarding NS fluxes and flux-axion polynomials. Appendix \ref{ap:curvature} develops the computations motivating our F-term Ansatz. Appendix \ref{ap:Hessian} contains the computation of the Hessian for geometric flux extrema.


\section{Type IIA orientifolds with general fluxes: bilinear formalism}
\label{s:IIAorientifoldgeom}

Let us consider again  type IIA string theory compactified on an orientifold of $X_4 \times X_6$ with $X_6$ a compact Calabi--Yau three-fold. As we have already seen in section \ref{sec:4deffaction}, this set-up can be described by a $\mathcal{N}=1$ supergravity theory in 4d. The presence of RR, NSNS and (non)-geometric generates a potential in the effective theory which this time has two pieces
\begin{align}
V=V_F+V_D\, ,
\end{align}
as we have explained in section \ref{generalsection}. Since all the details have been discussed in the previous chapters, we will directly give the explicit form of the potential in the bilinear formalism.

\subsection{The F-term flux potential}
As in \cite{Bielleman:2015ina,Herraez:2018vae}, one can show that the F-term potential displays a bilinear structure of the form
\begin{equation}\label{VF}
\kappa_4^2\, V_F  = {\rho}_\cA \, Z^{\cA\cB} \, {\rho}_\cB\, , 
\end{equation}  
where the matrix entries $Z^{\cA\cB}$ only depend on the saxions $\{t^a, n^\mu \}$, while the ${\rho}_\cA$ only depend on the flux quanta and the axions $\{b^a, \xi^\mu\}$. Indeed, one can easily rewrite the results in \cite{Gao:2017gxk} to fit the above expression, obtaining the following result.

The set of axion polynomials with flux-quanta coefficients are
\begin{equation}
    \rho_\cA=\{\rho_0,\rho_a,\tilde{\rho}^a,\tilde{\rho},\rho_\mu,\rho_{a\mu},\tilde{\rho}^a_\mu ,\tilde{\rho}_\mu \}\, ,
    \label{rhos}
\end{equation}
and are defined as
\bes
\label{RRrhos}
\begin{align}
  \ell_s  \rho_0&=e_0+e_ab^a+\frac{1}{2}\mathcal{K}_{abc}m^ab^bb^c+\frac{m}{6}\mathcal{K}_{abc}b^ab^bb^c+\rho_\mu\xi^\mu\, , \label{eq: rho0}\\
 \ell_s   \rho_a&=e_a+\mathcal{K}_{abc}m^bb^c+\frac{m}{2}\mathcal{K}_{abc}b^bb^c+\rho_{a\mu}\xi^\mu \, ,  \label{eq: rho_a}\\
  \ell_s  \tilde{\rho}^a&=m^a+m b^a + \tilde{\rho}^a_\mu\xi^\mu \, ,  \label{eq: rho^a}\\
 \ell_s   \tilde{\rho}&=m+\tilde{\rho}_\mu\xi^\mu \, ,   \label{eq: rhom}
\end{align}   
\ees 
and
\bes
\label{NSrhos}
\begin{align}    
\ell_s    \rho_\mu&=h_\mu+f_{a\mu}b^a+\frac{1}{2}\mathcal{K}_{abc}b^bb^cQ_\mu^a+\frac{1}{6}\mathcal{K}_{abc}b^ab^bb^cR_\mu \, , \\
 \ell_s   \rho_{a\mu}&=f_{a\mu}+\mathcal{K}_{abc}b^bQ^c_\mu+\frac{1}{2}\mathcal{K}_{abc}b^bb^cR_\mu \, ,  \label{eq: rho_ak} \\
 \ell_s   \tilde{\rho}^a_\mu &=Q^a_\mu+b^aR_\mu \, , \\
 \ell_s   \tilde{\rho}_\mu &=R_\mu \, .
\end{align}
\ees
The polynomials \eqref{NSrhos} are mostly new with respect to the previous case with only $p$-form fluxes, as they highly depend on the presence of geometric and non-geometric fluxes. As in \cite{Herraez:2018vae}, both \eqref{RRrhos} and \eqref{NSrhos} have the interpretation of invariants under the discrete shift symmetries of the combined superpotential $W = W_{\rm RR} + W_{\rm NS}$. This invariance is more manifest by writing $\ell_s \rho_\cA  = {\cal R}_\cA{}^\cB q_\cB$, where $q_\cA = \left\{e_0, \, e_b, \,  m^b, \, m, \, h_\mu, \, f_{b\mu}, \, Q^b{}_\mu, \, R_\mu \right\}$ encodes the flux quanta of the compactification and
\be
\label{eq:invRmat}
{\cal R} = \begin{bmatrix}
    {\cal R}_0  \quad & \qquad {\cal R}_0 \, \, \, \xi^\mu \,  \\
  0  \quad  & \qquad {\cal R}_0 \,\, \, \delta_\nu^\mu
\end{bmatrix}\, , \quad 
 {\cal R}_0 = \begin{bmatrix}
 1 & \quad b^b & \quad \frac{1}{2} \, {\cal K}_{abc} \, b^a \, b^c & \quad \frac{1}{6}\, {\cal K}_{abc} \, b^a \, b^b \, b^c \\
 0 & \quad \delta_a^b & \quad {\cal K}_{abc} \, b^c & \quad \frac{1}{2} \, {\cal K}_{abc} \, b^b \, b^c \\
 0 & \quad 0 & \quad \delta_b^a & \quad b^a\\
 0 & \quad 0 & 0 & \quad 1 \\
\end{bmatrix}\, ,
\ee
is an axion-dependent upper triangular  matrix, see Appendix \ref{ap:convm} for details. Including curvature corrections will modify ${\cal R}_0$, such that discrete shift symmetries become manifest, and shifting an axion by a unit period can be compensated by an integer shift of $q_\cA$ \cite{Escobar:2018rna}.

As for the bilinear form $Z$, one finds the following expression
\be
\label{eq:Z-matrix}
Z^{{\cal A}{\cal B}} =  e^K \, \begin{bmatrix}
   {\bf G}  \quad & \, \, {\cal O} \\
   {\cal O}^{\, t} \quad  & \, \, {\bf C}
\end{bmatrix}\, ,
\ee
where
\begin{equation}
    {\bf G} =\left(\begin{array}{cccc}
         4 & 0 & 0 & 0 \\
         0 & g^{ab} & 0 & 0\\
         0 & 0 & \frac{4\mathcal{K}^2}{9}g_{ab} & 0 \\
         0 & 0 & 0 & \frac{\mathcal{K}^2}{9}
    \end{array}
    \right)\, , \quad 
        \mathcal{O} =\left(\begin{array}{cccc}
         0 & 0 & 0 & -\frac{2\mathcal{K}}{3}u^\nu \\
         0 & 0 &  \frac{2\mathcal{K}}{3}u^\nu \delta^a_b & 0\\
         0 & -\frac{2\mathcal{K}}{3}u^\nu\delta^b_a & 0 & 0 \\
         \frac{2\mathcal{K}}{3}u^\nu & 0 & 0 & 0
    \end{array}
    \right)\, ,
    \label{eq:GOmatrix}
\end{equation}
\begin{equation}
    {\bf C} =\left(\begin{array}{cccc}
         c^{\mu\nu} & 0 & -\tilde{c}^{\mu\nu}\frac{\mathcal{K}_b}{2} & 0 \\
         0 & \tilde{c}^{\mu\nu}t^at^b+ g^{ab}u^\mu u^\nu  &  0 & -\tilde{c}^{\mu\nu}t^a\frac{\mathcal{K}}{6}\\
         -\tilde{c}^{\mu\nu}\frac{\mathcal{K}_a}{2} & 0 & \frac{1}{4}\tilde{c}^{\mu\nu}\mathcal{K}_a\mathcal{K}_b+\frac{4\mathcal{K}^2}{9}g_{ab}u^\mu u^\nu  & 0 \\
         0 & -\tilde{c}^{\mu\nu}t^b\frac{\mathcal{K}}{6} & 0 & \frac{\mathcal{K}^2}{36}c^{\mu\nu}
    \end{array}
    \right)\, .
\end{equation}
Here $K = K_K + K_Q$, $g_{ab} = \frac{1}{4} \partial_{t^a} \partial_{t^b} K_K\equiv \frac{1}{4}\p_a\p_b K_K$, and $c_{\mu\nu} = \frac{1}{4} \partial_{u^\mu}\partial_{u^\nu} K_Q\equiv \frac{1}{4}\p_\mu\p_\nu K_Q$, while upper indices denote their inverses. Also $u^\mu = \IM U^\mu$ stands for the complex structure saxions, and we have defined  $\mk_{a}=\mk_{abc}t^bt^c$ and $\tilde{c}^{\mu\nu}=c^{\mu\nu}-4u^\mu u^\nu $. 

Compared to the previous chapter, with only RR and NSNS fluxes, the matrices {\bf C} and ${\cal O}$ are more involved, again due to the presence of geometric and non-geometric fluxes. Interestingly, the off-diagonal matrix ${\cal O}$ has the same source as in the previous case, namely the contribution from the tension of the localised sources after taking into account  tadpole cancellation. Indeed, the contribution of background fluxes to the D6-brane tadpole is given by \cite{Aldazabal:2006up}
\be
{\cal D} F_{RR} = - \left(m h_\mu  -  m^a f_{a \mu} +  e_aQ^a{}_\mu  -  e_0 R_\mu  \right)\, \beta^\mu \, ,
\label{DFtadpole}
\ee
which can be easily expressed in terms of the $\rho_\cA$.
The corresponding absence of D6-branes needed to cancel such tadpole then translates into the following piece of the potential 
\be
\kappa_4^2 V_{\rm loc} = \frac{4}{3} e^K {\cal K}\, u^\mu  \left( \tilde\rho  \rho_\mu -  \tilde\rho^a  \rho_{a\mu}  +  \rho_a \tilde\rho^a{}_\mu - \rho_0 \, \tilde\rho_\mu \right)\, ,
\ee 
which is nothing but the said off-diagonal contribution.

Putting all this together, the final expression for the F-term potential reads
\begin{align}
   \kappa_4^2 V_F =\, &e^K\left[4\rho_0^2+g^{ab}\rho_a\rho_b+\frac{4\mathcal{K}^2}{9}g_{ab}\tilde{\rho}^a\tilde{\rho}^b+\frac{\mathcal{K}^2}{9}\tilde{\rho}^2+c^{\mu\nu}\rho_\mu\rho_\nu+\left(\tilde{c}^{\mu\nu}t^at^b+g^{ab}u^\mu u^\nu \right)\rho_{a\mu}\rho_{b\nu}\right.\nonumber\\
    &+\left(\tilde{c}^{\mu\nu}\frac{\mathcal{K}_a}{2}\frac{\mathcal{K}_b}{2}+\frac{4\mathcal{K}^2}{9}g_ {ab}u^\mu u^\nu \right)\tilde{\rho}^a_\mu\tilde{\rho}^b_\nu+\frac{\mathcal{K}^2}{36}c^{\mu\nu}\tilde{\rho}_\mu\tilde{\rho}_\nu-\frac{4\mathcal{K}}{3}u^\nu \rho_0\tilde{\rho}_\nu+\frac{4\mathcal{K}}{3}u^\nu \rho_a\tilde{\rho}^a_\nu\nonumber\\
    &\left.-\frac{4\mathcal{K}}{3}u^\nu \tilde{\rho}^a\rho_{a\nu}+\frac{4\mathcal{K}}{3}u^\nu \tilde{\rho}\rho_\nu-\tilde{c}^{\mu\nu}\mathcal{K}_a\rho_\mu\tilde{\rho}^a_\nu-\tilde{c}^{\mu\nu}t^a\frac{\mathcal{K}}{3}\rho_{a\mu}\tilde{\rho}_\nu\right]\, .
    \label{eq:potential}
\end{align}

\subsection{The D-term flux potential}

The potential derived in section \ref{generalsection} can be rewritten  in a bilinear form similar to \eqref{eq:potential} by defining the following flux-axion polynomials
\be
\label{eq:NSorbitsNew2}
\ell_s \hat{\rho}_\alpha{}^\mu =  f_\alpha{}^\mu + \hat{\cal K}_{a\alpha\beta}\, b^a \,  Q^{\beta \mu}\,,\qquad
 \ell_s\tilde\rho^{\alpha \mu} = Q^{\alpha \mu} \, ,
\ee
so that one has 
\bea\nonumber
 \kappa_4^2 V_D& = &\frac{1}{4}
\begin{bmatrix}
\hat{\rho}_\alpha{}^\mu & \, \,  \tilde\rho^{\alpha \mu} \\
\end{bmatrix} . \begin{bmatrix}
  \frac{3}{2\mk} g^{\alpha\beta}\, \partial_\mu K \partial_\nu K  & \quad 0 \\
0 &\frac{2\mk}{3}  g_{\alpha\beta} \partial_\mu K \partial_\nu K  \\
\end{bmatrix}
 . \begin{bmatrix}
\hat{\rho}_\beta{}^\nu \\
\tilde\rho^{\beta \nu} \\
\end{bmatrix}
\\ 
& = &\frac{1}{4}
\partial_\mu K \partial_\nu K  \left(  \frac{3}{2\mk}g^{\alpha\beta}  \hat{\rho}_\alpha{}^\mu \hat{\rho}_\beta{}^\nu +\frac{2\mk}{3} \, g_{\alpha\beta} \,\, \tilde\rho^{a \mu} \tilde\rho^{\beta \nu} \right)\, , 
\label{eq:D-terms-new}
\eea
with $g_{\a\b} = -\frac{3}{2\mk}{\rm Im}\, \hat{\cal K}_{\alpha\beta}$ and $g^{\a\b}$ its inverse. It is then easy to see that the full flux potential $V = V_F + V_D$ can be written of the bilinear form \eqref{VF}, by simply adding \eqref{eq:NSorbitsNew2} to the polynomials \eqref{rhos} and enlarging $Z$ accordingly. 


\section{Analysis of the potential}
\label{s:fluxpot}

While axion polynomials allow for a simple, compact expression for the flux potential, finding its vacua in full generality is still quite a formidable task. In this section we discuss some general features of this potential that, in particular, will lead to a simple Ansatz for the search of vacua. In the following section we will implement these observations for the case of compactifications with geometric fluxes. As the D-term piece of the potential will not play a significant role, in this section we will neglect its presence by considering compactifications such that $h_+^{1,1} = 0$.  Nevertheless, the whole discussion can be easily extended to a more general case.

\subsection{Stability and F-terms}\label{ss:fterms}

Given the F-term potential \eqref{eq:potential}, one may directly compute its first derivatives to find its extrema and, subsequently, its second derivatives to check their perturbative stability. However, as (meta)stability may be rather delicate to check for non-supersymmetric vacua, it is always desirable to have criteria that simplify the stability analysis. 

A simple criterium to analyse vacua metastability for F-term potentials in 4d supergravity was developed in \cite{GomezReino:2006dk,GomezReino:2006wv,GomezReino:2007qi,Covi:2008ea,Covi:2008zu}, with particular interest on de Sitter vacua. As argued in there, the sGoldstino direction in field space is the one more likely to become tachyonic in generic de Sitter vacua. Therefore, a crucial necessary condition for metastability is that such a mass is positive. Interestingly, the stability analysis along the sGoldstino direction can essentially be formulated in terms of the K\"ahler potential, which allows analysing large classes of string compactifications simultaneously. 

Following the general discussion in \cite{GomezReino:2006dk,GomezReino:2006wv,GomezReino:2007qi,Covi:2008ea,Covi:2008zu}  the sGoldstino masses can be estimated by
\be
m^2 = (3m_{3/2}^2 + \kappa_4^2 V)\, \hat{\sig}-  \frac{2}{3} \kappa_4^2 V\, ,
\label{sgoldmass}
\ee
where $m_{3/2} = e^{K/2} |W|$ is the gravitino mass, and
\be
\hat{\sig} = \frac{2}{3}-R_{A\bar B C \bar D} f^{A} f^{\bar B} f^{C} f^{\bar D}\, ,
\label{sigma}
\ee
is a function of the normalised F-terms $f_A = \frac{G_A}{(G^AG_A)^{1/2}}$ with $G_A = D_{A} W$, and the Riemann curvature tensor $R_{A\bar{B}C\bar{D}}$. Therefore, if $V$ is positive so must be $\hat{\sig}$, or else the extremum will be unstable. Reversing the logic, the larger $\hat{\sigma}$ is, the more favorable will be a class of extrema to host metastable  vacua. 

It is quite instructive to compute $\hat{\sigma}$ in our setup. Notice that because the Riemann curvature tensor only depends on the K\"ahler potential, the analysis can be done independently of which kind of fluxes are present. Moreover, because the moduli space metric factorises, $R_{A\bar B C \bar D} \neq 0$ only if all indices correspond to either K\"ahler or complex structure directions. As a consequence, the normalised F-terms can be expressed as
\be
f_A = \left({\rm cos}\, \b\, g_a, {\rm sin}\, \b\, g_\mu\right)\, 
\ee
where $g_a = \frac{G_a}{(G^aG_a)^{1/2}}$, $g_\mu  = \frac{G_\mu}{(G^\mu G_\mu)^{1/2}}$ are the normalised F-terms in the K\"ahler and complex structure sectors, respectively, and ${\rm tan}\, \b = \frac{(G^\mu G_\mu)^{1/2}}{(G^aG_a)^{1/2}}$. Therefore we have that
\be
\hat{\sig} = \frac{2}{3}-  \left({\rm cos}\, \b\right)^4 R_{a\bar b c \bar d}\, g^{a} g^{\bar b} g^{c} g^{\bar d} -  \left({\rm sin}\, \b\right)^4 R_{\mu\bar{\nu} \sigma\bar{\rho}}\, g^{\mu} g^{\bar \nu} g^{\sigma} g^{\bar \rho} \, .
\label{sigma2}
\ee
Following the discussion of Appendix \ref{ap:curvature}, one finds that the terms $R_{a\bar b c \bar d}\, g^{a} g^{\bar b} g^{c} g^{\bar d}$ and $R_{\mu\bar{\nu} \sigma\bar{\rho}}\, g^{\mu} g^{\bar \nu} g^{\sigma} g^{\bar \rho}$ are respectively minimized  by
\be
{g}_a = \frac{\g_K}{\sqrt{3}} K_a\, , \quad  {g}_\mu = \frac{\g_Q}{2} K_\mu\, ,
\label{partialmax}
\ee
where $\g_K, \g_Q \in \C$ are such that $|\g_K|^2 = |\g_Q|^2 = 1$.  In this case we have that
\be
\hat{\sigma} = \frac{2}{3} - \left({\rm cos}\, \b\right)^4 \frac{2}{3} - \left({\rm sin}\, \b\right)^4 \frac{1}{2}\, ,
\label{sigma3}
\ee
and it is positive for any value of $\b$. The choice \eqref{partialmax} corresponds to F-terms of the form
\be
G_A =\left\{G_a, G_\mu \right\}=\left\{\a_K K_a,\a_Q K_\mu \right\}\, ,
\label{solsfmax}
\ee
with  $\a_K, \a_Q \in \C$, the maximum value of \eqref{sigma3} being attained for $\a_K = \a_Q$ or equivalently $\tan\beta=2/\sqrt{3}$. Remarkably, the explicit branches of vacua obtained in the previous chapter  have this F-term pattern.\footnote{More precisely, {\bf S1} vacua branches in chapter \ref{ch:rrnsnsvacua} are of the form \eqref{solsfmax}. The solutions found within the branches {\bf S2} correspond to cases where the complex structure metric factorises in two, and so their F-terms are specified in terms of a third constant $\a$. Finally, F-terms for Minkowski vacua with D6-brane moduli also have a similar structure, except that \eqref{solsfmax} should be written in terms of contravariant F-terms \cite{Escobar:2018tiu}.} In the following we will explore type IIA flux vacua whose  F-terms are of the form \eqref{solsfmax}, assuming that they include a significant fraction of perturbatively stable vacua. It would be interesting to extend our analysis to other possible maxima of  $\hat\sigma$ not captured by \eqref{partialmax}.

\subsubsection*{An F-term Ansatz}

As it turns out, \eqref{solsfmax} can be easily combined with the bilinear formalism used in the previous section.
Indeed, as pointed out in \cite{Herraez:2018vae}, F-terms can be easily expressed in terms of the axion polynomials $\rho_\cA$. The expressions in \cite{Herraez:2018vae} can be  generalised to the more involved flux superpotential \eqref{eq:Wgen} and \eqref{eq:WgenNS}, obtaining that
\begin{align}
G_a =&\left[\rho_a-\mk_{ab}\tr^b_\mu u^\mu-\frac{3}{2}\frac{\mk_a}{\mk}\left(t^a\rho_a+u^\mu\rho_\mu-\frac{1}{2}\mathcal{K}_b\tilde{\rho}^b_\mu u^\mu+\frac{1}{6}\mk\tr\right)\right]\nonumber\\ +&i\left[\mk_{ab}\tr^b+\rho_{a\mu}u^\mu+\frac{3}{2}\frac{\mk_a}{\mk}\left(\rho_0-t^au^\mu\rho_{a\mu}-\frac{1}{2}\mk_b\tr^b-\frac{1}{6}\mk \tilde{\rho}_\mu u^\mu\right)\right]\, \label{eq: F-Ta}\, ,\\
G_\mu =&\left[\rho_\mu-\frac{1}{2}\mk_a\tr^a_\mu+\frac{\p_\mu K}{2}\left(t^a\rho_a+u^\mu\rho_\mu-\frac{1}{2}\mathcal{K}_b\tilde{\rho}^b_\mu u^\mu-\frac{1}{6}\mk\tr\right)\right]\nonumber \\ +&i\left(t^a\rho_{a\mu}-\frac{1}{6}\mk \tilde{\rho}_\mu-\frac{\p_\mu K}{2}\left(\rho_0-t^au^\mu\rho_{a\mu}-\frac{1}{2}\mk_b\tr^b+\frac{1}{6}\mk \tilde{\rho}_\mu u^\mu\right)\right)\, .
\label{eq: F-Umu}
\end{align}

Therefore, to realise \eqref{solsfmax}, one needs to impose the following on-shell conditions
\bes
\label{proprho}
\begin{align}
   \rho_a-\mk_{ab}\tr^b_\mu u^\mu & = \ell_s^{-1} {\mathcal P}\, \partial_a K \label{eq: f-term prop rho_a}\, ,\\
       \mathcal{K}_{ab}\tilde{\rho}^b+\rho_{a\mu}u^\mu & =  \ell_s^{-1} {\mathcal Q}\, \partial_a K \label{eq: f-term prop rho^a}\, ,\\
    \rho_\mu-\frac{1}{2}\mk_a\tr^a_\mu & =  \ell_s^{-1} \cM\, \partial_\mu K\, ,\\
   t^a\rho_{a\mu}-\frac{1}{6}\mk \tilde{\rho}_\mu  & =  \ell_s^{-1}\cN\, \partial_\mu K\, , \label{eq: f-term prop rhoak}
\end{align}
\ees
where ${\mathcal P}$, ${\mathcal Q}$, $\cM$, $\cN$ are real functions of the moduli. In the next section we will impose these conditions for compactifications with geometric fluxes, obtaining a simple Ansatz for the search of type IIA flux vacua.

\subsection{Moduli and flux invariants}
\label{ss:invariants}

If instead of the above Ansatz we were to apply the more standard strategy of the previous chapter, we would compute the first and second derivatives of the potential \eqref{eq:potential}, to classify its different families of extrema and determine the perturbative stability of each of them. As pointed out in \cite{Herraez:2018vae} for the Calabi--Yau case, the derivatives of the axion polynomials \eqref{RRrhos} and \eqref{NSrhos} are themselves combinations of axion polynomials, see Appendix \ref{ap:convm} for the expressions in our more general setup. As a result, all the derivatives of the potential are functions of the saxions $\{t^a, u^\mu\}$ and the $\rho_\cA$, and in particular the extrema conditions $\p V|_{\rm vac} =0$ amount to algebraic equations involving both:
\be
\left(\p_{\a} V\right) (t^a, u^\mu, \rho_\cA)|_{\rm vac} = 0 \, ,
\label{extrema}
\ee
where $\a$ runs over the whole set of moduli $\{b^a, \xi^\mu, t^a, u^\mu\}$. The fact that the extrema equations depend on the quantised fluxes $q_\cA$ only through the $\rho_\cA$ is not surprising, as these are the gauge invariant quantities of the problem  \cite{Bielleman:2015ina,Carta:2016ynn}. In addition, because in our approximation the axions $\{b^a, \xi^\mu\}$ do not appear in the K\"ahler potential and in the superpotential they appear polynomially, they do not appear explicitly in \eqref{extrema}, but only through the $\rho_\cA$ as well. Therefore, finding the extrema of the F-term potential amounts to solve a number of algebraic equations on $\{t^a, u^\mu, \rho_\cA\}$. 

This simplifying picture may however give the impression that the more fluxes that are present, the less constrained the system of equations is. Indeed, \eqref{extrema} always amounts to $2 (1+ h^{1,1}_- + h^{2,1})$ equations, while the number of unknowns is $1 + h^{1,1}_- + h^{2,1} + n_q$, with $n_q$ the number of different $\rho$'s, which depends on the fluxes that we turn on. For Calabi--Yau with $p$-form fluxes $n_q = 3 + 2h^{1,1}_- + h^{2,1}$, while by including geometric and non-geometric fluxes we can increase it up to $n_q = 2 (2 + h^{2,1}) (1 + h^{1,1}_-)$. From this counting, it would naively seem that the more fluxes we have, the easier it is to solve the extrema equations. This is however the opposite of what is expected for flux compactifications. 

The solution to this apparent paradox is to realise that the $\rho_\cA$ are not fully independent variables, but are constrained by certain relations that appear at linear and quadratic order in them. Such relations turn out to be crucial to properly describe the different branches of vacua. In the following we will describe them for different cases in our setup.

\subsubsection*{Calabi--Yau with $p$-form fluxes}

Let us consider the case where only the fluxes $F_{2n}$, $H$ are turned on, while $f = Q = R = 0$. The moduli stabilisation analysis reduces to that in the previous chapter, and the extrema conditions reduce to $2 h^{1,1}_- + h^{2,1} + 2$ because only one linear combination $h_\mu\xi^\mu$ of complex structure axions appears in the F-term potential. In this case the vector of axion polynomials $\rho_\cA=(\rho_0,\rho_a,\tilde{\rho}^a,\tilde{\rho},\rho_\mu)$ has $3 + 2h^{1,1}_- + h^{2,1}$ entries, but several are independent of the axions. Indeed, at the linear level 
\be 
 \tilde{\rho}=\ell_s ^{-1} m\, ,\qquad \quad    \rho_\mu=  \ell_s^{-1} h_\mu\, ,
 \label{invCYl}
\ee
are axion-independent, while at the quadratic level
\be
 \tilde{\rho}\rho_{a}-\frac{1}{2}\mathcal{K}_{abc}\tilde{\rho}^b\tilde{\rho}^c\, =\, \ell_s^{-2}\left(m e_a -  \frac{1}{2}\mathcal{K}_{abc} m^bm^c\right)\, ,
  \label{invCYq}
\ee
is also independent of the axions. If we fix the flux quanta $q_\cA = (e_0,  e_b,   m^b,  m,  h_\mu)$, the value of \eqref{invCYl} and \eqref{invCYq} will be fixed, and $\rho_\cA$ will take values in a $(1 + h^{1,1}_-)$-dimensional orbit. This orbit corresponds to the number of axions that enter the F-term potential, and so taking these constraints into account allows to see \eqref{extrema} as a determined system. 

Interestingly, the quadratic invariant \eqref{invCYq} was already identified in \cite{DeWolfe:2005uu} as the quantity that determines the value of the K\"ahler saxions in supersymmetric vacua of this kind. In fact, this is also true for non-supersymmetric vacua. One has that
\be
m e_a -  \frac{1}{2}\mathcal{K}_{abc} m^bm^c = \tilde{A} \CK_a\, ,
\ee
with $\tilde{A} \in \mathbb{R}$ fixed for each branch of vacua. Moreover, for the branches satisfying \eqref{solsfmax}, the complex structure saxions are fixed in terms of the fluxes as $h_\mu = \hat{A} \cK \p_\mu K$, with $\hat{A}$ constant. Therefore the fluxes fix both the saxions and the allowed orbit for the $\rho_{\cA}$. Finding the latter in terms of \eqref{extrema} is equivalent to finding the values of $b^a$ and $h_\mu\xi^\mu$.

\subsubsection*{Adding geometric fluxes}

Let us now turn to compactifications with fluxes $F_{2n}$, $H$, $f$, while keeping $Q = R = 0$. The number of axions $\xi^\mu$ that enter the scalar potential now corresponds to the dimension of the vector space spanned by $\langle h_\mu, f_{a\mu}\rangle$, for all possible values of $a$. If we see $f_{a\mu}$ as a $h^{1,1}_-  \times (h^{2,1}+1)$ matrix of rank $r_f$, the number of relevant entries on $\rho_\cA=(\rho_0,\rho_a,\tilde{\rho}^a,\tilde{\rho},\rho_\mu, \rho_{a\mu})$ is $2 + (2+ r_f)h^{1,1}_- + (1+ r_f)(1 +h^{2,1})- r_f^2$. At the linear level the invariants are
\be 
 \tilde{\rho}=\ell_s ^{-1} m\, ,\qquad \quad    \rho_{a\mu}=  \ell_s^{-1} f_{a\mu}\, ,
 \label{invmetl}
\ee
while at the quadratic level we have
\be
 \tilde{\rho}\rho_\mu-\tilde{\rho}^a\rho_{a\mu} = \ell_s^{-2} \left(mh_\mu - m^a f_{a\mu}\right)  \, , \qquad  c^{a} \left(\tilde{\rho}\rho_{a}-\frac{1}{2}\mathcal{K}_{\bar{abc}}\tilde{\rho}^b\tilde{\rho}^c\right)\, .
 \label{invmetq}
\ee
Here the $c^a \in \mathbb{Z}$ are such that $c^a \rho_{a\mu} = 0$ $\forall \mu$, so there are $h^{1,1}_- - r_f$ of this last class of invariants. Taking all these invariants into account we find that $\rho_\cA$ takes values in a $(1 + h^{1,1}_- + r_f)$-dimensional orbit,\footnote{If $d^a f_{a\mu} = h_\mu$ for some $d^a \in \R$, then the $\rho_\cA$ draw a  $(h^{1,1}_- + r_f)$-dimensional orbit, and one less axion is stabilised. As a result one can define an additional flux invariant.  See next section for an example.} signalling the number of stabilised axions. In other words, with the inclusion of metric fluxes the orbit of allowed $\rho_\cA$ increases its dimension, which implies that more moduli, in particular more axions $\xi^\mu$ are fixed by the potential. As in the CY case, the saxions are expected to be determined in terms of these invariants.

\subsubsection*{Adding non-geometric fluxes}

The same kind of pattern occurs when non-geometric fluxes are included. If one sets $R = 0$, the invariants at the linear level are $\tilde{\rho}$ and $\tilde{\rho}_\mu^a$, as well the combinations $c^a d^\mu \rho_{a\mu}$ with $c^a, d^\mu \in \mathbb{Z}$ such that $c^a d^\mu \cK_{abc}Q^c_\mu = 0$, $\forall b$. At the quadratic level, the first invariant in \eqref{invmetq} is replaced by 
\be
\tilde{\rho}\rho_\mu-\tilde{\rho}^a\rho_{a\mu} + \rho_a \tilde{\rho}^a_\mu \, ,
   \label{invngq}
\ee
where we have taken into account the Bianchi identity $f_{a[\mu}\, Q^a{}_{\nu]} = 0$.
Additionally, the second invariant  in \eqref{invmetq} may also survive if there are choices of $c^a \in \mathbb{Z}$ such that $c^a \rho_{a\mu} \xi^\mu = 0$ $\forall \xi^\mu$.
Finally, when all kind of fluxes are nonvanishing, the only invariant at the linear level is $R_\mu$, and some particular choices of $\tilde{\rho}^a_\mu$ and $\rho_{a\mu}$. At the quadratic level we have the generalisation of \eqref{invngq}
\be
\tilde{\rho}\rho_\mu-\tilde{\rho}^a\rho_{a\mu} + \rho_a \tilde{\rho}^a_\mu -\rho_0\tilde{\rho}_\mu\, ,
   \label{invngr}
\ee
where we have imposed the Bianchi identity $\rho_{[\mu} \, \tilde{\rho}_{\nu]} - \rho_{a[\mu}\, \tilde{\rho}^a{}_{\nu]} = h_{[\mu} \, R_{\nu]} - f_{a[\mu}\, Q^a{}_{\nu]} = 0$, see Appendix \ref{ap:conv}. Notice that this invariant and its simpler versions are nothing but  the D6-brane tadpole \eqref{DFtadpole} induced by fluxes. We also have the new invariants
\be
\tilde{\rho}_{[\mu}^a \tilde{\rho}_{\nu]}\, ,\qquad \qquad   \rho_{a(\mu}\tilde{\rho}_{\nu)}-\mathcal{K}_{abc}\tilde{\rho}^b_\mu\tilde{\rho}^c_\nu\, ,
\label{finalNGinv}
\ee
where as above $[\ ]$ and $(\ )$ stand for (anti-)symmetrisation of indices, respectively. Finally, if the second invariant in \eqref{finalNGinv} vanishes, or in other words if $f_{a(\mu}Q_{\nu)}=\mathcal{K}_{abc}Q^b_\mu Q^c_\nu$, we have that
\be
\rho_{a(\mu}\tilde{\rho}_{\nu)}^a - 3 \rho_{(\mu} \tilde{\rho}_{\nu)}\, ,
\label{ffinalNGinv}
\ee
is also an invariant.\footnote{Remarkably, both \eqref{ffinalNGinv} and the second invariant in \eqref{finalNGinv} vanish if the ``missing" Bianchi identities $f_{a(\mu}Q_{\nu)}=\mathcal{K}_{abc}Q^b_\mu Q^c_\nu$ and $f_{a(\mu}Q_{\nu)}^a - 3 h_{(\mu} R_{\nu)}$ proposed in \cite{Gao:2018ayp} turn out to hold generally.}

\section{Geometric flux vacua}
\label{s:geovacua}

In this section we would like to apply our previous results to the search of vacua in type IIA flux compactifications. For concreteness, we focus on those configurations with $p$-form and geometric fluxes only, leaving the systematic search of non-geometric flux vacua for future work. As we will see, for geometric flux vacua the Ansatz formulated in the last section, which amounts to impose on-shell F-terms of the form \eqref{solsfmax}, forbids de Sitter solutions. In contrast, we find two branches of AdS extrema corresponding to our Ansatz, one supersymmetric and one non-supersymmetric. The perturbative stability of the latter will be analysed in the next section. 

\subsection{The geometric flux potential}

Let us first of all summarise our previous results and restrict them to the case of $p$-form and geometric fluxes. The scalar potential reads $V = V_F + V_D$, with
\begin{align}
   \kappa_4^2 V_F =\, &e^K\left[4\rho_0^2+g^{ab}\rho_a\rho_b+\frac{4\mathcal{K}^2}{9}g_{ab}\tilde{\rho}^a\tilde{\rho}^b+\frac{\mathcal{K}^2}{9}\tilde{\rho}^2\right.\nonumber\\
    + &\left.c^{\mu\nu}\rho_\mu\rho_\nu+\left(\tilde{c}^{\mu\nu}t^at^b+g^{ab}u^\mu u^\nu \right)\rho_{a\mu}\rho_{b\nu}-\frac{4\mathcal{K}}{3}u^\nu \tilde{\rho}^a\rho_{a\nu}+\frac{4\mathcal{K}}{3}u^\nu \tilde{\rho}\rho_\nu\right]\, ,
    \label{eq:potentialgeom}\\
\kappa_4^2 V_D = \, &\frac{3}{8\mathcal{K}} \partial_\mu K \partial_\nu K \,  g^{\alpha\beta} \, \hat{\rho}_\alpha{}^\mu \hat{\rho}_\beta{}^\nu\, .\label{eq: D-potentialgeom}
\end{align}
The definitions for $g^{ab}$,  ${c}^{\mu\nu}$, $\tilde{c}^{\mu\nu}$ and $g^{\alpha\beta}$ are just as in section \ref{s:IIAorientifoldgeom}, while the $\rho_\cA$ simplify to
\bes
\label{RRrhosgeom}
\begin{align}
  \ell_s  \rho_0&=e_0+e_ab^a+\frac{1}{2}\mathcal{K}_{abc}m^ab^bb^c+\frac{m}{6}\mathcal{K}_{abc}b^ab^bb^c+\rho_\mu\xi^\mu\, , \label{eq:rho0g}\\
 \ell_s   \rho_a&=e_a+\mathcal{K}_{abc}m^bb^c+\frac{m}{2}\mathcal{K}_{abc}b^bb^c+\rho_{a\mu}\xi^\mu \, ,  \label{eq:rho_ag}\\
  \ell_s  \tilde{\rho}^a&=m^a+m b^a \, ,  \label{eq:rho^ag}\\
 \ell_s   \tilde{\rho}&=m \, ,   \label{eq:rhomg}\\
 \ell_s    \rho_\mu&=h_\mu+f_{a\mu}b^a \, , \\
 \ell_s   \rho_{a\mu}&=f_{a\mu} \, ,  \label{eq: ho_ak} \\
 \ell_s   \hat{\rho}_{\a}^{\mu}&=\hat{f}_{\a}^{\mu} \, . 
\end{align}   
\ees 

Using these explicit expressions one may compute the first order derivatives of the scalar potential with respect to the axions $\{\xi^\mu, b^a\}$ and saxions $\{u^\mu, t^a\}$ of the compactification. As expected the extrema conditions are of the form \eqref{extrema}, with 

\vspace*{.5cm}

\textbf{Axionic directions}

\bes
\label{paxions}
\begin{equation}
\label{paxioncpx}
  e^{-K}\frac{\partial V}{\partial \xi^\mu}=8\rho_0\rho_\mu +2g^{ab}\rho_a\rho_{b\mu} \, ,
\end{equation}
\begin{equation}
\label{paxionk}
 e^{-K}\frac{\partial V}{\partial b^a}= \ 8\rho_0\rho_a+\frac{8}{9}\mathcal{K}^2g_{ac}\tilde{\rho}\tilde{\rho}^c  +2\mathcal{K}_{abd}g^{bc}\rho_c\tilde{\rho}^d+2c^{\mu\nu}\rho_{a\mu}\rho_\nu\, ,
\end{equation}
\ees

\vspace*{.5cm}

\textbf{Saxionic directions}

\bes
\label{psaxions}
\begin{eqnarray}
\label{psaxioncpx}
 e^{-K}  \frac{\partial V}{\partial u^\mu} & =  & e^{-K} V_F\partial_\mu K+\frac{4}{3}\mathcal{K}\tilde{\rho} \rho_\mu +\partial_\mu c^{\kappa\sigma}\rho_\kappa\rho_\sigma - \frac{4}{3}\mathcal{K}\tilde{\rho}^a\rho_{a\mu} +2g^{ab}\rho_{a\mu}\rho_{b\nu}u^\nu\\ \nonumber
    & & + t^at^b(\partial_\mu c^{\kappa\sigma}\rho_{a\kappa}\rho_{b\sigma}-8\rho_{a\mu}\rho_{b\nu}u^\nu )
    + \frac{3}{4\mathcal{K}}e^{-K}\partial_\mu \partial_\sigma K \partial_\nu K \,  g^{\alpha\beta} \, \hat{\rho}_\alpha{}^\sigma \hat{\rho}_\beta{}^\nu\, ,
\end{eqnarray}
\begin{eqnarray}
\label{psaxionk}\nonumber
e^{-K} \frac{\partial V}{\partial t^a} &= & e^{-K} V_F\partial_{a}K+\partial_{a}\left(\frac{4}{9}\mathcal{K}^2\tilde{\rho}^b\tilde{\rho}^c
 g_{bc}\right)+\partial_{a}g^{cd}\rho_c\rho_d+\mathcal{K}_a\tilde{\rho}\left(\frac{2}{3}\mathcal{K}\tilde{\rho}+4u^\mu{\rho}_\mu \right)\\
 & & - 4 \mathcal{K}_a \tilde{\rho}^b\rho_{b\nu}u^\nu +2\tilde{c}^{\mu\nu}t^c\rho_{a\mu}\rho_{c\nu}+\partial_a g^{bc}\rho_{b\mu}u^\mu \rho_{c\nu}u^\nu\nonumber\\  & & +\frac{3}{8\mathcal{K}}e^{-K} \partial_\mu K \partial_\nu K \,  \p_a g^{\alpha\beta} \, \hat{\rho}_\alpha{}^\mu \hat{\rho}_\beta{}^\nu\,-\frac{9\mathcal{K}_a}{8\mathcal{K}^2}e^{-K} \partial_\mu K \partial_\nu K \,   g^{\alpha\beta} \, \hat{\rho}_\alpha{}^\mu\hat{\rho}_\beta^\nu\, .
\end{eqnarray}
\ees

\subsection{de Sitter no-go results revisited}
\label{subsec: no-go's}

From \eqref{psaxions} one can obtain the following off-shell relation
\begin{align}
\nonumber
    & u^\mu \partial_{u^\mu} V +x \, t^a \partial_{t^a} V=- (4+3x)V_F - (2+x) V_D + 4e^K\left[x\left(\frac{1}{2}g^{bc}\rho_b\rho_c+\frac{4\mathcal{K}^2}{9}g_{bc}\tilde{\rho}^b \tilde{\rho}^c +\frac{\mathcal{K}^2}{6}\tilde{\rho}^2\right) \right. \\
    & +\left. \frac{1}{2}c^{\mu\nu}\rho_\mu\rho_\nu  +\left(\frac{1}{3}+x\right)\mathcal{K}u^\nu \left(\tilde{\rho}\rho_\nu -\tilde{\rho}^b\rho_{b\nu}\right)  +\frac{1}{2}(1+x)(\tilde{c}^{\mu\nu}t^bt^c+g^{bc}u^\mu u^\nu )\rho_{b\mu}\rho_{c\nu}\right]\, ,
   \label{eq: arbitrary combination of partial derivatives}
\end{align}
with $x \in \R$ an arbitrary parameter. Different choices of $x$ will lead to different equalities by which one may try to constrain the presence of extrema with positive energy, in the spirit of  \cite{Hertzberg:2007wc,Flauger:2008ad}. In practice it is useful to rewrite this relation as
\begin{equation}
    u^\mu \partial_{u^\mu} V +x t^a \partial_{t^a} V = - 3V + \Xi_x\, ,
\end{equation}
where, for instance, the choice $x=1/3$ leads to 
\begin{equation}
    \label{eq: cosmo const relation}
   \Xi_{1/3} = \frac{2}{3} V_D  +  4e^K\left[-2\rho_0^2-\frac{1}{3}g^{bc}\rho_b\rho_c-\frac{2}{27}\tilde{\rho}^b\tilde{\rho}^c\mathcal{K}^2g_ {bc}+\frac{1}{6}(t^at^b\tilde{c}^{\mu\nu}+g^{ab}u^\mu u^\nu )\rho_{a\mu}\rho_{a\nu}\right] ,
\end{equation}
while the choice $x=1$ gives
\begin{equation}
 \Xi_1 = 4e^K\left[\frac{\mathcal{K}^2}{18}\tilde{\rho}^2-4\rho_0^2-\frac{1}{2}g^{ab}\rho_a\rho_b-\frac{1}{2}c^{\mu\nu}\rho_\mu\rho_\nu\right]\, .
    \label{eq: cosmo cons relation 2}
\end{equation}
Extrema of positive energy require $\p V =0$ and $V >0$, and so necessarily both \eqref{eq: cosmo const relation} and \eqref{eq: cosmo cons relation 2} should be positive. It is easy to see that this requires that both the Romans' parameter $\tilde{\rho}$ and geometric fluxes (either $\rho_{a\mu}$ or $\hat{\rho}_\alpha^\mu$)  are present, in agreement with previous results in the literature \cite{Haque:2008jz,Caviezel:2008tf,Flauger:2008ad,Danielsson:2009ff,Danielsson:2010bc,Danielsson:2011au}. In that case, it is unlikely that the potential satisfies an off-shell inequality of the form proposed in \cite{Obied:2018sgi}, at least at the classical level. 

In our formulation one can make more precise which kind of fluxes are necessary to attain de Sitter extrema. For this, let us express the last term of \eqref{eq: cosmo const relation} as
\be
(t^at^bc^{\mu\nu}+g^{ab}u^\mu u^\nu -4t^at^bu^\mu u^\nu )\rho_{a\mu}\rho_{a\nu} = \left[t^at^bc_{\rm P}^{\mu\nu}+u^\mu u^\nu g_{\rm P}^{ab} -\frac{5}{3} t^at^bu^\mu u^\nu \right]\rho_{a\mu}\rho_{a\nu}\, ,
\label{ccgeomterm}
\ee
where $g_{\rm P}^{ab}$, $c_{\rm P}^{\mu\nu}$ are the primitive components of the K\"ahler and complex structure metric, respectively. That is 
\be
\label{eq: primitive metric}
g_{\rm P}^{ab} = \frac{2}{3}\left(t^at^b - \mathcal{K}\mathcal{K}^{ab}\right)\, , \qquad \qquad c_{\rm P}^{\mu\nu} = \frac{1}{3}u^\mu u^\nu - 4G_Q G_Q^{\mu\nu}\, ,
\ee
where $G_Q = e^{-K_Q}$ and $G_Q^{\mu\nu}$ is the inverse of $\p_\mu \p_\nu G_Q$. These metric components have the property that they project out the K\"ahler potential derivatives along the overall volume and dilaton directions, namely $g_{\rm P}^{ab} \p_b K = c_{\rm P}^{\mu\nu} \p_\nu K =0$. So in order for the bracket in \eqref{eq: cosmo const relation} to be positive, the geometric fluxes $\rho_{a\mu}$ not only must be non-vanishing, but they must also be such that
\be
t^a \rho_{a\mu}\, t^b \rho_{a\nu}\, c_{\rm P}^{\mu\nu}+  \rho_{a\mu} u^\mu \, \rho_{a\nu} u^\nu\, g_{\rm P}^{ab} \neq 0\, .
\label{primcond}
\ee
In other words, either the vector $\rho_{a\mu} u^\mu$ is not proportional to $\p_a K$ or the vector $t^a \rho_{a\mu}$ is not proportional to $\p_\nu K$. The condition is likely to be satisfied at some point in field space, but in order to allow for a de Sitter extremum it must be satisfied on-shell as well. 

Remarkably, we find that the F-term Ansatz of section \ref{ss:fterms} forbids de Sitter extrema. Indeed, if we impose that the on-shell relations \eqref{proprho} are satisfied with the non-geometric fluxes turned off (cf. \eqref{proprhog} below) we obtain that, on-shell
\be
t^a \rho_{a\mu}\, t^b \rho_{a\nu}\, c_{\rm P}^{\mu\nu}+  \rho_{a\mu} u^\mu \, \rho_{a\nu} u^\nu\, g_{\rm P}^{ab} = \frac{4}{9} \cK^2 g^{\rm P}_{ab} \tilde{\rho}^a \tilde{\rho}^b \, ,
\label{primvalue}
\ee
with $g^{\rm P}_{ab}$ the inverse of $g_{\rm P}^{ab}$ in the primitive sector. Even if this term is positive, it can never be bigger than the other negative contributions within the bracket in \eqref{eq: cosmo const relation}. In fact, after plugging \eqref{primvalue} in \eqref{eq: cosmo const relation} there is a partial cancellation between the third and fourth term of the bracket, that then becomes semidefinite negative:
\be
4e^K\left[-2\rho_0^2-\frac{1}{3}g^{ab}\rho_a\rho_b-\frac{2}{27}\tilde{\rho}^a\tilde{\rho}^b\mathcal{K}^2g_ {ab}^{\rm NP}-\frac{5}{18}t^at^bu^\mu u^\nu \rho_{a\mu}\rho_{b\nu}\right]\, ,
\ee
with $g_ {ab}^{\rm NP} = g_{ab} -  g^{\rm P}_{ab} = \frac{3}{4}\frac{\cK_a \cK_b}{\cK^2} $ the non-primitive component of the K\"ahler moduli metric.

Even if the bracket in \eqref{eq: cosmo const relation} is definite negative,  there is still the contribution from the piece $\frac{2}{3}V_D$, which is positive semidefinite. However, one can see that with the Ansatz \eqref{solsfmax} this contribution vanishes. Indeed, using the Bianchi identity $f_{a\mu}\, \hat{f}_\alpha{}^\mu = 0$ and \eqref{eq: f-term prop rhoak geom}, one can see that for $f_{a\mu}\neq 0$ the D-term $D_a = i\p_\mu K \hat{f}_\alpha{}^\mu$ vanishes, and so does $V_D$.

To sum up, for type IIA geometric flux configurations, in any region of field space in which the F-terms are of the form \eqref{solsfmax} we have that the F-term potential satisfies
\be
u^\mu \partial_{u^\mu} V+\frac{1}{3}t^a \partial_{t^a} V \leq -3V\, , 
\label{eq: no-go geom inequality}
\ee
and so de Sitter extrema are excluded. In other words:
\begin{center}
{\em In type IIA geometric flux compactifications, classical  de Sitter extrema \\  are incompatible with F-terms of the form \eqref{solsfmax}.}
\end{center}
Here geometric flux compactifications refer to those with $f_{a\mu} \neq 0$, while for Calabi--Yau compactifications the no-go follows from \cite{Hertzberg:2007wc}. It would be interesting to extend this discussion to non-geometric flux compactifications, along the lines of \cite{deCarlos:2009fq,Shukla:2019dqd}.

\subsection{Imposing the Ansatz}

Besides the cosmological constant sign, let us see other constraints that the on-shell condition \eqref{solsfmax} leads to. By switching off all non-geometric fluxes, \eqref{proprho} simplifies to
\bes
\label{proprhog}
\begin{align}
   \rho_a & = \ell_s^{-1} {\mathcal P}\, \partial_a K \label{eq: f-term prop rho_a geom}\, ,\\
       \mathcal{K}_{ab}\tilde{\rho}^b+\rho_{a\mu}u^\mu & =  \ell_s^{-1} {\mathcal Q}\, \partial_a K \label{eq: f-term prop rho^a geom}\, ,\\
    \rho_\mu & =  \ell_s^{-1}\cM\, \partial_\mu K\, ,\\
   t^a\rho_{a\mu } & =  \ell_s^{-1} \cN\, \partial_\mu K\, , \label{eq: f-term prop rhoak geom}
\end{align}
\ees
where again ${\mathcal P}$, ${\mathcal Q}$, $\cM$, $\cN$ are real functions of the moduli. Such functions and other aspects of this Ansatz are constrained by the extrema conditions \eqref{paxions} and \eqref{psaxions} with which they must be compatible. Indeed, plugging \eqref{proprhog} into \eqref{paxions} one obtains  
\bes
\label{paxionsA}
\begin{equation}
\label{paxioncpxA}
8 \left(\rho_0\cM -  {\mathcal P}\cN\right) \partial_\mu K = 0 \, ,
\end{equation}
\begin{equation}
\label{paxionkA}
\left[ 8  {\mathcal P} (\rho_0 -  {\mathcal Q}) - \frac{1}{3} \tilde{\rho}  \cK \left( 10 {\mathcal Q} - 8 \cN \right)   \right]  \partial_a K   + \left[ \CK\tilde{\rho} + 8  {\mathcal P} - 8 \cM \right] \rho_{a\mu} u^{\mu} = 0 \, ,
\end{equation}
\ees
which must be satisfied on-shell. Even when both brackets in \eqref{paxionkA} vanish, this equation implies that on-shell
\be
\rho_{a\mu} u^{\mu} \propto \p_a K\, , \qquad {\rm and} \qquad \tilde{\rho}^a \propto t^a\, ,
\ee
simplifying the Ansatz. More precisely, we are led to the following on-shell relations
\bes
\label{Ansatz}
\begin{align}
   \ell_s \rho_0&= A \CK\, , \label{eq: ans rho0}\\
    \ell_s\rho_a&= B \cK \partial_a K  \, ,  \label{eq: ans rho_a}\\
    \ell_s\tilde{\rho}^a&= C t^a  \, ,  \label{eq: ans rho^a}\\
    \ell_s\tilde{\rho}&= D \, ,  \label{eq: ans rhotilde}\\
    \ell_s\rho_\mu&=E\cK \partial_\mu K  \, ,   \label{eq: nsns}\\
    \ell_s\rho_{a\mu} t^a &= \frac{F}{4}  \cK\partial_\mu K  \, ,  \label{eq: geoma}\\
     \ell_s\rho_{a\mu} u^\mu &= \frac{F}{3}  \CK \partial_a K   \, , \label{eq: geomu}
\end{align}
\ees
where $A, B, C, D, E, F$ are functions of the saxions. We have extracted a factor of $\CK$ in some of them so that the expression for the on-shell equations simplifies. In terms of \eqref{Ansatz} we have that the vanishing of \eqref{paxions} amounts to
\bes
\label{paxionsAA}
\begin{equation}
\label{paxioncpxAA}
 4 AE -  BF =  0 \, ,
\end{equation}
\begin{equation}
\label{paxionkAA}
 3AB  - \frac{1}{12} CD  + B C  -  EF = 0 \, ,
\end{equation}
\ees
assuming that at each vacuum $\p_\mu K \neq 0 \neq \p_a K$. Similarly, the vanishing of \eqref{psaxions} implies
\bes
\label{psaxionsAA}
\begin{equation}
\label{psaxioncpxAA}
 4A^2 + 12B^2 +\frac{1}{3} C^2 + \frac{1}{9}D^2 + 8 E^2 -\frac{5}{6} F^2 +CF -4DE = 0\, ,
\end{equation}
\begin{eqnarray}
\label{psaxionkAA}
4A^2 + 4 B^2 - \frac{1}{9} C^2 - \frac{1}{9}D^2 + 16E^2 -\frac{5}{9}F^2  = 0\, ,
\end{eqnarray}
\ees
where we have used the identities in \ref{app:relations}.

Expressing the extrema equations in terms of the Ansatz \eqref{Ansatz} has the advantage that we recover a system of algebraic equations. Nevertheless, eqs.\eqref{paxionsAA} and \eqref{psaxionsAA} may give the wrong impression that we have an underdetermined system, with four equations and six unknowns $A, B, C, D, E, F$. Notice, however, that these unknowns are not all independent, and that relations among them arise when the flux quanta are fixed. Indeed, let us first consider the case without geometric fluxes, which sets $F=0$. In this case, AdS vacua require that the Roman's parameter $m$ is non-vanishing so we may assume that $D \neq 0$. Because the lhs of \eqref{paxionsAA} and \eqref{psaxionsAA}  are  homogeneous polynomials of degree two, we may divide each of them by $D^2$ to obtain four equations on four variables: $A_D = A/D$, $B_D = B/D$, $C_D = C/D$, $E_D = E/D$. The solutions correspond to $A_D= 0$ and several rational values for $B_D, C_D, E_D$, which reproduce the different {\bf S1} branches found in chapter \ref{ch:rrnsnsvacua}.\footnote{To compare with chapter \ref{ch:rrnsnsvacua} one needs to use the  dictionary: $B = - C^{\text{Previous Chapter}} /3$, $C = B^{\text{PC}}$, $E = A^{\text{PC}}$ since in we are denoting differently the constants appearing in the ansatz.} Finally, the variable $D = m$ is fixed when the flux quanta are specified. 

The analysis is slightly more involved in the presence of geometric fluxes. Now we may assume that $F \neq 0$, since otherwise we are back to the previous case. Our Ansatz implies that the first flux invariant in \eqref{invmetq} is a linear combination of the vectors $(f_a)_\mu = f_{a\mu}$, as
\be
m\hat{h}_\mu \equiv m h_\mu - m^a f_{a\mu} = \left(DE - \frac{CF}{4}\right)  \cK \p_\mu K  =   \left( \frac{4DE}{F}  - C \right)  t^a  f_{a\mu} \, ,
\label{hath}
\ee
where $\cK$, $ \p_\mu K$, $\langle t^a\rangle$ correspond to the value of the K\"ahler saxions in the corresponding extremum, etc.  One can write the above relation as
\be
m\hat{h}_\mu = d^a f_{a\mu}\, ,
\label{condhhat}
\ee
where the constants $d^a$ are fixed once that we specify the fluxes $m$, $h_\mu$, $m^a$, $f_{a\mu}$. As a consequence, the number of stabilised complex structure axions $\xi^\mu$ is $r_f = {\rm rank}\, f_{a\mu}$, while the rest may participate in St\"uckelberg mechanisms triggered by the presence of D6-branes \cite{Camara:2005dc}.\footnote{Microscopically, \eqref{condhhat} means that $h_\mu$ is in the image of the matrix of geometric fluxes $f_{a\mu}$, and as such it is cohomologically trivial. Macroscopically, it means that the number of independent complex structure axions entering the scalar potential are $ {\rm dim} \langle h_\mu, f_{1\mu}, f_{2\mu}, \dots \rangle= {\rm rank} f_{a\mu} \equiv r_f $, and not $r_f+1$.} Strictly speaking, $d^a$ is only fixed up to an element in the kernel of $f_{a\mu}$, but this is irrelevant for our purposes. Indeed, notice that due to our Ansatz
\bea\nonumber
m\hat{e}_a \equiv  me_a -  \frac{1}{2}\mathcal{K}_{abc} m^bm^c  &= & \left(BD + \frac{C^2}{6} \right)  \cK \p_a K   - m f_{a\mu} \xi^\mu \\ 
 & = & \left[\left(\frac{3BD}{F} + \frac{C^2}{2F} \right)  u^\mu   -  D \xi^\mu  \right] f_{a\mu}  \, ,
 \label{hatea}
\eea
where again $\cK$, $u^\mu$, $\xi^\mu$ stand for the vevs at each extremum. This implies several things. First, the second set of invariants in \eqref{invmetq} vanish identically. Second, the combination $md^a \hat{e}_a$ is fully specified by the flux quanta, without any ambiguity.  Finally in terms of 
\begin{equation} 
m^2 \hat{e}_0 \equiv m^2 e_0 - m m^a e_a  + \frac{1}{3} \mathcal{K}_{abc}m^am^bm^c  \, ,
\label{hate0}
\end{equation} 
we can define the following cubic flux invariant
\be
m^2 \hat{e}_0 -  m d^a  \hat{e}_a  =  \cK  \left[AD^2  + 3BCD + \frac{C^3}{3} + \left( \frac{4DE}{F}  - C\right) \left(3BD + \frac{C^2}{2} \right)  \right] \, .
\label{extrainv}
\ee
The existence of this additional invariant is expected from the discussion of section \ref{ss:invariants}. As we now show, $\cK$ is fixed at each extremum by the choice of the flux quanta and the Ansatz' variables. Therefore \eqref{extrainv} and $D=m$ provide two extra constraints on these variables, which together with \eqref{paxionsAA} and \eqref{psaxionsAA}  yield a determined system of algebraic equations. 

To show how  $\cK$ is specified, let us first see how the saxionic moduli are determined. First \eqref{hath} determines $(4DE-CF) \cK  \partial_\mu K $ in terms of the flux quanta, which is equivalent to determine $(4DE-CF)^{-1}  u^\mu/\cK$. Plugging this value into \eqref{eq: geomu} one fixes $(4DE/F-C)^{-1}  \partial_a K$ in terms of the fluxes, which is equivalent to fix $(4DE/F-C)  t^a$. Therefore at each extremum we have that
\be
\left(\frac{4DE}{F}-C\right)^3  \cK  \, ,
\label{vevK}
\ee
is specified by the flux quanta. Notice that this is compatible with \eqref{hath}, and we can actually use this result to fix the definition of $d^a$, by equating \eqref{vevK} with $\cK_{abc}d^ad^bd^c$.

\subsection{Branches of vacua}\label{branchvacu}

\begin{center}    
\begin{figure}[h]
    \centering
    \includegraphics[scale=0.75]{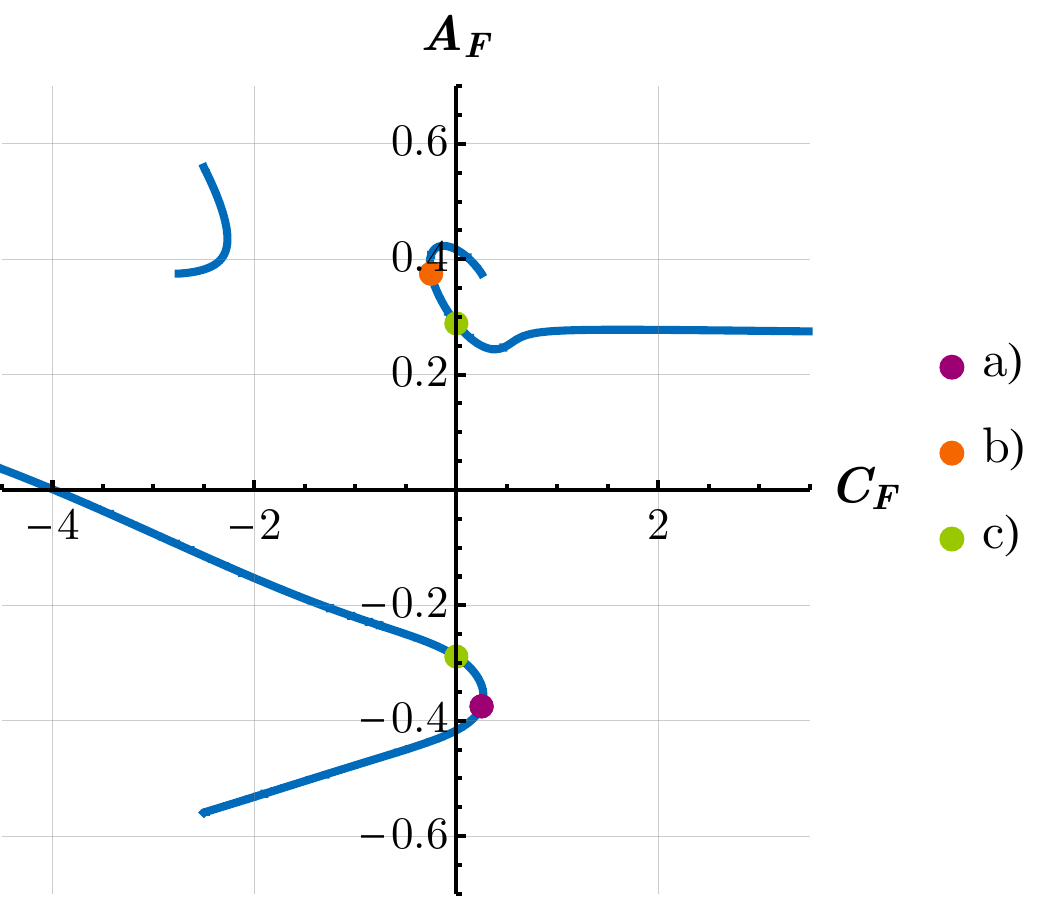}
    \caption{Set of points that verify \eqref{eq: megaeqF} (blue curve) and have $E_F^2\geq 0$. The coloured dots correspond to the particular solutions \eqref{delta=0sols}. Both curves tend asymptotically to $A_F=1/4$ for $C_F\rightarrow\pm \infty$. }
    \label{fig: generalsol}
\end{figure}
\end{center}

Let us analyse the different solutions to the algebraic equations \eqref{paxionsAA} and \eqref{psaxionsAA}. Following the strategy of the previous subsection, we assume that $F \neq 0$ and define $A_F = A/F$, $B_F = B/F$, $C_F = C/F$, $D_F = D/F$, $E_F = E/F$. Then, from \eqref{paxioncpxAA} we obtain
\begin{equation}
    B_F=4A_FE_F\, ,
    \label{eq: B_F}
\end{equation}
which substituted into \eqref{paxionkAA} gives the following relation
\begin{equation}
  C_FD_F= 12 E_F(12A_F^2+4A_FC_F-1)\, .
    \label{eq: C_FD_F}
\end{equation}
Then, multiplying \eqref{psaxionkAA} by $C_F^2$ and using \eqref{eq: C_FD_F} we obtain
\be
144 E_F^2 \Delta_F = C_F^2\left[36 A_F^2 - C_F^2 - 5 \right]  \, ,
  \label{eq: E_F Delta_F}
\ee
where
\be
\Delta_F =(12A_F^2+4A_FC_F-1)^2 - 4 A_F^2 C_F^2 - C_F^2\, .
   \label{eq: DeltaF def}
\ee
We have two possibilities, depending on whether $\Delta_F = 0$ or not. Let us consider both:

\begin{itemize}

\item $\Delta_F = 0$

In this case, from \eqref{eq: E_F Delta_F} and \eqref{eq: DeltaF def}, we find four different real solutions for $(A_F, C_F)$:
\bes
\label{delta=0sols}
   \begin{align}
   \label{SUSYsol}
            A_F=-\frac{3}{8}\, , \ \ \ & \ \ \ C_F=\frac{1}{4}\, ,\\
            A_F=\frac{3}{8}\, , \ \ \ & \ \ \ C_F=-\frac{1}{4}\, ,
            \label{nonSUSYdelta=0}\\
            A_F=\pm \frac{1}{2\sqrt{3}}\, , \ \ \ & \ \  \ C_F= 0 \, .
            \label{nonSUSYdelta=0C=0}
        \end{align}
\ees
Given the solution \eqref{SUSYsol}, one can solve for $D_F$ in \eqref{eq: C_FD_F} and check that \eqref{psaxioncpxAA} and \eqref{psaxionkAA} are automatically satisfied. We then find that:
\be
\eqref{SUSYsol} \ \raw \ B_F = -\frac{3}{2} E_F\, , \qquad D_F = 15E_F\, ,
\label{SUSYbranch}
\ee
with $E_F$ unfixed. Thus, at this level $(E,F)$ are free parameters of the solution. As we will see below, this case corresponds to the supersymmetric branch of solutions. The remaining solutions can be seen as limiting cases of the following possibility:

\item $\Delta_F \neq 0$

Under this assumption we can solve for $E_F$ in \eqref{eq: E_F Delta_F}:
     \begin{equation}
            E_F^2=\frac{C_F^2}{144\Delta_F} \left[36 A_F^2 - C_F^2 - 5 \right] 
            \label{eq: E_F delta neq0}
        \end{equation}
Then we see that \eqref{psaxioncpxAA} and \eqref{psaxionkAA} amount to solve the following relation:
      \begin{align}
       &\frac{8A_F^{2}C_F^{4}}{3}+4A_FC_F^{4}-\frac{7C_F^{4}}{6}+64A_F^{3}C_F^{3}+48A_F^{2}\,C_F^{3}-\frac{16A_FC_F^{3}}{3}-4C_F^{3} +576A_F^{4}C_F^{2}
       \nonumber\\ \nonumber 
           & +144A_F^{3}C_F^{2}-\frac{296A_F^{2}C_F^{2}}{3}-4A_FC_F^{2}+\frac{7C_F^{2}}{3}+2304A_F^{5}C_F -592A_F^{3}C_F+24A_F^{2}C_F\\
          & +\frac{100A_FC_F}{3}-2C_F+3456A_F^{6}-1176A_F^{4}+124A_F^{2}-\frac{25}{6} = 0\, ,
         \label{eq: megaeqF}
        \end{align}
which selects a one-dimensional family of solutions in the $(A_F,C_F)$-plane. We only consider those such that  \eqref{eq: E_F delta neq0} is non-negative, see figure \ref{fig: generalsol}.
One can check that all values in \eqref{delta=0sols} are also solutions of \eqref{eq: megaeqF}. Even if for them $\Delta_F =0$, we have that 
\be
D_F^2 = \left(1+ \frac{C_F^2(4A_F^2+1)}{\Delta_F}\right) \left[ 36A_F^2 -C_F^2-5\right]\, ,
 \label{eq: D_F delta neq0}
\ee
as well as \eqref{eq: E_F delta neq0}, attain regular limiting values that solve the equations of motion. Because \eqref{eq: megaeqF} constrains one parameter in terms of the other, we have two free parameters, say $(C,F)$, unfixed by the equations \eqref{paxionsAA} and \eqref{psaxionsAA}. 

\end{itemize}

\subsection{Summary}\label{sec:summary}

Let us summarise our results so far. Given the on-shell F-terms \eqref{solsfmax}, we find two branches of vacua, summarised in table \ref{vacuresul}. Each branch has two continuous parameters. However,  when taking specific values for the fluxes and taking into account that $D=m$ and \eqref{extrainv}, these two degrees of freedom are fixed. Then, as we scan over different flux quanta, we will obtain a discretum of values for the parameters of the Ansatz, within the above continuous solutions. 

\begin{table}[h]
\begin{center}
\scalebox{1}{%
    \begin{tabular}{| c ||     c | c | c | c |}
    \hline
  Branch & $A_F$  & $B_F$  & $C_F$  & $D_F$  \\
  \hline \hline
  \textbf{SUSY}  & $-\frac{3}{8}$ &  $-\frac{3}{2}E_F$  & $\frac{1}{4}$  & $15E_F$     \\ \hline
  \textbf{non-SUSY}   & eq.\eqref{eq: megaeqF}  & $4A_FE_F$  & eq.\eqref{eq: megaeqF}  & $ \sqrt{\frac{\Delta_F}{C_F^2}  + (4A_F^2+1)} \, 12E_F $       \\
      \hline
    \end{tabular}}      
\end{center}
\caption{Branches of solutions in terms of the quotients $A_F = A/F$, etc. of the parameters of the Ansatz \eqref{Ansatz}. In the SUSY branch $E_F$ is a free parameter, while in the non-SUSY extrema it is given by \eqref{eq: E_F delta neq0}. Moreover $\Delta_F$ is given by \eqref{eq: DeltaF def}, being always zero in the supersymmetric branch.  \label{vacuresul}}
\end{table}

As we show below, the branch where  $A_F = -3/8$, $C_F =1/4$ and $E_F$ is unconstrained corresponds to supersymmetric vacua, while the other branch contains non-supersymmetric ones. Remarkably, both branches intersect at one point. The non-supersymmetric branch splits into three when imposing the physical condition $E_F^2 \geq 0$, as can be appreciated from figure \ref{fig: generalsol}. Each point in the green physical region contains two solutions, corresponding to the two values $E_F = \pm \frac{C_F}{12} \sqrt{\Delta_F^{-1}(36A_F^2 - C_F^2 - 5)}$.

\bigskip

\noindent
\textbf{F-terms}
\medskip

\noindent
One can recast the F-terms for each of these extrema as
\bes
\begin{align}
    G_{a}&=\left[\left(-\frac{1}{2}B_F-2E_F+\frac{1}{12}D_F\right)+i\left(-\frac{1}{12}C_F-\frac{1}{2}A_F-\frac{1}{6}\right)\right]F \, \mathcal{K}^2\partial_a K\, ,\\
   G_{\mu}&=\left[\left(-\frac{3}{2}B_F-\frac{1}{12}D_F-E_F\right)+i\left(-\frac{1}{4}-\frac{1}{2}A_F+\frac{1}{4}C_F\right)\right]F\, \mathcal{K}^2\partial_\mu K\, ,
\end{align}
\ees
and one can see that requiring that they vanish is equivalent to impose \eqref{SUSYsol} and \eqref{SUSYbranch}.  Therefore, the branch \eqref{SUSYsol} corresponds to supersymmetric vacua, while general solutions to \eqref{eq: megaeqF} represent non-supersymmetric extrema of the potential. 

\bigskip
\noindent
\subsection{Vacuum energy and KK scale: AdS distance conjecture}
\medskip

\noindent
Using \eqref{eq: cosmo const relation} and imposing the extremisation of the potential, one can see that the vacuum energy has the following expression in the above branches of solutions:
\begin{equation}
   4\pi \kappa_4^4  V|_{\rm vac}= - \frac{4}{3}e^K\mathcal{K}^2F^2 \left(2A_F^2+64A_F^2E_F^2+\frac{1}{18}C_F^2+\frac{5}{18}\right)\, .
\end{equation}
In the supersymmetric branch this expression further simplifies to
\begin{equation}
    4\pi \kappa_4^4  V|_{\rm vac}^{\rm SUSY}=-  e^K\mathcal{K}^2F^2\left(12E_F^2+\frac{3}{4}\right)\, .
\end{equation}
So essentially we recover that the AdS$_4$ scale in Planck units is of order
\be
\frac{\Lambda_{\rm AdS}^2}{M_{\rm P}^2}  \sim e^{4D} V_{X_6}   F^2  \sim \frac{t^3}{u^4}  F^2 \chi\, ,
\ee
where in the last step we have defined  $\chi \equiv 2A_F^2+64A_F^2E_F^2+\frac{1}{18}C_F^2+\frac{5}{18}$. This is to be compared with the KK scale
\be
\frac{M_{\rm KK}^2}{M_{\rm P}^2} \sim   e^{2D} V_{X_6}^{-1/3} \sim  t^{-1} u^{-2} \, ,
\ee
obtaining the quotient
\be
\frac{\Lambda_{\rm AdS}^2}{M_{\rm KK}^2} \sim  e^{2D} V_{X_6}^{4/3} F^2 \sim \frac{t^{4}}{u^2} F^2 \chi \, .
\label{quot}
\ee
Scale separation will occur when this quotient is small, which seems hard to achieve parametrically, unlike in the case with only RR and NSNS fluxes. Indeed, unless some fine tuning occurs, at large $\{t,u\}$ one expects that $e^K|W|^2 \sim e^K|W_{\rm RR}|^2 + e^K |W_{\rm NS}|^2$, which in supersymmetric vacua dominates the vacuum energy. If both terms are comparable, then in type IIA setups with bounded geometric fluxes and Romans mass $m$, $u$ goes like $u \sim t^2$, and there is no separation due to the naive modulus dependence in \eqref{quot}. If one term dominates over the other the consequences are even worse, at least for supersymmetric vacua.\footnote{With specific relations between flux quanta parametric scale separation at the 4d level is possible \cite{Font:2019uva}. Remarkably, it was there found that this naive 4d scale separation did not occur at the 10d level.} Because $\chi$ is at least an order one number, the most promising possibility for achieving scale separation is that $F$ scales down with $t$. While this is compatible with \eqref{hath}, we have not been able to find examples where this possibility is realised. In any event, even if $F$ does not scale with the moduli, it would seem that generically $F \lesssim \cO(0.1)$  is a necessary condition to achieve a vacuum at large volume, weak coupling, and minimal scale separation. This is perhaps to be expected because in the limit $F \raw 0$ we recover the analysis of the previous chapter, where parametric scale separation occurs, at least from the 4d perspective considered here. It would be also interesting to relate to results to \cite{Cribiori:2021djm}, where vacua with scale separation and geometric fluxes (and no Roman's mass) are obtained after performing two T-dualities.

\subsection{Relation to previous results}

In order to verify the validity of our formalism and the results we have obtained, we proceed to compare them with some of the existing results in the literature. We will therefore focus on examples.

\bigskip
\noindent
\textbf{Comparison with Camara et al. \cite{Camara:2005dc}}
\medskip

\noindent
    This reference studies RR, NS and metric fluxes on a $T^6/(\Omega(-1)^{F_L}I_3)$ Type IIA orientifold. We are particularly interested in section 4.4, where $\CN=1$ AdS vacua in the presence of metric fluxes are analysed. One can easily use our SUSY branch (see table \ref{vacuresul}), the definitions of the flux polynomials \eqref{RRrhosgeom} and our Ansatz \eqref{Ansatz} to reproduce their relations between flux quanta and moduli fixing. We briefly discuss the most relevant ones.
    
    In \cite{Camara:2005dc} they study the particular toroidal geometry in which all three complexified K \"ahler moduli are identified. This choice greatly simplifies the potential and the flux polynomials.  To reproduce the superpotential in \cite[eq.(3.15)]{Camara:2005dc} we consider the case $T^a=T$,  $\forall a$, so that there is only one K\"ahler modulus and the K\"ahler index $a$ can be removed. The flux quanta $\{e_0, e_a, m^a,m,h_\mu,\rho_{a\mu}\}$ are such that  $e_a=3c_1$, $m^a=c_2$ and 
    \begin{equation}
    \rho_{a\mu}=
    \begin{cases}
    3a&\hspace{1cm} \mu=0\, ,\\
    b_\mu &\hspace{1cm} \mu\neq 0\, , 
    \end{cases}
    \qquad a, b_\mu \in \mathbb{Z}\, .
\end{equation}

This relation provides a constraint for the fluxes in order for this family of solutions to be realised (cf. \cite[eq.(4.32)]{Camara:2005dc}).
The complex structure saxions are instead determined in terms of  $\rho_{a\mu}$:
Imposing the constraint $D=m$ on the SUSY Ansatz we have
\begin{align}
    A&=-\frac{3}{8}F\, ,   &    B&=-\frac{m}{10}\, , &    C&=\frac{1}{4}F\, ,    &   D&=m=15E\, .
\end{align}
 The first step is to use the invariant combinations of fluxes and axion polynomials together with the Ansatz to fix the value of the saxions. Notice that because we only have one K\"ahler modulus, $\rho_{a\mu}$ has necessarily rank one, and so \eqref{hath} fixes $t$ as function of the fluxes and the parameter $F$:
\begin{equation}
\left(\frac{4 E D}{F}-C\right) \rho_{a \mu} t^{a}=m h_{\mu}-\rho_{a \mu} m^{a} \longrightarrow \begin{cases}\left(\frac{4 m^{2}}{15 F}-\frac{1}{4} F\right) 3 a t=m h_{0}-3 a c_{2} & \text { if } \mu=0 \\ \left(\frac{4 m^{2}}{15 F}-\frac{1}{4} F\right) b_{\mu} t=m h_{\mu}-b_{\mu} c_{2} & \text { if } \mu \neq 0\end{cases}
\end{equation}
This relation provides a constraint for the fluxes in order for this family of solutions to be realised (cf. \cite[eq.(4.32)]{Camara:2005dc}).
The complex structure saxions are instead determined in terms of  $\rho_{a\mu}$:
\begin{equation}
    \rho_{a\mu} t^a=\frac{F}{4}\mathcal{K}\partial_\mu K\longrightarrow \begin{cases}3a t=-\frac{F\mathcal{K}}{4 u^0}\, ,\\
    b_\mu t=-\frac{F\mathcal{K}}{4 u^\mu}\, ,
    \end{cases}
    \label{eq: camara u fix}
\end{equation}
which reproduces the relation  \cite[eq.(4.31)]{Camara:2005dc}.

To obtain the remaining relations of \cite[section 4.4]{Camara:2005dc}, we take into account that $\mathcal{K}=6t^3$ and take advantage of the particularly simple dependence of our Anstaz when considered on an isotropic torus. Using that $F=4C$ we can go back to \eqref{eq: camara u fix} to eliminate the $F$ dependence of the complex structure moduli. 
\begin{equation}
     \rho_{a\{\mu=0\}} t^a=F\mathcal{K}\partial_{\mu=0} K=-\frac{6t^3F}{4u^0}=-C\frac{6t^3}{u^0}=-\frac{6t^2}{u^0}\tilde{\rho}^a \longrightarrow 3atu^0=-6t^2(c_2+vm)\, ,
     \label{eq: camara u fix 2}
\end{equation}
which, up to redefinition of the parameters, is just relation  \cite[eq.(4.34)]{Camara:2005dc}. Similarly, we have
\begin{equation}
    \rho_{\mu=0}=E\mathcal{K}\partial_{\mu=0}K\longrightarrow h_0+3av=-\frac{m}{15}\frac{6t^3}{u^0}\, .
    \label{eq: camara t fix}
\end{equation}
Replacing $u_0$ using \eqref{eq: camara u fix 2} in the above expression leads to
\begin{equation}
    t^2=\frac{5(h_0+3av)(c_2+mv)}{am}\, ,
    \label{eq: camara u-t fix}
\end{equation}
which is equivalent to \cite[eq.(4.41)]{Camara:2005dc} and provides an alternative way to fix the K\"ahler moduli $t$. 

To fix the complex structure axions $\xi^\mu$ we note that
\begin{equation}
    \rho_a=B\mathcal{K}\partial_a K =-\frac{3}{2}E\mathcal{K}\partial_a K= \frac{3u^0}{2}\rho_{\mu=0}\partial_a K \longrightarrow \rho_at^a=-\frac{9}{2}(h_0+3av)u^0\, .
\end{equation}
Expanding $\rho_a$ and replacing $t$ using \eqref{eq: camara u fix 2} we arrive at
\begin{equation}
    3c_1+6c_2v+3mv^2+3a\xi^0+\sum_\mu b_\mu \xi^\mu=\frac{9}{a}(c_2+mv)(h_0+3av)\, ,
    \label{eq: complex axion combination camara}
\end{equation}
and hence we derive an analogous relation to  \cite[eq.(4.33)]{Camara:2005dc}. We observe that it  only fixes one linear combination of complex structure saxions. This was to be expected, since by construction the geometric fluxes are of rank one. Finally, we can fix the  K\"ahler axion $b$ using the flux polynomial $\rho_0$
\begin{equation}
    \rho_0=A\mathcal{K}=-\frac{3C}{2}\mathcal{K}=-\frac{3}{2t}\tilde{\rho}^a\mathcal{K}\longrightarrow\rho_0=-9(c_2+mv)t^2\, ,
\end{equation}
which after replacing the complex axions using \eqref{eq: complex axion combination camara} and substituting $t$ using \eqref{eq: camara t fix} and \eqref{eq: camara u-t fix} leads to the same equation for the K\"ahler axion as the one shown in \cite[eq.(4.40)]{Camara:2005dc}.

\bigskip
\noindent
\textbf{Comparison with Dibitetto et al. \cite{Dibitetto:2011gm}}
\medskip

\noindent
In this reference the vacuum structure of isotropic $\mathbb{Z}_2\times\mathbb{Z}_2$ compactifications is analysed, combining algebraic geometry and supergravity techniques. We are particularly interested in the results shown in \cite[section 4]{Dibitetto:2011gm}, where they consider a setup similar to \cite[section 4.4]{Camara:2005dc}, but go beyond supersymmetric vacua.\footnote{It is worth noting that in order to solve the vacuum equations, \cite{Dibitetto:2011gm} follows a complementary approach to the standard one. Typically, one starts from the assumption that the flux quanta have been fixed and then computes the values of the axions and saxions that minimise the potential. Ref.\cite{Dibitetto:2011gm} instead fixes a point in field space, and reduces the problem to find the set of consistent flux backgrounds compatible with this point being an extremum of the scalar potential. Both descriptions should be compatible.} More concretely, in this section they study type IIA orientifold compactifications on a $\mathbb{T}^6/(\mathbb{Z}_2\times \mathbb{Z}_2)$ isotropic orbifold in the presence of metric fluxes. Hence, they have an $STU$ model with the axiodilaton $S$, the overall K\"ahler modulus $T$ and the overall complex structure modulus $U$.

They obtain sixteen critical points with one free parameter and an additional solution with two free parameters. This last case is not covered by our Ansatz, since the associated geometric fluxes do not satisfy \eqref{eq: geoma} and \eqref{eq: geomu}. Therefore it should correspond to a non-supersymmetric vacuum with F-terms different from \eqref{solsfmax}. The remaining sixteen critical points are grouped into four families and summarised in \cite[table 3]{Dibitetto:2011gm}. Taking into account their moduli fixing choices, we can relate their results for the flux quanta with the parameters of our Ansatz as follows:

\begin{itemize}
    \item When $s_2=1$, solution $1$ from  \cite[table 3]{Dibitetto:2011gm} corresponds to a particular point of the SUSY branch in our table \ref{vacuresul}, with $E_F=\pm \frac{1}{4\sqrt{15}}$ (sign given by $s_1$). 
    \item When $s_2=-1$, solution $1$ of  \cite[table 3]{Dibitetto:2011gm} corresponds to the limit solution \eqref{nonSUSYdelta=0} of the non-SUSY branch (point (b) in figure \ref{fig: generalsol}). We confirm the result of \cite{Dibitetto:2011gm} regarding stability: similarly to the SUSY case, this is a saddle point with tachyonic mass $m^2=-8/9|m^2_{BF}|$ (for a detailed analysis on stability check section \ref{s:stabalidity} and Appendix \ref{ap:Hessian}).
    
    \item Solution $2$ from \cite[table 3]{Dibitetto:2011gm} corresponds to a limit point $C_F=0$ of the non-SUSY branch with $\Delta_F\neq0$ and $A_F=\pm5/12$. Such solution was not detailed in our analysis of section \ref{branchvacu} since, despite being a limit point, it still verifies \eqref{eq: E_F delta neq0}, \eqref{eq: megaeqF} and \eqref{eq: D_F delta neq0}. In \cite[table 4 ]{Dibitetto:2011gm} it is stated that this solution is perturbatively unstable, in agreement with our results below (see figure \ref{fig: excludsol}).
    
    \item Solution $3$ from \cite[table 3]{Dibitetto:2011gm} is a particular case of the non-SUSY branch, corresponding to $A_F=s_1/4$ and $C_F=s_1/2$ (with $s_1=\pm1$). This specific point falls in the stable region of figure \ref{fig: excludsol}. The analysis of section \ref{s:stabalidity} reveals that the mass spectrum has two massless modes, confirming the results of \cite{Dibitetto:2011gm}.

    \item Solution $4$ of \cite[table 3]{Dibitetto:2011gm} is not covered by our ansatz since, similarly to the two-dimensional solution, our parameter $F$ is not well-defined under this combination of geometric fluxes. We then expect F-terms not of the form \eqref{solsfmax}.
\end{itemize}

Hence, the results of \cite{Dibitetto:2011gm} provide concrete examples of solutions for both the supersymmetric and non-supersymmetric branches of table \ref{vacuresul}.

\bigskip
\noindent
\textbf{Examples of de Sitter extrema}
\medskip

\noindent
In \cite{Caviezel:2008tf}, the authors study the cosmological properties of type IIA compactifications on orientifolds of manifolds with geometric fluxes. They apply the no-go result of \cite{Flauger:2008ad} to rule out de Sitter vacua in all the scenarios they consider except for the manifold $SU(2)\times SU(2)$, where they find a de Sitter extremum, albeit with tachyons. One can check that the fluxes considered in section 4.2 of \cite{Caviezel:2008tf} do not satisfy condition \eqref{condhhat}. Therefore, this example lies outside of our Ansatz and so relation \eqref{eq: no-go geom inequality} does not hold. 


\section{Perturbative stability}
\label{s:stabalidity}

Given the above set of 4d AdS extrema some questions arise naturally. First of all, one should check which of these points are \textit{actual} vacua, meaning stable in the perturbative sense. In other words, we should verify that they do not contain tachyons violating the BF bound \cite{Breitenlohner:1982bm}. As it will be discussed below, for an arbitrary geometric flux matrix $f_{a\mu}$ it is not possible to perform this analysis without the explicit knowledge of the moduli space metric. Nevertheless, the problem can be easily addressed if we restrict to the case in which $f_{a\mu}$ is a rank-one matrix, which will be the case studied in section  \ref{s:stabalidity}. On the other hand, one may wonder if these 4d solutions have a 10d interpretation. We relegate this second question to chapter \ref{ch:review10d}.

Following the approach of the previous chapter we will compute the physical eigenvalues of the Hessian by decomposing the K\"ahler metrics (both for the complex structure and K\"ahler fields) into their primitive and non-primitive pieces. This decomposition together with the Ansatz \eqref{Ansatz} reduces the Hessian to a matrix whose components are just numbers and whose eigenvalues are proportional to the physical masses of the moduli. The explicit computations and details are given in Appendix \ref{ap:Hessian}, whose main results we will summarise in here. To simplify this analysis we will initially ignore the contribution of the D-term potential, that is, we will set $\hat{\rho}_\alpha{}^\mu=0$. We will briefly discuss its effect at the end of this section.

As mentioned above, we will consider the case in which $f_{a\mu} = \ell_s \rho_{a\mu}$ has rank one, since the case with a higher rank cannot be solved in general. Let us see briefly why. One can show that the Ansatz \eqref{Ansatz} implies:
\begin{align}
\label{rhoproblem}
    f_{a\mu}&=-\frac{F K}{12}\p_a K\p_\mu K+\tilde f_{a\mu},    &   &\text{with}    &    & t^a \tilde f_{a\mu}=0= u^\mu \tilde f_{a\mu}\, ,
\end{align}
and so $\tilde f_{a\mu}$ must be spanned by $ t_a^\bot\otimes u_\mu^\bot$, where the  $\left\{t_a^\bot\right\}$ form a basis of the subspace orthogonal to $ t^a$, and similarly for  $ u_\mu^\bot$. The contribution of the first term of \eqref{rhoproblem} to the Hessian can be studied in general. The contribution of the second term depends, among other things, on  how both the $ t_a^{\bot}$ and $ u_\mu^{\bot}$ are stabilised, which can only be studied if the explicit form of the internal metric is known. Therefore, in the following we will set  $\tilde f_{a\mu}=0$. Notice that, for this case, our Ansatz implies that just one linear combination of axions is stabilised, since from \eqref{Ansatz} it follows that $\rho_\mu \propto \rho_{a\mu}, \forall a$.

\vspace*{.5cm}
\noindent
\textbf{SUSY Branch}
\\
As expected, the SUSY case is perturbatively stable. The results can be summarised as: 

\begin{table}[h]
\begin{center}
\scalebox{1}{%
    \begin{tabular}{| c ||     c | c | c | c |}
    \hline
  Branch & Tachyons (at least)  & Physical eigenvalues  & Massless modes (at least)  \\
  \hline \hline
  \textbf{SUSY}  & $h^{2,1}$ &  $m^2_{tach}=\frac{8}{9} m_{BF}^2$  & $h^{2,1}$  \\
  \hline
    \end{tabular}}      
\end{center}
\caption{Massless and tachyonic modes  for the supersymmetric minimum\label{susytach}.}
\end{table}
Let us explain the content of the table and especially the meaning of ``at least''. All the details of this analysis are discussed in appendix \ref{ap:Hessian}

\begin{itemize}
    \item Since the potential only depends on a linear combination of complex structure axions and the dilaton, the other $h^{2,1}$ axions of this sector are seen as flat directions. Their saxionic partners, which pair up with them into complex fields,  are tachyonic directions with mass $\frac{8}{9}m_{BF}^2$. Both modes are always present for any value of $E_F$ so we refer to them with the ``at least" tag. This is expected form general arguments, see e.g. \cite{Conlon:2006tq}. 
    
    \item For $E_F\lesssim 0.1$ there appear new tachyons with masses above the BF bound, in principle different from $\frac{8}{9}m_{BF}^2$. The masses of these modes change continuously with $E_F$, and so they become massless before becoming tachyonic.
    \item Finally, there are also modes which have a positive mass for any $E_F$.
\end{itemize} 

\vspace*{.5cm}
\noindent
\textbf{Non-SUSY branch}
\\
This case presents a casuistry that makes it difficult to summarise in just one table. As discussed in section \ref{branchvacu}, the non-SUSY vacuum candidates are described  by  the physical solutions of eq.\eqref{eq: megaeqF}, represented in figure \ref{fig: generalsol}. On top of this curve one can represent the regions that are excluded  at the perturbative level: 
\label{ss:stability}
\begin{center}    
\begin{figure}[h]
    \centering
    \includegraphics[scale=0.8]{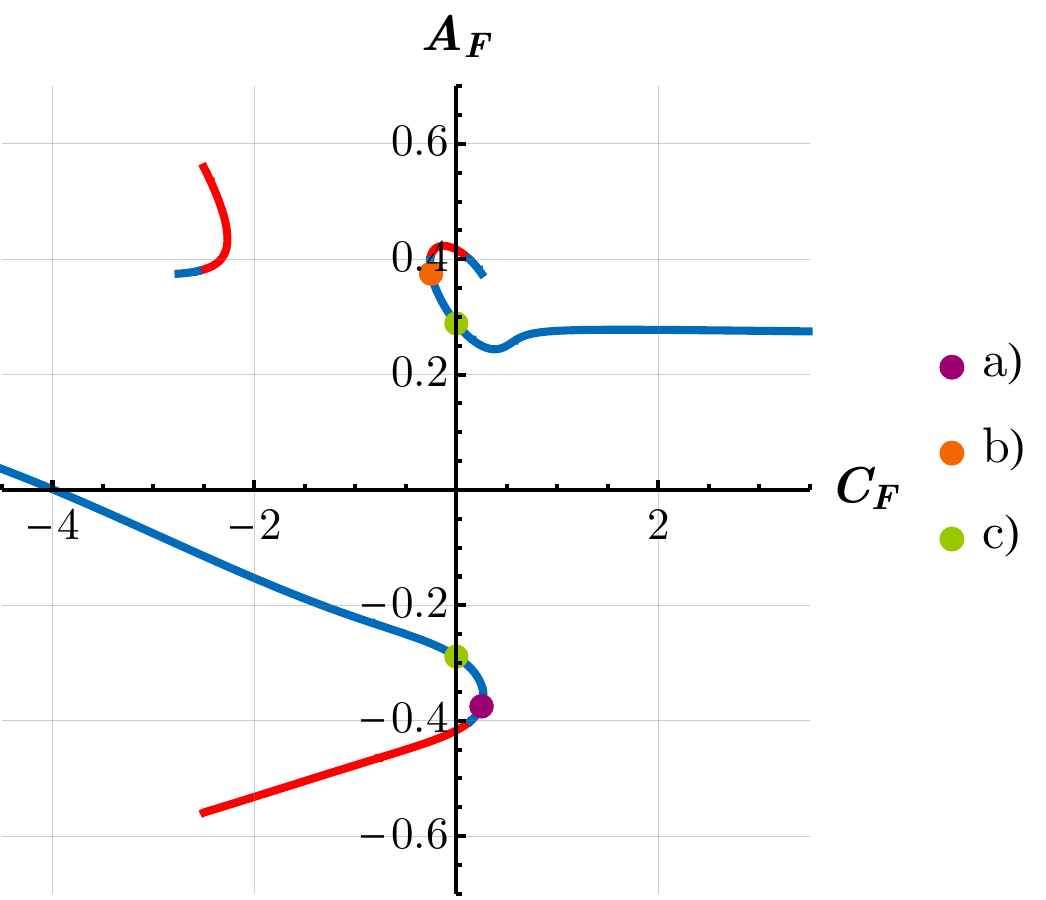}
    \caption{Set of points that verify \eqref{eq: megaeqF} with $E_F^2\geq 0$ and:  have no tachyons violating the BF bound and therefore are perturbatively  stable (blue curve);  have tachyons violating the BF bound and therefore are perturbatively unstable (red curve). The colored dots correspond to the particular solutions \eqref{delta=0sols}.}
    \label{fig: excludsol}
\end{figure}
\end{center}
Some comments are in order regarding the behaviour of the modes:
\begin{itemize}
    \item In the regions with $|A_F|\gtrsim 0.4$ there is always a tachyon whose mass violates the BF bound. This corresponds to the red pieces of the curves in figure \ref{fig: excludsol}.
    \item On the blue region of the curves, tachyons appear only in the vicinity of the red region, while away from it all the masses are positive. For instance, in the curve stretching to the right  there are no tachyons for $C_F\gtrsim 1.5$.
\end{itemize}

The explicit computation of the modes and their masses is studied in appendix \ref{ap:Hessian}.

\vspace*{.5cm}
\noindent
\textbf{D-term contribution}

As announced in the introduction, let us finish this section by commenting on the effect of the D-terms on stability. The first thing one has to notice is that, although $V_D=0$ once we impose the ansatz $\eqref{Ansatz}$, the Hessian $H_D$ associated to the D-terms is generically different from zero -see \eqref{dhessian}-. Indeed one can show that the matrix $H_D$ is a  positive semidefinite matrix. Therefore, splitting the contribution of  $V_F$ and  $V_D$ to the Hessian into $H=H_F+H_D$ and using the inequalities collected in \cite{bhatia2013matrix}, one can prove that the resulting eigenvalues of the full Hessian $H$ will always be equal or greater than the corresponding $H_F$ eigenvalues. Physically, what this means is that the D-terms push the system towards a more stable regime. In terms of the figure \eqref{fig: excludsol} and taking into account the directions affected by $H_D$ -see again \eqref{dhessian}-, one would expect that, besides having no new unstable points (red region), some of them do actually turn into stable ones (blue points) once the D-terms come into play.

\section{Conclusions}
\label{s:concluu}

In this chapter we have continued our systematic approach towards moduli stabilisation in 4d type IIA orientifold flux compactifications. The first step has been to rewrite the scalar potential, including both the F-term and D-term contributions, in a bilinear form, such that the dependence on the axions and the saxions of the compactification is factorised. This bilinear form highlights the presence of discrete gauge symmetries on the compactification, which correspond to simultaneous discrete shifts of the axions and the background fluxes. This structure has been already highlighted for the F-term piece of the potential in Calabi-Yau compactifications with $p$-form fluxes \cite{Bielleman:2015ina,Carta:2016ynn,Herraez:2018vae}, and in here we have seen how it can be extended to include general geometric and non-geometric fluxes as well. 

Besides a superpotential, these new fluxes  generate a D-term potential, which displays the same bilinear structure. The D-term potential arises from flux-induced  St\"uckelberg gaugings of the U(1)'s of the compactification by some axions that do not appear in the superpotential, and that generate conventional discrete gauge symmetries arising from $B \wedge F$ couplings. Such discrete symmetries are unrelated to the ones in the F-term potential. However, the D-term potential itself depends on the B-field axions $b^a$, because they appear in the gauge kinetic function $f_{\alpha\beta}$, and these axions do appear as well in the F-term potential, participating in its discrete symmetries. It would be interesting to understand the general structure of discrete shift symmetries that one can have in flux compactifications with both F-term and D-term potentials. In addition, it would be interesting to complete the analysis by including the presence of D6-branes with moduli and curvature corrections, along the lines of the previous chapter.

As in \cite{Bielleman:2015ina,Carta:2016ynn,Herraez:2018vae}, it is the presence of discrete shift symmetries that is behind the factorisation of the scalar potential into the form \eqref{VF}, where $ Z^{\cA\cB}$ only depends on the saxionic fields, and ${\rho}_\cA$ are gauge invariant combinations of flux quanta and axions. With the explicit form of the ${\rho}_\cA$ one may construct combinations that are axion independent, and therefore invariant under the discrete shifts of the compactification. In any class of compactifications, some of the fluxes are invariant by themselves, while others need to be combined quadratically to yield a flux invariant. We have analysed the flux invariants that appear in Calabi--Yau, geometric and non-geometric flux compactifications, their interest being that they determine the vev of the saxions at the vacua of the potential. Therefore, in practice, the value of these flux invariants will control whether the vacua are located or not in regions in which the effective field theory is under control. 

Another important aspect when analysing flux vacua is to guarantee their stability, at least at the perturbative level. Guided by the results of \cite{GomezReino:2006dk,GomezReino:2006wv,GomezReino:2007qi,Covi:2008ea,Covi:2008zu}, we have analysed the sGoldstino mass estimate in our setup, imposing that it must be positive as a necessary stability criterium to which de Sitter extrema are particularly sensitive. Our analysis has led us to the simple Ansatz \eqref{solsfmax} for the F-terms on-shell, which can be easily translated to relations between the $\rho_\cA$ and the value of the saxions at each extremum, cf. \eqref{proprho}.

The next step of our approach has been to find potential extrema based on this Ansatz, a systematic procedure that we have implemented for the case of geometric flux compactifications. This class of configurations is particularly interesting because they contain de Sitter extrema and are therefore simple counterexamples of the initial de Sitter conjecture \cite{Obied:2018sgi}, although so far seem to satisfy its refined version \cite{Garg:2018reu,Ooguri:2018wrx}. In this respect, we have reproduced previous de Sitter no-go results in the literature \cite{Hertzberg:2007wc,Flauger:2008ad} with our bilinear expression for the potential, but with two interesting novelties. First, when imposing that the F-terms are of the form \eqref{solsfmax} either on-shell or off-shell, we recover an inequality of the form \eqref{eq: no-go geom inequality} that forbids de Sitter extrema. We find quite amusing that this result is recovered after imposing an Ansatz inspired by de Sitter metastability. Second, our analysis includes a flux-induced D-term potential, and so the possibility of D-term uplifting, typically considered in the moduli stabilisation literature, does not seem to work in the present setting. We see our result as an interesting product of integrating several de Sitter criteria, and it would be interesting to combine it with yet other no-go results in the literature, like for instance those in \cite{Andriot:2018ept,Andriot:2019wrs,Grimm:2019ixq}. 

As is well known, type IIA  orientifold compactifications with geometric fluxes provide a non-trivial set of AdS$_4$ vacua, which we have analysed from our perspective. We have seen that, by imposing the on-shell Ansatz \eqref{solsfmax}, the equations of motion translate into four algebraic equations. By solving them, we have found two different branches of vacua, one supersymmetric and one-non-supersymmetric, and we have shown how both of them include most of the vacua found in the geometric flux compactification literature. This link with previous results can be made both with references that perform a 4d analysis and those that solve the equations of motion at the 10d level which is particularly interesting for non-supersymmetric solutions, which are scarce. 

All these results demonstrate that analysing the bilinear form of the scalar potential provides a systematic strategy to determine the vacua of this class of compactifications, overarching previous results in the literature. Needless to say, to obtain a clear overall picture it would be important to generalise our analysis in several directions. First, it would be interesting to consider other on-shell F-term Ansatz beyond  \eqref{solsfmax} that also guarantee vacua metastability. Indeed, our analysis of the Hessian shows that, for certain geometric flux compactifications, perturbative stability occurs for a very large region of the parameter space of our F-term Ansatz, and it would be important to determine how general this result is. Second, a natural extension of our results would be to implement our  approach to compactifications with non-geometric fluxes, a task that we leave for the future. In this case it would be particularly pressing to characterise the potential corrections to the effective flux-potential, and in particular to the K\"ahler potential that we have assumed throughout our analysis. For the case of geometric fluxes these corrections should be suppressed for those vacua that sit at large volume and weak coupling, which generically corresponds with the set of solutions with a small value for the Ansatz parameter $F$. Remarkably, it is through the same small parameter that it seems to be possible to control the separation between the AdS$_4$ scale and the cut-off scale of the theory. This is in agreement with that in the limit $F \raw 0$ such scale separation may, a priori, be realised parametrically. 

In any event, we hope to have demonstrated that with our systematic approach one may be able to obtain an overall picture of classical type IIA flux vacua. Our strategy not only serves to find and characterise different metastable vacua, but also to easily extract the relevant physics out of them, like the F-terms, vacuum energy and light spectrum of scalars. A global picture of this sort is essential to determine what the set of string theory flux vacua is and it is not, and the lessons that one can learn from it. Hopefully, our results will provide a non-trivial step towards this final picture. 
\clearpage


\chapter{Type IIA orientifold compactifications: a (smeared) 10d perspective}
\label{ch:review10d}
Up to this moment, we have been studying type IIA (CY) orientifold compactifications just focusing on the 4d effective action, without worrying about the EOM in the 6d internal manifold. As repeated several times during the previous chapters, the issue in these scenarios is that the back-reaction of the intersecting O6-planes is not taken into account. Technically, this would require going beyond the usual Calabi-Yau manifolds and considering $SU(3)\times SU(3)$ structure manifolds, which are far less understood. For that reason, one works usually in the \emph{smeared} approximation, where the punctual sources are \emph{smeared}. That is, they are distributed uniformly along all the internal dimensions while keeping the total amount of charge fixed. Though this is not describing truly the physical situation, the idea is that at low energies -so in the long-wavelength approximation- both solutions should be similar, and so the smearing approximation should be valid in this regime\footnote{An example in which this also happens, though for parallel O6 planes, can be found in \cite{Baines:2020dmu}.}.  

This approach is not exempt from criticism, as we have already mentioned. It could happen that some of the nice properties of the smeared solution, such as scale separation, are lost when the localised nature of the sources and the full 10d picture is taken into account\footnote{See for example \cite{Font:2019uva} for a case in which something similar happens.}. Indeed, we know for sure that the smeared solution cannot be capturing all the properties of the true solution since gradients for the dilaton and the warp factor -which in the smearing approximation are constant- are expected to appear. The question is whether the information hidden by the smearing approximation changes drastically when one considers the fully back-reacted background.

In this same spirit, the full 10d picture is also needed to study the non-perturbative stability of the non-SUSY solutions. Only in this way one can compute the spectrum of the theory and check if there is some membrane triggering an instability, as predicted by the refined version of the WGC. Therefore, we see that to have better control and understanding of the vacua derived so far, studying the 10d equations of motion, and not only the 4d effective action is an indispensable requirement

In this chapter we will initiate the study of the 10d picture of the 4d vacua derived so far, focusing here on the SUSY solutions and its uplift to $SU(3)$ structure manifolds. We will start by reviewing the 10d equations of motion equations in section \ref{sec:10dequa} and recalling how the SUSY scenario can be recast in a very elegant way using polyforms in section \ref{sec:susyback}. We will be short and concise, referring the reader to \cite{Grana:2005jc, Zabzine:2006uz, Koerber:2007hd, Tsimpis:2016bbq, Tomasiello:2022dwe} -which we will be following- to a detailed introduction and further references. After having presented the framework, we will finalise in section \ref{sec:smearingup} presenting the smeared uplift of the SUSY vacua derived in chapters \ref{ch:rrnsnsvacua} and \ref{ch:geometricflux}.

\section{10d equations of motion and Bianchi identities}
\label{sec:10dequa}

Let us start by listing here all the equations that a consistent background has to satisfy. Though we will see afterwards that the SUSY case can be expressed in a much simpler form, when we study in chapter \ref{ch:nonsusy} the stability of some of the non-SUSY vacua derived in chapter \ref{ch:rrnsnsvacua}, we will need to work with these equations directly. We will use the expressions derived in \cite{Junghans:2020acz} adapted to our conventions. Since we will study (non-)SUSY AdS$_4$ vacua, we take an ansatz for the metric of the form
\begin{align}
ds^2=e^{2A}ds^2_{\text{AdS}_4}+ds_6^2\, .
\end{align}

\subsection{Bianchi identities}
\label{sec:bie}
Already introduced  in \eqref{IIABI},  let us now write them explicitly. In the presence of $D6$-branes and $O6$-planes they read
\be
dG_0 = 0\, , \qquad d G_2 = G_0 H -4 \d_{\rm O6} +   N_\a \d_{\rm D6}^\a \, ,  \qquad d G_4 = G_2 \wedge H\, , \qquad dG_6 = 0\, ,
\label{BIG0}
\ee
where  we have defined $\d_{\rm D6\rm /O6}\equiv \ell_s^{-2}  \d(\Pi_{\rm D6/\rm O6})$. Recall that $\Pi_\alpha$ hosts a D-brane source, and $\delta(\Pi_\alpha)$ is the bump $\delta$-function form with support on $\Pi_\alpha$ and indices transverse to it, such that $\ell_s^{p-9} \d(\Pi_\a)$ lies in the Poincar\'e dual class to $[\Pi_\a]$. This in particular implies that
\be
{\rm P.D.} \left[4\Pi_{\rm O6}- N_\a \Pi_{\rm D6}^\a\right] = m [\ell_s^{-2} H] \, ,
\label{tadpolewo}
\ee
constraining the quanta of Romans parameter and NS flux.

\subsection{RR and NSNS equations}
\label{subs:rrnseq}
In our conventions, the EOM for the field strengths -in the string frame- derived from \eqref{10daction} are
\begin{subequations}
\label{eq:rrnsns}
\begin{align}
0&=\mathrm{d}\left(\star_{10} G_{2}\right)+H_{3} \wedge \star_{10} G_{4}\, , \\
0 &=\mathrm{d}\left(\star_{10} G_{4}\right)+H_{3} \wedge \star_{10} G_{6}\, , \\
0 &=\mathrm{d}\left(\star_{10} G_{6}\right)\, , \\ 
0 &=\mathrm{d}\left(e^{-2\phi} \star_{10} H_{3}\right)+\star_{10} G_{2} \wedge G_{0}+\star_{10} G_{4} \wedge G_{2}+\star_{10} G_{6} \wedge G_{4}\label{problem}\, .
\end{align}
\end{subequations} 

\subsection{Einstein and dilaton equation}
\label{subsec:einseq}

Finally, we will also write these equations explicitly for completeness
\begin{align}
\label{eq:einstein1}
0=& 12 \frac{e^{-2\phi}}{e^{2A}}+12 \frac{e^{-2\phi}}{e^{2A}}(\partial e^A)^{2}+4 \frac{e^{-2\phi}}{e^A} \nabla^{2} e^A+12 \frac{e^{-\phi}}{e^A}(\partial e^A)(\partial e^{-\phi})+e^{-\phi} \nabla^{2} e^{-\phi}+(\partial e^{-\phi})^{2}\nonumber \\
&-\frac{1}{2} e^{-2\phi}\left|H_{3}\right|^{2}-\sum_{q=0}^{6} \frac{q-1}{4}\left|F_{q}\right|^{2}+\frac{1}{2} e^{-\phi} \sum_{i} \left(-4\delta_{\mathrm{O} 6}^i+\delta_{\mathrm{D}6 }^i\right)\, ,
\end{align}
\begin{align}
0=&-e^{-2\phi} R_{m n}+4 \frac{e^{-2\phi}}{e^A} \nabla_{m} \partial_{n} e^A+\frac{e^{-\phi}}{e^A} g_{m n}(\partial e^A)(\partial e^{-\phi})+\frac{1}{4} g_{m n} e^{-\phi} \nabla^{2} e^{-\phi}+\frac{1}{4} g_{m n}(\partial e^{-\phi})^{2}\nonumber \\
&+2 e^{-\phi} \nabla_{m} \partial_{n} e^{-\phi}-2\left(\partial_{m} e^{-\phi}\right)\left(\partial_{n} e^{-\phi}\right)+\frac{1}{2} e^{-2\phi}\left(\left|H_{3}\right|_{m n}^{2}-\frac{1}{4} g_{m n}\left|H_{3}\right|^{2}\right) \\
&+\frac{1}{2} \sum_{q=0}^{6}\left(\left|F_{q}\right|_{m n}^{2}-\frac{q-1}{8} g_{m n}\left|F_{q}\right|^{2}\right)+\sum_{i}\left(\Pi_{i, m n}-\frac{7}{8} g_{m n}\right) e^{-\phi} \left(-4\delta_{\mathrm{O} 6}^i+\delta_{\mathrm{D}6 }^i\right)\nonumber\, ,
\end{align}

\begin{equation}
\label{eq:einstein3}
\begin{aligned}
0=&-8 \nabla^{2} e^{-\phi}-24 \frac{e^{-\phi}}{e^{2A}}-\frac{32}{e^A}(\partial e^A)(\partial e^{-\phi})-24 \frac{e^{-\phi}}{e^{2A}}(\partial e^A)^{2}-16 \frac{e^{-\phi}}{e^A} \nabla^{2} e^A \\
&+2 e^{-\phi} R_{m n} g^{m n}-e^{-\phi}\left|H_{3}\right|^{2}+2 \sum_{i}\left(-4\delta_{\mathrm{O} 6}^i+\delta_{\mathrm{D}6 }^i\right)\, ,
\end{aligned}
\end{equation}
where the subindex $i$ sums over all the $O6$-planes and $D6$-brane sources and the $\delta_{\mathrm{D} 6 / \mathrm{O} 6}^i$ was defined in section \ref{sec:bie}. $\Pi_{i,mn}=-\frac{2}{\sqrt{g_{\pi_i}}}\frac{\delta\sqrt{g_{\pi_i}}}{\delta g^{mn}}$ is the projector of the stress-energy tensor of the $i$th O6-planes/D6-brane. The conventions for the hodge star and in general the conventions in 10d used in the next chapters are relegated to appendix \ref{ap:10dconv}

\section{SUSY backgrounds}
\label{sec:susyback}

As is often the case, asking for the solution to preserve some amount of supersymmetry makes life easier. Without going into the details, it was shown first in \cite{Lust:2004ig}  that the equations of motion (which are second-order equations)  are automatically satisfied by backgrounds solving the vanishing of the supersymmetric variations of the fermions and the Bianchi identities (which are all first-order equations). In other words, SUSY$+$Bianchi identities$\rightarrow$EOM. Then, in \cite{Grana:2005sn} the SUSY equations were written in a very elegant and simple way, using the language of polyforms. Let us review here briefly the main tools needed. We will be following \cite{Koerber:2008rx,Tomasiello:2022dwe} mostly. 

Having supersymmetry in the 4d theory constrains severely the internal space $\mathcal{M}_6$, since it requires the existence of a constant   spinor defined on the manifold. The compactification ansatz for the $10d$ supergravity generators  $\epsilon^{1,2}$ in IIA is
\begin{align}
\epsilon^{1} &=\zeta_{+} \otimes \eta_{+}^{(1)}+\zeta_{-} \otimes \eta_{-}^{(1)}\, ,	&	 \epsilon^{2} &=\zeta_{+} \otimes \eta_{-}^{(2)}+\zeta_{-} \otimes \eta_{+}^{(2)}\, ,
\end{align}
where $\zeta_{+}$ are constant complex Weyl spinors in the 4d theory -yielding $\mathcal{N}=1$ supersymmetry-, whereas $\eta_{\pm}$ are two internal spinors of positive and negative chirality respectively. The existence of these spinors reduces the structure group of $\mathcal{M}_6$ to $SU(3)\times SU(3)$.

\begin{tcolorbox}[breakable, enhanced,every float=\centering]
\begin{small}
\textbf{Structure groups}
\\
Consider the internal manifold $\mathcal{M}_d$ and its tangent bundle $T\mathcal{M}_d$, that is, the bundle over the manifold $\mathcal{M}_6$ with fiber in each point $p\in\mathcal{M}_d$ the set of bases of the tangent space $T_p\mathcal{M}_6$. Locally, on a patch $U_\alpha$ of $\mm_d$, we have $\left(p, e_a\right)$ where $p\in U_\alpha$ and $e_a=e^i_a\partial_{x_i}, \,\,  a,i=1,\dots d$ is a local basis of $T_p\mm_d$. 

Let us now consider two different patches $\{U_\alpha$, $U_\beta\}$ with some region in common, where they are described by $\left(p, e_a\right)$ and $\left(p, \hat{e}_a\right)$ respectively. On the this overlap, there exists the following relation inherited from the associated tangent bundle
\begin{align}
 \hat{e}^{i}_a=e^i_b\left(t_{\beta\alpha}\right)^b_a\, ,
\end{align}
where $t_{\beta\alpha}\in \text{GL}\left(d,\mathds{R}\right)$ are called the transition functions and must satisfy some consistency conditions -see \cite{Koerber:2008rx} for more details-. In practice the $t_{\beta\alpha}$  tell us how a local basis of the tangent space of the manifold transform when moving between different patches. This group of transformations is called the \emph{structure group} of the manifold, and will be in general  $\text{GL}\left(d,\mathds{R}\right)$.

\begin{center}    
    \centering
    \includegraphics[width=12cm]{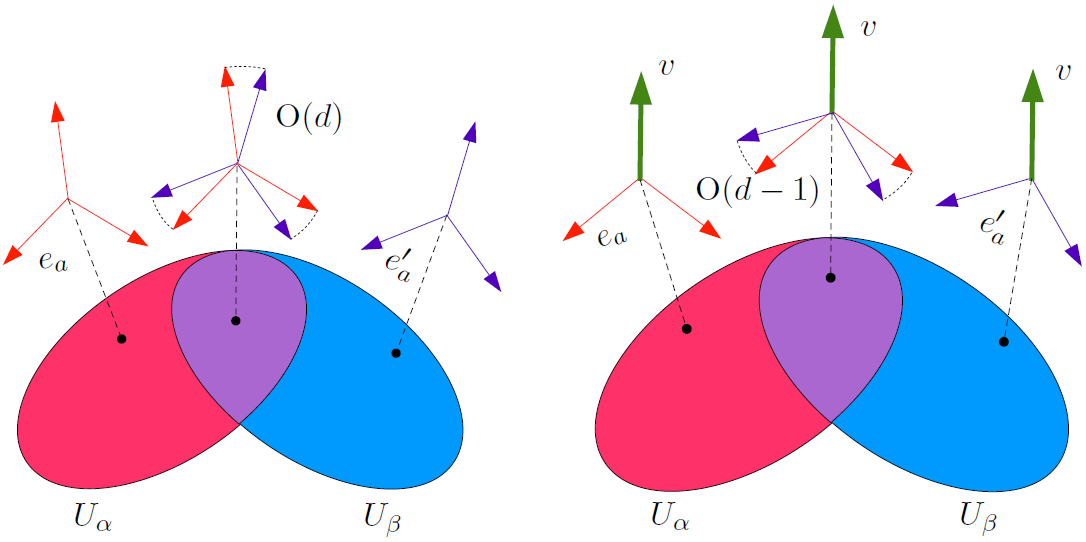}
    \captionsetup{font={footnotesize}}
     \captionof{figure}{Extracted from \cite{Koerber:2008rx}. On the left, the structure group is assumed to be reduced to $O(d)$. On the right: a globally defined non-vanishing vector $v$ is introduced and the structure group is reduced to $O(d-1)$.} \label{fig:structuregrou}
\end{center}
 We will be interested in the cases in which the structure group is reduced to some subgroup $G\subset \text{GL}$, which depends on the topological properties of the manifold. This happens for instance when there are structures globally defined  on the manifold. For example, if the manifold admits an everywhere non-vanishing vector $v$, the structure group is reduced to $O(d-1)$ -see figure \ref{fig:structuregrou}-; if it admits a metric, it reduces $O\left(d,\mathds{R}\right)$ and if there is a globally defined volume form associated with it, to $SO\left(d,\mathds{R}\right)$
\end{small}
\end{tcolorbox}

From the internal spinors, one can build a pair of bispinors\footnote{Indeed, one could build 3 different pairs of bispinors, but only the one considered in the main text will play a significant role.}, $\Phi_\pm \equiv \eta^1_+ \otimes  \eta^{2\,\dagger}_\pm$. In turn, these bispinors can be expressed, using Fierz identities, as
\begin{equation}
\Phi_\pm \equiv \eta^1_+ \otimes  \eta^{2\,\dagger}_\pm=\frac{1}{8}\sum_{k=0}^{6} \frac{1}{ k !}\left( \eta^{2\,\dagger}_\pm\gamma_{n_{k} \ldots n_{1}}\eta^1_+\right) \gamma^{n_{1} \ldots n_{k}}\, ,
\end{equation}
where the gamma matrices are defined in the usual way $\{\gamma_a,\gamma_b\}=2g_{ab}\, a=1, \dots, 6$;  $\gamma_{n_{k} \ldots n_{1}}$ is the antisymmetric product of gammas, $\gamma_{n_{k} \ldots n_{1}}\equiv \gamma_{[n_k}\dots\gamma_{n_1]}$ -for instance $\gamma_{mn}=\frac{1}{2}\left[\gamma_m,\gamma_n\right]$-  and the same for $ \gamma^{n_{1} \ldots n_{k}}$ with the indices raised by the metric. Under the Clifford map, the previous expression can be interpreted as a polyform in the internal space
\begin{align}
\Phi_\pm=\frac{1}{8}\sum_{k=0}^{6} \frac{1}{ k !}\left( \eta^{2\,\dagger}_\pm\gamma_{n_{k} \ldots n_{1}}\eta^1_+\right) \gamma^{n_{1} \ldots n_{k}}\leftrightarrow \frac{1}{8}\sum_{k=0}^{6} \frac{1}{ k !}\left( \eta^{2\,\dagger}_\pm\gamma_{n_{k} \ldots n_{1}}\eta^1_+\right) dx^{n_1}\wedge\dots \wedge dx^{n_k}\, .
\end{align}  
\begin{tcolorbox}[breakable, enhanced,every float=\centering]
\begin{small}
\textbf{The Clifford map}
\\
The \textit{Clifformd map} -see\cite{Tomasiello:2022dwe} for a more detailed discussion- is an isomorphism between  polyforms and operators acting on spinors, defined as
\begin{align}
\alpha_k=\frac{1}{k!}\alpha_{m_1\dots m_k} dx^{m_1}\wedge\dots\wedge dx^{m_k}\quad \longleftrightarrow\quad \cancel{\alpha}_k =\frac{1}{k!}\alpha_{m_1\dots m_k}\gamma^{m_1\dots m_k}\, .
\end{align}
The action of $\gamma^m$  on the bispinor $\cancel{\alpha}_k$ translates to the polyform formalism $\alpha_k$ as the operator
\begin{align}
\overrightarrow{\gamma^{m}}&\longleftrightarrow\mathrm{d} x^{m} \wedge+g^{m n} \iota_{n}\, ,	&	\overleftarrow{\gamma^{m}}&\longleftrightarrow\left(\mathrm{d} x^{m} \wedge-g^{m n} \iota_{n}\right)(-1)^{\operatorname{deg}}\, ,
\label{eq:bispinorform}
\end{align}
where $\overrightarrow{}\left(\overleftarrow{}\right)$ indiactes that the gamma matrix is acting on the bispinor on the left (right).$\iota_{n}$ is the \textit{contraction} operator defined as $$\iota_{m}\left(\mathrm{~d} x^{n_{1}} \wedge \ldots \wedge \mathrm{d} x^{n_{k}}\right) \equiv k \delta_{m}^{\left[n_{1}\right.} \mathrm{d} x^{n_{2}} \wedge \ldots \wedge \mathrm{d} x^{\left.n_{k}\right]}\label{eq:contrac}\, ,$$ for instance $\iota_{m} \mathrm{~d} x^{n} \wedge \mathrm{d} x^{p}=\delta_{m}^{n} \mathrm{~d} x^{p}-\delta_{m}^{p} \mathrm{~d} x^{n}$- Finally, the operator $\operatorname{deg}$ counts the degree of a form, $\operatorname{deg}\alpha_k=k\alpha_k$, where $\alpha_k$ is a $k$-form. 
\end{small}
\end{tcolorbox}

Depending on how the internal spinors $\eta^i$ are -parallel, orthogonal or generic-, the structure group of the manifold can be reduced further. We will give the explicit expressions for $\Phi_{\pm}$ for each case in the next sections. Then in \ref{secc:susyeq}, we will directly give the differential constraints that the SUSY variations impose on $\Phi_\pm$.

\subsection{SU(3)-structure}

If the two spinors are proportional, $\eta^1_\pm\propto \eta^2_\pm$, the structure group reduces to $SU(3)$. If we choose a normalisation for the spinors  such that  $\left(\eta^2_+\right)^\dagger\eta_+^1\equiv 8 e^{3A-\phi}e^{i\theta}$, then
\begin{align}
\label{eq:su3phi}
\Phi_+&= e^{3A-\phi}e^{i\theta}e^{-iJ}\, ,	&	\Phi_-&= e^{3A-\phi}\Omega\, ,
\end{align}
where we have introduced the warp factor $A$ and the dilaton $\phi$ in the normalisation (they can change along the internal manifold) since they will appear naturally when we consider the SUSY equations. $\{J,\Omega\}$ are a globally defined real $2$-form and a globally defined complex 3-form respectively, satisfying
\begin{align}
J\wedge\Omega&=0\,	,	&	d\vol_6=-\frac{1}{6}J\wedge J\wedge J=-\frac{i}{8}\Omega\wedge\bar\Omega\, .
\end{align} 
Moreover, $\Omega$ has to be \textit{decomposable}: at every point, it should be possible to write it as the wedge product of three one-forms. This constraint allows to reconstruct $\IM \Omega$ from $\RE \Omega$, or vice-versa; explicit formulas for this were given in \cite{Hitchin:2000jd} and reviewed for example in \cite[Sec.3.1]{Tomasiello:2007zq}.

We can decompose $\left(d J, d\Omega\right)$ in terms of $SU(3)$-representations \cite{Chiossi:2002tw}:
\begin{align}
\label{eq:torsionclasses}
d J & = -\frac{3}{2} \IM (\bar{W}_1 \Omega) +W_4 \wedge J +W_3 \, , \\
d \Omega & =W_1 J^2 +W_2 \wedge J + \bar{W}_5 \wedge \Omega \, ,
\end{align}
where the $W_i$ are the {\em torsion classes}: $W_1$ is a complex scalar, $W_2$ is a complex primitive (1,1)-form, $W_3$ is a real
primitive $(1,2)+(2,1)$-form, $W_4$ is a real one-form and $W_5$ a complex (1,0)-form. Depending on which of these torsion classes vanishes, manifolds with $SU(3)$ structure receive the classification shown in table \ref{ta:manifolds}. When all of them vanish, the manifold is Calabi-Yau, and $\{J,\Omega\}$ become respectively the K\"ahler and complex form introduced in the first chapter.

As we will see in section \ref{sec:smearingup}, $SU(3)$ manifolds are enough only to accomodate the smeared uplift of the vacua studied in chapters \ref{ch:rrnsnsvacua} and \ref{ch:geometricflux}, but we need $SU(3)\times SU(3)$ structure to describe the localised nature of the sources.
\begin{table}
\begin{center}
\begin{tabular}{|c|c|}
\hline
{\textbf{Manifold}} & {\textbf{Vanishing torsion class}}\\
\hline
Complex & $W_1=W_2=0$ \\ 
\hline
Symplectic & $W_1=W_3=W_4=0$ \\ 
\hline
Half-flat & $\IM W_1=\IM W_2=W_4=W_5=0$ \\
\hline
Special Hermitean & $W_1=W_2=W_4=W_5=0$ \\ 
\hline
Nearly K\"ahler & $W_2=W_3=W_4=W_5=0$ \\
\hline
Almost K\"ahler & $W_1=W_3=W_4=W_5=0$ \\
\hline
K\"ahler & $ W_1=W_2=W_3=W_4=0$ \\
\hline
Calabi-Yau & $ W_1=W_2=W_3=W_4=W_5=0$ \\
\hline
``Conformal'' Calabi-Yau & $ W_1=W_2=W_3= 3 W_4-2 W_5=0$ \\
\hline
\end{tabular}
\caption{ \label{ta:manifolds}
\text{Extracted from \cite{Granalectures}. Manifolds with $SU(3)$ structure with vanishing torsion classes.}}
\end{center}
\end{table}

\subsection{(Static) SU(2) structure}
\label{sec:su2static}
If the two spinors are orthogonal $\left(\eta^2_+\right)^{\dagger}\eta_+^1=0$ the structure group of the manifold is $SU(2)$ and the polyforms, choosing again an appropriate normalisation, are
\begin{align}
\label{eq:su2phi}
\Phi_+&= e^{3A-\phi}e^{\frac{1}{2}v\wedge\bar v}\wedge j\, ,	&	\Phi_-=- e^{3A-\phi}v\wedge e^{ij}\, ,
\end{align}
where $v$ is a complex 1-form and $j,\omega$ a real and complex two-forms respectively satisfying
\begin{align}
\label{su(2)prop}
j\wedge\omega&=\omega\wedge\omega=0\, , &	\omega\wedge\bar{\omega}&=2j^2\, ,\nonumber\\
\iota_v j&=\iota_{\bar v} j=0\, ,	&		\iota_v \omega&=\iota_{\bar v} \omega=0\, ,
\end{align}
with the contraction $\iota_{\bar v}$ defined in \eqref{eq:contrac}. The volume form is in this case is $d\mathrm{vol}_6 = -\frac i4 v \wedge \bar v \wedge j^2$. An equivalent definition is obtained by taking an $SU(3)$-structure $(J, \Omega)$ and adding a complex one-form $v$. The forms $j$ and $\omega$ are then obtained as
	\begin{equation}
		j= J - \frac i2 v \wedge \bar v \, ,\qquad \omega = \frac12 \iota_v \Omega \equiv \frac12 v \cdot \Omega\,.
	\end{equation}
	In other words, $\Omega= v \wedge \omega$.

 We will not use manifolds of this kind, since SUSY AdS$_4$ compactifications of classical IIA SUGRA in static $SU(2)$ manifolds do not exist \cite{Koerber:2010bx, Bovy:2005qq, Koerber:2008rx}. Nevertheless, $j,v$ and $\omega$ will appear when we study $SU(3)\times SU(3)$ manifolds (or dynamic $SU(2)$ manifolds).

\subsection{$SU(3)\times SU(3)$ structure or dynamic $SU(2)$ structure}

In the most generic case, and choosing an appropriate normalisation, $$\left(\eta^2_+\right)^{\dagger}\eta_+^1=8\cos\left(\psi\right) e^{3A-\phi}e^{i\theta}\, ,$$ where $\{\theta, \psi\}$ can change along the internal manifold.  The polyforms are
\begin{equation}
	\label{su3phi0}
		\Phi_+ = e^{3A-\phi}\mathrm{e}^{\mathrm{i} \theta} \cos\psi \exp[-i J_\psi] \, ,\qquad \Phi_- =e^{3A-\phi} \cos\psi\, v \wedge \exp[i \omega_\psi]\,,
	\end{equation}
with
\begin{equation}\label{eq:Jpsi0}
	J_\psi \equiv \frac1{\cos(\psi)}j + \frac i{2 \tan^2(\psi)} v\wedge \bar v \,,\qquad 
	\omega_\psi \equiv \frac1{\sin(\psi)} \left({\rm Re} \omega + \frac i{\cos(\psi)} {\rm Im} \omega\right)\, ,
\end{equation}   
and $\{j,v,\omega\}$  defined in the previous section \ref{sec:su2static}. We will work with this kind of manifolds in the next chapter. 

To close this section, it is important to point out that this and the other two previous cases can be described in a unifying and more natural way using the language of generalised complex geometry \cite{Hitchin:2003cxu, Gualtieri:2003dx}. We will not talk about it in this thesis.

\subsection{SUSY equations}
\label{secc:susyeq}

As we have commented when we introduced the whole set of 10d equations in section \ref{sec:10dequa}, looking for SUSY solutions is simpler, since one only has to solve the Bianchi identities and the vanishing of the SUSY variations of the fermions (the gravitino and dilatino). Bianchi identities were introduced in section \ref{sec:bie}, whereas we will present now  SUSY equations.

SUSY equations impose some differential constraints in the spinors $\eta^{1,2}_\pm$. These constraints can be re-expressed in a very compact way using the polyforms $\Phi_\pm$ introduced in the previous sections. We will skip here  the technical details and refer the reader to the original reference \cite{Grana:2005sn} -or to the reviews previously mentioned- to see the complete derivation. In our conventions, the differential equations imposed by supersymmetry on the $\Phi_\pm$ can be written as 
\begin{subequations}\label{eq:psp0}
	\begin{align}
		\label{eq:psp+0}
		d_H \Phi_+ &= - 2 \mu e^{-A} \mathrm{Re} \Phi_-\, , \\
		\label{eq:psp-0}
		d_H \left(e^{A} \im{ \Phi_-}\right) &= -3 \mu \mathrm{Im} \Phi_+ + e^{4A}\star \lambda \mathbf{G}\,.
	\end{align}
\end{subequations}
Here $\star$ is the internal Hodge dual,  $\lambda$ is a sign reversal operation defined on a $k$-form as $\lambda(\alpha_k)= (-1)^{\lfloor k/2 \rfloor} \alpha_k$ and $d_H=d-H\wedge$. For our purposes, it is  convenient to recall that  the second equation admits an alternative expression \cite{Tomasiello:2007zq}
\begin{equation}\label{eq:calJ0}
	{\mathcal J}_+ \cdot d_H \left( e^{-3A} \im{\Phi_-}\right) = -5 \mu e^{-4A} \mathrm{Re} \Phi_+ + \mathbf{G}\,.
\end{equation}
The new operator ${\mathcal J}_+\cdot$ is associated in a certain way to the form $\Phi_+$; we will see in explicit examples what it reduces to.

To find a SUSY solution, one needs to solve equations \eqref{eq:psp0} together with the Bianchi identities for the fluxes \eqref{BIG0}. The specific form of $\Phi_\pm$, -\eqref{eq:su3phi}, \eqref{sec:su2static} or \eqref{su3phi0}- will depend on the ingredients we consider since they will restrict the manifold in which we have to do the compactification. Having presented the tools needed, we are ready to look for solutions to these equations and see which of them correspond to the 4d vacua obtained in the first chapters. 

\section{SUSY (smeared) uplift}
\label{sec:smearingup}

We are ready finally to start the study of the uplift of the SUSY vacua found in the previous chapters. Before discussing the solutions, let us start by explaining what the smearing approximation is since we have not presented it properly yet. Originally introduced in \cite{Acharya:2006ne}, it replaces the localised sources with a homogeneous distribution of the charge over the internal manifold. In other words, it \textit{spreads} the charge. Mathematically, the idea is the following. Consider the Bianchi identity for $G_2$
\begin{align}
\label{eq:problematicbianchi}
dG_2=G_0 H+\delta_{\text{D}6/\text{O}6}\, ,
\end{align}
with $\delta_{\text{D}6/\text{O}6}$ taking into account the presence of O6 and D6 sources. In the smearing approximation, one replaces
\begin{align}
\delta_{\text{D}6/\text{O}6}\rightarrow -G_0H\, ,
\end{align}
so that now equation \eqref{eq:problematicbianchi} only imposes that $G_2$ must be closed. In this section, we will see that the uplift of the previous vacua to manifolds with $SU(3)$ structure -which are simpler than the ones with $SU(3)\times SU(3)$ structure- does not work, unless one work in the smearing approximation, which is not describing the true physical situation. Chapter \ref{ch:uplift10d} will be entirely dedicated to studying if one can go beyond this approximation, focusing on the case with only RR and NSNS fluxes.

\subsection{RR and NSNS fluxes}
\label{sec:susynsnsrr}
Massive type IIA 10d supergravity solutions of the form AdS$_4 \times X_6$ are relatively well understood in several instances, like when the internal manifold $X_6$ is endowed with a SU(3)-structure underlying 4d $\CN=1$ supersymmetry \cite{Behrndt:2004km,Behrndt:2004mj,Lust:2004ig}. In that case, using the expressions introduced in the previous sections, the 10d background supersymmetry equations reduce to
\bes
\label{adssu3}
\begin{equation}
\label{gauge}
d_{H} \IM \Omega \, =\, e^\phi *_6 \left(G_0 - G_2 + G_4 - G_6 \right) -3 \IM (w_0e^{-iJ})\, ,
\end{equation}
\begin{equation}
\label{DW}
d_{H} e^{-iJ} \, =\, -2 \bar{w}_0 \RE \Omega\ ,
\end{equation}
\begin{equation}
\label{string}
d_{H}  \RE \Omega \, =\, 0\, ,
\end{equation}
\ees
where $\phi$ is constant, $d_{H} = d - H \wedge$ and $w_0=e^{i\th}\mu = e^{i\th}/R\in \mathbb{R}$ is the constant entering the Killing equation of an AdS$_4$ of radius $R\ell_s$. $e^A$ does not appear in the equations because they imply $dA=0$, so $A$ is constant and we have already fixed its value to $e^ A=1$ -any other number can be reabsorbed in $\mu$-. For $w_0 \in  \mathbb{R}$ the solution to the above equations can be parametrised as
\be\label{solutionsu30}
G_6= 0\, , \quad \quad H = \frac{2}{5} e^{\phi} G_0 \,  \RE \Omega\, , \quad \quad G_2 =- e^{-\phi} W_2 \, , \quad \quad G_4 = \frac{3}{10} G_0 J \wedge J \, , \quad \quad dJ=0\, ,
\ee
where $G_0 =  5e^{-\phi}\, \RE w_0$ is a constant and $W_2$ is a real primitive $(1,1)$-form, namely a SU(3) torsion class of $\Omega$ -recall expressions \eqref{eq:torsionclasses}. Notice that because of \eqref{gauge} $W_2$ cannot have a harmonic component. One may now express the RR background fluxes as
\be\label{expflux}
G_0 = \tilde{\rho}\, ,  \quad G_2 = \tilde{\rho}^a\, \ell_s^{-2} \omega_a + \alpha_2 \, ,  \quad G_4 = \rho_a\, \ell_s^{-4}  \tilde{\omega}^a  +\alpha_4 \, , \quad G_6 = \rho_0\, \ell_s^{-6} \frac{d{\rm vol}_{X_6}}{{\rm vol}_{X_6}} + \alpha_6\,, 
\ee
where $\omega_a$ and $\tilde{\omega}^a$ are a basis of harmonic two- and four-forms of $X_6$ introduced in table \ref{base}, and $\alpha_{2p}$ are globally well-defined forms with no harmonic component. We also expand the NS-flux as in \eqref{Hflux}. One can then see that \eqref{solutionsu30} amounts to apply the Ansatz \eqref{solutions} with the choice of constants $A, B, C$ corresponding to the supersymmetric {\bf A1-S1} branch, together with $\alpha_2 =- e^{-\phi} W_2$ and $\alpha_4 = \alpha_6 =0$. Even if a $W_2 \neq 0$ signals that the metric on $X_6$ is not Calabi-Yau, supersymmetry requires that $W_2$ has no harmonic component, just as in type IIA compactifications to 4d Minkowski  \cite{Marchesano:2014iea}. As such, its presence can be considered as a deformation of the Calabi-Yau metric similar to a warp factor, rather than a discrete deformation or genuine geometric flux carrying topological information, see e.g. \cite{Tomasiello:2005bp,Marchesano:2006ns}. 

Despite these suggestive features, one can see that the \eqref{adssu3} is too simple to describe an actual 10d background corresponding to a type IIA compactification with fluxes, O6-planes and D6-branes. First, it features a constant dilaton and warp factor, which are in tension with the backreaction of such localised sources. Second, 
it is incompatible with the Bianchi identity for $G_2$. This reads
\be\label{BIG2}
d G_2 =HG_0 +\delta_{\text{D}6/\text{O}6}\quad \rightarrow \quad dW_2 = - e^{\phi} \left[G_0 H +\delta_{\text{D}6/\text{O}6}\right]\, ,
\ee
where $\delta_{\text{D}6/\text{O}6}$ are bump functions localised on the 3-cycles wrapped by the D6-branes and O6-planes and include their relative charge. As usual, RR tadpole cancellation amounts require that the quantity in brackets vanishes in cohomology, so that $G_2 = -e^{-\phi}W_2$ can be globally well-defined. However, \eqref{BIG2} together with $\Omega \wedge G_2 = 0$ implies a flux density $|G_4|^2$ which is negative in the bulk and singular on top of any localised source. A proposal to circumvent these problems, as explained at the beginning of this section, is to modify the Bianchi identity \eqref{BIG2} by replacing the localised sources with smeared ones \cite{Acharya:2006ne}, so that one can take $G_2 \equiv 0$. 

Instead of modifying the Bianchi identity, one may try to embed the above solution into a type IIA AdS$_4$ compactification based on a background with $SU(3) \times SU(3)$ structure, which is compatible with a non-trivial dilaton and warp factor \cite{Lust:2008zd, Lust:2009zb, Lust:2009mb}. We will study this possibility in the next chapter, where we will also revisit the \textit{smearing problem} problem in more detail.

\subsection{Geometric fluxes}
\label{sec:upliftgeom}

For those geometric vacua that fall in the large-volume regime, one may try to infer a microscopic description in terms of a 10d background AdS$_4 \times X_6$. In this section we will do so by following the general philosophy of the previous section, by interpreting our 4d solution in terms of an internal manifold $X_6$ with SU(3)-structure. As in the previous case, it could be that the actual 10d background displays a more general $SU(3)\times SU(3)$-structure that is approximated by an $SU(3)$-structure in some limit.  This is in fact to be expected for type IIA supersymmetric backgrounds with localised sources like O6-planes since the problem with the Bianchi identity for $G_2$ discussed in the anterior section always appears.  One should be able to describe the 4d vacua from a 10d SU(3)-structure perspective if the localised sources are smeared so that the Bianchi identities amount to the tadpole conditions derived from \eqref{DFtadpole}, already taken into account by our analysis.

Following section \ref{sec:susynsnsrr}, one may translate our Ansatz into 10d backgrounds in terms of the gauge-invariant combination of fluxes
\be
G_{RR} \,=\, d_HC_{RR} + e^{B} \wedge F_{RR} \, ,
\label{bfG}
\ee
where $d_H = d - H \wedge$. From here one reads
\begin{align}
\label{solutionsu3s0}
\ell_s G_6= 6 A\, d{\rm vol}_{X_6}\, , \quad \ell_s G_4 = 3 B\, J \wedge J \, , \quad \ell_s G_2 = C\, J \, , \quad \ell_s H = -6 E \,  g_s \IM (e^{-i\theta} \Omega)\, ,
\end{align}
and $ \ell_s G_0 = - D$, with the constants $\{A,\dots, F\}$ defined in \eqref{eq: ans rho0}-\eqref{eq: geomu}. Moreover, a vanishing D-term $D_\alpha = \frac{1}{2} \partial_\mu K \hat{f}_\alpha{}^\mu$ implies no torsion classes, as in the setup in \cite{Camara:2011jg}. In this case from  \eqref{eq: geoma} and \eqref{eq: geomu} it follows that
\be
dJ =  \frac{3}{2}  F g_s  \ell_s \IM (e^{-i\theta} \Omega)\, , \qquad d \RE (e^{-i\theta} \Omega) = - F g_s \ell_s J \wedge J\ ,
\label{geomsu3}
\ee
which translate into the following  $SU(3)$ torsion classes
\be
W_1 = - \ell_s g_s e^{i\theta} F \, , \qquad W_2 = W_3 = W_4 = W_5 = 0\, .
\label{torsionsu3}
\ee
Therefore, in terms of an internal SU(3)-structure manifold, our vacua correspond to nearly-K\"ahler compactifications.  

With this dictionary, it is easy to interpret our SUSY branch of solutions in terms of the general SU(3)-structure solutions for ${\cal N} = 1$ AdS$_4$ type IIA  vacua \cite{Behrndt:2004km,Lust:2004ig}. Taking for instance the choice $\theta = - \pi/2$, we can compare with the parametrisation of \cite[eq.(4.24)]{Koerber:2010bx}, and see that the relations \eqref{SUSYsol} and \eqref{SUSYbranch} fit perfectly upon identifying
\be
\ell_s |W_0|  e^{- A - i\hat{\theta}} =  3g_s\left(E + i \frac{F}{4}\right) \, ,
\label{SUSYdict}
\ee
where $|W_0|$ is the AdS$_4$ scale from the 10d frame, and $\hat{\theta}$ a phase describing the solution. 

One can in fact use this dictionary to identify some solutions in the non-supersymmetric branch with 10d solutions in the literature, like e.g. those in \cite{Lust:2008zd}. Indeed, let us in particular consider \cite[section 11.4]{Lust:2008zd}, where $\mathcal{N}=0$ AdS$_4$ compactifications are constructed by extending integrability theorems for 10d supersymmetric type II backgrounds. We first observe that the second Bianchi identity in \cite[eq.(11.29)]{Lust:2008zd} describes our first vacuum equation \eqref{paxioncpxAA}. Similarly \cite[eqs.(11.31),(11.35),(11.36)]{Lust:2008zd} are directly related to \eqref{psaxioncpxAA}, \eqref{psaxionkAA} and \eqref{paxionkAA} respectively.

Using these relations three classes of solutions are found in \cite[section 11.4]{Lust:2008zd}:
\begin{enumerate}
    \item The first solution \cite[(11.38)]{Lust:2008zd} is a particular case of the non-SUSY branch, corresponding to $A_F=\pm1/4$ and  $C_F=\pm1/2$, with $A_FC_F>0$.
    \item The second solution \cite[(11.38)]{Lust:2008zd} corresponds  the limit solution of the non-SUSY branch with $C_F=0$ and $\Delta_F\neq0$.
    \item The third solution \cite[(11.40)]{Lust:2008zd} describes a point in the SUSY branch characterised by $E_F=\pm \frac{1}{4\sqrt{15}}$. 
\end{enumerate}

To sum up, the results of \cite{Lust:2008zd} provide concrete 10d realisation of solutions for both the supersymmetric and non-supersymmetric branches of table \ref{vacuresul}.

Finally, this 10d picture allows us to understand our no-go result of section \ref{subsec: no-go's} from a different perspective. Indeed, given the torsion classes \eqref{torsionsu3} the Ricci tensor of the internal manifold $X_6$ reads \cite{BEDULLI20071125,Ali:2006gd}
\be
{\cal R}_{mn} = \frac{5}{4} g_{mn} |W_1|^2\, ,
\ee
and so it corresponds to a manifold of positive scalar curvature, instead of the negative curvature necessary to circumvent the obstruction to de Sitter solutions \cite{Silverstein:2007ac}.

\section{Summary}

In this chapter we have initiated the study of the 10d picture of the 4d vacua derived in chapters \ref{ch:rrnsnsvacua} and \ref{ch:geometricflux}. In other words, the study of the 10d uplift of type IIA orientifold flux compactifications.

We started by recapping the 10d equations of motion and recalling how, for SUSY backgrounds, they can be rewritten in a very elegant way using the language of polyforms. We also pointed out that SUSY AdS$_4$ flux compactifications of (classical) type IIA (SUGRA) only admit two kinds of internal manifolds, either  $SU(3)$ or $SU(3)\times SU(3)$ structure manifolds.

We then tried to embed IIA orientifold compactifications on these backgrounds, first only considering NSNS and RR fluxes and then including geometric fluxes. In general, this is important to understand if some of the nice properties the 4d vacua have (scale separation, stability of the non-SUSY solutions...) are preserved when we look at them with the 10d glasses.

Regarding orientifold compactifications with NSNS and RR fluxes, we recovered the results of \cite{Acharya:2006ne}. We started by checking if $SU(3)$ manifolds are enough to accommodate these vacua. We saw that, unfortunately, they are not: the Bianchi identity for $G_2$ cannot be solved unless one smears the O6/D6 sources. 

This same problem happens when geometric fluxes are included. We have seen that our Ansatz corresponds to a nearly-K\"ahler geometry in the limit of smeared sources. On the other hand,  AdS$_4$ SUSY vacua with $W_2\neq 0$  have been recently derived in \cite{Cribiori:2021djm}, which seem to be missed by our ansatz. It would be nice to understand how these solutions are recovered in our formalism. 

We limited ourselves to discussing the uplift of the previously derived 4d vacua in $SU(3)$ structure manifolds. We explained why this is not possible, but this is not the end of the story. The natural next step is to check if $SU(3)\times SU(3)$ manifolds can accommodate these solutions. This question will be addressed in the next chapter \ref{ch:uplift10d}, where we will look carefully at NSNS and RR flux compactifications and their possible uplift. With the results derived there, we will finish this thesis in chapter \ref{ch:nonsusy} by studying the uplift of the non-SUSY solutions of that same scenario, focusing on the stability of the vacua.

\clearpage
\chapter{Type IIA orientifold vacua beyond the smeared uplift}
\label{ch:uplift10d}

\section{Introduction}

String theory is known to support many Anti-de Sitter (AdS) vacua,  solutions of the form ${\rm AdS}_d \times {\cal M}_p$ where all fields are invariant under the AdS isometries. 
Strikingly, for the vast majority of AdS vacua, the Kaluza--Klein (KK) scale is comparable to the scale of the cosmological constant: one often says that there is no ``scale separation''. This means that the solutions are not really $d$-dimensional in any physical sense: physics looks ten- or eleven-dimensional to a hypothetical observer. There have been many studies on the property of scale separation in string theory, see in particular \cite{Tsimpis:2012tu, Gautason:2015tig, Gautason:2018gln, Blumenhagen:2019vgj, Font:2019uva, Apruzzi:2019ecr}.

Recently, the feature of scale separation was revisited as part of the Swampland program, as we reviewed in section \ref{sec:adsdc}. As we explained there, it was suggested that a Swampland condition could be that the value of the cosmological constant sets the mass scale of an infinite tower of states. The AdS Distance Conjecture (ADC) states that this mass scale $m$ is related to the cosmological constant as 
\be
m \sim \Lambda^{\alpha} \;,
\ee
with $\alpha \sim {\cal O}(1)$. The conjecture was motivated by examples in string theory, but also by the fact that the $\Lambda \rightarrow 0$ limit is in the infinite distance in the space of metrics. Further, a Strong version of this conjecture was also proposed which states that for supersymmetric vacua $\alpha = \frac12$. This stronger form would be satisfied in any AdS vacuum which has no separation of scales.\footnote{The absence of scale separation was actually shown for general classical supersymmetric AdS$_7$ vacua in \cite{Apruzzi:2019ecr}.} 

We have seen in chapter \ref{ch:rrnsnsvacua}, section \ref{s:validity4d}, that type IIA compactified on a CY orientifold with RR and NSNS fluxes violates this conjecture, yielding $\alpha=7/18$ for both  SUSY and non-SUSY vacua. This was first noticed in \cite{DeWolfe:2005uu} for the SUSY case, to which we will refer as DGKT vacua -we saw in section \ref{sec:comp} how this solution is included in our formalism-. All these vacua have orientifold singularities and indeed O-planes are supposed to be necessary for scale separation \cite{Gautason:2015tig}.

This four-dimensional possible counter-example to the Strong ADC was already discussed in \cite{Lust:2019zwm}, where it was argued that since there is no known ten-dimensional uplift of this vacuum its features are not established and therefore may not be trustable. This chapter aims to take some initial steps towards improving our understanding of ten-dimensional solutions which are based on these 4d proposals.\footnote{Another approach towards establishing their validity would be to construct a dual CFT, which would have the so far unrealised features of a parametric hierarchy between the central charge and the scaling dimensions of an infinite number of operators. An initial search for the dual CFT to \cite{DeWolfe:2005uu} was carried out in \cite{Aharony:2008wz, Conlon:2021cjk, Apers:2022zjx, Apers:2022tfm, Quirant:2022fpn}.} 

Over the years, several attempts have been made to lift the four-dimensional DGKT construction to a ten-dimensional solution. The main difficulty, as we have seen in section \ref{sec:smearingup}, lies in the presence of the O-plane sources. There exist several AdS solutions with back-reacted O-planes (but without scale separation: see for example \cite{Apruzzi:2019ecr} for a discussion in AdS$_7$). But in this case, the most concrete examples proposed in \cite{DeWolfe:2005uu} involve \emph{intersecting} O-planes, whose back-reaction isn't even known in flat space. In \cite{Acharya:2006ne} it was proposed to simply smear the O-planes; with this trick, an uplift to ten dimensions can indeed be found using $SU(3)$ structure manifolds. Other similar solutions were found with the same trick in \cite{Petrini:2013ika}.

Smearing O-planes is not physically sensible -though this approximation can capture the main physical properties \cite{Baines:2020dmu}-, so the next step was to investigate whether a similar solution could be found, where the O-planes could be localised. In \cite{Saracco:2012wc}, a local solution was found as a candidate for the behaviour near the individual O6-planes, with a resolved singularity and large-distance asymptotics to the smeared solution of \cite{Acharya:2006ne}. However, it was not clear whether it could be made global; this partially motivated scepticism about the solution \cite{McOrist:2012yc}. 

Our approach to looking for a solution is to utilise the supersymmetry equations introduced in section \ref{secc:susyeq}. First, we restrict to the equations at the two-derivative level, so neglecting higher-order $\alpha'$ corrections. The supersymmetry is related to the structure group of the manifold. The Ricci-flat metric on a Calabi--Yau has $SU(3)$-structure, and it is known that there are no $SU(3)$-structure solutions with localised $O$-planes \cite{Acharya:2006ne}. Therefore, any solution must deform the metric away from the Ricci flat one. For this deformation to be supersymmetric it should exhibit $SU(3)\times SU(3)$ structure, which is the most general possibility. We, therefore, study whether there are ten-dimensional solutions with $SU(3)\times SU(3)$-structure that exhibits the properties of DGKT. Even though these would not be compactifications on the Ricci-flat Calabi--Yau metric, they may morally be considered the uplifts to DGKT. 

We find an approximate solution to the supersymmetry equations. Specifically, in \cite{Saracco:2012wc} it was proposed that one could look for solutions that are perturbations of the smeared solution controlled by an expansion parameter related to the value of the cosmological constant. Following this approach, we find a solution to the supersymmetry equations and the Bianchi identities with localised sources, at leading order in this expansion parameter. The solution is very different to the one considered in \cite{Saracco:2012wc}, specifically we have an exact, rather than approximately, vanishing Freund--Rubin flux.\footnote{It should be noted that AdS solutions without a Freund--Rubin flux exist, notably in cases where they are forbidden by the dimensionality of spacetime, such as for AdS$_5$ in M-theory \cite{lin-lunin-maldacena} or AdS$_7$ in type IIA \cite{afrt}, but also not, such as in \cite{Couzens:2016iot}.}  However, the methodology is the same.\footnote{An argument against exactly vanishing Freund--Rubin flux was suggested in \cite[Sec.~7.5]{Saracco:2012wc}. However, we have found a mistake in that argument (which does not influence the rest of the paper).} 

Note that this approach was also recently utilised in a closely related paper \cite{Junghans:2020acz}. Our work focuses on the supersymmetry equations, which were not considered in \cite{Junghans:2020acz}, but at least so far as the existence of a first-order solution our results agree with those of \cite{Junghans:2020acz}. 

Returning to the question of separation of scales, our results show that DGKT vacua have passed a first non-trivial test. However, we do not claim that our results show conclusively that DGKT really does uplift to a full exact solution of string theory, nor that if such a solution exists it exhibits separation of scales. We discuss the remaining open questions 
in section \ref{sec:conclu}.

The rest of the chapter is organised as follows. In section \ref{sec:IIA} we review again the basics ingredients that are important for a 10d description, and the most general class of 10d supersymmetric backgrounds that they can correspond to. In section \ref{sec:nogosol} we discuss how the 4d features of DGKT constrain such 10d supersymmetric vacua, narrowing down the search for solutions. In section \ref{sec:approx} we present a large volume/weak coupling approximation of the supersymmetry equations compatible with DGKT. In section \ref{sec:Bianchi} we solve exactly the Bianchi identities corresponding to DGKT in a generic Calabi--Yau. In section \ref{sec:solution} we present our solution to the supersymmetry equations and Bianchi identities in the large volume approximation. We express such a solution in terms of Calabi--Yau quantities, and discuss its general features. We finally draw our conclusions in section \ref{sec:conclu}.

The most technical details of the chapter have been relegated to the Appendices. Appendix \ref{ap:SU33} reviews type IIA supersymmetry solutions from the viewpoint of $SU(3)\times SU(3)$ structures -the equations were introduced in sec \ref{sec:susyback}-, and appendix \ref{ap:SBEproof} contains the proof of what we dub the source balanced equation, see \eqref{preGOE}.

\section{Supersymmetric type IIA flux vacua}
\label{sec:IIA}

In this section we review the setup considered in \cite{DeWolfe:2005uu} and in chapter \ref{ch:rrnsnsvacua}, and in particular the features that should appear in a 10d description. Since the vacua in which we will be interested are supersymmetric from a 4d viewpoint, one expects their corresponding 10d backgrounds to solve the 10d supersymmetry equations with four supercharges. These equations can be efficiently encoded in the language of compactifications with $SU(3)\times SU(3)$ structures -see section \ref{sec:susyback}- which we also review. As we will show in the next section, the results of \cite{DeWolfe:2005uu} imply that only a specific class of $SU(3)\times SU(3)$-structure compactifications can describe the global aspects of these vacua. Most of the concepts presented here have already been introduced along the pages of this thesis, but we will repeat some of them to guarantee the internal coherence of this chapter and to make it as self-contained as possible.

\subsection{4d description of type IIA AdS$_4$ orientifold vacua}
Let us summarize here all the concepts introduced in the previous sections that we will need in this chapter. Consider again type IIA string theory compactified in an orientifold of $X_4 \times X_6$ with $X_6$ a compact real six-manifold with a Calabi--Yau metric, and therefore a K\"ahler 2-form $J_{\rm CY}$ and a holomorphic 3-form $\Omega_{\rm CY}$. The orientifold action is generated by $\Omega_p (-1)^{F_L}\cR$ and the fixed locus $\Pi_{\rm O6}$ of $\CR$ is one or several 3-cycles of $X_6$ in which O6-planes are located. In a consistent compactification, the RR charge of such O6-planes must be cancelled by a combination of D6-branes wrapping three-cycles of $X_6$ and background fluxes. 

In the presence of only O6-planes the Bianchi identities for the RR fluxes read
\be
dG_0 = 0\, , \qquad d G_2 = G_0 H -4 \d_{\rm O6}\, ,  \qquad d G_4 = G_2 \wedge H\, , \qquad dG_6 = 0\, ,
\label{BIG00}
\ee
where  we have defined $\d_{\rm O6}\equiv \ell_s^{-2} \d(\Pi_{\rm O6})$. This in particular implies that
\be
4{\rm P.D.} [\Pi_{\rm O6}] = m [\ell_s^2 H] \, ,
\label{tadpole1}
\ee
constraining the quanta of Romans parameter and NS flux. 

In chapter \ref{ch:rrnsnsvacua}, as in the original approach in \cite{DeWolfe:2005uu}, we performed a 4d analysis of this scenario,  finding an infinite discretum of $\cN=1$ AdS$_4$ vacua.  Indeed, one finds that 
\be
\langle g_s^{-1}\rangle [ H ]  = \frac{2}{5} G_0 [\re \Om_{\rm CY} ] \, , \qquad \langle G_2\rangle  =  0\, ,  \qquad \langle G_4\rangle  =   \frac{3G_0}{10}J\wedge J \, , \qquad \langle G_6\rangle  =  0\, , 
\label{intflux1}
\ee
where we have defined
\be
\langle g_s^{-1} \rangle = \frac{\int_{X_6} e^{-\phi}  J_{\rm CY}^3 }{\int_{X_6} J_{\rm CY}^3}\, , \ 
\langle G_2 \rangle = \frac{\int_{X_6} G_2 \wedge J_{\rm CY}^2 }{\int_{X_6} J_{\rm CY}^3}\, , \ \langle G_4 \rangle = \frac{\int_{X_6} G_4  \wedge J_{\rm CY}}{\int_{X_6} J_{\rm CY}^3 }\, , \ \langle G_6 \rangle = \frac{\int_{X_6} G_6}{-\int_{X_6} J_{\rm CY}^3}\, ,
\ee
with $\phi$ the 10d dilaton\footnote{Remarkably, similar relations hold when adding curvature corrections \cite{Palti:2008mg,Escobar:2018rna} and mobile D6-branes \cite{Escobar:2018tiu}.}. It was also obtained that the Calabi--Yau volume $\ell_s^6 \cV_{\rm CY} (X_6) = -\frac{1}{6} {\int_{X_6} J_{\rm CY}^3}$ is controlled by the four-form and two-form flux quanta, more precisely by the combination $\hat{e}_a = e_a - \oh \frac{\cK_{abc} m^am^b}{m}$, with $\cK_{abc}$ the triple intersection numbers of $X_6$ \cite{DeWolfe:2005uu}. As such, one may arbitrarily increase the Calabi--Yau volume by increasing the value of $\hat{e}_a$, while the density of four-form flux remains constant, as captured by \eqref{intflux1}. On the other hand, because of \eqref{tadpole1} the quanta of $H$ and $m$ are bounded and in practice be considered to be fixed. This implies that the average value of the (inverse) dilaton scales as
\be
\langle g_s^{-1}\rangle \, \sim \, \cV_{\rm CY}^{1/2} \, \sim\, \hat{e}^{3/4}\, .
\label{scaling0}
\ee
Finally, the AdS$_4$ radius $R_{\rm AdS}$ scales like 
\be
R_{\rm AdS} M_{\rm P} \sim \hat{e}^{9/4}\, .
\label{4dRads}
\ee
Therefore the 4d EFT considered in \cite{DeWolfe:2005uu} suggests that as we increase $\hat{e}$ along the infinite family of solutions we go to a limit of weak coupling, large volume and large AdS radius.

\subsection{10d description and supersymmetry equations} 
\label{sub:10d-susy}

We now review the conditions for ten-dimensional supersymmetry directly in the pure spinor formalism.  A ten-dimensional AdS$_4$ vacuum has a metric of the form
\begin{equation}\label{eq:warped-product1}
	ds^2 = e^{2A}ds^2_{\mathrm{AdS}_4} + ds^2_6\,.
\end{equation}

Preserved supersymmetry imposes differential equations on the internal part of the supersymmetry parameters $\eta^a_\pm$. From these one can build a bispinor $\Phi_\pm \equiv \eta^1_+ \otimes  \eta^{2\,\dagger}_\pm$, which can be interpreted as a polyform in the internal space by the Clifford map $\gamma^m \to d x^m$. This form obeys some algebraic constraints, that follow from its definition in terms of spinors, and some differential equations that follow from supersymmetry. 

Preserved supersymmetry  allows several types of solutions. Only two classes are relevant for us. The first class is made of the $SU(3)$-structure solutions, and are enough to describe the smeared uplift -\cite{Acharya:2006ne} and section \ref{sec:smearingup}-; they depend on a two-form $J$ and a three-form $\Omega$. The second class, which is the generic solution, comprises the $SU(3)\times SU(3)$-structure solutions; this is the one relevant for this chapter. Both classes were reviewed in \ref{sec:susyback}; here we only need to know that the $SU(3)\times SU(3)$ class depends on the following data:
\begin{itemize}
	\item Two functions $\psi$, $\theta$, 
	\item A complex one-form $v$,
	\item A real two-form $j$,
	\item A complex two-form $\omega$. 
\end{itemize}
The function $\psi$ measures the departure from the $SU(3)$ class; $\psi\to 0$ makes one fall to that case. In that limit, the data reassemble in those of an $SU(3)$-structure as
\begin{equation}\label{eq:JO-jo}
	J = j + \frac i{2\tan^2\psi} v \wedge \bar v \, ,\qquad \Omega = \frac i{\tan\psi}\, v \wedge \omega \,.
\end{equation}
On the other hand, $\theta$ is the phase of $\eta^{2\,\dagger}_+\eta^1_+$.   Finally, the forms $v$, $j$, $\omega$ are the same data that define an $SU(2)$-structure in six dimensions, see \ref{sec:susyback} for more details.

The differential equations imposed by supersymmetry on the $\Phi_\pm$ were introduced in section \ref{sec:susyback} and read
\begin{subequations}\label{eq:psp}
	\begin{align}
		\label{eq:psp+}
		d_H \Phi_+ &= - 2 \mu e^{-A} \mathrm{Re} \Phi_-\, , \\
		\label{eq:psp-}
		d_H \left(e^{A} \im{ \Phi_-}\right) &= -3 \mu \mathrm{Im} \Phi_+ + e^{4A}\star \lambda \mathbf{G}\, ,
	\end{align}
\end{subequations}
with the second equation admitting the following equivalent formulation \cite{Tomasiello:2007zq}
\begin{equation}\label{eq:calJ}
	{\mathcal J}_+ \cdot d_H \left( e^{-3A} \im{\Phi_-}\right) = -5 \mu e^{-4A} \mathrm{Re} \Phi_+ + \mathbf{G}\,.
\end{equation}
Recall that $\star$ is the internal Hodge dual, $\lambda$  is defined as $\lambda(\alpha_k)= (-1)^{\lfloor k/2 \rfloor} \alpha_k$ and $ \sqrt{-\Lambda/3}$ is the AdS$_4$ radius seen from the 10d string frame perspective. The mean value  of $e^{-A}$ can be absorbed into the definition of $ \sqrt{-\Lambda/3}$ so we can fix it to one. Regarding the  operator ${\mathcal J}_+\cdot$,  we will see in explicit examples what it reduces to. 

Let us now present in some detail what one gets by plugging in (\ref{eq:psp+}), (\ref{eq:calJ}) the solutions to the algebraic constraints for $\Phi_\pm$. Here we will focus on those classes of solutions that are more relevant for the computations of the following sections, leaving the rest for the more detailed discussion of App.~\ref{ap:SU33}. 

\subsubsection{$SU(3)$-structure} 
\label{ssub:su3}

For an $SU(3)$-structure we have seen that the pure spinors have the form
\be \label{eq:Phi-su3}
\Phi_+ \, =\, e^{3A-\phi} e^{i\theta} e^{-i J}\, , \qquad \qquad \Phi_-\,  =\, e^{3A-\phi}\Omega \, .
\ee
where $J$ and $\Omega$ do not need to be closed, allowing for $SU(3)$-structure torsion classes. From here one finds that 
\begin{equation}
 d\th =0\,\qquad 3dA =d\phi   
\end{equation}
and the following expression for the fluxes \cite{Koerber:2010bx}:
\begin{subequations}	
\label{su3flux}
\begin{align}
H & = 2 \mu e^{-A} {\rm cos}\, \th\, \re \Om\, ,\\
\label{su3fluxG0}
G_0 & = 5 \mu e^{-\phi-A}  {\rm cos}\, \th\, , \\
\label{su3fluxG2}
G_2 & = \frac{1}{3} \mu e^{-\phi-A}  {\rm sin}\, \th\, J - J \cdot d \left( e^{-\phi}  \im  \Om \right)\, , \\
G_4 & =  \frac{3}{2} \mu e^{-\phi-A}  {\rm cos}\, \th\, J \wedge J \, , \\
G_6 & = 3 \mu e^{-\phi-A}  {\rm sin}\, \th\, d\mathrm{vol}_{X_6}\, .
\end{align}
\end{subequations}     
The operation $J\cdot$ is defined as $J^{-1} \llcorner:$ one inverts the two-form $J$ to obtain a bivector, and one contracts this bivector with the forms that follow it. We will see more precisely how that works in the solutions below.

\subsubsection{$SU(3)\times SU(3)$ with $\theta=0$} 
\label{ssub:t0}

Here we consider the special case $\theta=0$, since, as we will argue in section \ref{sec:nogosol}, this case is the one suitable for a microscopic description of the DGKT vacua. $SU(3)\times SU(3)$ backgrounds with $\theta\neq 0$ have a similar but slightly more involved description; we defer their discussion to App.~\ref{sub:tneq0}.

In the case $\theta=0$ the pure spinors have the form
\begin{equation}
	\label{su3phi}
		\Phi_+ = e^{3A-\phi} \cos\psi \exp[-i J_\psi] \, ,\qquad \Phi_- =e^{3A-\phi} \cos\psi\, v \wedge \exp[i \omega_\psi]\,,
	\end{equation}
where
\begin{equation}\label{eq:Jpsi}
	J_\psi \equiv \frac1{\cos(\psi)}j + \frac i{2 \tan^2(\psi)} v\wedge \bar v \,,\qquad 
	\omega_\psi \equiv \frac1{\sin(\psi)} \left({\rm Re} \omega + \frac i{\cos(\psi)} {\rm Im} \omega\right)\,.
\end{equation}    
Details about $v$, $j$, $\omega$ are reviewed in section \ref{sec:susyback}.

There are first some equations that do not involve the fluxes:
\begin{subequations}	
\begin{align}
	\label{eq:revt0}    
	&{\rm Re} v= - \frac{e^A}{2 \mu}\left(3 dA - d \phi - \tan \psi d \psi\right) = -\frac{e^A}{2\mu} d\log(\cos\psi e^{3A-\phi}) \, ,\\
    \label{eq:dJpsit0}
    &d(e^{3A-\phi} \cos \psi J_\psi)=0 \, .
\end{align}
\end{subequations}

To arrive to \eqref{su3phi} one needs to perform a B-field transformation on the pure spinors and the fluxes \cite{gualtieri,Grana:2006kf,Saracco:2012wc}, with $b_{\Phi_\pm} = {\rm tan}\, \psi\, \im \om$. The physical fluxes are obtained by undoing it:
\begin{equation}\label{eq:GF-t0}
	H= \hat H + d(\tan\psi \mathrm{Im} \omega) \, ,\qquad \mathbf{G}= e^{\tan \psi \mathrm{Im} \omega \wedge} \mathbf{F}\,,
\end{equation}
where
\begin{subequations}\label{eq:flux-t0}
\begin{align}
    \label{eq:Ht0}
 	& \hat H= 2\mu e^{-A} {\rm Re} (i v \wedge \omega_\psi) \, , \\
	\label{eq:F0t0}
	& F_0 = -J_\psi\cdot
    d (\cos \psi e^{-\phi}{\rm Im} v)
    + 5 \mu \cos \psi e^{-A-\phi} \,,\\  
	\label{eq:F2t0}
	& F_2 = -J_\psi\cdot d\, {\rm Im} (i \cos \psi e^{-\phi} v \wedge \omega_\psi) - 2 \mu \frac{\sin^2 \psi}{\cos \psi} e^{-A-\phi} {\rm Im} \omega_\psi\,,\\
	\label{eq:F4t0}
& F_4 = J_\psi^2\left[ \frac12 F_0 - \mu \cos \psi e^{-A-\phi}\right] + J_\psi \wedge d\, {\rm Im} (\cos \psi e^{-\phi} v) \,,\\
& F_6 = 0 \label{f6su3su3}\,.
\end{align}
\end{subequations}
In the limit $\psi \rightarrow 0$ and upon making the replacements \eqref{eq:JO-jo} one recovers \eqref{su3flux} with $\theta=0$.

\section{Constraining the solution}
\label{sec:nogosol}

As expected for data obtained from the 4d EFT, the relations \eqref{intflux1} correspond to integrated quantities, and so there could be an infinite number of 10d backgrounds that correspond to them. Nevertheless, when combined with supersymmetry they result in some stringent constraints on the microscopic description of DGKT vacua. In this section we develop such  constraints by using the machinery of $SU(3)\times SU(3)$-structure compactifications. The result is quite simple to state: DGKT vacua should correspond to 10d backgrounds such that {\it i)} the internal flux $G_6$ vanishes pointwise and {\it ii)} it corresponds to a genuine $SU(3)\times SU(3)$ structure with $\theta=0$. 

\subsection{The Freund--Rubin flux}
\label{sec:susFR}

A key characteristic of type IIA flux compactifications studied so far is their Freund--Rubin flux. Given the compactification Ansatz \eqref{eq:warped-product1}, this flux is of the form
\be
G^{\rm{10d}}_4 =  c\,  d\mathrm{vol}_{X_4} + G_4 \, ,
\ee
where $c$ is defined by
\be
c = e^{4A} \star G_6\, ,
\ee
and $G_4$, $G_6$ are the four- and six-form components of the internal flux \eqref{bfG}. The equations of motion imply that $c$ is a constant, since
\be
d\left(\star_{10} G_6\right) =  d\mathrm{vol}_{X_4} \wedge d \left(e^{4A} \star G_6 \right) =  d\mathrm{vol}_{X_4} \wedge d c = 0 \, ,
\ee
or in other words that the internal six-form takes the expression
\be
G_6 = c\, e^{-4A} d\mathrm{vol}_{X_6} \, ,
\ee
with $c$ constant.
Notice that the volume form $d\mathrm{vol}_{X_6}$ need not be $-\frac{1}{3} J_{\rm CY}^3$, because the actual internal metric of the solution is not supposed to be Calabi--Yau, even if $X_6$ admits a Calabi--Yau metric. In any case the last relation in \eqref{intflux1} reads
\be
c \int_{X_6} e^{-4A} d\mathrm{vol}_{X_6} = 0 \, .
\label{inte4a}
\ee
This in principle leads to two possibilities: either $c$ or the integral vanishes. Notice however that the integrand is positive definite -- excluding perhaps regions very close to the O6-planes where the supergravity approximation breaks down -- and so should be its integral.\footnote{In practice one may shifts the warp factor by a constant that is absorbed into the AdS$_4$ scale $\mu$, to fix the value of the integral to a certain positive value. We take the simple choice $\langle e^{-4A}\rangle = 1$ in our solution of section \ref{sec:solution}.} Therefore, sensible 10d uplifts of DGKT vacua are those in which the six-form flux $G_6$ (and dual Freund--Rubin flux) must vanish point-wise on $X_6$ 
\be
G_6 = 0 \, .
\label{f6ha0}
\ee
As follows from the discussion of Appendix \ref{ap:SU33}, this condition has a straightforward implication for the poly-forms describing the $SU(3)\times SU(3)$ structure. Namely
\be
\left. \im \Phi_+ \right|_{\mathrm{0-form}} =0\, .
\label{imphi0}
\ee
This simple constraint rules out several candidates for DGKT 10d vacua.

\subsection{No $SU(3)$-structure solution}
\label{sec:nosu3}

As noticed in \cite{Acharya:2006ne}, the relations \eqref{intflux1} are very suggestive from the viewpoint of type IIA flux backgrounds with $SU(3)$ structure, whose most general solution was found in \cite{Behrndt:2004km,Lust:2004ig}. Nevertheless, this particular subcase of $SU(3)\times SU(3)$-structure compactification cannot accommodate a 10d uplift of \cite{DeWolfe:2005uu} unless the orientifold sources are smeared, as we already discussed in the previous chapter. While this is a well-known obstruction, it will prove useful to review it from the present perspective. 

Recall the $SU(3)$-structure solutions (\ref{su3flux}). It is easy to see that the choice $\theta=0$ is reminiscent of the relations \eqref{intflux1}, and in particular that it is compatible with the constraints \eqref{f6ha0} and \eqref{imphi0}. However, this choice is not allowed in the present setup, unless the Bianchi identity is modified by smearing the O6-plane sources. Indeed, it follows from the Bianchi identity for $G_0$ that there both the warp factor and dilaton are constant, from where one obtains that 
\be
d  \im  \Om = i  W_2 \wedge J\, ,
\ee
with $W_2$ a real, primitive (1,1)-form. Then, the Bianchi identity for $G_2$ becomes \cite{Acharya:2006ne,Koerber:2007jb}
\be
e^{-\phi} \left[ \frac{1}{4} |W_2|^2 + e^{-2A} \mu^2 \left( 10\,  {\rm cos}^2 \th - \frac{2}{3}  {\rm sin}^2 \th\right)  \right]  \re \Om = - \delta_{O6}\, .
\label{tf2loc}
\ee
Away from the O6-plane locus, the LHS of \eqref{tf2loc} needs to vanish, which necessarily imposes that $\theta \neq 0$ and a non-vanishing internal flux $G_6$. Therefore, by the requirement \eqref{f6ha0} this cannot be a 10d realisation of \cite{DeWolfe:2005uu}. If $\d_{O6}$ is replaced with a smeared three-form source in the appropriate cohomology class
\be 
\label{eq:smearing}
- \delta_{O6}\, \rightarrow \, G_0H = 10 e^{-\phi-2A} \mu^2   {\rm cos}^2 \th\,  \re \Om 
\ee
such obstruction is gone, and one find solutions with $W_2 = \theta =0$. This would-be solution would have $dJ = d\Omega =0$, and would correspond to a Calabi--Yau metric. The fluxes would read
\begin{equation}\label{eq:SU3th0}
	\begin{split}
		H&= 2 \mu {\rm Re}\, \Omega\, , \\
		G_0&= 5 \mu e^{- \phi} \, ,\qquad  \quad \ G_2 =0\ ,\\
		G_4&= \frac32 \mu e^{- \phi} J^2 \, ,\qquad G_6 = 0 \, . 
	\end{split}	
\end{equation}
Since $A$ is constant, we have set it to zero, reabsorbing it in $\mu$. To see how things scale, assume as in \cite{DeWolfe:2005uu} that $F_0\sim O(1)$ and that the internal space has volume $\cV(X_6) \sim R^6$ in string units. We know already that $\delta \propto {\rm Re}\,  \Omega$; if we take $\delta_{O6} \sim - \frac 1 {R^3} {\rm Re}\, \Omega$, integrating $\delta$ along a 3-cycle gives $O(1)$, as it should. From all this we read
\begin{equation}\label{eq:scales}
	g_s = \frac{5}{m} \hat{\mu} \sim R^{-3}\,,
\end{equation}
with $\hat{\mu} = \mu\ell_s$, in agreement with \eqref{scaling0}.

It has been recently proposed in \cite{Junghans:2020acz} that this Calabi--Yau solution with smeared sources can be seen as the leading order contribution to an expansion in the flux quantity $\hat{e}$ controlling the volume of the compactification. As we will see in section \ref{sec:solution}, this is manifest in the solution that we find for the $SU(3)\times SU(3)$-structure supersymmetry equations, approximated in the large volume regime. Before deriving such equations, let us constrain which kind of $SU(3)\times SU(3)$ structure can describe DGKT microscopically.

\subsection{Setting $\theta = 0$}
\label{sec:nothetaneq0}

Massive type IIA supergravity backgrounds solving the $SU(3)\times SU(3)$-structure supersymmetry equations have been analysed in \cite{Gaiotto:2009mv,Saracco:2012wc}. In particular, in \cite{Saracco:2012wc} two different branches of solutions were identified, as reviewed in section \ref{sub:10d-susy} and Appendix \ref{ap:SU33}. These two branches are described in terms of the function $\theta$ defined in section \ref{sub:10d-susy}, which  appears in the pure spinors as in \eqref{su3phi}. One branch has $\theta=0$ (see section \ref{sub:10d-susy}) and the other one has non-vanishing, varying $\theta$ (see section \ref{sub:tneq0}).

Our discussion above implies that the branch with $\theta=0$ should be more suitable to describe DGKT. Indeed, given \eqref{su3phi} one can rewrite \eqref{imphi0} as $\theta = 0$ or $\pi$. Accordingly, one can show that after the B-field transformation \eqref{eq:GF-t0} one obtains $G_6=0$ from \eqref{eq:flux-t0}, see Appendix \ref{ap:SU33}.  The compatibility of this branch with the presence of O6-planes seemed unlikely from the symmetry arguments used in \cite{Saracco:2012wc}. However, in the following sections we will see that supersymmetry equations for the case $\theta=0$ are rich enough to host localised and smeared sources at the same time. In this sense, what our results of the following sections suggest is that the 10d description of \cite{DeWolfe:2005uu} consists of a $SU(3)\times SU(3)$-structure background with $\theta=0$ that at large volumes can be approximated by an $SU(3)$-structure background with $\theta=0$. Indeed, we will see that the background that we find can be organised as a perturbative expansion on the small parameter $g_s \sim \cV_{X_6}^{-1/2}$. The zeroth order contribution is nothing but the background \eqref{eq:SU3th0}.

The branch in which $\theta$ is non-vanishing is a priori not suitable to describe the 10d uplift of DGKT. First, as reviewed in section \ref{sub:tneq0}, such a solution has a varying $G_6$ flux, which prevents it to satisfy the point-wise constraint \eqref{f6ha0}. In addition, this sort of backgrounds are characterised by an NS three-form flux $H$ which is exact. As this implies vanishing $H$-flux quanta, it can never describe the global features of a DGKT vacuum. Finally, a crucial aspect of this branch is that $\left. \im \Phi_+ \right|_{\mathrm{0-form}} \neq 0$, and in fact it is not even constant. The aim of \cite{Saracco:2012wc} was to find a solution that only asymptotes to $\left. \im \Phi_+ \right|_{\mathrm{0-form}} = 0$, but we have shown that this must hold locally anywhere on $X_6$ where a 10d supergravity description is reliable. Therefore it seems unlikely the solution in \cite{Saracco:2012wc} can be part of a 10d description of  DGKT.

This being said, let us stress that the approximate solution that we find in section \ref{sec:solution} breaks down near the O6-plane loci. In particular in those regions we find the same metric singularities associated with O6-planes in flat space, featuring a divergent negative warp factor $e^{-4A}$. In the case of flat space it is known that the divergent negative warp factor around the O6-plane is resolved by string theory corrections, uplifting the solution to M-theory on a $G_2$ manifold with an Atiyah-Hitchin metric on the former O6-plane region \cite{Gibbons:1986df, Sen:1997kz, Seiberg:1996nz, Hanany:2000fw}. In the case at hand we are dealing with massive type IIA string theory and therefore we lack an M-theory description, so it would be very interesting to understand how the theory resolves such a singularity. One possibility could be that the full solution with $\theta =0$ does not have any singularity. This would be quite analogous to the result found in \cite{Saracco:2012wc} for the $\theta \neq0$ branch. Indeed, there it was shown that such massive type IIA solutions with O6-planes can resolve the O6-plane singularity  without resorting to an M-theory description. As these belong to a different branch of $SU(3)\times SU(3)$-structure backgrounds, we will take an agnostic approach towards this possibility.

\subsection{The source balanced equation}
\label{sec:SBE}

If the obstruction for $SU(3)$-structure solutions can be circumvented by $SU(3)\times SU(3)$-structure backgrounds with $\theta=0$ a natural question is how the equation \eqref{tf2loc} leading to the obstruction is modified. In the following we would like to present a generalisation of eq.\eqref{tf2loc}, valid for any $SU(3)\times SU(3)$-structure background, which we dub {\em source balanced equation}. 

For this we first need to introduce the Mukai pairing
\be
\left< \omega_1 , \omega_2 \right> \equiv \left. \omega_1 \wedge \lambda \left( \omega_2 \right) \right|_{\mathrm{top}}\, ,
\ee
for the poly-forms $\omega_1$ and $\omega_2$, where $|_{\rm top}$ indicates that we only extract the top form of the product. The source balanced equation then reads
\be
3\mu^2 e^{-4A} \left< \mathrm{Re\;}\Phi_+ , \mathrm{Im\;}\Phi_+ \right>  - e^{4A} \sum_k G_k \wedge \star G_k + dX_5 
=  \left< \delta^{(3)}_{\mathrm{source}} , e^A  \mathrm{Im\;}\Phi_- \right> \; ,
 \label{preGOE}
\ee
where $\delta^{(3)}_{\mathrm{source}} = \sum_\a \delta (\Pi_\alpha)$ contains all the localised sources wrapping three-cycles, both O6-planes and D6-branes. In addition
\be
X_5 \equiv \left<  e^A \mathrm{Im\;}\Phi_-, {\bf G} \right> =   -G_2 \wedge \left( e^A \mathrm{Im\;}\Phi_- \right)_{3}  + G_4 \wedge \left( e^A \mathrm{Im\;}\Phi_- \right)_{1} + G_0 \left( e^A \mathrm{Im\;}\Phi_- \right)_{5}   \;,
\label{X5}
\ee
where the subscripts denote the degree of the form to be picked out. 

Eq.\eqref{preGOE} is derived in Appendix \ref{ap:SBEproof} by using the Bianchi identities and the supersymmetry equations. Notice that it takes a similar form to \eqref{tf2loc} in the sense that the left-hand side is supported over the whole manifold, while the right-hand side is localised. One can see this relation as a generalisation of (\ref{tf2loc}), in which case one had $X_5=0$. Indeed, recall that in the SU(3)-structure case $ \left( \mathrm{Im\;}\Phi_- \right)_{1} = \left( \mathrm{Im\;}\Phi_- \right)_{5} = 0$ and that $G_2$ is a (1,1)-form. In section \ref{ss:comparison} we will test our solution  with this equation, to see in which way \eqref{tf2loc} is modified to allow for a consistent $SU(3)\times SU(3)$-structure solution.

\section{The large volume approximation}
\label{sec:approx}

We now consider the BPS equations in a limit where the cosmological constant is small, aiming for a solution similar to (\ref{eq:SU3th0}) but \emph{without smearing}. We will do this by taking the parameter $\hat\mu = \mu\ell_s$ in (\ref{eq:psp}), (\ref{eq:calJ}) to be small; recalling (\ref{eq:scales}) $g_s$ will then also be small and $R$ large. 

\subsection{Defining the limit} 
\label{sub:def-lim}

As we have seen in section \ref{sec:nogosol}, the smeared solution comes from an $SU(3)$-structure, but the solution with localised O6-planes that we are looking for cannot. As discussed around (\ref{eq:JO-jo}), the function $\psi$ interpolates between $SU(3)$ and $SU(3)\times SU(3)$. So in the limit we also have to take the function $\psi$ to be at least of order $\hat\mu \sim R^{-3}\sim g_s$ at leading order, recalling (\ref{eq:scales}). In addition, from (\ref{eq:SU3th0}) we see that for $F_0$ to stay non-zero in the limit we need $e^\phi \to 0$. On the other hand, since we are already making $\hat\mu\to0$, $e^A$ should not scale. For simplicity in the following we will  fix $\langle e^{-4A}\rangle = 1$.

A limit with all these features was originally devised in \cite{Saracco:2012wc}, exactly for the solution at hand. As we commented earlier, there the focus was on the local behaviour, and we have argued above that solution cannot capture the global solution, essentially because $\theta\neq 0$ was taken there. In the limit, the problem presented itself already in eq.\cite{Saracco:2012wc}; it was noted below eq.(6.12) in that paper that $\mathrm{Re}\,  \Omega$ has to be exact at leading order, and that this could be an obstruction to finding a global solution.

Nevertheless, we can still apply the same ideas of \cite{Saracco:2012wc} to the $\theta=0$ case. Notably, it was decided there to expand in $\mu$, but with a subleading behaviour that is either an even or odd function of $\mu$. This was found to simplify the equations significantly, and it was inspired in turn by a similar limit in \cite{Gaiotto:2009yz} where $\psi \to 0$ but the cosmological constant remained fixed. 

Implementing this strategy in our case leads us to taking $g_s\to 0$, with the following Ansatz: 
\begin{equation}\label{eq:limit}
\begin{split}
	&\hat{\mu} = \frac m5 g_s\, ,\qquad \psi = g_s \psi_1 +  \cO(g_s^3) \, ,\qquad \theta=0 \, ,\\
	&e^\phi = g_s e^{\delta\phi_0 + g_s^2 \delta\phi_2 + \cO(g_s^4)}   \, ,\qquad e^A = e^{A_0 + g_s^2 A_2 + \cO(g_s^4)}\,.
\end{split}	 
\end{equation}

It is important to stress that the equations will fix the coefficients of the expansion as a function of the coordinates, in such a way that some extra powers of the parameter $g_s\sim R^{-3}$ will appear. For example we will find below that
\begin{equation}
    e^{A_0}\sim a_0 + a_1 R^{-4} \sim a_0 + a_1 g_s^{4/3}\,.
\end{equation}
This might look confusing, but the method is sensible as long as these ``hidden'' powers of $g_s$ are not smaller with respect to terms we have ignored in (\ref{eq:limit}). The same comment applies to the expansion of the forms, to which we now turn.

\subsection{Forms and fluxes} 
\label{sub:lim-forms}

We now have to decide how to scale the forms. $\mathrm{Re}\, v$ is already determined by (\ref{eq:revt0}). Due to the $\mu$ in the denominator of that equation,  (\ref{eq:limit}) would imply that $\mathrm{Re} v \sim -\frac{5e^A}{2g_s m}d(3A_0- \delta \phi_0)$, whereas as we explained above we would like $\mathrm{Re}\, v\to 0$ in the limit. For this reason we take 
\begin{equation}
    \delta \phi_0 = 3 A_0\,.
\end{equation}
Now we obtain
\begin{equation}\label{eq:Rev1-t0}
	\mathrm{Re}\, v= g_s\mathrm{Re}\, v_1 + \cO(g_s^3) \, ,\qquad \mathrm{Re}\, v_1=\frac12 e^{A_0} d f_\star \,,\qquad  f_\star \equiv -\frac5m \left(3A_2 - \delta\phi_2 - \frac12 \psi_1^2\right)\,.
\end{equation}
As for $j$ and $\omega$, we want them to reconstruct in the limit an $SU(3)$-structure $(J,\Omega)$. We will simply assume here the latter to be fixed, and $(j,\omega)$ to be determined by (\ref{eq:JO-jo}). 
We don't know whether this assumption is really warranted at higher orders in our expansion, but up to the order of our computations we will see no difference. All this leads to 
\begin{equation}\label{eq:lim-JO}
	J_\psi = J + \cO(g_s^2)  \, ,\qquad \Omega= \frac i{\psi_1} v_1 \wedge \omega_0 + \cO(g_s^2)\,. 
\end{equation}
In fact we will be able to write everything in terms of $v$ and the fixed $(J, \Omega)$. 
It should be remarked that we are not assuming these to be those of the underlying Calabi--Yau, since we are aiming at removing the smearing. 

$\Omega$ still defines an \emph{almost} complex structure $I$: we mentioned in section \ref{sub:10d-susy} that it is at every point the wedge product of three one-forms $h^a$, which are then defined to be the $(1,0)$-forms of $I$. In fact we see from (\ref{eq:lim-JO}) that $v_1$ is one of these $(1,0)$-forms, and we can use this to determine $\mathrm{Im}\, v_1$ in the expansion $\mathrm{Im}\, v\sim g_s \mathrm{Im}\, v_1 + O(g_s^3)$.  But we are not assuming $I$ to be integrable; this would be implied by $d \Omega=0$, which is not part of the equations we found in section \ref{ssub:t0}. On the other hand, at leading order (\ref{eq:dJpsit0}) simply becomes 
\begin{equation}\label{eq:dJ0}
	dJ=0\,.
\end{equation}

The metric is not really needed to find a solution; it is determined by the forms of the $SU(3)\times SU(3)$-structure. The procedure comes originally from generalised complex geometry \cite{gualtieri}, and was explained in detail in \cite[Sec.~2.2.2]{Saracco:2012wc}. Fortunately at the leading order we are working with, the procedure reduces to the simpler one for $SU(3)$-structures, which we will illustrate in an example later on. In terms of this metric, one can invert the relation for $\Omega$ in (\ref{eq:lim-JO}) with a contraction:
\begin{equation}
	\omega_0=-\frac i{2 \psi_1} \bar v_1\cdot \Omega\,. 
\end{equation}

One last comment about the geometric forms: we are taking the volume of the internal space $X_6$ to be $\mathrm{Vol}(X_6)\sim R^6$, but we are taking care of that by scaling coordinates rather than the metric and forms. So for example for the torus cases below, the periodicities of the internal coordinates will scale like
\begin{equation}
	\Delta y \sim R\,.
\end{equation}
One can of course easily always switch to another point of view, where the coordinates don't rescale and forms do; this would lead to $J\sim R^2 J_0$, $\Omega\sim R^3 \Omega_0$. We take this viewpoint in the explicit example of section \ref{ss:torus}.

Applying the above procedure to \eqref{eq:flux-t0} we obtain the following relations between the fluxes and the $SU(3) \times SU(3)$-structure forms:
\begin{subequations}
\label{eq:leading-fluxes}
\begin{align}
	\label{Hleading}
	\hat H&= \frac25 F_0 g_s e^{-A_0} \mathrm{Re}\, \Omega + \cO(g_s^3)\,,\\
	\label{F0leading}
	F_0 &=  F_0 e^{-4A_0} - J\cdot d(e^{-3A_0}\mathrm{Im}\, v_1) + \cO(g_s^2) \,,\\
	\label{F2leading}
	F_2 &= -\frac{1}{g_s}  J \cdot d(e^{-3A_0} \mathrm{Im}\, \Omega) + \cO(g_s)\,,\\
	\label{F4leading}
	 F_4 &= F_0 J^2 \left(\frac12 - \frac{1}{5} e^{-4A_0}\right) +  J \wedge d(e^{-3A_0} \mathrm{Im}\, v_1) + \cO(g_s^2)\,,\\
	 F_6 &= 0\,.
\end{align}
\end{subequations}
Where recall that $F_0 = \ell_s^{-1} m$. Using (\ref{eq:GF-t0}) we find the physical fluxes
\begin{equation}
	H= \hat H + g_s d(\psi_1 \mathrm{Im}\, \omega_0) + \cO(g_s^3)\, ,\qquad
	\mathbf{G} = e^{ (g_s\psi_1 \mathrm{Im}\, \omega_0 + \cO(g_s^3))\wedge} \mathbf{F}\,.
\end{equation}
Notice that $(\mathrm{Im}\, \omega_0)^3=0$, so the exponential truncates. 

To summarise, in order to find a solution in this limit we need to find an $SU(3)$-structure $(J,\Omega)$, a $(1,0)$-form $v_1$, and a funcntion $A_0$, such that $e^{A_0}{\rm Re}\, v_1$ is exact, $J$ is closed  ((\ref{eq:Rev1-t0}),  (\ref{eq:dJ0})). When plugged into (\ref{eq:leading-fluxes}) these should provide an expression for the fluxes that solves the Bianchi identities, up the order of the approximation.


\section{Solving the Bianchi identities}
\label{sec:Bianchi}

In this section we solve exactly the Bianchi identities for the internal sources that correspond to \cite{DeWolfe:2005uu} and to the vacua derived in chapter \ref{ch:rrnsnsvacua}. For this we consider a manifold $X_6$ that admits a Calabi--Yau metric, namely a metric of $SU(3)$ holonomy, so that we can have a 10d interpretation of the sources. Looking at the first relation in \eqref{intflux1} and the expression for $H$ in terms of $SU(3)\times SU(3)$ structures with $\theta=0$ one infers that it must be of the form
\be
H =  2\mu \re \Omega_{\rm CY} + d\tilde{B}\, ,
\label{Hini}
\ee
with $5 \mu  \langle g_s^{-1}\rangle= G_0$.  Let us for now set $\tilde{B} = 0$ and solve the Bianchi identities in this case, and then recover the general solution by applying a $B$-field transformation. For the particular solution we denote the RR fluxes by $\tilde{F}_{2p}$. 

The Bianchi identity for the two-form flux reads
\be
  \ell_s^2 d\tilde{F}_2 =  2 m \hat{\mu} \re \Omega_{\rm CY} -4  \d(\Pi_{\rm O6}) \, , 
  \label{BIF2ini}
\ee
with $\hat{\mu} = \mu\ell_s$. By Hodge decomposition the most general solution is of the form
\be
 \tilde{F}_2 =   d^\dag_{\rm CY} K + \tilde{F}_2^{\rm h}  + dC_1\, ,
\label{F2K}
\ee
with $dC_1$ exact, $\tilde{F}_2^{\rm h}$ Calabi--Yau harmonic, and $d^\dag_{\rm CY}$ constructed with the Calabi--Yau metric. 
Finally, $K$ is a 3-form current that always exists, as it satisfies the following Laplace  equation
\be
\ell_s^2 \Delta_{\rm CY} K =  2m\hat{\mu}\re \Omega_{\rm CY} -4 \d(\Pi_{\rm O6})\, ,
\label{defK1}
\ee
where $\Delta_{\rm CY} = d^\dag_{\rm CY} d + d d^\dag_{\rm CY}$ is constructed from the CY metric. Indeed, following  \cite[sec.~3.4]{Hitchin:1999fh} notice that $d\re \Omega_{\rm CY} = d \, \d(\Pi_{\rm O6}) = 0$, $\Delta_{\rm CY} d K =0$ and $dK$ is harmonic. Because it is also exact, then necessarily $dK=0$. We conclude that $\Delta_{\rm CY} K = dd^\dag_{\rm CY} K$ from where \eqref{F2K} follows.  One can then constrain $K$ by using that $J_{\rm CY}$, $\Omega_{\rm CY}$ are covariantly constant with respect to $\Delta_{\rm CY}$: 
\begin{align}
\Delta_{\rm CY} K \wedge J_{\rm CY} = 0 &  \Rightarrow  \Delta_{\rm CY}(K \wedge J_{\rm CY}) = 0 \ \Rightarrow \  K \wedge J_{\rm CY} = 0\, , \\
\Delta_{\rm CY} K \wedge \re \Om_{\rm CY} = 0  & \Rightarrow  \Delta_{\rm CY}(K \wedge \re \Om_{\rm CY}) = 0  \Rightarrow   K = \varphi \re \Om_{\rm CY} + c \im \Om_{\rm CY} + \re k ,
\label{formK1}
\end{align}
with $\varphi$ a real function, $c$ a constant that we will take to be zero, and $k$ a (2,1) primitive current. Here we have used that there are no harmonic 5-forms in the CY metric. One then obtains that
\be
\label{F2H}
 d^\dag_{\rm CY} K  =   \star_{\rm CY} \left( d\varphi \wedge \im  \Om_{\rm CY} \right) -  \star_{\rm CY} d \im   k  = - J_{\rm CY} \cdot d \left(2 \varphi \im  \Om_{\rm CY} \right) - V_2  \, ,
\ee
where $V_2$ is a primitive (1,1)-form in the CY sense. One can check that this implies that
\be
\tilde{F}_2 = - J_{\rm CY} \cdot d( 4 \varphi \im \Om_{\rm CY} - \star_{\rm CY} K)  + \tilde{F}_2^{\rm h}  + dC_1 \, .
\label{keyimOm}
\ee

As for the remaining fluxes, it is easy to see that
\be
\tilde{F}_4 =   \tilde{F}_4^{\rm h} - 4 \mu \varphi \, J_{\rm CY} \wedge J_{\rm CY} +  2\mu \re \Omega_{\rm CY} \wedge C_1 + dC_3  \, ,
\label{F4H}
\ee
with $dC_3$ exact, $\tilde{F}_4^{\rm h}$ Calabi--Yau harmonic, satisfies the Bianchi identity $d\tilde{F}_4 = 2\mu \re \Omega_{\rm CY} \wedge \tilde{F}_2$. As for the six-form flux, we can set $\tilde{F}_6 = dC_5$ to be an exact form. 

Finally, whenever $\tilde{B}$ in \eqref{Hini} is not trivial, the solution for the fluxes will be given by
\be
\mathbf{G}= e^{\tilde{B} \wedge} \mathbf{\tilde{F}}\, ,
\ee
with $\tilde{F}_0 = G_0$ and the remaining $\tilde{F}_{2p}$ as specified.

\section{Solving the supersymmetry equations}
\label{sec:solution}

Thanks to our previous results, in this section we will be able to give a 10d supersymmetric background describing the  relations \eqref{intflux1} for any manifold $X_6$ that admits a Calabi--Yau metric. Our strategy will be simple: we will  provide expressions for $\Om$, $J$, $e^{A_0}$ and $v_1$ in terms of Calabi--Yau quantities, such that when plugged in \eqref{eq:leading-fluxes} provide backgrounds fluxes solving the Bianchi identities up to the appropriate order of the expansion. Because of that, our background can only be thought of as an approximation to an actual supersymmetric solution describing a 10d counterpart of \cite{DeWolfe:2005uu}. This approximation becomes more accurate in the limit of large volume and weak coupling, approaching the $SU(3)$-structure smeared solution in that limit.

\subsection{General Calabi--Yau manifolds}

Since by assumption $X_6$ admits a Calabi--Yau metric, we can profit from the discussion in section \ref{sec:Bianchi} as a guiding principle to construct the $SU(3)\times SU(3)$-structure metric in $X_6$. First, as the two-form $J$ is closed, we will assume that
\be \label{eq:J0=JCY}
J = J_{\rm CY} +  \cO(g_s^2)\, ,
\ee
where recall that $g_s = 5\mu/m = 5 \cV_{X_6}^{-1/2} / m$. Then, one can guess the form of $\im \Om$ by comparing \eqref{F2leading} and \eqref{keyimOm}. Indeed, let us consider the following expression
\be 
e^{-3A_0} \im \Om = \left(1 +  g_s 4 \varphi \right)  \im \Omega_{\rm CY} - g_s \star_{\rm CY} K + \cO(g_s^2)\, ,
\label{AimOm}
\ee
with $K$ an exact three-form defined by \eqref{defK1} and $\varphi$ defined by \eqref{formK1}. Plugging this into \eqref{F2leading} one obtains \eqref{keyimOm} with $\tilde{F}_2^{\rm h}= dC_1 =0$. Therefore with the choice \eqref{AimOm}, $F_2$ in \eqref{F2leading} satisfies  \eqref{BIF2ini}. Finally, such an $F_2$ also satisfies the Bianchi identity up to $\cO(g_s^2)$ terms  if we assume that $\tilde{B} \sim \cO(g_s^2)$ in \eqref{Hini}, as we will do in the following.

From here, one may construct the rest of $\Omega$. In general its real and imaginary parts are related by a method in \cite{Hitchin:2000jd} and reviewed in \cite[Sec.~3.1]{Tomasiello:2007zq}. In particular here we can consider ${\rm Im}\,\Omega$ as a perturbation over ${\rm Im}\,\Omega_{\rm CY}$, and apply the perturbation formulas \cite[(3.8)--(3.10)]{Tomasiello:2007zq}. One finds that
\be
e^{-3A_0} \re \Om  =\left( 1 + g_s 2\varphi \right) \re \Om_{\rm CY} + g_s K +\cO(g_s^2)\, ,
\ee
which satisfies the $SU(3)$-structure relation $\re \Om \wedge \im \Om = \frac{2}{3} J^3$ provided that
\be 
e^{-4A_0} =  1 + g_s 4 \varphi + \cO(g_s^2)\, .
\label{Awarp}
\ee
From the definition of $\varphi$ it is easy to see that $\int_{X_6} \varphi = 0$ and therefore $\langle e^{-4A} \rangle = 1$ up to this order of approximation, as expected. This moreover leads to
\be
e^{-A_0} \re \Om  = \re \Om_{\rm CY} + g_s K + \cO(g_s^2)\, ,
\label{AreOm}
\ee
and so, since $dK=0$, the Bianchi identity $dH = 0$ is satisfied up to order $\cO(g_s^3)$. Finally, when plugging \eqref{AreOm} into \eqref{Hleading} we obtain that $d\tilde{B} =  2\mu g_s K +  \cO(g_s^2)$, consistently with our assumption. We finally obtain
\bea \label{eq:Omega0-def}
\Omega & = & 
\Omega_{\rm CY}  + g_s  k + \cO(g_s^2) \, ,
\eea
where $k$ the primitive  (2,1)-from $k$ defined by \eqref{formK1}. Notice that this expression is compatible with  SU(3)-structure torsion classes, since
\be
d\Om = g_s dk = - d\varphi \wedge \Om_{\rm CY} + i g_s V_2 \wedge J = dA_0 \wedge \im \Om + i W_2 \wedge J +\cO(g_s^2)\, ,
\ee
with $W_2 = g_s V_2 +\cO(g_s^2)$. So the leading correction to the pair $(J, \Omega)$ corresponds to a non-Ricci-flat, symplectic metric with SU(3)-structure.

With this choice of $(J, \Om)$ and warp factor it is easy to accommodate the remaining expressions in \eqref{eq:leading-fluxes} to satisfy the Bianchi identities with $\tilde{B} \sim  \cO(g_s^2)$. Indeed, to fit \eqref{F4leading} into \eqref{F4H} one simply needs to take
\be \label{eq:C3-pert}
\tilde{F}_4^{\rm h} = \frac{3}{10} F_0\, J_{\rm CY} \wedge J_{\rm CY} \, , \qquad C_3 = e^{-3A} J_{\rm CY}\wedge \im v_1 \, .
\ee
Finally, the Bianchi identity for $F_0$ is compatible with the RHS of \eqref{F0leading} if one takes $\im v_1$ to be the imaginary completion of 
\be \label{eq:v1pert}
\re v_1 = \oh e^{A_0} d f_\star \, ,
\ee
with
\be
\ell_s \Delta_{\rm CY} f_\star  = - g_s m 8 \varphi  +  \cO(g_s^2) \, .
\label{defstar}
\ee
In other words, $\im v_1 = I\cdot \re v_1$, with $I\cdot$ the action of the complex structure. At the leading order of the  expansion, this implies that $v_1 =  \p_{\rm CY} f_\star$. It follows from this result that the $B$-field transformation in \eqref{eq:GF-t0} is suppressed by $g_s^2$ and does not induce any change in the fluxes at the present order.
In the following we will discuss in detail how this approximate solution looks like in the case of a toroidal orbifold, where the above expressions can be made more explicit.

In summary, our solution is specified by the $SU(3)$-structure given in \eqref{eq:J0=JCY}, \eqref{eq:Omega0-def}; the one-form $v_1$ specified by \eqref{eq:v1pert}; and the warping function $A_0$ in \eqref{Awarp}. The Bianchi identity were shown to be solved for $F_2$ in \eqref{AimOm}; for $H$ in \eqref{eq:Omega0-def}; for $F_0$ in \eqref{defstar}; for $F_4$ in \eqref{eq:C3-pert}. By the general results of \cite{Koerber:2007hd}, once the supersymmetry equations and Bianchi identities are satisfied, the equations of motion for the fields are also solved.

\subsection{A toroidal orbifold example}
\label{ss:torus}

Let us consider the particular case where $X_6 = T^6/\IZ_2\times\IZ_2$, as in \cite{Camara:2005dc}. We consider the choice of discrete torsion that makes it T-dual to the closed string background in \cite{Berkooz:1996km}, so that all O6-planes have negative charge and tension. In the orbifold limit, the Calabi--Yau structure is essentially inherited from the covering space $T^6$, so we can write
\bea
J_{\rm CY} & = & 4\pi^2 t_i dx^i \wedge dy^i \, ,\\
\re \Om_{\rm CY} & = &  h\left(\tau_1\tau_2\tau_3 \b^0 - \tau_1 \b^1 - \tau_2 \b^2 - \tau_3 \b^3 \right) \, ,\\
\im \Om_{\rm CY} & = &   h\left( \a_0 - \tau_2\tau_3 \a_1 - \tau_1\tau_3 \a_2 - \tau_1\tau_2 \a_3 \right) \, ,
\eea
where
\be
t^i  =  R_{x^i} R_{y^i}\, , \qquad \tau_i = \frac{R_{y^i}}{R_{x^i}}\, , \qquad h =  8\pi^3 \sqrt{\frac{t_1 t_2 t_3}{\tau_1\tau_2\tau_3}} =  8\pi^3 R_{x^1}R_{x^2}R_{x^3 }\, ,
\ee
and we have the following basis of bulk three-forms
\bea\nonumber
\a_0 = dx^1 \wedge dx^2 \wedge dx^3\, , & \quad & \b^0 = dy^1 \wedge dy^2 \wedge dy^3 \, ,\\ \nonumber
\a_1 = dx^1 \wedge dy^2 \wedge dy^3\, , & \quad & \b^1 = dy^1 \wedge dx^2 \wedge dx^3 \, ,\\ \nonumber
\a_2 = dy^1 \wedge dx^2 \wedge dy^3\, , & \quad & \b^2 = dx^1 \wedge dy^2 \wedge dx^3 \, ,\\ \nonumber
\a_3 = dy^1 \wedge dy^2 \wedge dx^3\, , & \quad & \b^3 = dx^1 \wedge dx^2 \wedge dy^3 \, .
\eea
In principle one can consider partially cancelling the charge of the O6-planes with D6-branes on top of them, and so different choices of $H$-flux that will cancel the corresponding generalisation of the tadpole condition \eqref{tadpole1}. For simplicity, we will consider those cases where the $H$-flux is of the form 
\be
\ell_s [H] = 8 q\left([\b^0] - [\b^1] - [\b^2] -[\b^3] \right) \, ,
\ee
for some choice of $q \in \IZ$, with $q m = 4$ in the particular case where no D6-branes are present. Then supersymmetry requires that
\be
\tau_1=\tau_2=\tau_3 = 1\, , \quad \text{and} \quad \hat{\mu} = \frac{4q}{h}\, ,
\ee
from where it is clear that $\hat{\mu} \sim \cV^{-1/2}_{X_6}$. In this setup we find a solution for \eqref{defK1} of the form
\be
K =  qm  \left(B_0  \b^0 - B_1  \b^1 - B_2 \b^2 - B_3  \b^3 \right)\, ,
\ee
with
\bea
\label{Bs}
B_0 & = & - h^{2/3}  \sum_{\vec{\eta}} \sum_{\vec{0}\neq\vec{n}\in \IZ^3} \frac{e^{2\pi i  \vec{n}\cdot \left[(y^1, y^2,y^3)+\vec\eta\right]}}{4\pi^2\vec{n}^2} \, , \
 B_1  =  - h^{2/3} \sum_{\vec{\eta}} \sum_{\vec{0}\neq\vec{n}\in \IZ^3} \frac{e^{2\pi i  \vec{n}\cdot \left[(y^1, x^2,x^3)+\vec\eta\right]}}{4\pi^2\vec{n}^2} \, , \qquad   \\ \nonumber 
 B_2 & = & -  h^{2/3} \sum_{\vec{\eta}}  \sum_{\vec{0}\neq\vec{n}\in \IZ^3} \frac{e^{2\pi i  \vec{n}\cdot \left[(x^1, y^2,x^3)+\vec\eta\right]}}{4\pi^2\vec{n}^2} \, , \
 B_3  =  - h^{2/3} \sum_{\vec{\eta}} \sum_{\vec{0}\neq\vec{n}\in \IZ^3} \frac{e^{2\pi i  \vec{n}\cdot \left[(x^1, x^2,y^3)+\vec{\eta}\right]}}{4\pi^2\vec{n}^2} \, , \qquad 
\eea
where for simplicity we have set $R_{y^1} = R_{y^2} = R_{y^3} = R$,\footnote{Otherwise one should replace $\vec{n}^2/R^2$ by $|\vec{n}|^2 = \sum_i \left(n_i/R_{y^i}\right)^2$.} and $\vec\eta$ has entries which are either 0 or $1/2$. Notice that $d(B_i\b^i) = 0\, \forall i$ so that $K$ is closed, and in fact exact.

It is important to point out that the expansion for the $B_i$'s in terms of Fourier modes should be understood as a formal solution since the sum is not convergent. A regularised version of these Green functions using the Jacobi theta function was originally suggested in \cite{Shandera:2003gx} and some details have been recently studied in \cite{Andriot:2019hay}. For practical purposes, the regularised functions behave as standard Green functions in flat space when approximating the source and go to zero as we move away. 

Following the Calabi--Yau general discussion, can rewrite things as \eqref{formK1} with 
\be
\varphi  =  \frac{qm}{4h}  \sum_{i=0}^3 B_i \, ,\qquad \re k   =  K - \varphi \re \Om^{\rm CY} \, .
\ee
Notice that $\varphi \sim \cO(R^{-1})$ but it is not suppressed by an extra factor of $g_s$. Eq.\eqref{AimOm} becomes 
\be
e^{-3A_0} \im \Om = C_0 \a_0 - C_1 \a_1 - C_2 \a_2 - C_3 \a_3 + \cO(g_s^2)\, ,
\label{AimOmT}
\ee
with $C_i = h - g_s q m \left(B_i - \sum B_i\right)$. Stability techniques from \cite{Hitchin:2000jd} (reviewed for example in  \cite[Sec.~3.1]{Tomasiello:2007zq}) tell us when a three-form can be the imaginary part of a decomposable form $e^{-3A_0}\Omega$, and what the real part is. For \eqref{AimOmT} this tells us that $\Omega$  exists in regions where $C_0 C_1 C_2 C_3 > 0$, and determines
\bea
e^{-A_0} \re \Om &= & h^{2/3}  \left(C_0C_1C_2C_3\right)^{1/3} \left[C_0^{-1} \b^0  - C_1^{-1} \b^1  -C_2^{-1} \b^2  -C_3^{-1} \b^3   \right] + \cO(g_s^2)  \\ \nonumber
& = & \re \Om_{\rm CY} +  g_s K + \cO(g_s^2)\, .
\eea
where we have used that from imposing the relation $\re \Om \wedge \im \Om = \frac{2}{3} J_{\rm CY}^3$ one obtains
\be
e^{-4A_0} = h^{-4/3} \left(C_0C_1C_2C_3\right)^{1/3} =  1 + 4g_s \varphi +  \cO(g_s^2)\, ,
\ee
in agreement with \eqref{Awarp} and \eqref{AreOm}. By combining all these expressions we obtain
\be
\Om = i  e^{3A_0} C_0  \left[ dx^1 + i  \left(\frac{C_2C_3}{C_0C_1}\right)^{1/2}\hspace{-.3cm}  dy^1 \right] \wedge  \left[ dx^2 + i  \left(\frac{C_1C_3}{C_0C_2}\right)^{1/2}\hspace{-.3cm}  dy^2 \right] \wedge  \left[ dx^3 + i  \left(\frac{C_1C_2}{C_0C_3}\right)^{1/2}\hspace{-.3cm}  dy^3 \right] \, ,
\ee
which corresponds to the metric
\bea
ds^2 & = & h^{2/3} \left[\left(\frac{C_0C_1}{C_2C_3}\right)^{1/2}(dx^1)^2 + \left(\frac{C_0C_2}{C_1C_3}\right)^{1/2}(dx^2)^2  + \left(\frac{C_0C_3}{C_1C_2}\right)^{1/2}(dx^3)^2\right] \\ \nonumber
& + &  h^{2/3} \left[\left(\frac{C_2C_3}{C_0C_1}\right)^{1/2}(dy^1)^2 + \left(\frac{C_1C_3}{C_0C_2}\right)^{1/2}(dy^2)^2  + \left(\frac{C_1C_2}{C_0C_3}\right)^{1/2}(dy^3)^2\right] + \cO(g_s^2)\, .
\eea
Regarding the fluxes, we find that
\bea
H & = & \frac{2}{5}F_0 g_s \left(\re \Om_{\rm CY} + g_s K \right) - g_s\oh   d\re \left(\bar{v}_1 \cdot \Omega_{\rm CY} \right) + \cO(g_s^3)  \, , \\
 F_2 & = &  d^{\dag}_{\rm CY} K  + \cO(g_s)  \, , \\
F_4 & = & F_0 J_{\rm CY}^2 \left(\frac{3}{10}  - \frac{4}{5} g_s \varphi \right)  + J_{\rm CY} \wedge d \im v_1 + \cO(g_s^2) \, ,
\eea
where $\im v_1$ is the imaginary completion of
\be
2 e^{-A_0} \re v_1  =  df_\star + \cO(g_s^2) \, , \qquad {\rm with} \qquad 
f_\star  =  - \ell_s g_s \frac{2qm^2}{h} \sum_i \tilde{B}_i\, .
\ee
where $\tilde{B}_i$ stand for the functions $B_i$ in \eqref{Bs} with the replacement $R^2/\vec{n}^2 \rightarrow R^4/|\vec{n}|^4$. Note that, unlike the $B_i$, the $\tilde{B}_i$ can be shown to be convergent, so there is no need to regularise them.

\subsection{Comparison with the smeared solution}
\label{ss:comparison}

Let us summarise our approximate solution. We obtain that the background fluxes are given by $\ell_s G_0 = m$ and
\begin{subequations}
	\label{solutionflux1}
\begin{align}\nonumber
H & =  g_s \frac{2}{5}G_0 \left(\re \Om_{\rm CY} + g_s K \right) - \oh   d\re \left(\bar{v} \cdot \Omega_{\rm CY} \right) + \cO(g_s^{3})= g_s \frac{2}{5}G_0 \re \Om_{\rm CY}  \left( 1+ \cO(g_s^{4/3})\right) \, , \\
G_2 & =   d^{\dag}_{\rm CY} K  + \cO(g_s)  = \cO( g_s^{2/3}) \, , \\
G_4 & =  G_0 J_{\rm CY}^2 \left(\frac{3}{10}  - \frac{4}{5} g_s \varphi \right) + J_{\rm CY} \wedge g_s^{-1}d \im v  + \cO(g_s^2) = \frac{3}{10} G_0 J_{\rm CY}^2 \left( 1+ \cO(g_s^{4/3})\right)\, , \\
G_6 & = 0\, ,
\end{align}
\end{subequations}   
where  $g_s = 5 \cV^{-1/2}_{X_6}/m$, $K$ is defined by \eqref{defK1} and the proof of $G_6=\left( e^{b}\bf F\right)|_6= 0$ is given in appendix \ref{sub:teq0}. The warp factor, dilaton and internal metic are specified by
\begin{subequations}	
	\label{solutionsu31}
\begin{align}
e^{-A}  & = 1 + g_s \varphi + \cO(g_s^2) \,  =\, 1+ \cO(g_s^{4/3}) \, ,\\
e^{\phi}  & = g_s \left(1 - 3  g_s \varphi\right) + \cO(g_s^3) \, =\, g_s \left( 1 + \cO(g_s^{4/3}) \right) \, ,\\
\Om & = \Om_{\rm CY} + g_s k +  \cO(g_s^2) \, =\, \Om_{\rm CY} \left( 1 + \cO(g_s^{4/3}) \right)\, , \\
J & = J_{\rm CY} + \cO(g_s^2) \, =\, J_{\rm CY} \left( 1 + \cO(g_s^{4/3}) \right)\, , \\
v & = g_s \p_{\rm CY} f_\star + \cO(g_s^3) \, =\, \cO(g_s^{2})
\end{align}
\end{subequations}     
where recall that $\varphi$ and $k$ are defined by \eqref{formK1}, and $f_\star$ by \eqref{defstar}. When next to a $p$-form, the above scalings $\cO(g_s^k)$  are to be interpreted with respect to the natural scaling of the  $p$-form, so the total scaling of the object is $\cO(g_s^{k-p/3})$. 

We notice that the natural parameter of the expansion is $g_s^{4/3} \sim \cV_{\rm CY}^{-2/3}$, or in other words the quantum of four-form flux $G_4$.  We also notice that at leading order we recover precisely the Calabi--Yau background with fluxes \eqref{eq:SU3th0}. At the next order our solution is an $SU(3) \times SU(3)$ background, which contains an SU(3)-structure pair $(J, \Omega)$ with the following torsion classes 
\bea
\label{Omsu3}
d\Omega & = & i W_2 \wedge J + d(\phi-2A) \wedge \Omega\\
dJ & = & 0
\eea
and with $e^{\phi-3A} = g_s$. While this is the starting point for the analysis of SU(3)-structure backgrounds with $\theta=0$, the difference here is that a varying warp factor is allowed. This is thanks to the presence of a non-trivial one-form $v$.
This varying warp factor, and in general the three-form $K$ obtained from solving the Bianchi identity for $G_2$ at leading order, also modifies the fluxes $H$ and $G_4$ at this order, adding a non-CY-harmonic piece.

Given the no-go results for SU(3)-structure compactifications of section \ref{sec:nogosol}, one may wonder how this approximate background at $\cO(g_s)$ can overcome the obstructions therein. In particular, let us see how \eqref{tf2loc} is modified to allow for a non-smeared solution. First notice that to arrive at this equation one uses that \cite{Lust:2004ig}
\be
dG_2 \wedge \Omega + G_2 \wedge d \Omega = d (G_2 \wedge \Omega)\, .
\ee
In type IIA SU(3)-structure compactifications $G_2 \wedge \Omega \equiv 0$ iff the warp factor is constant, as one can show from \eqref{su3fluxG2} and the general expression for $d\Omega$. In the case of a SU(3)-structure background, this 
follows when we impose that the RHS of \eqref{su3fluxG0} is closed. In our more general background, this expression generalises to \eqref{F0leading} allowing for a non-constant, subleading piece or the warp factor, as the solution shows explicitly. This in turn implies that $d(G_2 \wedge \Omega) \neq 0$ adding the extra term to \eqref{tf2loc}. In our solution, this term is comparable to the terms in the LHS of \eqref{tf2loc} wedged with $\im \Omega$, which would then scale like $\Om$. Therefore the cancellation of this term is possible away from the localised sources and no smearing is needed. 

Notice that this the term $d (G_2 \wedge \Omega)$ is a non-trivial contribution to $X_5$ in the source balanced equation \eqref{preGOE}. So let us analyse how this more general equation can be satisfied for our approximate solution. Using the background in \eqref{solutionflux1} and \eqref{solutionsu31} one obtains
\begin{subequations}
	\label{balancedsol}
\begin{align}
 \label{dX5sol1}
3\mu^2 e^{-4A} \left< \mathrm{Re\;}\Phi_+ , \mathrm{Im\;}\Phi_+ \right>  \sim  -\frac{2}{25} G_0^2J^3  & \sim   \cO(g_s^0)\, d\mathrm{vol}_{X_6}\, , \\
 \label{dX5sol2}
 e^{4A}  \sum\nolimits_k G_k \wedge \star G_k  \sim \frac{26}{75}G_0^2J^3+\frac{ |G_2|^2}{6}J^3 & \sim   \cO(g_s^0)\,  d\mathrm{vol}_{X_6}+ \cO(g_s^{4/3})\,  d\mathrm{vol}_{X_6} \, ,\\
 -dX_5  \sim -\frac{4}{15}G_0^2 J^3-\delta^{(3)}_{\mathrm{source}}\wedge \frac{1}{g_s}\im\Omega_{\text{CY}} & \sim \cO(g_s^0)\,  d\mathrm{vol}_{X_6}\, ,
 \label{dX5sol}
\end{align}
\end{subequations} 
where, although $G_2\wedge \star G_2\sim \cO(g_s^{4/3})$ at leading order, we are writing explicitly this term to make easier the comparison with \eqref{tf2loc}.  Even if the sum of the first two terms gives a positive definite quantity --- recovering the case $\theta=0$ in \eqref{tf2loc} --- the terms coming from $dX_5$ are able to compensate this contribution. Indeed, for the case at hand one can check that the leading contribution to \eqref{dX5sol} comes from $d\left(G_2 \wedge \left( e^A \mathrm{Im\;}\Phi_- \right)_{3}\right)$ and that it cancels the other two contributions at order $\cO(g_s^0)$.
 In fact, one can easily check that this corresponds to the contribution $d (e^{-\phi} G_2 \wedge \Omega)$ that would allow to circumvent the obstruction related to \eqref{tf2loc}.

\section{Conclusions}
\label{sec:conclu}

In this chapter, we have found approximate solutions to the ten-dimensional supersymmetry equations which exhibit some key features of four-dimensional vacua derived in chapter \ref{ch:rrnsnsvacua} and found originally in \cite{DeWolfe:2005uu}, the so-called DGKT vacua. The solutions are first order in an expansion parameter corresponding to the average 10d dilaton $g_s$, or equivalently to the AdS$_4$ scale $\mu$ or $\cV_{X_6}^{-1/2}$ in string units.

The solutions are such that in the limit $g_s\rightarrow 0$ the background metric of the Calabi--Yau $X_6$ is the Ricci-flat one and the warp factor is constant. The non-vanishing fluxes are $G_0$ and $G_4 = \frac{3}{10} G_0 J_{\rm CY}^2$. This background corresponds to the smeared-O6-plane solution proposed in \cite{Acharya:2006ne} and reviewed in the previous chapter. For small but non-vanishing $g_s$, corrections to this background appear. The leading ones can be described in terms of the solution to the Bianchi identity $dG_2 = G_0H + \d_{\rm O6}$, which defines a function $\varphi$ and a (2,1)-form $k$. The first one corrects the warp factor and the dilaton, and the second one the three-form $\Omega$. Due to this metric deformation $X_6$ becomes a manifold with $SU(3)\times SU(3)$-structure.\footnote{This can also be thought of as $SU(3)$-structure with an additional 1-form. The $SU(3)$-structure part has the same torsion classes as type IIA Minkowski backgrounds with O6-planes.} Finally, $H$ and $G_4$ are also corrected in terms of $\varphi$ and $k$, no longer being harmonic forms in the Calabi--Yau sense.

Given that the solution was obtained in an expansion in the average string coupling $g_s$, one might wonder whether it competes with the genus expansion in string theory. Since we are at weak coupling, certainly the leading order part of the solution is under good control. The next order comes in at $g_s^{4/3}$. We expect that string loop corrections should appear at order $g_s^2$ or higher, and therefore the analysis should hold at least to first order. We leave a more detailed analysis of the magnitude of string corrections in this background for future work. 

Perhaps an even more delicate issue is the fact that we have solved the equations at the two-derivative level, so at leading order in $\alpha'$. Higher $\alpha'$ corrections are controlled by the curvature radius which is again related to our expansion parameter $g_s \sim R^{-3}$. In this case, a more accurate analysis of the magnitude and, importantly, the precise form of such corrections is needed to see whether their effect is substantial. 

Regarding the issue of scale separation and the Strong ADC, our results show that the DGKT proposal (type IIA Ad$_4$ orientifold compactifications) for scale separation has passed a first non-trivial test. There could have been an obstruction manifest already at first order in the supersymmetry equations, but we have shown that this is not the case (at least at the two-derivative level). 

However, they are still far from settling the issue. Before even asking about separation of scales we may ask whether a full ten-dimensional solution actually exists. At a technical level, a first possible obstruction may appear at the next order in the expansion parameter. Indeed, a crucial part of DGKT is that it involves intersecting sources, and the ten-dimensional solutions of such sources are poorly understood. At the first, linearised level of the expansion, the interactions between the sources drop out which is why we are able to find a solution relatively easily. The interactions only appear at the next level, where this feature of the construction is first tested. Note that if a solution does exist, it would be interesting to see if it also realises the picture proposed in section \ref{sec:smearingup} in which the pure spinors $\Phi_{\pm}$ differ from the Calabi--Yau ones only by non-harmonic forms.

There are also other, older and more general, open problems with any solutions of massive type IIA with O-planes. In our approximate solution, near an O6-plane we obtain an   warping behaviour of the form $e^{-4A} \sim 1-\frac{g_s\ell_s}{r}$ \cite{Andriot:2019hay}.  This defines a region in which we enter strong coupling and the supergravity approximation breaks down.
Typically the O-plane singularities may be resolved by uplifting to M-theory or F-theory. This is not possible here due to the mass parameter, and so the fate of these singularities remains an open question. It should be noted that the mass parameter, which obstructs an M-theory uplift \cite{Aharony:2010af}, is the crucial element to obtaining scale-separated vacua (it cannot be turned off, unlike some of the other fluxes). Practically, what this means is that we are simply not able to say anything about what happens near the O-planes in our solution. The hope is therefore that either we make progress on understanding the O-planes, or that we are able to settle the relevant questions without needing to worry about them.\footnote{An interesting possibility is that the singularities are removed already in the IIA supergravity description, similar to the ideas in \cite{Saracco:2012wc}.}

There are other open questions, some of which were raised already in \cite{DeWolfe:2005uu}, such as the control of higher derivative terms in the presence of large fluxes parameters. Further, even if a ten-dimensional solution of string theory can be firmly proven under controlled approximations, it still remains to be checked that it really does exhibit separation of scales. For example, as we have shown, the solution exhibits dilaton gradients and warp factors which must be accounted for in establishing the mass scales of the KK and string modes. On the other hand, one can use the formalism derived here to find the approximate uplift of the non-SUSY vacua derived in chapter \ref{ch:rrnsnsvacua} and study its non-perturbative stability. We will do this in the next (and last) chapter.

If the remaining open questions can be addressed and the property of scale separation proven, we would reach a significant result in string theory, and a counterexample to the Strong ADC (the normal ADC is of course satisfied in the scenarios studied here). In such a case it would be interesting if there is a possible refinement of the Strong ADC which may hold. A particularly interesting proposal was made in \cite{Buratti:2020kda} related to the presence of discrete symmetries. In any case, we find it exciting and encouraging that the recent activity, and progress reported in this thesis, suggests that at least the issue of DGKT and scale separation may be settled one way or the other in the not too distant future.

\clearpage
\chapter{Non-perturbative instabilities of non-SUSY AdS$_4$ orientifold vacua}
\label{ch:nonsusy}
\section{Introduction}
\label{s:intro}

Out of the different aspects of the Swampland Programme \cite{Vafa:2005ui,Brennan:2017rbf,Palti:2019pca,vanBeest:2021lhn,Grana:2021zvf} one of the most far-reaching is the interplay between quantum gravity and supersymmetry breaking. In the specific context of non-supersymmetric vacua, several proposals for Swampland criteria put severe constraints on their stability. In particular, the AdS Instability Conjecture \cite{Ooguri:2016pdq,Freivogel:2016qwc} -reviewed in section -\ref{sec:nonsusyconjecture}- proposes that all $\cN=0$ AdS$_d$ vacua are at best metastable, with  bubble nucleation always mediating some non-perturbative decay. The motivation for this proposal partially arises from a refinement of the Weak Gravity Conjecture (WGC) stating that the WGC inequality is only saturated in supersymmetric theories \cite{Ooguri:2016pdq}. Applied to $(d-2)$-branes, this implies a specific decay mechanism for $\cN=0$ AdS$_d$ vacua supported by $d$-form background fluxes, in which a probe superextremal $(d-2)$-brane nucleates and expands towards the AdS$_d$ boundary, as in \cite{Maldacena:1998uz}.

These proposals have been tested in different contexts, and in particular for type II setups in which the AdS solution is supported by fluxes \cite{Gaiotto:2009mv,Antonelli:2019nar,Apruzzi:2019ecr,Bena:2020xxb,Suh:2020rma,Guarino:2020jwv,Guarino:2020flh,Basile:2021vxh,Apruzzi:2021nle,Bomans:2021ara}. Remarkably, the kind of compactifications studied in this thesis, that is of the form AdS$_4 \times X_6$, where $X_6$ admits a Calabi--Yau metric \cite{DeWolfe:2005uu,Camara:2005dc}, remain elusive of the conjecture, because so far  the decays  observed for perturbatively stable $\cN=0$ vacua are marginal \cite{Narayan:2010em}, and the corresponding membranes saturate the WGC inequality. A better understanding of these constructions seems thus crucial to the Swampland Programme: their non-supersymmetric version challenges the AdS Instability Conjecture, and more precisely the WGC for membranes, while the supersymmetric settings challenge the strong version of the AdS Distance Conjecture \cite{Lust:2019zwm} -see chapter \ref{ch:uplift10d}-. As pointed out in \cite{Buratti:2020kda}, the tension with the AdS Distance Conjecture could be solved by taking into account the discrete symmetries related to 4d membranes, so the spectrum and  properties of 4d membranes seem to be at the core of both issues.  Finally, the constructions in \cite{DeWolfe:2005uu,Camara:2005dc} are particularly interesting phenomenologically, since besides supersymmetry breaking they incorporate key features like scale separation and chiral gauge theories supported on D6-branes wrapping intersecting three-cycles of $X_6$.

An important caveat of the constructions in \cite{DeWolfe:2005uu,Camara:2005dc} and that we have highlighted along this thesis, is that they do not solve the 10d equations of motion and Bianchi identities, unless localised sources like D6-branes and O6-planes are smeared over the internal dimensions, as we saw in chapter  \ref{ch:review10d}. Nevertheless, one may look for solutions with localised sources by formulating the problem as a perturbative expansion, of which the leading term is the smeared-source Calabi--Yau approximation, and where the expansion parameter is essentially the AdS$_4$ cosmological constant \cite{Saracco:2012wc}, recall chapter \ref{ch:uplift10d} for more details. The first-order correction to this expansion displays localised sources and a natural expansion parameter $R^{-4/3} \sim g_s^{4/3}$, where $R$ is the AdS$_4$ radius in string units and $g_s$ is the average 10d string coupling.\footnote{Another caveat surrounding these constructions is that they combine O6-planes and a non-vanishing Romans mass, which makes difficult to understand them microscopically. However, T-dual versions of the solutions in \cite{Junghans:2020acz,Marchesano:2020qvg} have been constructed in \cite{Cribiori:2021djm} with similar properties, vanishing Romans mass and an 11d description.}  

In this chapter we revisit the stability of the AdS$_4$ vacua obtained in chapter \ref{ch:rrnsnsvacua} and  \cite{DeWolfe:2005uu,Camara:2005dc,Narayan:2010em}, with the vantage point of the more precise 10d description of chapter \ref{ch:uplift10d}. We consider $\cN=1$ and $\cN=0$ vacua which, in the smearing approximation,  are related by an overall sign flip of the internal four-form flux $G_4$, considered firstly in \cite{Narayan:2010em} for the specific case $X_6 = T^6/ (\Z_3 \times \Z_3)$. For the supersymmetric backgrounds a rather explicit 10d solution was provided in chapter \ref{ch:uplift10d}, in terms of an $SU(3)\times SU(3)$-structure deformation of the Calabi--Yau metric. For their non-supersymmetric cousins we use the approach in \cite{Junghans:2020acz} to provide a solution at the same level of approximation. 
In this setup we consider 4d membranes that come from wrapping D$(2p)$-branes on $(2p-2)$-cycles of $X_6$. These membranes couple to  fluxes that support the AdS$_4$ background, more precisely to the dynamical fluxes of the 4d theory \cite{Lanza:2019xxg,Lanza:2020qmt}. Therefore, even if there could be other non-perturbative decay channels, the $\cN=0$ sharpening of the WGC suggests that at least one of these membranes or a bound state of them should be superextremal, and thus a candidate to yield an expanding bubble. Note that these AdS$_4$ backgrounds have  not been constructed as near-horizon limits of a backreacted black brane solutions, so it is a priori not clear which membrane is the most obvious candidate to fulfil the conjecture. 

It was argued in \cite{Aharony:2008wz,Narayan:2010em} that D4-branes wrapping either holomorphic or anti-holomorphic cycles of $X_6$ saturate a BPS bound for the $\cN=1$ and $\cN=0$ vacua mentioned above, while D2-branes and D6-branes wrapping four-cycles never do. By looking at each of their couplings to the fluxes supporting the AdS$_4$ background and their tension we recover the same result. Remarkably, we not only do so for the smeared-source Calabi--Yau  approximation considered in \cite{Aharony:2008wz,Narayan:2010em}, but also when the first-order corrections to this background are taken into account. It follows that at this level of approximation such (anti-)D4-branes give rise to extremal objects, that can at most mediate marginal decays. This extends to bound states of D6, D4 and D2-branes, in the sense that they do not yield any superextremal 4d membrane. 

We then turn to consider D8-branes wrapping $X_6$. Due to a Freed--Witten anomaly generated by the $H$-flux, D6-branes must be attached to the D8 worldvolume. From the 4d perspective, these are membranes that not only change the Romans mass flux $F_0$ when crossing them, but also the number of space-time filling D6-branes, so that the tadpole condition is still satisfied. It turns out that the presence of attached D6-branes acts as a force on the D8-branes, and exactly cancels the effect of their charge and tension in supersymmetric vacua, as it should happen for a BPS object. This provides a rationale  for the precise relation between $F_0$, $R$, $g_s$ found in \cite{DeWolfe:2005uu}. In $\cN=0$ vacua the energetics of D8-branes is more interesting, because curvature corrections induce D4-brane charge and tension on their worldvolume. The induced tension is in general negative, implying that the D8-brane is dragged towards the boundary of $\cN=0$ AdS$_4$.  As we argue, this corresponds to a superextremal 4d membrane that mediates a decay to another non-supersymmetric vacuum with larger $|F_0|$ and fewer D6-branes, in agreement with the sharpened Weak Gravity Conjecture.

This picture is however incomplete, since it relies on the smeared description. First-order corrections to the Calabi--Yau background modify the D8-brane action by terms comparable to an induced D4-brane tension. In fact, beyond the smearing approximation the D8/D6 system should be treated as a BIon-like solution, whose tension differs from the sum of D8 and D6-brane tensions. We compute this difference for $X_6 = T^6/(\Z_2 \times \Z_2)$, and find that this new correction is comparable to curvature-induced effects. Nevertheless, we find that it is also negative, and so the D8-branes are still dragged towards the $\cN=0$ AdS$_4$ boundary. We then argue that the same is true in more general setups, so that the combined effect of curvature and BIon-like corrections provide a non-perturbative instability for $\cN=0$ AdS$_4$ vacua with space-time filling D6-branes, in line with the AdS Instability Conjecture.

The chapter is organised as follows. In section \ref{s:memb} we discuss the energetics of membranes in AdS$_4$ backgrounds with four-form fluxes, which we then use as criterion for membrane extremality. In section \ref{s:dgkt} we review the $\cN=1$ AdS$_4$ Calabi--Yau orientifold vacua with fluxes in the smearing approximation, and classify BPS membranes that come from wrapped D-branes. Section \ref{s:nonsusy} does the same for non-supersymmetric AdS$_4$, finding superextremal membranes thanks to curvature corrections, and section \ref{s:insta} argues that they mediate actual decays in the 4d theory. Section \ref{s:nonsmeared} describes the 10d background with localised sources for $\cN=1$ and $\cN=0$ AdS$_4$ vacua, and shows that D4-branes saturate a BPS bound in both cases. Section \ref{s:bion} describes D8/D6-brane systems as BIons, and shows that they are BPS in $\cN=1$ but feel a net force in $\cN=0$ vacua. We finally present our conclusions in section \ref{s:conclu}.

Several technical details have been relegated to the appendices \ref{ap:end}.  Appendix \ref{ap:10deom} shows that the  backgrounds of section \ref{s:nonsmeared} satisfy the 10d equations of motion. Appendix \ref{ap:dbi} shows how the BIon profile of section \ref{s:bion} linearises the  DBI action. Appendix \ref{ap:IIBion} relates this profile to 4d strings in type IIB warped Calabi--Yau compactifications and to SU(4) instantons in Calabi--Yau four-folds. Appendix \ref{ap:torus} computes the  BIonic D8-branes excess tension for $X_6 = T^6/(\Z_2 \times \Z_2)$.


\section{Membranes in AdS$_4$}
\label{s:memb}

In a 4d Minkowski background with $\cN=1$ supersymmetry, simple examples of static BPS membranes are  3d hyperplanes of $\pr^{1,3}$ including the time-like direction. Analogous objects in Anti-de-Sitter can be described by considering the Poincar\'e patch of AdS$_4$, whose metric reads
\be
ds^2_4 =e^{\frac{2z}{R}} (-dt^2 + d\vec{x}^2) + dz^2\, ,
\label{PPatch}
\ee
with $R$ the AdS length scale, $\vec{x} = (x^1, x^2)$ and all coordinates range over $\pr$. In these coordinates, the AdS$_4$ boundary is located at $z = \infty$. Similarly to the Minkowski case, one may consider a membrane that spans the coordinate $t$ and a surface within $(x^1, x^2, z)$. Particularly simple is the case where the surface is the plane $z=z_0$, with $z_0 \in \pr$ fixed. While this object may look like the BPS membranes of Minkowski, the tension of such a membrane decreases exponentially as we take $z_0 \to -\infty$. Therefore, if we place such an object in AdS$_4$ and take the probe approximation, it will inevitably be driven away from the boundary and it cannot be BPS. 

This can be avoided if on top of the AdS$_4$ metric we consider a four-form flux background $F_4$, to whose three-form potential $C_3$ the membrane couples as $-\int C_3$. Indeed, if we have
\be
\langle F_4 \rangle = -\frac{3Q}{R} {\rm vol}_4
\qquad \Longrightarrow \qquad \langle C_3 \rangle = Q\, e^{\frac{3z}{R}} dt \wedge dx^1 \wedge dx^2 \, ,
\label{3form}
\ee
and $Q$ coincides with the tension of the membrane $T$, then the variation of the tension when moving in the $z$ coordinate is compensated by the potential energy $-\int \langle C_3 \rangle$ gained because of its charge. Moving along this coordinate is then a flat direction and the membrane may be BPS. If $Q > T$ one may still find BPS membrane configurations, but they cannot be parallel to the boundary. We instead have that force cancellation occurs for embeddings of the form
\be
\left\{t, x^1, x^2 = \pm \frac{R}{\sqrt{\frac{Q^2}{T^2}-1}} \, e^{-\frac{z}{R}} + c\right\}\, , \qquad c \in \pr\, .
 \label{QnotT}
\ee

Four-form flux backgrounds are ubiquitous in AdS$_4$ backgrounds obtained from string theory, and in particular in those with 4d $\cN=1$ supersymmetry or $\cN=0$ spontaneously broken. The membrane profiles $z=z_0$ and \eqref{QnotT} were found in \cite{Koerber:2007jb} in the context of $\cN=1$ AdS$_4$ backgrounds obtained from type II string theory, but from the above discussion it follows that they can also be present in backgrounds with supersymmetry spontaneously broken by fluxes.  One can in fact see that  the set of 4d fluxes arising from the compactification is directly related to the spectrum of BPS branes, as well as to the internal data specifying the supersymmetry generators. 

In the following we will be chiefly concerned with those membranes whose profile is given by $z=z_0$. As argued in \cite{Koerber:2007jb}, for $Q=T$ and at $z \to \infty$ they capture the BPS bound of a spherical membrane in global coordinates at asymptotically large radius. It is precisely the domain walls that correspond to spherical membranes near the AdS boundary that determine if the non-perturbative decay of one vacuum to another with lower energy is favourable or not. Thus, by considering the energetics of membranes in the Poincar\'e patch with $z = z_0 \to \infty$ we may detect if there could be some domain wall triggering such a decay. If all membranes satisfy $T > Q$ such a decay should not occur, if $Q=T$ it should be marginal, and  if $T<Q$ the AdS background may develop a non-perturbative instability. According to the conjectures in \cite{Ooguri:2016pdq,Freivogel:2016qwc}, any $\cN=0$ AdS background of this sort should have at least one non-perturbative instability towards a new vacuum, and therefore a membrane with $T<Q$. In the following sections we will consider the membranes that appear from wrapping D-branes on internal cycles in backgrounds of the form AdS$_4 \times X_6$, where $X_6$ admits a Calabi--Yau metric, and compute $T$ and $Q$ for them. In particular we will consider the $\cN=1$ vacua  and some of the non-supersymmetric vacua found in chapter \ref{ch:rrnsnsvacua}, which are stable at the perturbative level. We will not only consider the Calabi--Yau approximation of these references, but also the solutions with localised sources, using the tools of chapter \ref{ch:uplift10d}. As we will see, for non-supersymmetric vacua the answer is not the same once this more precise picture is taken into account.


\section{Supersymmetric AdS$_4$ orientifold vacua}
\label{s:dgkt}

Examples of membranes satisfying $Q=T$ are typically found in supersymmetric AdS$_4$ backgrounds, where the equality follows from saturating a BPS bound. In this section we analyse for which membranes this condition is met for the supersymmetric type IIA flux compactifications of \cite{DeWolfe:2005uu}, for an arbitrary Calabi--Yau geometry $X_6$, in the approximation of smeared sources \cite{Acharya:2006ne}. With the simple criterion $Q=T$ one can reproduce the results of \cite{Aharony:2008wz} for membranes arising from D2, D4 and D6-branes wrapping internal cycles of $X_6 = T^6/(\Z_3 \times \Z_3)$, and extend them to any Calabi--Yau manifold. Furthermore, one may detect an additional set of BPS membranes, namely those coming from D8-branes wrapping $X_6$, to which space-time filling D6-branes are attached. This last feature makes such membranes quite special, particularly when one considers them for non-supersymmetric AdS$_4$ backgrounds and beyond the smearing approximation.

\subsection{10d background in the smearing approximation}
\label{ss:smeared}

Let us consider again type IIA string theory compactified in an orientifold of $X_4 \times X_6$, where $X_6$ is a compact Calabi--Yau three-fold, with RR and NSNS fluxes. This scenario has been presented several times in this thesis, and so we will just repeat the essential expressions to make this chapter self-contained, referring to chatper \ref{ch:review} for more details.

The internal RR flux quanta  were defined in terms of the following integer numbers
\begin{equation}
m \, = \,  \ell_s G_0\, ,  \quad  m^a\, =\, \frac{1}{\ell_s^5} \int_{X_6} \bar{G}_2 \wedge \tilde \omega^a\, , \quad  e_a\, =\, - \frac{1}{\ell_s^5} \int_{X_6} \bar{G}_4 \wedge \omega_a \, , \quad e_0 \, =\, - \frac{1}{\ell_s^5} \int_{X_6} \bar{G}_6 \, ,
\label{RRfluxes1}
\end{equation}
with $\omega_a$, $\tilde \omega^a$ defined in table \ref{base}, in terms of which we can expand the K\"ahler form as $J_{\rm CY} = t^a \omega_a$ and $- J_{\rm CY} \wedge J_{\rm CY} = {\cal K}_a \tilde{\omega}^a$. Recall that in our conventions $-\frac{1}{6}J_\text{CY}^3=-\frac{i}{8}\Omega_{\text{CY}}\wedge\bar{\Omega}_{\text{CY}}$ is the volume form\footnote{Due to this choice of volume form the triple intersection numbers must be defined with an additional minus sign compared to the more standard definition in the literature so that, whenever $\{[\ell_s^{-2}\omega_a]\}_a$ is dual to a basis of Nef divisors, ${\cal K}_{abc} \geq 0$. The same observation applies to the curvature correction term $K_a^{(2)}$ defined in \eqref{Kcurv}.}. In the presence of D6-branes and O6-planes the Bianchi identities for the RR fluxes read
\be
dG_0 = 0\, , \qquad d G_2 = G_0 H - 4 \d_{\rm O6} +   N_\a \d_{\rm D6}^\a \, ,  \qquad d G_4 = G_2 \wedge H\, , \qquad dG_6 = 0\, ,
\label{BIG}
\ee
implying that
\be
{\rm P.D.} \left[4\Pi_{\rm O6}- N_\a \Pi_{\rm D6}^\a\right] = m [\ell_s^{-2} H] \, ,
\label{tadpole}
\ee
constraining the quanta of Romans parameter and NS flux. Let us in particular choose P.D.$[\ell_s^{-2}H] = h [\Pi_{\rm O6}] = h [\Pi_{\rm D6}^\a]$, $\forall \a$. We then find the constraint
\be
mh +N = 4\, ,
\label{tadpole2}
\ee
with $N$ the number of D6-branes wrapping $\Pi_{\rm O6}$. Supersymmetry in addition implies that $mh$ and $N$ are non-negative, yielding a finite number of solutions.\footnote{In several instances (e.g., toroidal orbifolds) $[\Pi_{\rm O6}]$ may be an integer multiple $k$ of a three-cycle class. In those cases $h, N$ need not be integers, but instead $kh, kN \in \mathbb{Z}$, allowing for a richer set of solutions to \eqref{tadpole2}.} 

The constraint on sign$(mh)$ can be seen by means of a 4d analysis of the potential generated by background fluxes, following chapter \ref{ch:rrnsnsvacua}. As derived there and discussed in section \ref{sec:susynsnsrr}, the internal fluxes satisfy
\be
\ell_s [ H ]  = \frac{2}{5} m g_s [\re \Omega_{\rm CY} ] \, , \qquad  G_2  =  0\, ,  \qquad  \ell_s G_4  = -\hat{e}_a  \tilde{\omega}^a  = \frac{3}{10} G_0 J \wedge J   \, . \qquad  G_6 =  0\, , 
\label{intfluxsm}
\ee
with
\be
\hat{e}_a = e_a - \oh \frac{\cK_{abc} m^am^b}{m}\, .
\label{hate}
\ee

Care should however be taken when interpreting such relations from the viewpoint of the actual 10d supergravity solution, since the presence of fluxes and localised sources will deform the internal geometry away from the Calabi--Yau metric, and a $G_2$ and $G_4$ of the above form will never satisfy the Bianchi identities \eqref{BIG}. The standard way to deal with both issues is to see \eqref{intfluxsm} as a formal solution in which all localised sources have been smeared \cite{Acharya:2006ne}. This so-called smeared solution is then the leading term in a perturbative series that should converge to the actual background \cite{Saracco:2012wc}, with expansion parameter $g_s^{4/3}$, and where sources are localised.  Instead of \eqref{intfluxsm}, the relations that this background must satisfy are
\be
[ H ]  = \frac{2}{5} G_0 g_s  [\re \Omega_{\rm CY} ] \, , \quad \int_{X_6} G_2 \wedge \tilde{\omega}^a =  0\, ,  \quad \frac{1}{\ell_s^6} \int_{X_6} G_4  \wedge \omega_a  =  - \frac{3}{10} G_0 {\cal K}_a \, , \quad  G_6  =  0\, , 
\label{intflux}
\ee
where $g_s$ is the average value of $e^{\phi}$, with $\phi$ a varying 10d dilaton. This value determines the AdS$_4$ length scale in the 10d string  frame $R$, from the following additional relation
\be
\frac{\ell_s}{R} = \frac{1}{5} |m| g_s  \, .
\label{Rads}
\ee
There is in addition a non-trivial warp factor, and the Calabi--Yau metric on $X_6$ is deformed to an $SU(3) \times SU(3)$-structure metric, as explained in chapter \ref{ch:uplift10d}. We will discuss this more accurate background in section \ref{ss:mpqt}, and for now focus on the smearing approximation. 

It follows from such a description that the Calabi--Yau volume ${\cal V}_{\rm CY}  =  \frac{1}{6} {\cal K}$ depends on $m$ and $\hat{e}_a$, growing large when we increase their absolute value. One can then for instance see that $1/R$ grows as we increase $h$ or $m$, and decreases as we increase $\hat{e}_a$ The more precise result can be obtained from the 4d analysis of chapter \ref{ch:rrnsnsvacua}, which yields the following 4d Einstein frame vacuum energy 
\be
\Lambda = - \frac{16\pi}{75\kappa_4^4} e^K {\cal K}^2 m^2 \, ,
\label{lambda}
\ee
where $K$ is the K\"ahler potential, given by \eqref{eq:kpotential}. One can then see that $\Lambda$ scales like $|m|^{5/2}$, as in the explicit toroidal solutions in \cite{DeWolfe:2005uu,Camara:2005dc}. Recall however that the allowed values for $m$ are bounded by the tadpole condition \eqref{tadpole2}.

Finally, one can include the effect of curvature corrections to the 4d analysis, following \cite{Palti:2008mg,Escobar:2018rna}. We will only include those corrections dubbed $K^{(1)}_{ab}$ and $K_a^{(2)}$ in \cite{Palti:2008mg,Escobar:2018rna}, given by
\be
K^{(1)}_{ab} = \frac{1}{2} {\cal K}_{aab} \, , \qquad K_a^{(2)} = -\frac{1}{24} \int_{X_6} c_2(X_6) \wedge \omega_a\, ,
\label{Kcurv}
\ee
and that respectively correspond to $\cO(\alpha')$ and $\cO(\alpha'^2)$ corrections, since higher orders will be beyond the level of accuracy of our analysis. If $\{[\omega_a]\}_a$ is dual to a basis of Nef divisors, then $K_a^{(2)} \geq 0$ \cite{Miyaoka1987}. The effect of such corrections is to redefine the background flux quanta as follows
\be
e_0 \to e_0 - m^a K_a^{(2)} \, , \qquad \quad e_a \to e_a  - K_{ab}^{(1)} m^b + m K_a^{(2)} \, ,
\label{curvflux}
\ee 
so in particular they modify the flux combinations \eqref{hate} that determine the K\"ahler moduli vevs. This modification makes more involved the scaling of $\Lambda$ with $m$, but since in the regime of validity we have that ${\cal K}_a \gg K_a^{(2)}$, it turns out that $\Lambda \sim |m|^{5/2}$ is still a good approximation. 

\subsection{4d BPS membranes}
\label{ss:4dmem}

Given a type II flux compactification to $\CN=1$ AdS$_4$, one may study the spectrum of BPS D-branes via $\kappa$-symmetry or pure spinor techniques, as in \cite{Aharony:2008wz,Koerber:2007jb}, and in particular determine those D-branes that give rise BPS membranes from the 4d perspective. In the following we will take the more pedestrian viewpoint of section \ref{s:memb} to identify such BPS membranes. This criterion will also be useful when considering non-supersymmetric AdS$_4$ vacua. 

An analysis of 4d BPS membranes parallel to the AdS$_4$ boundary in the Poincar\'e patch was carried out in \cite{Aharony:2008wz}, for the particular case $X_6 = \mathbb{T}^6/(\Z_3 \times \Z_3)$ of \cite{DeWolfe:2005uu}, in the smearing approximation. It was found that D4-branes wrapping holomorphic cycles are BPS, while D2 and D6 branes cannot be so. Let us see how to recover such results and extend them to general Calabi--Yau geometries using the picture of section \ref{s:memb}. For this we recall that in the type IIA democratic formulation the RR background fluxes take the form
\be
{\bf G} = {\rm vol}_4 \wedge \tilde{G} + \hat{G}\, ,
\label{demoflux}
\ee
where ${\rm vol}_4$ is the AdS$_4$ volume form and $\tilde{G}$ and $\hat{G}$ only have internal indices, satisfying the relation $\tilde{G} = - \lambda ( *_6 \hat{G})$. Therefore from \eqref{intfluxsm} and \eqref{Rads} we find the following fluxes that translate into a 4d four-form background 
\be
G_6 =  - \frac{3\eta}{Rg_s} {\rm vol}_4 \wedge J_{\rm CY}\, , \qquad G_{10} =  - \frac{5\eta}{6 R g_s} {\rm vol}_4 \wedge J^3_{\rm CY}\, ,
\label{610fluxes}
\ee
with $\eta =  {\rm sign }\, m$. Contrarily, no component of ${\rm vol}_4$ appears in $G_4$ or $G_8$. We hence deduce the following  couplings for 4d membranes arising from D(2$p$)-branes wrapping (2$p-$2)-cycles of $X_6$:
\be
Q_{D2} = 0 \, , \qquad Q_{D4} =  e^{K/2} \eta \int_\Sigma J_{\rm CY} \, , \qquad Q_{D6} = 0\, , \qquad Q_{D8} =  -\frac{5}{3} e^{K/2} \eta\, q_{\rm D8} {\cal V}_{\rm CY}\, ,
\label{QDGKT}
\ee
expressed in 4d Planck units. Here $\Sigma$ is the two-cycle wrapped by the D4-brane, and $q_{\rm D8} = \pm 1$ specifies the orientation with which the D8-brane wraps $X_6$. This implies that for $\eta=1$ a BPS D4-brane must wrap a holomorphic two-cycle with vanishing worldvolume flux $\cF = B + \frac{\ell_s^2}{2\pi} F$ to be BPS, so that $e^{K/2} \int_\Sigma J_{\rm CY} = e^{K/2} {\rm area}(\Sigma) \equiv T_{\rm D4}$, while for $\eta = -1$ the fluxless two-cycle must be anti-holomorphic. This choice of orientation for $\Sigma$ can be understood from looking at how the four-form varies when crossing the D4-brane from $z = \infty$ to $z = -\infty$. In both cases, due to \eqref{IIABI} and the choice of orientation for $\Sigma$ one decreases the absolute value of the four-form flux quanta $\hat{e}_a$, and therefore the vacuum energy. This is consistent with our expectations, as it permits to have a BPS domain-wall solution mediating a marginal decay from a vacuum with higher energy (at $z = \infty$) to one with lower energy (at $z = -\infty$). Considering this set of BPS membranes allows us to scan over the set of vacua with different four-form flux quanta.  Differently, D6-branes wrapping four-cycles of $X_6$ and D2-branes can never yield 4d BPS membranes. This indeed reproduces and generalises the results found in \cite{Aharony:2008wz}, adapted to our conventions. 

It however remains to understand the meaning of $Q_{\rm D8}$, which naively does not seem to allow for BPS membranes that come from wrapping (anti-)D8-branes on $X_6$. On general grounds one would expect that such BPS membranes exist as well, in order to scan over the different values of $m$. In particular, one would expect that for $\eta=1$ D8-branes $(q_{\rm D8} =1)$ wrapping $X_6$ are BPS, while for $\eta=-1$ the same occurs for anti-D8-branes $(q_{\rm D8} = -1)$. Indeed, when crossing the corresponding domain wall from $z = \infty$ to $z = -\infty$ the value of $|m|$ increases and the vacuum energy decreases in both setups, paralleling the case for D4-branes. However, the factor of $5/3$ and a sign prevent achieving the necessary BPSness condition $Q_{\rm D8}=T_{\rm D8} \equiv e^{K/2} {\cal V}_{\rm CY}$.

\begin{figure}[h!]
    \centering
    \includegraphics[width=10cm]{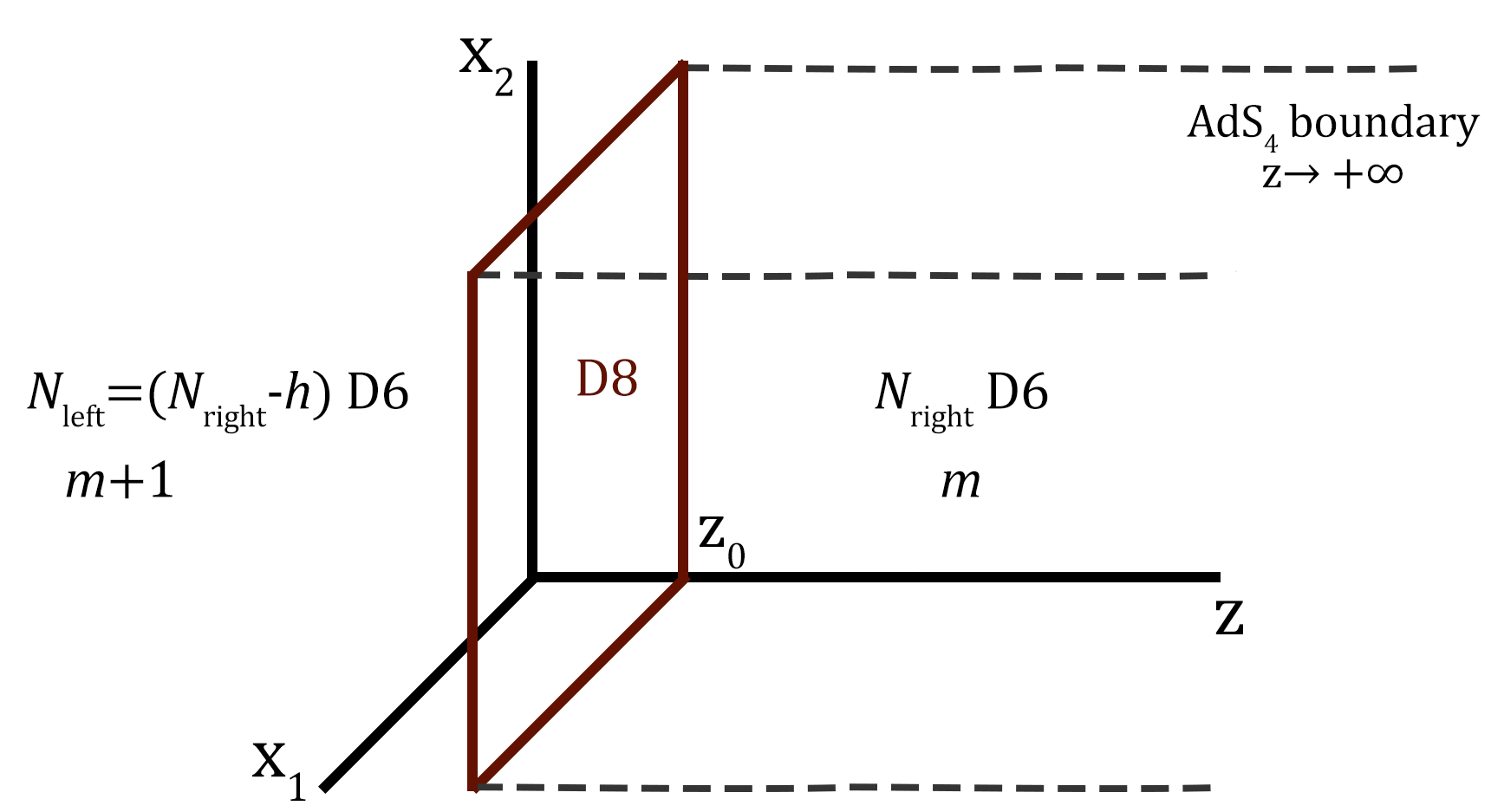}
    \caption{To cure the Freed--Witten anomaly induced by the $H$-flux on the D8-brane worldvolume, an excess of $|h|$ space-time filling D6-branes must be attached from its position to the AdS$_4$ boundary. We take $m, h >0$ in the figure.}
    \label{fig:D8D6}
\end{figure}

The resolution to this puzzle comes from realising that D8-branes wrapping $X_6$ cannot be seen as isolated objects. Instead, D6-branes must be attached to them, to cure the Freed--Witten anomaly generated on the (anti-)D8-brane by the NS flux background $H$. In the above setup the D6-branes will be wrapping a three-cycle of $X_3$ on the Poincar\'e dual class to $\eta [\ell_s^{-2} H] = |h| [\Pi_{\rm O6}]$, and extend along the 4d region of AdS$_4$ $(t,x^1,x^2) \times [z_0 , \infty)$ that is bounded by the 4d membrane. More generally, we need an excess of space-time filling D6-branes wrapping $\Pi_{\rm O6}$ on the interval $[z_0 , \infty)$ to the right of the (anti-)D8-brane, as compared to the ones in the left-interval $(-\infty, z_0]$ to cancel the said Freed--Witten anomaly: 
\be
N_{\rm right} - N_{\rm left} = |h| \, ,
\label{excessD6}
\ee
see figure \ref{fig:D8D6}. Since $m$ jumps by $\eta$ when crossing the membrane from right to left, $mh$ jumps by $|h|$, and so \eqref{excessD6} guarantees that the tadpole condition \eqref{tadpole2} is satisfied at both sides. 

Since the number of space-time filling D6-branes is different at both sides of the D8-brane, their presence will induce an energy dependence in terms of the D8-brane position. Indeed, if we decrease $z_0$ and move the D8-brane away from the AdS$_4$ boundary the region of AdS$_4$ filled by $N_{\rm right}$ D6-branes will grow, and so will the total energy of the system. As a result, the D6-brane jump induced by the Freed--Witten anomaly pulls the D8-branes towards the boundary of AdS$_4$. It turns out that this effects precisely cancels the effect of the tension $T_{\rm D8}$ and coupling $Q_{\rm D8}$ of the D8-brane, which both drag the 4d membrane away from the AdS boundary. 

One can derive such a cancellation via a microscopic calculation of the DBI+CS action for the D8/D6 system, dimensionally reduced to 4d. Of course, from the viewpoint of the 4d membrane the tension of space-time filling D6-branes extended along $(-\infty, z_0]$ and $[z_0, \infty)$ is infinite. Nevertheless, one may compute how the energy of the system varies  as we modify the D8-brane position $z_0$. Indeed, the DBI contribution to the action is given by the sum of the following two terms:
\begin{align}
 \label{DBID8}
    S_{\rm DBI}^{\rm D8} =& -\frac{1}{g_s} {\cal V}_{\rm CY}e^{\frac{3z_0}{R}} \frac{2\pi}{\ell_s^3} \int  dt dx^1 dx^2\, , \\
    S_{\rm DBI}^{\rm D6} = &
    -\frac{1}{g_s}  {\cal V}_{\Pi_{\rm O6}} \left(N_{\rm left} \int_{-\infty}^{z_0} dz'e^{\frac{3z'}{R}} +N_{\rm right} \int_{z_0}^\infty dz'e^{\frac{3z'}{R}}\right) \int  dt dx^1 dx^2\, ,
    \label{DBID6}
    \end{align}
with
\be
{\cal V}_{\Pi_{\rm O6}} = \frac{1}{\ell_s^3} \int_{\Pi_{\rm O6}} \im\, \Omega_{\rm CY} = \frac{1}{h \ell_s^5}  \int_{X_6}  \im \Omega_{\rm CY} \wedge H  = \frac{8}{5} \frac{m}{h} g_s {\cal V}_{\rm CY} = \frac{8\ell_s}{|h|R}{\cal V}_{\rm CY} \, ,
\ee
where we have used that in our conventions O6-planes and BPS D6-branes are calibrated by $\im\, \Omega_{\rm CY}$, and then the relations \eqref{intflux} and \eqref{Rads}. Let us now consider an infinitesimal variation $z_0 \to z_0 + \ell_s \epsilon$.  The variation of these actions is
\begin{align}
\label{DBID8var}
\delta_\epsilon  S_{\rm DBI}^{\rm D8} & = -\frac{3}{Rg_s } {\cal V}_{\rm CY}\, e^{\frac{3z_0}{R}}   \frac{2\pi}{\ell_s^2} \int   dt dx^1 dx^2\, ,\\
\delta_\epsilon S_{\rm DBI}^{\rm D6} & = -\frac{8}{Rg_s} \frac{N_{\rm left} - N_{\rm right}}{|h|} {\cal V}_{\rm CY}\, e^{\frac{3z_0}{R}}   \frac{2\pi}{\ell_s^2} \int   dt dx^1 dx^2 =  \frac{8}{Rg_s} {\cal V}_{\rm CY}\, e^{\frac{3z_0}{R}}   \frac{2\pi}{\ell_s^2} \int   dt dx^1 dx^2\, .
\label{DBID6var}
\end{align}
That is, the dragging effect of the D6-branes ending on the D8-brane overcomes the effect of its tension, acting like an additional coupling $Q^{\rm eff}_{\rm D6} =  \frac{8}{3} e^{K/2} {\cal V}_{\rm CY}$.  This precisely compensates the coupling of the 4d membrane made up from a D8-brane in the case $\eta =1$ and from an anti-D8-brane in the case $\eta =-1$, as claimed. Microscopically, this cancellation is seen from the variation of the (anti-)D8-brane Chern-Simons action. By evaluating the coupling to the RR potential $C_9$ that corresponds to \eqref{610fluxes} and integrating over $X_6$ one obtains:
\be
S_{\rm CS}^{D8} = q_{\rm D8} \frac{2\pi}{\ell_s^9} \int C_9  =  -q_{\rm D8}\,  \eta \frac{5}{3} g_s^{-1}  {\cal V}_{\rm CY}  e^{\frac{3z_0}{R}} \, \frac{2\pi}{\ell_s^3} \int  dt dx^1 dx^2 \, .
\label{CSD8}
\ee
It is then easy to see that for $q_{\rm D8} = \eta$ the variation $\delta_\epsilon  S_{\rm CS}^{\rm D8}$ precisely cancels \eqref{DBID8var}+\eqref{DBID6var}. Therefore, the effect of the D6-branes can be understood as generating an effective coupling $Q_{\rm D8/D6}^{\rm eff} = Q_{\rm D8} + Q^{\rm eff}_{\rm D6} =  \eta q_{\rm D8}  e^{K/2} {\cal V}_{\rm CY}$. Indeed, notice that if one chose $q_{\rm D8} = - \eta$ then the Freed--Witten anomaly would be opposite and the D6-branes would be extending along $z \in (-\infty, z_0]$. This would result into $Q_{\rm D8/D6}^{\rm eff} = - e^{K/2} {\cal V}_{\rm CY}$, destabilising the system towards $z_0 \to -\infty$. 

\subsubsection*{Considering bound states}

In general, the Chern-Simons action of a D8-brane reads
\be
S_{\rm CS}^{\rm D8} =  \frac{2\pi}{\ell_s^9} \int P\left[ {\bf C} \wedge e^{-B}\right] \wedge e^{- \frac{\ell_s^2}{2\pi}  F} \wedge \sqrt{\hat{A}({\cal R})}\, ,
\ee
where ${\bf C} = C_1 + C_3 + C_5 + C_7 + C_9$ and $\hat{A}({\cal R}) = 1 + \frac{1}{24} \frac{\Tr R^2}{8\pi^2} + \dots $ is the A-roof genus. These couplings encode that in the presence of a worldvolume flux and/or curvature, we actually have a bound state of a D8 with lower-dimensional D-branes. If the bound state is BPS, then its tension will be a sum of D8 and D4-brane tensions. Taking also into account the effect of the D6-branes ending on it we have that
\be
T_{\rm D8}^{\rm total} = T_{\rm D8} + \left(K^F_a - K_a^{(2)}\right)  T_{\rm D4}^a\, ,
\label{TD8tot}
\ee
where $T_{\rm D4}^a = e^{K/2} t^a$, $K^F_a = \frac{1}{2} \int_{X_6} \cF \wedge \cF \wedge \omega_a$ and $K_a^{(2)}$ has been defined in \eqref{Kcurv}. Similarly, the  Chern-Simons action of this bound state will give, upon dimensional reduction 
\be
q_{\rm D8} Q_{\rm D8}^{\rm total} = Q_{\rm D8/D6}^{\rm eff} + \left(K^F_a - K_a^{(2)}\right)  Q_{\rm D4}^a = \eta T_{\rm D8}^{\rm total}\, ,
\label{QD8tot}
\ee
where $ Q_{\rm D4}^a = \eta e^{K/2}   t^a$. Hence, again for $\eta=1$ a D8-brane will satisfy the BPS condition $Q=T$, while for $\eta =-1$ this will occur for an anti-D8-brane. One important aspect of these corrections is that the induced D4-brane tension in \eqref{TD8tot} is in general negative. Indeed, the curvature term $K_a^{(2)} t^a =- \frac{1}{24} \int_{X_6} c_2(X_6) \wedge J$ is positive in the interior of the K\"ahler cone for Calabi--Yau geometries, inducing a negative D4-brane tension. In the present case this is compensated by an induced negative D4-brane charge in \eqref{QD8tot}. However, such a compensation will no longer occur for the non-supersymmetric AdS$_4$ flux backgrounds that we now turn to discuss.


\section{Non-supersymmetric AdS$_4$ vacua}
\label{s:nonsusy}

The type IIA flux potential obtained in  \cite{Grimm:2004ua} has, besides the supersymmetric vacua found in \cite{DeWolfe:2005uu}, further non-supersymmetric families of vacua. This can already be seen by the toroidal analysis of \cite{DeWolfe:2005uu,Camara:2005dc}, and we generalised to arbitrary Calabi--Yau geometries in chapter \ref{ch:rrnsnsvacua}. A subset of such vacua was analysed in \cite{Narayan:2010em} in terms of perturbative and non-perturbative stability, for the particular case of $X_6 = \mathbb{T}^6/(\Z_3 \times \Z_3)$. It was found that one particular family of vacua, dubbed type 2 in \cite{Narayan:2010em}, was stable both at the perturbative and non-perturbative level.\footnote{As we pointed out in chapter \ref{ch:rrnsnsvacua} the remaining non-supersymmetric families (type 3 - type 8) found in \cite{Narayan:2010em} are not actual extrema of the flux potential, and only seem so when the potential is linearised as in \cite{Narayan:2010em}.} In the following we will extend this analysis to general Calabi--Yau geometries in the smearing approximation, and to new membranes like those arising from the D8/D6 configuration considered above. 

The non-supersymmetric vacua dubbed type 2 in \cite{Narayan:2010em} are in one-to-one correspondence with supersymmetric vacua, by a simple sign flip of the internal four-form flux $G_4 \to - G_4$. Because $G_4$ enters quadratically in the 10d supergravity Lagrangian, the energy of such a vacuum is similar to its  supersymmetric counterpart and, as argued in \cite{Narayan:2010em}, one expects it to share many of its nice properties. It was indeed found in chapter \ref{ch:rrnsnsvacua} that such non-supersymmetric vacua (dubbed {\bf A1-S1} therein) exist for any Calabi--Yau geometry, and that they are  stable at the perturbative level. Instead of the (smeared) supersymmetric relations \eqref{intfluxsm} we now have
\be
\ell_s [ H ]  = \frac{2}{5} m g_s [\re \Omega_{\rm CY} ] \, , \qquad  G_2  =  0\, ,  \qquad  \ell_s G_4  = \hat{e}_a  \tilde{\omega}^a  = \frac{3}{10} m \, \cK_a \tilde{\omega}^a  \, , \qquad  G_6 =  0\, , 
\label{intfluxsmnosusy}
\ee
and most features are analogous to the supersymmetric case. In particular, the AdS$_4$ radius and vacuum energy are also given by \eqref{Rads} and \eqref{lambda}, respectively. 

Because the energy dependence with the flux quanta is the same, one should be looking for similar non-perturbative transitions that jump to a vacuum of lower energy: Those that decrease $|\hat{e}_a|$ and those that increase $|m|$ or $|h|$. The objects that will implement such jumps will again be 4d membranes that come from (anti-)D4-branes and (anti-)D8-branes. Because of the sign flip in $G_4$, the role of the D4-branes will be exchanged with that of anti-D4-branes with respect to the supersymmetric case. 

Indeed, the relations \eqref{intfluxsmnosusy} imply that \eqref{610fluxes} is replaced by
\be
G_6 =   \frac{3\eta}{Rg_s} {\rm vol}_4 \wedge J\, , \qquad G_{10} =  - \frac{5\eta}{6 R g_s} {\rm vol}_4 \wedge J^3\, ,
\label{610fluxesnosusy}
\ee
with no further external fluxes. As a result we find the following 4d membrane couplings:
\be
Q_{D2} = 0 \, , \qquad Q_{D4}^{\rm ns} =  - \eta e^{K/2} \int_\Sigma J \, , \qquad Q_{D6} = 0\, , \qquad Q_{D8} =  -\frac{5}{3}  \eta\, q_{\rm D8} e^{K/2} {\cal V}_{\rm CY}\, .
\label{QDGKTnosusy}
\ee
By analogy with the supersymmetric case, we now find that the equality $Q =T$ is realised by D4-branes wrapping anti-holomorphic two-cycles, for $\eta =1$, and holomorphic two-cycles for $\eta=-1$. This essentially amounts to exchanging the roles of D4-brane and anti-D4-brane, as advanced. If we chose the object with opposite charge (e.g. a D4-brane wrapping a holomorphic two-cycle for $\eta=1$) then we would have that $Q = -T$ and the effects of the tension and the coupling to the flux background would add up, driving the membrane away from the boundary. In general, it is not possible to find a D4-brane such that $Q >T$, just like it is not possible to find it in supersymmetric vacua. This reproduces the result of  \cite{Narayan:2010em} that D4-brane decays are, at best, marginal. Regarding D8-branes, the naive story is essentially the same as for ${\cal N} =1$ vacua. Since $Q_{\rm D8}$ remains the same, $Q^{\rm eff}_{\rm D8/D6}$ will compensate $T_{\rm D8}$ for $\eta=q_{\rm D8}$. 

Now, the interesting case occurs when we consider bound states of D8 and D4-branes, by introducing the effect of worldvolume fluxes and/or curvature corrections. In a D8-brane configuration similar to the one in the supersymmetric case the tension is the same:
\be
T_{\rm D8}^{\rm total} = T_{\rm D8} + \left(K^F_a - K_a^{(2)}\right)  T_{\rm D4}^a\, .
\label{TD8totnosusy}
\ee
In the large volume approximation $T_{\rm D8} \gg T_{\rm D4}^a$, and so just like in the supersymmetric case we need to consider a D8-brane whenever $\eta =1$, or else $T>Q$. The coupling of the corresponding 4d membranes is now different from \eqref{QD8tot}, and reads
\be
q_{\rm D8} Q_{\rm D8}^{\rm total} = Q_{\rm D8/D6}^{\rm eff} + \left(K^F_a - K_a^{(2)}\right)  Q_{\rm D4}^{{\rm ns}, a} = \eta \left[ T_{\rm D8} -  \left(K^F_a - K_a^{(2)}\right)  T_{\rm D4}^a\right]  \, .
\label{QD8totnosusy}
\ee
As a result we find that
\be
Q_{\rm D8}^{\rm total} - T_{\rm D8}^{\rm total} = 2 \left(K_a^{(2)} -K^F_a\right)  T_{\rm D4}^a\, ,
\label{QTsmnosusy}
\ee
where we have imposed $\eta=q_{\rm D8}$. On the one hand, by assumption the D8-brane worldvolume flux induces pure D4-brane charge, which means that $K^F_aT_{\rm D4}^a >0$.\footnote{If we consider fluxes that induce pure anti-D4-brane charge, their contributions would cancel in \eqref{QTsmnosusy}.} On the other hand,  generically $K_a^{(2)} T_{\rm D4}^a > 0$, since for a Calabi--Yau \cite{Miyaoka1987}
\be
-\int_{X_6} c_2(X_6) \wedge J_{\rm CY} \geq 0\, ,
\label{condcurv}
\ee
with the equality occurring only at the boundary of the K\"ahler cone. This means that the curvature corrections are inducing negative D4-brane charge and tension on the D8-brane. The effects of such negative tension and charge add up in the present non-supersymmetric background, and they drag the D8-brane towards the AdS$_4$ boundary\footnote{Notice that this mechanism is analogous to the one in  \cite{Maldacena:1998uz}, in which a D5-branes wraps the $K3$ in AdS$_3 \times S^3 \times K3$.}. So if the worldvolume fluxes are absent or give a smaller contribution, we will have that $Q_{\rm D8}^{\rm total} > T_{\rm D8}^{\rm total}$ and the energy of the configuration will be minimised at $z_0 \to \infty$. As such, these D8/D4 bound states are clear candidates to realise the AdS instability conjecture of \cite{Ooguri:2016pdq,Freivogel:2016qwc}. In the next section we will argue that this is indeed the case.

While a remarkable result, one must realise that it does not apply to all non-supersymmetric vacua of this sort. We need that the flux vacua contain space-time filling D6-branes, or in other words that $N>0$ in \eqref{tadpole2}. If $N=0$ it means that we cannot consider a transition like the above in which $m$ increases its absolute value, or in other words that the D8-brane configuration described above cannot exist.\footnote{Or it could at the expense of introducing anti-D6-branes, which would introduce a whole new set of instabilities.} These are precisely the kind of vacua considered in \cite{Narayan:2010em} which, even with these new considerations, would a priori remain marginally stable. Moreover, if \eqref{condcurv} vanished at some boundary of the K\"ahler cone, there would be a priori no instability triggered by D8/D4-brane bound states, which would be marginal.  In fact, this last statement is not true, but only a result of the smearing approximation. As we will see, when describing the same setup but in terms of a background that admits localised sources, corrections to the D8-brane tension will appear, modifying the above computation.

\section{AdS$_4$ instability from the 4d perspective}
\label{s:insta}

The results of the previous section suggest that non-supersymmetric AdS$_4 \times X_6$ vacua with a flux background of the form \ref{intfluxsmnosusy} develop non-perturbative instabilities if they contains space-time filling D6-branes. From the 4d perspective such an instability would be mediated by a membrane that arises from wrapping a D8-brane on $X_6$, since it becomes a membrane with $Q > T$ upon dimensional reduction. However, the link between the inequality $Q > T$ and a non-perturbative gravitational instability typically follows an analysis similar to \cite{Maldacena:1998uz}, implicitly relying on the thin-wall approximation. As pointed out in \cite{Narayan:2010em}, D8-branes are not in the thin-wall approximation unless the value of $|m|$ is very large, which is not generically true. Therefore in this section we would like to provide an alternative argument of why these vacua are unstable. 

For this we will make use of the symmetry between supersymmetric and non-supersymmetric vacua mentioned in section \ref{s:nonsusy}. That is, for the same value of the fluxes $m$, $h$ and $|\hat{e}_a|$ the saxion vevs are stabilised at precisely the same value in both supersymmetric and non-supersymmetric vacua, and the vacuum energy \eqref{lambda} is also the same. For simplicity let us consider a pair of supersymmetric and non-supersymmetric vacua in which $e_0 = m^a=0$ and
\be
m^{\rm susy} = m^{\cancel{\rm susy}} > 0\, , \qquad  h^{\rm susy} = h^{\cancel{\rm susy}} > 0\, , \qquad \hat{e}_a^{\rm susy} = - \hat{e}_a^{\cancel{\rm susy}} > 0\, .
\ee
In both backgrounds, a D8-brane without worldvolume fluxes will induce the following shift of flux quanta as we cross it from $z = \infty$ to $z = -\infty$ as
\begin{align}
m^{\rm susy} \to m^{\rm susy} + 1\, , \qquad & |\hat{e}_a^{\rm susy}| \to |\hat{e}_a^{\rm susy}  + K_a^{(2)}| \, , \\
m^{\cancel{\rm susy}} \to m^{\cancel{\rm susy}} + 1\, , \qquad & |\hat{e}_a^{\cancel{\rm susy}}| \to  |\hat{e}_a^{\cancel{\rm susy}} + K_a^{(2)}| = |\hat{e}_a^{{\rm susy}}  - K_a^{(2)}|\, .
\end{align}
Because the absolute value of the four-form flux quanta $\hat{e}_a$ are different after the jump for the supersymmetric and the non-supersymmetric case, so are the vevs of the K\"ahler moduli and the vacuum energy. To fix this, let us add to the supersymmetric setup a D4-brane wrapping a holomorphic two-cycle in the Poincar\'e dual class to $2K_a^{(2)}[\tilde{\omega}^a]$. The resulting 4d membrane can create a marginal bound state with the one coming from the D8-brane, implementing the combined jump
\begin{align}
m^{\rm susy} \to m^{\rm susy} + 1\, , \qquad  |\hat{e}_a^{\rm susy}| \to |\hat{e}_a^{\rm susy}  - K_a^{(2)}| \, .
\end{align}
Now both supersymmetric and non-supersymmetric jumps are identical, in the sense that the variation of the scalar fields from the initial to the final vacuum is the same, and so are the initial and final vacuum energies. As a result, the energy stored in the field variation of both solutions should be identical. What is different is the tension of the membranes. We have that
\be
T_{\rm susy} = T_{\rm D8} +  K_a^{(2)}  T_{\rm D4}^a > T_{\rm D8} -  K_a^{(2)}  T_{\rm D4}^a = T_{\cancel{\rm susy}}\, ,
\ee
assuming as before that \eqref{condcurv} is met. Therefore, because the supersymmetric decay is marginal, the non-supersymmetric one should be favoured energetically, rendering the non-supersymmetric vacuum unstable. 


\section{Beyond the smearing approximation}
\label{s:nonsmeared}

The Calabi--Yau flux backgrounds of section \ref{s:dgkt} and \ref{s:nonsusy} can be thought of as an approximation to the actual 10d solutions to the equations of motion and Bianchi identities, in which O6-planes and D6-branes are treated as localised sources. More precisely, the smeared Calabi--Yau solution can be recovered from the actual solution in the limit of small cosmological constant, weak string coupling and large internal volume, as derived in chapter \ref{ch:uplift10d}. Any of these quantities can be used to define an expansion parameter, so that the actual 10d solution can be described as a perturbative series, of which the smeared solution is the leading term. While a solution for the whole series (i.e. the actual 10d background) has not been found yet, the next-to-leading term of the expansion was found in chapter \ref{ch:uplift10d} for the case of the supersymmetric vacua. In the following we will review the main results of that chapter, and then use the approach of \cite{Junghans:2020acz} to construct, at the same level of accuracy, a similar background with localised sources for the non-supersymmetric vacua of section \ref{s:nonsusy}. As we will see, these more precise backgrounds do not affect significantly the energetics of 4d membranes made up from D4-branes. However, as it will be discussed in the next section, they yield non-trivial effects for membranes that correspond to D8/D6 systems.

\subsection{Supersymmetric AdS$_4$}
\label{ss:mpqt}

To incorporate localised sources to the type IIA flux compactification of section \ref{ss:smeared} one must first consider a warped metric of the form
\begin{equation}\label{eq:warped-product}
	ds^2 = e^{2A}ds^2_{\mathrm{AdS}_4} + ds^2_{X_6}\, ,
\end{equation}
 with $A$ a function on $X_6$. Then, as pointed out in chapter \ref{ch:uplift10d}, the Calabi--Yau metric on $X_6$ must be deformed to a metric that corresponds to an $SU(3) \times SU(3)$-structure solution with $G_6 =0$. Assuming as before that P.D.$[\ell_s^{-2}H] = h [\Pi_{\rm O6}] = h [\Pi_{\rm D6}]$,  the first-order correction to the smearing approximation can be described in terms of the following equation
\be
\ell_s^2 \Delta_{\rm CY} K =  \frac{2}{5} m^2 g_s \re \Omega_{\rm CY} + (N-4)\d(\Pi_{\rm O6})\, ,
\label{defK}
\ee
where $\Delta_{\rm CY} = d^\dag_{\rm CY} d + d d^\dag_{\rm CY}$ is constructed from the CY metric. The solution is of the form
\be
 K = \varphi \re \Omega_{\rm CY}  + \re k \, ,
\label{formK}
\ee
with $k$ a (2,1) primitive current and $\varphi$ is a real function that satisfies $\int_{X_6} \varphi = 0$ and
\be
\Delta_{\rm CY}  \varphi = \frac{mh}{4}\left(\frac{{\cal V}_{\Pi_{\rm O6}}}{{\cal V}_{\rm CY}} - \delta^{(3)}_{\Pi_{\rm O6}}\right)  \ \implies \ \varphi \sim \cO(g_s^{1/3})\, ,
\ee
where $\delta^{(3)}_{\Pi_{\rm O6}} \equiv *_{\rm CY} (\im \Omega_{\rm CY} \wedge \d(\Pi_{\rm O6}))$. In term of these quantities we can describe the metric background and the varying dilaton profile as
\begin{subequations}	
	\label{solutionsu3}
\begin{align}
J & = J_{\rm CY} + \cO(g_s^2) \, , \qquad   \qquad  \Omega  = \Omega_{\rm CY} + g_s k +  \cO(g_s^2)\, , \\
e^{-A}  & = 1 + g_s \varphi + \cO(g_s^2) \, , \qquad e^{\phi}   = g_s \left(1 - 3  g_s \varphi\right) + \cO(g_s^3)\, .
\end{align}
\end{subequations}   
where we have taken $g_s$ as the natural expansion parameter. Notice that $\varphi \sim -\frac{mh}{4r}$ near $\Pi_{\rm O6}$, and so as expected the 10d string coupling blows up and the warp factor becomes negative near that location. The function $\varphi$ indicates the region $\tilde{X}_6 \equiv \{p \in X_6 | g_s |\varphi(p)| \ll 1\}$ in which the perturbative expansion on $g_s$ is reliable; beyond that point one may use the techniques of \cite{DeLuca:2021mcj} to solve for the 10d supersymmetry equations. The background fluxes are similarly expanded as
\begin{subequations}
	\label{solutionflux}
\begin{align}
 H & =   \frac{2}{5} \frac{m}{\ell_s} g_s \left(\re \Omega_{\rm CY} + g_s K \right) - \oh   d\re \left(\bar{v} \cdot \Omega_{\rm CY} \right) + \cO(g_s^{3}) \label{H3sol} \, , \\
 \label{G2sol}
 G_2 & =     d^{\dag}_{\rm CY} K  + \cO(g_s)  = - J_{\rm CY} \cdot d(4 \varphi \im \Omega_{\rm CY} - \star_{\rm CY} K) + \cO(g_s) \, , \\
G_4 & =  \frac{m}{\ell_s} J_{\rm CY} \wedge J_{\rm CY} \left(\frac{3}{10}  - \frac{4}{5} g_s \varphi \right)+   J_{\rm CY} \wedge g_s^{-1} d \im v + \cO(g_s^2) \, , \\
G_6 & = 0\, .
\end{align}
\end{subequations}   
Here $v$ is a (1,0)-form whose presence indicates that we are in a genuine $SU(3)\times SU(3)$ structure, as opposed to an $SU(3)$ structure. It is determined by
\be
v  = g_s \p_{\rm CY} f_\star + \cO(g_s^3)\, , \qquad \text{with} \qquad \ell_s \Delta_{\rm CY} f_\star  = - g_s 8 m \varphi \, .
\ee 
It is easy to see that \eqref{solutionsu3} and \eqref{solutionflux} reduce to the smeared solution in the limit $g_s \to 0$.  Moreover, as shown in chapter \ref{ch:uplift10d}, this background satisfies the supersymmetry equations and the Bianchi identities up to order $\cO(g_s^2)$.  As a cross-check of this result, we discuss in Appendix \ref{ap:10deom} how the 10d equations of motion are satisfied,  to the same level of accuracy.

Given this new background, one may reconsider the computation of the tension and coupling made in the smearing approximation. Let us for instance consider a D4-brane wrapping a two-cycle $\Sigma$. Instead of the expression for $G_6$ in \eqref{610fluxes} we obtain
\bea\nonumber
G_6 & = & -  {\rm vol}_4 \wedge \left[ J_{\rm CY}  \frac{m}{5\ell_s } \left(3 - 8 g_s \varphi \right) - \oh *_{\rm CY} d \left(J_{\rm CY} \wedge d^c f_\star \right)  \right] e^{4A}+ \cO(g_s^2) \\ \nonumber
& =& -  {\rm vol}_4 \wedge \left[ J_{\rm CY}  \frac{m}{5\ell_s } \left(3 - 20 g_s \varphi \right) - \oh \left( \Delta_{\rm CY} - dd^\dag_{\rm CY}\right) \left(f_\star  J_{\rm CY}  \right)  \right] + \cO(g_s^2) \\
& =&  -  {\rm vol}_4 \wedge \left[ J_{\rm CY} \frac{3\eta}{Rg_s} + \oh dd^\dag_{\rm CY} \left(f_\star  J_{\rm CY}  \right)  \right] + \cO(g_s^2)\, .
\label{6fluxesloc}
\eea
where $d^c \equiv i(\bar{\p}_{\rm CY} - \p_{\rm CY})$ and we have used that $J_{\rm CY} \wedge d^c f  = *_{\rm CY} d(J_{\rm CY} f)$. Since the only difference with respect to the smearing approximation is an exact contribution, the membrane coupling $Q_{D4}$ remains unchanged, and it is still given by $Q_{D4} =  \eta e^{K/2} \int_\Sigma J_{\rm CY}$. As before, D4-branes  wrapping holomorphic ($\eta=1$) and anti-holomorphic ($\eta=-1$) two-cycles will be BPS, and will feel no force in the above AdS$_4$ background, as expected from supersymmetry.

\subsection{Non-supersymmetric AdS$_4$}
\label{ss:nonsmearednonsusy}

Just like for supersymmetric vacua, one would expect a 10d description of the non-supersymmetric vacua of section \ref{s:nonsusy} compatible with localised sources. Again, the idea would be that the smeared background is the leading term of an expansion in powers of $g_s$. In the following we will construct a 10d background with localised sources which can be understood as a first-order correction to the smeared Calabi--Yau solution \eqref{intfluxsmnosusy} in the said expansion. 

The main feature of the non-supersymmetric background \eqref{intfluxsmnosusy} is that it flips the sign of the RR four-form flux $G_4$, while it leaves the remaining fluxes, metric, dilaton and vacuum energy invariant. This means that the Bianchi identities \eqref{BIG} do not change at leading order, and in particular the leading term for two-form flux $G_2$ should have the same form \eqref{G2sol} as in the supersymmetric case. Moreover, the localised solution is likely to be described in terms of the quantities $\varphi$ and $k$ that arise from the Bianchi identity of $G_2$, at least at the level of approximation that we are seeking. Because of this, it is sensible to consider a 10d metric and dilaton background similar to the supersymmetric case, namely \eqref{solutionsu3}. 

Regarding the background flux $G_4$, there should be a sign flip on its leading term, but it is clear that this cannot be promoted to an overall sign flip, because the co-exact piece of $G_4$, that contributes to the Bianchi identity, must be as in the supersymmetric case. Since the harmonic and co-exact pieces of the fluxes are fixed by the smearing approximation and the Bianchi identities, the question is then how to adjust their exact pieces to satisfy the equations of motion. Using the approach of \cite{Junghans:2020acz}, we find that the appropriate background reads
\begin{subequations}
	\label{solutionfluxnnosusy}
\begin{align}
 H & =   \frac{2}{5} \frac{m}{\ell_s} g_s \left(\re \Omega_{\rm CY} - 2g_s K \right) + \frac{1}{10}   d\re \left(\bar{v} \cdot \Omega_{\rm CY} \right) + \cO(g_s^{3})  \, , \\
 G_2 & =     d^{\dag}_{\rm CY} K  + \cO(g_s)   \, , \\
G_4 & =  -\frac{m}{\ell_s} J_{\rm CY} \wedge J_{\rm CY} \left(\frac{3}{10}  + \frac{4}{5} g_s \varphi \right) -\frac{1}{5}   J_{\rm CY} \wedge g_s^{-1} d \im v + \cO(g_s^2) \, , \\
G_6 & = 0\, ,
\end{align}
\end{subequations} 
with the same definition for the (1,0)-from $v$. In Appendix \ref{ap:10deom} we show that this background satisfies the 10d equations of motion up to order $\cO(g_s^2)$, just like the supersymmetric case.

With this solution in hand, one may proceed as in the supersymmetric case and recompute the 4d membrane couplings and tensions. If the result is different from the one in the smearing approximation the difference could be interpreted as a $g_s$ correction. To begin, let us again consider a D4-brane wrapping a two-cycle $\Sigma$. The coupling of such a brane can be read from the six-form RR flux with legs along AdS$_4$, which reads
\bea\nonumber
G_6 & = & {\rm vol}_4 \wedge \left[ J_{\rm CY}  \frac{m}{5\ell_s } \left(3 + 8 g_s \varphi \right) + \frac{1}{10} *_{\rm CY} d \left(J_{\rm CY} \wedge d^c f_\star \right)  \right] e^{4A}+ \cO(g_s^2) \\ \nonumber
& =& {\rm vol}_4 \wedge \left[ J_{\rm CY}  \frac{m}{5\ell_s } \left(3 - 4 g_s \varphi \right) + \frac{1}{10} \left( \Delta_{\rm CY} - dd^\dag_{\rm CY}\right) \left(f_\star  J_{\rm CY}  \right)  \right] + \cO(g_s^2) \\
& =&  {\rm vol}_4 \wedge \left[ J_{\rm CY} \frac{3\eta}{Rg_s} - \frac{1}{10} dd^\dag_{\rm CY} \left(f_\star  J_{\rm CY}  \right)  \right] + \cO(g_s^2)\, .
\label{6fluxeslocnosusy}
\eea
Remarkably, we again find that the first non-trivial correction to the smearing approximation is an exact form, and so it vanishes when integrating over $\Sigma$. As a result, the 4d membrane couplings $Q_{D4}^{\rm ns} =  - \eta e^{K/2} \int_\Sigma J$ remain uncorrected at this level of the expansion, and there is a force cancellation for D4-branes wrapping anti-holomorphic ($\eta=1$) and holomorphic ($\eta=-1$) two-cycles, just like in our discussion of section  \ref{s:nonsusy}. Presumably, by looking at higher-order corrections one may find one that violates the equality $Q_{\rm D4}^{\rm ns} = T_{\rm D4}^{\rm ns}$ in one way or the other, which would be a non-trivial test of the conjecture in \cite{Ooguri:2016pdq}. Such a computation is however beyond the scope of the present work. Instead, we will focus on membranes whose coupling and tension departure from the smeared result already at this level of approximation, namely those membranes arising from D8/D6 systems. To see how this happens, one must first take into account that beyond the smearing approximation such systems are described by BIonic configurations, as we now discuss. 


\section{BIonic membranes}
\label{s:bion}

A D$p$-brane ending on a D$(p+2)$-brane to cure a Freed--Witten anomaly constitutes a localised source for gauge theory on the latter. When going beyond the smearing approximation one should take this into account, and describe the combined system as a BIon-like solution \cite{Gibbons:1997xz}. In this section we do so for the D8/D6-brane system, and compute the  tension and flux coupling of the associated 4d membrane for both the supersymmetric and non-supersymmetric backgrounds of the last section. As we will see, the BIonic nature of the membrane will modify their coupling and tension of the membrane with respect to the smeared result.

\subsection{Supersymmetric AdS$_4$}
\label{ss:bionsusy}

Let us consider a D8-brane wrapping $X_6$ with orientation $q_{\rm D8} = \pm 1$ and extended along the plane $z = z_0$ in the Poincar\'e patch of AdS$_4$. As pointed out above, due to the non-trivial $H$-flux background we must have an excess of $h$ D6-branes wrapping $\Pi_{\rm O6}$ and extended to the right of the D8-brane, namely along $(t,x^1,x^2) \times [z_0 , \infty) \subset$ AdS$_4$. This setup implies a Bianchi identity for the D8-brane worldvolume flux $\cF = B + \frac{\ell_s^2}{2\pi} F$ of the form
\be
d{\cal F} = H - \frac{h}{\ell_s} \delta(\Pi_{\rm O6})\, .
\label{dFD8}
\ee
Because by construction the rhs is trivial in cohomology, this equation always has a solution. Moreover, if we are in the smearing approximation, we have that the rhs of \eqref{dFD8} vanishes, and so $\cF$ must be closed. The energy-minimising configurations then correspond to solving the standard F-term and D-term-like equations for $\cF$ \cite{Koerber:2007jb}, which in our setup means that $\cF$ is a harmonic (1,1)-form of $X_6$ such that $3\cF \wedge J_{\rm CY}^2  = \cF^3$. When we see such a D8-brane as a membrane in 4d, this harmonic worldvolume flux is the one responsible for the contribution $K_a^F Q^a_{\rm D4}$ to their flux coupling and tension. 

If we describe our system beyond the smearing approximation, the D8-brane worldvolume flux can no longer be closed. Instead, it must satisfy a Bianchi identity that is almost identical to the one of the RR two-form flux. Even when the harmonic piece of $\cF$ vanishes, we find that
\be
\cF =  \frac{G_2}{G_0} = \frac{\ell_s}{m} d^{\dag}_{\rm CY} K  + \cO(g_s) \, .
\label{cfsol}
\ee
BPS configurations with D$p$-branes ending on D$(p+2)$-branes, inducing a non-closed worldvolume flux on the latter are usually described by BIon-like solutions \cite{Gibbons:1997xz}, in which the D$(p+2)$-brane develops a spike along the direction in which the D$p$-branes are extended. A relatively simple configuration of this sort is given by the D5/D3 system in type IIB flux compactifications, that was analysed in \cite{Evslin:2007ti} from the viewpoint of calibrations. In this setup a D5-brane wraps a special Lagrangian three-cycle $\Lambda$ of a warped Calabi--Yau compactification, and extends along the plane $x^3 = x^3_0$ of $\pr^{1,3}$. If $\int_{\Lambda} H = - N$, then $N$ space-time filling D3-branes must end on the D5-brane, stretched along $(t,x^1,x^2) \times [x^3_0 , \infty) \subset \pr^{1,3}$ and located at a point $p \in \Lambda$. This induces an internal worldvolume flux on the D5-brane, solving the equation $d\cF = N\left(\delta(p) - \frac{d{\rm vol}_\Lambda}{{\rm Vol}(\Lambda)}\right)$. To render the configuration BPS it is necessary to give a non-trivial profile to the D5-brane position field $X^3$, such that $dX^3 = *_{\Lambda} \cF$. The resulting profile features a spike $X^3 \sim \frac{N}{r}$ around the point $p$, which represents the $N$ D3-branes ending on the D5. Therefore, the D5-brane BIon configuration accounts for the whole energy of the D5/D3 system. 

Our D8/D6 setup can be seen as a six-dimensional analogue of the D5/D3 system. The presence of the worldvolume flux \eqref{dFD8} can be made compatible with a BPS configuration if we add a non-trivial profile for the D8-brane transverse field $Z$. The relation with the worldvolume flux is now given by
\be
*_{\rm CY} dZ = q_{\rm D8} \im \Omega_{\rm CY} \wedge \cF + \cO(g_s) \, .
\label{BIonrel}
\ee
This expression can be motivated in a number of ways. In Appendix \ref{ap:dbi} we show that upon imposing it the DBI action is linearised at the level of approximation that we are working, as required by a BPS configuration. In Appendix \ref{ap:IIBion} we describe a similar configuration in type IIB flux compactifications, that can then be mapped to the BPS Abelian SU(4) instantons of Calabi--Yau four-folds \cite{Donaldson:1996kp}. Finally, notice that \eqref{BIonrel} implies that
\be
\Delta_{\rm CY} Z = \ell_s q_{\rm D8} h \left( \delta^{(3)}_{\Pi_{\rm O6}} -\frac{{\cal V}_{\Pi_{\rm O6}}}{{\cal V}_{\rm CY}} \right)\, ,
\ee
and so whenever $q_{\rm D8} h = |h|$ we recover a spike profile of the form $Z \sim \frac{|h|\ell_s}{r}$ near $\Pi_{\rm O6}$, as expected. In fact we can draw the more precise identification
\be
Z = z_0 - \frac{4\ell_s \varphi}{|m|}\, ,
\ee
where we have imposed the BPS relation $q_{\rm D8} = \eta \equiv {\rm sign}\, m$. Notice that this identifies the spike profile of the BIon solution towards the AdS$_4$ boundary with the strong coupling region near the O6-plane location, where our perturbative  expansion on $g_s$ is no longer trustable. 

\begin{figure}[h!]
    \centering
    \includegraphics[width=10cm]{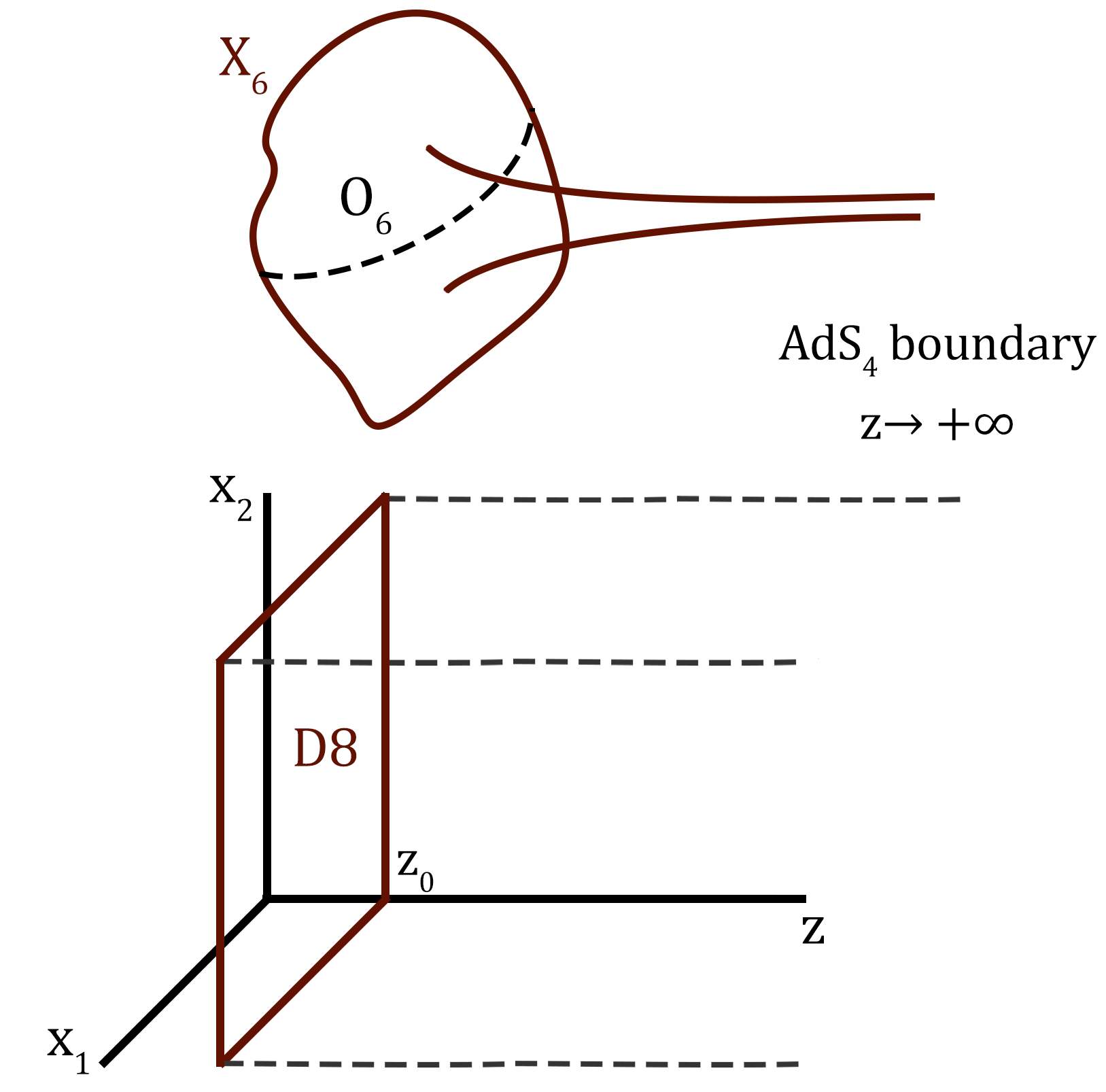}
    \caption{Beyond the smearing approximation, the D8/D6 system of figure \ref{fig:D8D6} becomes a BIon-like solution for the D8-brane, with a BIon profile that peaks at the O6-plane location.}
    \label{fig:D8D6bion}
\end{figure}

The relation \eqref{BIonrel} implies that the DBI action of the BIon can be computed in terms of calibrations. Indeed, ignoring  curvature corrections, the calibration for a D8-brane wrapping $X_6$ and with worldvolume fluxes is given by 
\be
- \im \Phi_+ = - g_s^{-1} q_{\rm D8} \im e^{-iJ_{\rm CY}} + \cO(g_s) = g_s^{-1} q_{\rm D8}\left(- \frac{1}{6} J_{\rm CY}^3 + J_{\rm CY}\right) + \cO(g_s)\, ,
\label{imPhi+}
\ee
while that for D6-branes wrapping a three-cycle of $X_6$ is 
\be
\im \Phi_- =  g_s^{-1} \left( \im v + \Im \Omega - \oh \psi \im \omega \wedge \im \Omega  \right) + \cO(g_s)  = g_s^{-1} \Im \Omega_{\rm CY} + \cO(g_s^0)\, ,
\label{imPhi-}
\ee
at leading order in our expansion.  Here $\psi$ and $\omega_0$ are a complex function and 2-form which describe the $SU(3)\times SU(3)$ structure, and such that $\Omega = \frac{i}{\psi} v \wedge \omega + \cO(g_s^2)$, see  chapter \ref{ch:review10d} for details. Applying the general formulas of \cite{Evslin:2007ti}, we find that the BIon DBI action reads 
\bea
\label{psp}
dS_\text{DBI}^{\rm D8} &= & dt \wedge dx^1 \wedge dx^2 \wedge e^{\frac{3Z}{R}} q_{\rm D8} \left(  \im \Phi_+ - dZ \wedge e^A \im \Phi_-  \right) \wedge e^{-\cF} \nonumber\\
& \simeq & dt \wedge dx^1 \wedge dx^2 \wedge g_s^{-1} e^{\frac{3z_0}{R}}  \left( \frac{1}{6}J_{\rm CY}^3  - \oh J_{\rm CY} \wedge \cF^2 + q_{\rm D8}dZ \wedge \im \Omega_{\rm CY} \wedge \cF \right) \nonumber \\
& = &   - dt \wedge dx^1 \wedge dx^2 \wedge g_s^{-1} e^{\frac{3z_0}{R}}  \left(- \frac{1}{6}J_{\rm CY}^3  + \oh J_{\rm CY} \wedge \cF^2 + *_{\rm CY} dZ \wedge dZ \right)\, .
\label{finaldbi}
\eea
The last line coincides with our result of Appendix \ref{ap:dbi}, and with what is expected for a BIon solution. Indeed, the first two terms of \eqref{finaldbi} correspond to the DBI action of the magnetised D8-brane, while the third one corresponds to the D6-branes that stretch towards the AdS$_4$ boundary. Nevertheless, notice that the middle term $\oh J_{\rm CY} \wedge \cF^2$ gives an extra contribution to the DBI action compared to the smearing approximation of section \ref{ss:4dmem}. Indeed, when $\cF$ is a harmonic form this term accounts for the contribution $K_a^FT_{\rm D4}^a$ in \eqref{TD8tot}. When going away from the smearing approximation $\cF$ will also have a co-exact piece, given by \eqref{cfsol}, that will contribute to the DBI even if $\cF^{\rm harm} =0$. Because it induces a non-trivial D4-brane charge, one may interpret this extra contribution to the D8-brane tension as a curvature correction induced by the non-trivial BIon profile, as opposed to D6-branes sharply ending on the D8-brane, although it would be interesting to derive this expectation from first principles. As we will see, this additional contribution to the tension does not play much of a role in the present supersymmetric setup, but it is crucial for the dynamics of Bionic membranes in non-supersymmetric backgrounds. 

Eq.\eqref{psp} suggests how to generalise \eqref{BIonrel} to a relation describing the BIon profile to all orders in $g_s$. The natural choice is 
\be
  *_6 dZ =  -  q_{\rm D8} e^{\phi-2A} \left. \im \Phi_- \wedge  e^{-\cF}\right|_5 \, ,
  \label{stardZ}
\ee
where the Hodge star is performed with the exact, non-Calabi--Yau metric of $X_6$, and $|_5$ means that we are only keeping the five-form component of the polyform on the rhs. With this choice the BIon DBI action would read
\be
dS_\text{DBI}^{\rm D8} =
 dt \wedge dx^1 \wedge dx^2 \wedge  e^{\frac{3Z}{R}} q_{\rm D8} \left(  \im \Phi_+  \wedge e^{-\cF} - e^{3A-\phi} *_{6} dZ \wedge dZ \right)\, ,
\label{finaldbiex}
\ee
as expected on general grounds. In addition, \eqref{psp} encodes the force cancellation observed for the D8/D6 system in the smearing approximation, which can now be derived for the single object which is the BIonic D8-brane, and in the exact background. For this, notice that the Chern-Simons part of the D8-brane action reads
\be  
dS_\text{CS}^{\rm D8}  = - dt \wedge   dx^1 \wedge dx^2 \wedge \frac{R}{3} e^{\frac{3Z}{R}} e^{4A} q_{\rm D8}  *_6 \lambda \hat{G}  \wedge e^{-\cF}  \, ,
\ee
where $\hat{G}$ is defined as in \eqref{demoflux}. Putting both contributions together and using the bulk supersymmetry equation
\be
 d_H  \left( e^A \im \Phi_- \right) +  \frac{3}{R} \im \Phi_+ =  e^{4A} *_6 \lambda \hat{G}\, ,
 \label{susy1}
\ee
and \eqref{dFD8} one finds that 
\begin{align}
\label{DBICS}
  dS_\text{DBI}^{\rm D8}  & + dS_\text{CS}^{\rm D8} = - dt \wedge dx^1 \wedge dx^2 \wedge \frac{R}{3} q_{\rm D8}  \left[ de^{\frac{3Z}{R}} \wedge  e^A \im \Phi_- +  e^{\frac{3Z}{R}}  d_H   \left( e^A \im \Phi_- \right)  \right]  \wedge e^{-\cF} \\ \nonumber
& =   - dt \wedge dx^1 \wedge dx^2 \wedge \frac{R}{3} q_{\rm D8}  \left[ d \left(e^{\frac{3Z}{R}}   e^A \im \Phi_- \wedge e^{-\cF} \right) + \frac{h}{\ell_s} e^{\frac{3Z}{R}}  \delta(\Pi_{\rm O6})  \wedge e^A \im \Phi_- \wedge  e^{-\cF} \right]  \, .
\end{align}
The first term of the second line is a total derivative that will vanish when integrating over $X_6$, while the second term is an infinite contribution to the action, that accounts for the DBI action of the $|h|$ D6-branes extending along $[z_0, \infty)$. Indeed, it is easy to see that the leading piece of this term is of the form $ |h| g_s^{-1} e^{\frac{3Z}{R}} \delta(\Pi_{\rm O6})  \wedge \im \Omega_{\rm CY} = |h|g_s^{-1} {\cal V}_{\Pi_{\rm O6}} e^{\frac{3Z_\infty}{R}}$, with $Z_\infty \equiv Z|_{\Pi_{\rm O6}} = \infty$. The relevant point is that $Z_{\infty}$ is independent of $z_0$, and therefore this second term is independent of the D8-brane transverse position. Therefore, the total energy of the BIonic 4d membrane will be independent of $z_0$, even if contains some infinite contributions. This matches the results obtained in the smearing approximation, and is equivalent to the BPS equilibrium relation $Q_{\rm D8}^{\rm BIon} = T_{\rm D8}^{\rm BIon}$. 

The above computation is quite general, and essentially follows from some  general observations made in \cite{Koerber:2007jb} applied to the present setup. It is nevertheless instructive to see how \eqref{susy1}, which is a key relation to achieve force cancellation for our BIonic D8-brane, is satisfied for the background \eqref{solutionsu3} and \eqref{solutionflux}, in preparation for the non-supersymmetric case. We have that
\begin{align}
&d_H  \left( e^A \im \Phi_- \right)  = \frac{1}{2}d d^c f_\star + *_{\rm CY} G_2 - \frac{2}{3} G_0 \left(\frac{2}{5}- g_s\varphi \right)J_{\rm CY}^3  +  \cO(g_s^{5/3})\, , \\
 &\frac{3}{R} \im \Phi_+  = \frac{3}{5} q_{\rm D8}  |G_0| \left( - J_{\rm CY} + \frac{1}{6} J_{\rm CY}^3\right) + \cO(g_s^{2})\, , \\ 
 & e^{4A} *_6 \lambda \hat{G}  = -\frac{1}{2} dd^\dag_{\rm CY} \left(f_\star  J_{\rm CY}  \right) -\frac{3}{5} G_0 J_{\rm CY} - *_{\rm CY} G_2 - \frac{1}{6} G_0 \left(1 - 4g_s \varphi\right) J_{\rm CY}^3 + \cO(g_s^{5/3})\, , 
\end{align}
and so one only has to impose $\eta=q_{\rm D8} $ and use that $d d^c f = - dd^\dag_{\rm CY} \left(f  J_{\rm CY}  \right)$  to show the equality. 

\subsection{Non-supersymmetric AdS$_4$}
\label{ss:bionnosusy}

Let us now consider the D8-brane BIon in the non-supersymmetric AdS$_4$ background of section \ref{ss:nonsmearednonsusy}. Notice that the metric and dilaton background are similar to  the supersymmetric case, and that the $H$-flux only changes by an exact piece at subleading order, so that \eqref{cfsol} remains intact. Because of this, the DBI action of the BIon should be identical to the supersymmetric case, at least to the level of approximation that we are working, and so should be the BIon profile \eqref{BIonrel}. One may thus run a very similar argument to \eqref{DBICS} to see whether the D8-brane is in equilibrium or not with the background. If not, the same computation will determine whether it is dragged towards the boundary or away from it. 

The key relation to look at is again the bulk supersymmetry equation \eqref{susy1}. If  satisfied, the BIonic membrane will be at equilibrium for any choice of transverse position $z_0$. In the smearing approximation we have already seen that there is no equilibrium whenever there is a non-trivial D4-brane charge induced by curvature or worldvolume fluxes, c.f. \eqref{QTsmnosusy}, so we do not expect \eqref{susy1} to be satisfied. Evaluating the background \eqref{solutionsu3} and \eqref{solutionfluxnnosusy} one finds that
\begin{align}
&d_H  \left( e^A \im \Phi_- \right)  = \frac{1}{2}d d^c f_\star + *_{\rm CY} G_2 - \frac{2}{15} G_0 \left(2 +  g_s\varphi \right)J_{\rm CY}^3  +  \cO(g_s^{4/3})\, , \\
 &\frac{3}{R} \im \Phi_+  = \frac{3}{5} q_{\rm D8}  |G_0| \left( - J_{\rm CY} + \frac{1}{6} J_{\rm CY}^3\right) + \cO(g_s^{2})\, , \\ 
 & e^{4A} *_6 \lambda \hat{G}  = \frac{1}{10} dd^\dag_{\rm CY} \left(f_\star  J_{\rm CY}  \right) + \frac{3}{5} G_0 J_{\rm CY} - *_{\rm CY} G_2 - \frac{1}{6} G_0 \left(1 - 4g_s \varphi\right) J_{\rm CY}^3 + \cO(g_s^{5/3})\, , 
\end{align}
which results in\footnote{In the language of \cite{Lust:2008zd,Held:2010az}, this corresponds to a background where gauge BPSness is not satisfied, and as a result some space-time filling D-branes may develop tachyons. One can however check that D6-branes wrapping special Lagrangians of $X_6$, and in particular those on top of the orientifold, do not develop any instability. It would be interesting to see if D8-branes wrapping coisotropic five-cycles \cite{Font:2006na} could develop them.}
\be
 d_H  \left( e^A \im \Phi_- \right) +  \frac{3}{R} \im \Phi_+ -  e^{4A} *_6 \lambda \hat{G} = - \frac{3}{5} dd^\dag \left(f_\star  J_{\rm CY}  \right) -\frac{6}{5} G_0 J_{\rm CY}  - \frac{4}{5}  G_0 g_s \varphi J_{\rm CY}^3  +\cO(g_s^{4/3})\, .
\ee
Plugged into the DBI and CS actions, and again ignoring curvature terms, this translates into
\begin{align}
\nonumber
  dS_\text{DBI}^{\rm D8}   + dS_\text{CS}^{\rm D8} &= - dt \wedge dx^1 \wedge dx^2 \wedge \frac{R}{3} q_{\rm D8}e^{\frac{3Z}{R}}  \left[\frac{3}{10} \left( dd^\dag \left(f_\star  J_{\rm CY}  \right) + 2 G_0 J_{\rm CY}\right) \wedge \cF^2  + \frac{4}{5}  G_0 g_s \varphi J_{\rm CY}^3 \right] + \dots\\
  & = - dt \wedge dx^1 \wedge dx^2 \wedge \frac{R}{3} e^{\frac{3z_0}{R}}   \left[\frac{3}{5} |G_0| J_{\rm CY} \wedge \cF^2 + \frac{4}{5}  |G_0| g_s \varphi J_{\rm CY}^3 \right] + \dots
  \label{DBICSns}
\end{align}
where we have neglected terms that do not depend on $z_0$, and in the second line we have only kept terms up to order $\cO(g_s^{4/3})$. Out of the two remaining terms, one of them will vanish when integrating over $X_6$, since  $\int_{X_6} \varphi = 0$. The other one finally gives
\be
 Q_{\rm D8}^{\rm BIon, ns} - T_{\rm D8}^{\rm BIon, ns} =  - e^{K/2} \int_{\rm X_6} J_{\rm CY} \wedge \cF^2 + \cO(g_s^2) \, .
\label{QTbionnosusy}
\ee
This result is perhaps not very surprising, because it reproduces the result \eqref{QTsmnosusy} of the smearing approximation when curvature corrections are omitted and $\cF$ is a harmonic form. However remember that in the present setup $\cF$ is always non-vanishing, even when the harmonic piece of $\cF$ is set to zero. Therefore, 
\be
2\Delta_{\rm D8}^{\rm Bion} \equiv - e^{K/2} \int_{\rm X_6} J_{\rm CY} \wedge \cF^2\qquad \text{with} \qquad  \cF = \frac{\ell_s}{m} d^{\dag}_{\rm CY} K 
\label{QTbionnosusyexp}
\ee
constitutes a correction to the previous result \eqref{QTsmnosusy}. Since a vanishing harmonic piece for $\cF$ is always a choice, there will always be some BIonic membrane whose charge-to-tension ratio will be fixed by the curvature term $2K_a^{(2)} T_{\rm D4}^a$ plus  \eqref{QTbionnosusyexp}. 

One may thus wonder what is the magnitude of $\Delta_{\rm D8}^{\rm Bion}$ compared to $2K_a^{(2)} T_{\rm D4}^a$, as well as its sign. For this notice that \eqref{cfsol} is suppressed as $\cO(g_s^{2/3})$ compared to a harmonic two-form. Therefore $\Delta_{\rm D8}^{\rm Bion}$ gets an relative suppression of $\cO(g_s^{4/3}) \sim {\cal V}_{\rm CY}^{-2/3}$, just like both terms in \eqref{QTsmnosusy}. In other words, $\Delta_{\rm D8}^{\rm Bion}$ and  $2K_a^{(2)} T_{\rm D4}^a$ scale similarly with the string coupling. As for the sign, it will be the result of two competing quantities, since
\be
2\Delta_{\rm D8}^{\rm Bion} =  e^{K/2} \int_{X_6}  *_{\rm CY} \cF_2 \wedge \cF_2 -   *_{\rm CY} \cF_1 \wedge \cF_1\, ,
\ee
where $\cF_1 \equiv  \cF^{(1,1)}$ and $\cF_2 \equiv \cF^{(2,0)+(0,2)}$. If we assume \eqref{cfsol} we obtain
\begin{align}
\label{cF1}
\cF_1 &\equiv  \cF^{(1,1)} = \frac{i}{2G_0} J_{\rm CY} \cdot \bar{\partial} k = G_0^{-1}J_{\rm CY} \cdot d\left(*_{\rm CY} K - 2 \varphi \im \Omega_{\rm CY} \right) \, , \\
\cF_2 & \equiv \cF^{(2,0)+(0,2)} = - G_0^{-1} J_{\rm CY} \cdot d\left( 2\varphi \im \Omega_{\rm CY} \right) \, .
\label{cF2}
\end{align}
Intuitively, a $(1,1)$ component of $\cF$ induces D4-brane charge on the BIon worldvolume, and drags it away from the boundary, while a $(2,0)+(0,2)$ component induces anti-D4-brane charge and therefore the opposite effect. So if the integrated norm of $\cF_2$ wins over that of $\cF_1$ the BIonic membrane suffers an additional force that draws it towards the boundary of AdS$_4$, providing a source of instability for the non-supersymmetric vacuum. 

Computing $\Delta_{\rm D8}^{\rm Bion}$ is in general non-trivial, but one may do so for toroidal or toroidal orbifold geometries, where vacua of this sort can be constructed explicitly \cite{Camara:2005dc,Narayan:2010em,Ihl:2006pp}. We have computed its value  in Appendix \ref{ap:torus}, for the particular case of $T^6/ (\Z_2 \times \Z_2)$.  The result is
\be
\Delta_{\rm D8}^{\rm Bion} (T^6/ (\Z_2 \times \Z_2)) = \frac{(8h)^2}{24} \sum_i T_{\rm D4}^i \, ,
\label{finalresult}
\ee
where $T_{\rm D4}^i = e^{K/2} \frac{1}{4} {\cal V}_{T^2_i}$ is the tension of a fractional D4-brane wrapping the $i^{\rm th}$ two-torus. As discussed in Appendix \ref{ap:torus} the factor $(8h)^2$ is related to the O6-plane intersections, while the $1/24$ is reminiscent of \eqref{Kcurv}. It is thus tempting to interpret \eqref{finalresult} as a curvature correction to the BIon action induced by D6-brane intersections. This intuition matches well with the positive sign in \eqref{finalresult} that adds up to the effect of the Calabi--Yau curvature corrections. Together, they mean that the induced charge and tension is negative compared to that of D4-branes, and this drags the BIonic membrane towards the boundary of AdS$_4$. Following our previous discussion, this will induce a non-perturbative instability towards non-supersymmetric vacua with larger values for $|m|$ and less space-time filling D6-branes, until none of the latter remain.  A more general analysis of $T^6/ (\Z_2 \times \Z_2)$ and other toroidal orientifolds will be carried in \cite{Casas:2022mnz}. Notice that even if we stabilise the K\"ahler moduli away from the orbifold limit as in \cite{DeWolfe:2005uu,Narayan:2010em}, in the trustable regime in the blow-up modes are significantly smaller than the untwisted K\"ahler moduli, and so the sign of \eqref{finalresult} still determines whether the BIonic contribution creates an excess charge for the membrane. If  the contribution $\Delta_{\rm D8}^{\rm Bion}$ is non-negative like $K_a^{(2)} T_{\rm D4}^a$ both effects will add up and, as a consequence, a non-perturbative instability would be induced for vacua with D6-branes.


\section{Conclusions}
\label{s:conclu}

In this chapter we have revisited the non-perturbative stability of type IIA $\cN=0$ AdS$_4 \times X_6$ orientifold vacua, where $X_6$ has a Calabi--Yau metric in the smeared-source approximation. For our analysis we have used the results of  \cite{Junghans:2020acz} and chapter \ref{ch:uplift10d},  which give a description of these backgrounds beyond the Calabi--Yau approximation. Such a description is quite accurate in the large-volume, weak-coupling regime, at least at regions of $X_6$ away from the O6-plane location. However, as already pointed out, we are still working with an approximate solution which will have further corrections at higher orders in the expansion. At such a higher level of accuracy, and specially in non-supersymmetric settings, there will be additional corrections that one should take into account, and which are beyond the scope of the present analysis. 

Given our results, there are several open questions to be addressed. First, we have unveiled a potential decay channel for $\cN=0$ AdS$_4$ vacua with space-time filling D6-branes, triggered by nucleating D8-branes that take the system to a new $\cN=0$ vacuum with larger $|F_0|$ and fewer D6-branes. There are two quantities that determine if this decay channel exist, namely the curvature correction term $K_a^{(2)} T_{\rm D4}^a$ to the D8-brane action and the BIon correction $\Delta_{\rm D8}^{\rm Bion}$ defined in \eqref{QTbionnosusyexp}. The sharpened WGC for membranes \cite{Ooguri:2016pdq} predicts that $K_a^{(2)} T_{\rm D4}^a+\Delta_{\rm D8}^{\rm Bion} > 0$, securing the decay channel. We have shown that this is the case for a simple D6-brane configuration in $X_6 = T^6/ (\Z_2 \times \Z_2)$, and it would be interesting to extend our analysis to other configurations, other toroidal orbifolds and more general Calabi--Yau geometries. It was shown recently in \cite{Casas:2022mnz} that this is indeed the case for most of the toroidal examples studied there, though there are configurations that seem to be in tensions with what was expected  from the WGC.  It would however remain to see if the two terms always add up to yield a positive quantity, the key question being how $\Delta_{\rm D8}^{\rm Bion}$ behaves in general. Because $\cF$ is a non-closed but nevertheless quantised two-form, it could be that $\Delta_{\rm D8}^{\rm Bion}$ is determined by the topological data of the problem, as the simple result obtained for toroidal geometries would suggest.

More generally, the instabilities that we have discussed only apply to vacua with space-time filling D6-branes. For instance, the explicit vacua discussed \cite{DeWolfe:2005uu,Narayan:2010em} were based on toroidal orbifolds, but the $H$-flux and $F_0$ quanta were chosen such that no D6-branes were present. For these vacua and others alike, our results find no superextremal membrane that could mediate the decay, since D4-branes saturate a BPS bound in the same sense that they do in the smeared-source approximation analysis. It would be interesting to see if pushing our analysis to the next term in the expansion one could find that $Q_{\rm D4} \neq T_{\rm D4}$ in $\cN=0$ backgrounds, or if some other kind of corrections sourced by supersymmetry-breaking effects creates an imbalance. If not, one may consider more exotic classes of processes where four-form flux is discharged, like decays involve a mixture of bubbles of nothing and D4-brane charge (see e.g. \cite{Bomans:2021ara}) to fully test the sharpened WGC for membranes. 

In any event, we believe that the decay processes that we have studied are interesting per se, and deserve further study. Notice for instance that after bubble nucleation the AdS$_4$ flux dual to the Romans mass is not discharged, as in \cite{Maldacena:1998uz}, but on the contrary it increases. And the same happens with the 4d four-form flux dual to $G_4$. From the 4d viewpoint there is nothing wrong with this fact, as we jump to a new $\cN=0$ vacuum with lower vacuum energy. Indeed, we have argued in \ref{s:insta} that these decays are favourable from the 4d viewpoint, even when we are away from the thin-wall approximation. It would however be interesting to carry a more detailed 4d analysis of this process, as well as to build the explicit 4d solution. Moreover, it would be important to analyse the superextremality of the membranes from a standard 4d viewpoint, like the analysis of the WGC for membranes carried out in \cite{Lanza:2020qmt}.

From the microscopic viewpoint, it would be interesting to see if our computations can be generalised to other string theory settings. To start with, an analysis analogous to the one performed in this chapter will be applied to the so-called A2-S1 branch of table \ref{vacuresulnsrr} in \cite{nuevo}. Other obvious candidates are the class of type IIA orientifold compactifications studied in \cite{Banks:2006hg,Cribiori:2021djm}, which share many similar properties with the ones considered in this paper. But one may also consider other compactifications which share key ingredients like scale separation and non-Abelian chiral gauge theories, and see if similar results are obtained. After all, our results hint that $\cN=0$ 4d EFTs with non-trivial gauge sectors are more susceptible to decay to vacua where such gauge sectors are absent. If true in general, this would have deep implications for string theory model building, and probably result into a new branch of implications of the Swampland Programme.

\clearpage
\chapter{Conclusions and final remarks}
\label{ch:conclusions}

In this thesis  we have explored the vacua structure of (massive) type IIA (CY) orientifold compactifications with fluxes. We have done it with one eye on the swampland program and one eye on the phenomenology applications. The constructions obtained here can serve both as an arena to test the swampland conjectures and as a first step in connecting string theory with our universe. We have been following the most natural reasoning, starting with the basics and adding ingredients to make more complex elaborations. We will close these pages by recapping what we have done and establishing what questions are still open.

After motivating the work and recalling the very fundamental concepts in chapter \ref{ch:introducion}, we started in chapter \ref{ch:review} by reviewing (massive) type IIA Calabi-Yau orientifold compactifications. Following mainly \cite{Grimm:2005fa}, we wrote the 4d $\mathcal{N}=1$ effective action that describes this scenario, adapted to our conventions and in the presence of fluxes. We first considered only NSNS and RR fluxes,  and then also added  (non)-geometric fluxes. We explained how the fluxes generate a potential for the moduli\footnote{Massless scalar fields appearing in the effective action when compactifying that determine the geometrical properties of the internal manifold.} that can be described in the SUGRA language in terms of a superpotential $W$ and a K\"ahler potential $K$. In the presence of only RR and NSNS, it was was shown in \cite{Herraez:2018vae} that the potential can be rewritten in full generality in a bilinear form, $V= \left(\vec{\rho}\right)^ {t} Z\vec{\rho}$, result  that was generalised in chapter \ref{ch:geometricflux} including also (non)-geometric fluxes. We finished with section  \ref{sec:swamplandcon},  listing the main swampland conjectures tested in the thesis.

Having an effective action and a potential, the next step was to analyse the extrema of the potential, to see what vacua the theory has. We started in chapter \ref{ch:rrnsnsvacua} by turning off (non)-geometric fluxes and taking only NSNS and RR fluxes into account. The bilinear form of the potential was very useful to perform a systematic search of vacua and to see that de Sitter (dS) extrema cannot be obtained in this setup. We obtained several families of SUSY and non-SUSY AdS$_4$ vacua -see table \ref{vacuresulnsrr}- generalising previous results  \cite{DeWolfe:2005uu,Camara:2005dc}, computed in toroidal examples, to any CY orientifold. We analysed the \text{validity} of the vacua obtained, verifying its perturbative stability and checking that they are in a regime (large volume, weak coupling) in which we can trust the theory. We also saw they all have separation of scales: the radius of the Calabi-Yau manifold can be made parametrically smaller than the AdS$_4$ radius, contradicting the strong form of the AdS distance conjecture \cite{Ooguri:2016pdq}.

Having studied the extrema of the potential generated by RR and NSNS fluxes, we added geometric fluxes and performed a similar analysis in chapter \ref{ch:geometricflux}. We took an ansatz for the fluxes in the vacuum -motivated by stability arguments- and exploited again the bilinear form of the potential. We derived several families of SUSY and non-SUSY AdS$_4$, vacua which are summarised in table \ref{vacuresul}. As in the previous chapter, we studied the validity of these new vacua, checking which of them are perturbatively stable. In this case we were not able to find a regime in which the radius of the internal manifold can be made parametrically smaller than the radius of the AdS$_4$

After focusing on the 4d point of view in chapters \ref{ch:review}-\ref{ch:geometricflux}, in chapter \ref{ch:review10d} we started the analysis of the 10d uplift of the vacua obtained (recall that massive type IIA in a CY orientifold does not solve all the internal equations). We began by reviewing the main tools needed for a 10d analysis. We recalled how the SUSY equations can be rewritten in a very elegant way using the language of polyforms and how SUSY constrains the internal manifold to be either a $SU(3)$ or a $SU(3)\times SU(3)$ structure manifold. Subsequently, we went over the \textit{smearing problem}, the fact that the 10d uplift cannot be in a $SU(3)$ structure manifold. For the vacua considered, with fluxes and O6-planes,  the Bianchi identities cannot be solved in $SU(3)$ manifolds unless one considers \emph{smeared} sources, which is not describing the true physical situation. We saw that this happens both in the case of only RR and NSNS fluxes and when geometric fluxes are added. We gave the smeared uplift for both cases, agreeing with the previous results in the literature \cite{Acharya:2006ne}.

From chapter \ref{ch:uplift10d} until the end of the thesis, we focused on the case with only RR and NSNS fluxes. In chapter \ref{ch:uplift10d} we studied the uplift of the SUSY vacua computed in chapter \ref{ch:rrnsnsvacua} going beyond the smeared uplift, and considering $SU(3)\times SU(3)$ structure manifolds. To do so, we expanded the equations in terms of $g_s$. At zeroth order the smeared uplift is recovered. At first order the localised nature of the sources is taken into account -though the intersection terms between different O6-planes decouple, since they appear at next order-. We solved there all the equations of motion and the Bianchi identities at first order. We did it in general and also explicitly for a particular toroidal example.

In the last chapter of the thesis, chapter \ref{ch:nonsusy}, we used the machinery developed in chapter \ref{ch:uplift10d} to study the non-perturbative stability of one of the families of non-SUSY vacua obtained in chapter \ref{ch:rrnsnsvacua}. Namely, the non-SUSY branch that is related to the SUSY one changing $G_4^{\text{non-SUSY}}=-G_4^{\text{SUSY}}$. According to a refined version of the WGC \cite{Ooguri:2016pdq}, there should be branes in the spectrum of this background with $Q>T$, triggering its decay. In the smearing approximation and both for the SUSY and non-SUSU vacua, we were able to find only BPS branes, that is, branes with $Q=T$ and branes with $Q<T$. Therefore, the decay of the non-SUSY vacua can only be at best marginal in this approximation. We had to go beyond the smeared uplift and generalise the results of chapter \ref{ch:uplift10d} to compute the solution of the EOM at first order in the $g_s$ expansion for the non-SUSY background. In this more accurate description, the SUSY spectrum continues to contain only BPS branes and branes with $Q<T$, as was expected. For the non-SUSY solution, D8 branes wrapping the internal manifold with D6 branes ending on them can have $Q>T$, sourcing the decay of the AdS$_4$, as predicted by the refined form of the WGC.

Having explored the vacua structure of type IIA orientifold compactifications, we have opened the door to many interesting questions. First of all, there is the problem of having dS vacua in string theory. We have seen that in type IIA orientifold compactifications they cannot be obtained with only RR and NSNS fluxes in the large volume, weak coupling regime (where the vanilla SUGRA description is valid). But there is still room to obtain them either in other regimes of string theory (in which nowadays we have less control) or maybe with (non)-geometric fluxes. In this sense, one can try to generalise the ansatz taken for the geometric fluxes in chapter \ref{ch:geometricflux}, and explore the extrema of the potential for more complicated proposals. Also, non-geometric fluxes are still poorly understood, so more research in this sense is needed.

Another very stimulating and urgent topic is the construction of scale separated vacua in string theory. As of today, the only models yielding scenarios with this property are (massive) type IIA orientifold compactifications. In the last years, the study of the would-be CFT duals to these theories has been initiated  \cite{Aharony:2008wz, Conlon:2021cjk, Apers:2022zjx, Apers:2022tfm, Quirant:2022fpn}, with the aim of shed light in this problem using a different approach. As explained along the pages of this thesis, the issue with the aforementioned constructions is that a full solution for the 10d equations of motion is not known unless the  O6-planes are smeared. In chapter \ref{ch:uplift10d} we made a non-trivial step in this direction, showing that a solution beyond the smeared uplift exists. But this is not the end of the story, since problems could appear at next orders. The existence of a full-controlled 10d set-up with separation of scales between the internal and the external manifold is another puzzle that we aim to solve in the future.

Finally,  related to the previous problem, there is also the question of building non-SUSY  (meta)stable vacua in string theory\footnote{Metastability is fine as long as the lifetime of the vacuum is large enough to accommodate our universe.}. This is crucial for string theory to describe our universe. In this respect, there are still some non-SUSY vacua derived in chapters \ref{ch:rrnsnsvacua} and \ref{ch:geometricflux} whose non-perturbative stability should be checked. Indeed, for the so-called branch A2-S1 in chapter  \ref{ch:rrnsnsvacua} we are working in this problem in \cite{nuevo}, in the lines of the non-SUSY instability conjecture \cite{Ooguri:2016pdq}.

So, as is usually the case in science, we conclude with the idea that though this thesis closes some questions, it opens up many new ones. This is precisely the beauty of this job. Hopefully, we can continue contributing to solving some of them in the not too distant future.

\chapter{Conclusiones y comentarios finales}
\label{ch:conclusionss}
\thispagestyle{empty}
En esta tesis hemos explorado la estructura de vacios de la teor\'ia de cuerdas tipo IIA (masiva) compactificada en \textit{orientifolds} (de variedades de Calabi-Yau) incluyendo flujos. Lo hemos hecho con  un ojo puesto en el programa de la ci\'enaga (el \textit{swampland program}) y otro en las aplicaciones fenomenol\'ogicas. Las construcciones que hemos obtenido pueden servir tanto como para testear las conjeturas del pantano (las \textit{swampland conjectures}) como un primer paso en la conexi\'on de la teor\'ia de cuerdas con nuestro universo. Hemos ido seguimiento un razonamiento \textit{natural}, empezando por lo m\'as simple y a\~{n}adiendo ingredientes para construir escenarios m\'as complejos. Cerraremos estas p\'aginas resumiendo el trabajo realizado y  comentando qu\'e preguntas quedan todav\'ia por responder.

Despu\'es de motivar la tesis y recordar los conceptos m\'as fundamentales en el cap\'itulo \ref{ch:introducion}, empezamos el cap\'itulo \ref{ch:review} repasando  la compactificaci\'on  de la teor\'ia de cuerdas de tipo IIA (masiva) en orientifolds de variedades de Calabi-Yau. Siguiendo principalmente \cite{Grimm:2005fa}, derivamos la acci\'on de supergravedad $\mathcal{N}=1$ en 4d que describe este escenario, adaptada a nuestros convenios y teniendo en cuenta la presencia de flujos. Primero consideramos solo flujos de tipo RR y NSNS y luego a\~{n}adimos tambi\'en flujos (no)-geom\'etricos. Explicamos c\'omo los flujos generan un potential para los \textit{moduli}\footnote{Campos escalares sin masa que aparecen en la acci\'on efectiva al compactificar y que determinan las propiedades geom\'etricas de la variedad interna.} que puede ser descrito en el lenguaje de supergravedad en t\'erminos de un superpotential $W$ y un potential K\"ahler $K$. En presencia de flujos  de tipo RR y NSNS, en \cite{Herraez:2018vae} demostraron que el potential puede ser reescrito, con total generalidad, como una expresi\'on bilinear, $V=\left(\vec{\rho}\right)^t Z \vec{\rho}$, resultado que nosotros generalizamos en el cap\'itulo \ref{ch:geometricflux} incluyendo tambi\'en flujos (no)-geom\'etricos. Acabamos este cap\'itulo con la secci\'on \ref{sec:swamplandcon}, haciendo un listado y repasando las principales conjeturas de la ci\'enaga analizadas en esta tesis.

Teniendo una acci\'on efectiva y un potencial, el siguiente paso fue analizar los exremos de este potential, para ver qu\'e vac\'ios tiene la teor\'ia. Empezamos esta tarea en el cap\'itulo \ref{ch:rrnsnsvacua} apagando los flujos (no)-geom\'etricos y considerando solo flujos de tipo NSNS y RR. La forma de escribir el potential como una expresi\'on bilinear fue muy \'util para llevar a cabo una b\'usqueda de vac\'ios sistem\'atica y para ver  que no se puede obtener vac\'ios de tipo de Sitter (dS) en estas construcciones. Obtuvimos varias familias de vac\'ios AdS$_4$ tanto supersim\'etricas como no-supersim\'etricas -ver tabla  \ref{vacuresulnsrr}- generalizando resultados previos \cite{DeWolfe:2005uu,Camara:2005dc} calculados en ejemplos toroidales, a \textit{orientifolds} de cualquier variedad de Calabi-Yau. Analizamos tambi\'en la validez de los vac\'ios obtenidos, verificando su estabilidad a nivel perturbativo y comprobando que est\'en en un r\'egimen (volumen grande, acoplo d\'ebil) en el que podamos confiar en la descripci\'on usada. Tambi\'en vimos que todos estos vac\'ios tienen separaci\'on de escalas: el radio de la variedad de Calabi-Yau puede hacerse param\'etricamente m\'as peque\~{n}o que el radio de AdS$_4$, contradiciendo la versi\'on fuerte de la llamada \textit{AdS distance conjecture} \cite{Ooguri:2016pdq}.

Habiendo estudiado los extremos del potencial cuando solo se incluyen flujos RR y NSNS, en el cap\'itulo \ref{ch:geometricflux} a\~{n}adimos flujos geom\'etricos para realizar un an\'aslisis an\'alogo. Asumimos  un cierto \textit{ansatz} para los flujos geom\'etricos en el vac\'io -motivado por argumentos de estabilidad- e hicimos uso de la forma bilinear del potential. Encontramos de nuevo varias familias de vacios AdS$_4$ tanto supersim\'etricas como no-supersim\'etricas -un resumen puede verse en la tabla \ref{vacuresul}-. Como en el cap\'itulo anterior, estudiamos la validez de estos vac\'ios, comprobando cu\'ales de ellos son estables a nivel perturbativo. En este caso no fuimos capaces de encontrar un r\'egimen en el que el radio de la variedad interna pueda hacerse param\'etricamente m\'as peque\~{n}o que el radio de AdS$_4$.

Despu\'es de centrarnos en el punto de vista 4-dimensional en los cap\'itulos  \ref{ch:review}-\ref{ch:geometricflux}, en el cap\'itulo \ref{ch:review10d} empezamos el estudio del \textit{uplift} 10-dimensional de los vacios obtenidos previamente (recordemos que la IIA masiva compactificada en un orientifold de un Calabi-Yau no resuelve todas las ecuaciones en las dimensiones internas). Empezamos repasando las herramientas principales necesarias para un an\'alisis 10-dimensional. Recordamos c\'omo las ecuaciones de supersimetr\'ia pueden reescribirse de una forma muy elegante utilizando el lenguaje de las poliformas y c\'omo el hecho de exigir que cierta cantidad de supersimetr\'ia se preserve al compactificar constri\~{n}e severamente la variedad interna: esta solo solo puede tener o bien una estructura $SU(3)$ o bien una esctructura $SU(3)\times SU(3)$. Seguidamente, recordamos tambi\'en el problema del \textit{smearing} -lo que traduciremos como el problema de la deslocalizaci\'on- el hecho de que el \textit{uplift} 10-dimensional no puede ser en una variedad con escturctura $SU(3)$. Para los vacios que estamos estudiando, con flujos y O6-planos, las identiadades de Bianchi no tienen soluci\'on en variedades de tipo $SU(3)$, a menos que consideremos que las fuentes est\'an deslocalizads, lo que no representa la verdadera situaci\'on f\'isica. Vimos que esto ocurre tanto en el caso en el que solo incluimos flujos RR y NSNS, as\'o como cuando a\~{n}adimos flujos geom\'etricos. Con esto en mente, obtivuimos el \textit{uplift} deslocalizado para ambos casos, de acuerdo con los resultados previamente derivados en la literatura.

Desde el cap\'itulo \ref{ch:uplift10d} hasta el final de la tesis, nos centramos en el caso en el que solo incluimos flujos de tipo RR y NSNS. En este mismo cap\'itulo, el cap\'itulo \ref{ch:uplift10d}, estudiamos el \textit{uplift} de los vac\'ios supersim\'etricos calculados en el cap\'itulo \ref{ch:rrnsnsvacua} yendo m\'as all\'a de la aproximaci\'on deslocalizada, considerando variedades con una estructura $SU(3)\times SU(3)$. Para ello, expandimos todas las ecuaciones en t\'erminos de $g_s$. A orden zero se recupera la aproximaci\'on deslocalizada. A primer orden la naturaleza localizada de las fuentes se hace manifiesta -aunque no los t\'erminos de intersecci\'on entre diferentes O6-planos, que vienen a siguiente orden-. Resolvimos tanto las ecuaciones de movimiento como las identiades de Bianchi a primer orden. Lo hicimos tanto en general como expl\'icitamente para un ejemplo toroidal.

En el \'ultimo cap\'itulo de la tesis, el cap\'itulo  \ref{ch:nonsusy}, usamos todo el formalismo desarrollado en el cap\'itulo \ref{ch:uplift10d} para estudiar la estabilidad no-perturbativa de una de las familias de vac\'ios no-supersim\'etricas obtenidas en el cap\'itulo  \ref{ch:rrnsnsvacua}. En concreto, la rama no-supersim\'etrica que est\'a relacionada con la rama supersim\'etrica mediante el cambio de signo $G_4^{\text{no-supersim\'etrica}}=-G_4^{\text{supersim\'etrica}}$. Seg\'un una versi\'on refinada de la conjetura de la gravedad d\'ebil (la \textit{WGC} por sus siglas en ingl\'es, \textit{Weak Gravity Conjecture}) \cite{Ooguri:2016pdq}, el espectro de la teor\'ia deber\'ia contener branas cuya carga satisfaciera $Q>T$, lo que provocar\'ia el decaimiento del vac\'io. En la aproximaci\'on deslocalizada y tanto para los vac\'ios supersim\'etricos como los no-supersim\'etricos, solo fuimos capaces de encontrar branas BPS, esto es, branas cuya carga y tensi\'on satisfacen $Q=T$ y branas con $Q<T$. Por tanto, el decaimiento de los vac\'ios no-supersim\'etricos en esta aproximaci\'on solo puede ser, como mucho, marginal. Tuvimos que ir m\'as all\'a del uplift deslocalizado y generalizar los resultados del cap\'itulo  \ref{ch:uplift10d}, calculando tambi\'en para el caso no-supersim\'etrico la soluci\'on de las ecuaciones de movimiento a primer orden en la expansi\'on en $g_s$. En esta descprici\'on m\'as precisa, el espectro supersim\'etrico continua contenieno solo branas de tipo BPS y branas con $Q<T$, como era de esperar. Sin embargo, para la soluci\'on no-supersim\'etrica, D8 branas envolviendo la variedad interna y con D6 branas acabando en ella pueden tener $Q>T$, provocando una inestabilidad del vac\'io AdS$_4$, como predice la forma refinada de la \textit{WGC}.

Habiendo explorado la estructura de vac\'ios de la tipo IIA compactificada en \textit{orientifolds}, hemos abierto la puerta a muchas preguntas muy interesantes. En primer lugar est\'a el problema de construir vac\'ios dS en teor\'ia de cuerdas. Hemos visto que en compactificaciones en \textit{orientifolds} de la IIA no pueden obtenerse si solo incluimos flujos RR y NSNS y nos quedamos en el r\'egimen de volumen grande y acopldo d\'ebil (donde la descripci\'on m\'as simple, en t\'erminos de una teor\'ia de supergravedad sin correcciones, es suficiente). Esto no significa que no puedan obtenerse en general. Podr\'ian aparecer en otros r\'egimenes de la teor\'ia de cuerdas (en los que hoy en d\'ia tenemos menos control) o puede que incluyendo flujos (no)-geom\'etricos. En este sentido, uno podr\'ia tratar de generalizar el \textit{ansatz} que asumimos para los flujos geom\'etricos en el cap\'itulo  \ref{ch:geometricflux}, buscando otros extremos del potential en  propuestas m\'as complejas. As\'i mismo, los flujos no-geom\'etricos todav\'ia no se entienden completamente, por lo que se necesita m\'as investigaci\'on en esta direcci\'on.

Otro tema muy estimulante, a la vez que urgente, es la construcci\'on de vac\'ios con separaci\'on de escalas en teor\'ia de cuerdas. A d\'ia de hoy, los \'unicos modelos con esta propiedad que se conocen son las compactificaciones en \textit{orientifolds} de la IIA (masiva). En los \'ultimos a\~{n}os se ha iniciado el estudio de las que ser\'ian las teor\'ias CFT duales a estos escenarios \cite{Aharony:2008wz, Conlon:2021cjk, Apers:2022zjx, Apers:2022tfm, Quirant:2022fpn}, con la idea de arrojar luz usando un enfoque distino. Como se ha explicado a lo largo de las p\'aginas de esta tesis, el problema de estas construcciones es que solo se conoce una soluci\'on completa a las ecuaciones 10-dimensionales cuando los O6-planos se deslocalizan en las dimensiones internas. En el cap\'itulo  \ref{ch:uplift10d}  dimos un paso no trivial en esta direcci\'on, demostrando que una primera soluci\'on m\'as all\'a de la aproximaci\'on deslocalizada existe. Pero esto no cierra el asunto, ya que podr\'ian aparecer obst\'aculos en los siguientes \'ordenes de la expansi\'on que realizamos. La existencia de una soluci\'on 10d que est\'e totalmente bajo control y en la que haya separaci\'on de escalas entre la dimensi\'on interna y la externa es otro puzzle que esperamos que se resuelva en el futuro.

Finalmente, relacioando con el problema anterior, est\'a la cuesti\'on de construir vac\'ios no-supersim\'etricos (meta)estables en teor\'ia de cuerdas\footnote{La metaestabilidad puede ser satisfactoria siempre que la vida media del vac\'io sea lo suficientemente grande para acomodar la historia de nuestro universo.}. Esto es crucial si aspiramos a que la teor\'ia de cuerdas sea capaz de describir nuestro universo. En este respecto, todav\'ia queda comprobar la estabilidad no-perturbativa de algunas de las familias de vac\'ios no-supersim\'etricas derivadas en los cap\'itulos \ref{ch:rrnsnsvacua} y \ref{ch:geometricflux}. De hecho, para la llamada rama A2-S1 en el cap\'itulo  \ref{ch:rrnsnsvacua}, estamos actualmente trabajando en ello \cite{nuevo}, siguiendo el esp\'iritu de la \textit{non-SUSY instability conjecture} \cite{Ooguri:2016pdq}.

Con todo ello concluimos, como es el caso habitual en la ciencia,  con la idea de que aunque esta tesis cierra algunas preguntas, abre muchas nuevas. Esta es precisamente la belleza de este trabajo. Con suerte, esperamos poder seguir contribuyendo a resolver algunas de ellas en un futuro no muy lejano.

\thispagestyle{empty}
\phantom{lala}

\part*{\scshape  \textcolor{MyDarkRed}{Appendices}} \label{part:appendices}
\appendix  

\renewcommand{\thesection}{\Alph{chapter}.\arabic{section}}
\chapter{Basic relations and conventions}
\label{app:relations}
On the one hand, we derive in this appendix some relations regarding the metrics accompanying the kinetic terms of the K\"ahler and complex moduli. On the other hand, we define the operators used for the (non)-geometric fluxes in the main text and list the Bianchi identities in the presence of these fluxes.

\section{Relations for the K\"ahler metrics}\label{ap:relations}
As reviewed  in section \ref{generalsection}, type IIA compactifications on Calabi-Yau orientifolds come with moduli spaces parameterised by K\"ahler moduli and complex structure moduli. These moduli spaces are endowed with a K\"ahler geometry with the K\"ahler metric being proportional to the second derivative of the  K\"ahler potential:
\begin{align}
\label{eq:kpotential}
K=K_K+K_Q&=-\log (\mathcal{G}_T\mathcal{G}_Q^2)=	-\log (\mathcal{G})	&	 &\longrightarrow 	&	&K_{AB}=\frac{1}{4}\p_A\p_B K\, ,
\end{align}
where, following our notation, $A=\{t^a,u^\alpha\}$, $\mathcal{G}_T=\frac{4}{3}\mk$ is a homogeneous function of degree three on the $t^a$ and $\mathcal{G}_Q^2$ is a homogeneous function of degree four on the $u^\alpha$. These properties, along with the fact that $K_{a\mu}=0$, allow to compute some useful relations regarding the K\"ahler potential and the  K\"ahler metric. Let us start by noting that:
\begin{align}
t^a\p_{t^a} &\mathcal{G}=3\mathcal{G}	\, ,&	u^\mu\p_{u^\mu} &\mathcal{G}=4\mathcal{G}\, .
\end{align}
It will be  convenient to write the explicit form of the metric in each sector:
\begin{align}
&K_{ab}=\frac{3}{2\mathcal{K}}\left(\frac{3\mathcal{K}_a\mathcal{K}_b}{2\mathcal{K}}-\mathcal{K}_{ab}\right)	\, ,&	
&K_{\mu\nu}=\frac{1}{4}\left(\frac{\p_\mu G\p_\nu G}{G^2}-\frac{\p_\mu\p_\nu G}{G}\right)\, ,
\end{align}
with $\mk_{ab}=\mk_{abc}t^c$, $\mk_{a}=\mk_{abc}t^bt^c$, and $G= {\cal G}_Q^2$. Then, it is  straightforward to check that:
\begin{multicols}{2}
\begin{itemize}
\item $u^\mu \partial_\mu K=-4$,
\item $K^{\mu\nu}\partial_\mu K=-4u^\nu$,
\item $K^{\mu\nu}\partial_\mu K\partial_\nu K=16$,
\item $\p_\alpha K^{\mu\nu}\partial_\mu K=-8\delta^\nu_\alpha$,
\item $K^{\mu\nu}\p_\alpha\p_\mu K\p_\nu K=4\p_\alpha K$,
\item $\p_\alpha\p_\beta K^{\mu\nu}\p_\mu K\p_\nu K=8\p_\alpha\p_\beta K$,
\item $u^\alpha\partial_\alpha K_{\mu\nu}=-2K_{\mu\nu}$,
\item $u^\alpha\partial_\alpha K^{\mu\nu}=2K^{\mu\nu}$,
\item $u^\mu\p_\mu\p_\alpha K=-\p_\alpha K$;
\end{itemize}
\end{multicols}
and:
\begin{multicols}{2}
\begin{itemize}
\item $K^{ab}\mathcal{K}_a\mathcal{K}_b=\frac{4}{3}\mathcal{K}^2$,
\item $\mathcal{K}^{cd}\mathcal{K}_d=t^c$,
\item $K^{cd}\mathcal{K}_d=\frac{4}{3}\mathcal{K}t^c$,
\item $\mathcal{K}^{bd}K_{ab}=-\frac{1}{4\mathcal{K}}\left(6\delta^d_a-\frac{9t^d\mathcal{K}_a}{\mathcal{K}}\right)$,
\item $t^cK_{ca}=\frac{3}{4}\frac{\mathcal{K}_a}{\mathcal{K}}$,
\item $K^{cd}\mathcal{K}_{da}=\mathcal{K}\left(-\frac{2}{3}\delta^c_a+\frac{2t^c\mathcal{K}_a}{\mathcal{K}}\right)$,
\item $t^a\partial_aK_{bc}=-2K_{bc}$,
\item $t^a\partial_aK^{bc}=2K^{bc}$,
\item $\p_a\mk^{cb}\mk_c=-\delta^b_a$,
\item $\p_a K^{bc}\mk_b=\frac{8}{3}\mk \delta^c_a$;
\end{itemize}
\end{multicols}
\noindent
where implicitly we have defined $\p_\alpha\equiv \p_{u^\alpha}$,  $\p_a\equiv \p_{t^a}$,  $K^{\mu\nu}K_{\mu\eta}=\delta^\nu_\eta$, $K^{ab}K_{bc}=\delta^a_c$ and $\mk^{ab}\mk_{bc}=\delta^a_c$.

\section{(Non)-geometric fluxes}
\label{ap:conv}

In type IIA orientifold compactifications, geometric and non-geometric fluxes are defined in terms of their action on the basis of $p$-forms of table \ref{base}, that correspond to the harmonic representatives of $p$-form cohomology classes of a would-be Calabi--Yau manifold $X_6$. In this framework, and following the conventions in  \cite{Ihl:2007ah}, the action of the different NS fluxes on each $p$-form is determined as 

\begin{align}
\label{eq:fluxActions0}
H \wedge {\bf 1} &= -h_\mu \beta^\mu \, , &  H \wedge \alpha_\mu &= -  h_\mu \Phi_6 \nonumber\, , \nonumber\\
f \triangleleft \om_a &= -f_{a \mu}\, \beta^\mu \, , & f \triangleleft \omega_\alpha &= \hat{f}_{\alpha}{}^\mu\, \alpha_\mu\, , \nonumber\\
f \triangleleft \alpha_\mu &= -f_{a \mu} \, \tilde\om^a\, ,& f \triangleleft \beta^\mu &= -\,f_{\alpha}{}^\mu \, \tilde{\omega}^\alpha \,,\nonumber\\
& & \\
Q \triangleright \tilde\om^a &= Q^{a}{}_{\mu}\, \beta^\mu\, , & Q \triangleright  \tilde{\omega}^\alpha &=  Q^{\alpha \mu}\, \alpha_\mu\, , \nonumber\\
Q \triangleright \alpha_\mu &= -\, Q^{a}{}_{\mu} \, \om_a\, , & Q \triangleright \beta^\mu &=  Q^{\alpha\, \mu} \, \omega_\alpha\,, \nonumber\\
R \bullet {\Phi_6} &= R_\mu\, \beta^\mu \, ,& R \bullet \alpha_\mu &= R_\mu   {\bf 1} \,, \nonumber
\end{align}
and we also have that $H\wedge \beta^\mu  = R \bullet \beta^\mu = 0$. The NS flux quanta are $h_\mu, f_{a\, \mu},f_{\a}{}^\mu, Q^a{}_\mu, Q^{\a\, \mu},  R_\mu \in \bZ$. This specifies the action of the twisted differential operator \eqref{eq:twistedD} on each $p$-form, and in particular the superpotential \eqref{eq:WgenNS} and the RR potential transformation \eqref{eq:C3change} leading to the D-term potential.

\subsubsection*{Constraints from Bianchi identities}

On compactifications with geometric and non-geometric fluxes, one important set of consistency constraints are the flux Bianchi identities. In our setup, these can be obtained by imposing that the twisted differential ${\cal D}$ in \eqref{eq:twistedD} satisfies the idempotency constraint ${\cal D}^2 = 0$ when applied on the $p$-form basis of table \ref{base}  \cite{Robbins:2007yv}. Applying the definitions \eqref{eq:fluxActions0}, one obtains\footnote{Compared to \cite{Robbins:2007yv}, in our setup the flux components $h^\mu$, $R^\mu$, $f_a{}^\mu, Q^{a\mu}, \hat{f}_{\alpha \mu}$ and $\hat{Q}^\alpha{}_\mu$ are projected out. }
\bea
\label{eq:bianchids2}
& & h_\mu\, \hat{f}_\alpha{}^\mu = 0\, , \quad h_\mu\, \hat{Q}^{\alpha \mu} = 0\, , \quad f_{a\mu}\, \hat{f}_\alpha{}^\mu = 0\, , \quad f_{a\mu}\, \hat{Q}^{\alpha \mu} =0\, , \nonumber\\
& & R_\mu\, \hat{Q}^{\alpha \mu} = 0\, , \quad R_\mu \, \hat{f}_\alpha{}^\mu = 0\, , \quad Q^a{}_\mu \, \hat{Q}^{\alpha \mu} = 0\, , \quad \hat{f}_\alpha{}^{\mu} \,Q^a{}_\mu=0\, ,\\
& & \hat{f}_\alpha{}^{[\mu}\, \hat{Q}^{\alpha \nu]} = 0\, , \quad h_{[\mu} \, R_{\nu]} - f_{a[\mu}\, Q^a{}_{\nu]} = 0\,. \nonumber
\eea

\newpage
\chapter{Analysis of the Hessian in the presence of RR and NSNS fluxes}
\label{ap:Hessiannsrr}

In this appendix we analyse the properties of the matrix of second derivatives of the potential, or Hessian. As discussed in section \ref{ss:hessian}, due to our Ansatz \eqref{Ansatz} the Hessian can be written as
\begin{align}
H_{\a\b}&=\p_\alpha\p_\beta V\rvert_{\rm vac}=2\left(\p_\alpha\vec{\g}^t\right) \hat{\bf Z}_{1} \left(\p_\beta\vec{\g}+\vec{\eta_\beta}\right)\,.
\label{ap:Hfin}
\end{align}
For the solutions in table \ref{vacuresul} within the branches {\bf A1-S1} and {\bf A2-S1} (or equivalently for the solutions of the form \eqref{solutions} with $\hat{\epsilon}^p_\mu=0$) one can write and explicit expression for {\bf H} in terms of the parameters $A,B, C \in \mathbb{R}$. Ordering the derivatives as $\left(\p_{\xi^\mu},\p_{b^a},\p_{u^\alpha}, \p_{t^a} \right)$ one finds that:
\begin{align}
\label{gZ1g}
&\left(\p_\alpha\vec{\g_r}^t\right) \hat{\bf Z}_1 \left(\p_\beta\vec{\g_r}\right)=\nonumber\\&e^K\left(\begin{matrix}
4A^2\mk^2\tr^2\p_\nu K\p_\mu K & 4 AC\mk\tr^2\partial_\nu K\mk_a	&0 &	0\\
\\
4 AC\mk\tr^2\partial_\mu K\mk_b&-C_1^2\mk\mk_{ab}+C_2^2\mk_a\mk_b &0 &	B\left(C_3\mk \mk_{ab} -C_4\mk_a\mk_b\right)\\
\\
0 &0 &C_5^2\mk^2K_{\alpha\beta}  &	\frac{3C_5^2}{4}\mk\partial_\alpha K\mk_b\\ 
\\
 0 &B\left(C_3\mk \mk_{ab} -C_4\mk_a\mk_b\right) &\frac{3C_5^2}{4}\mk\partial_\alpha K\mk_a&	-C_6^2\mk\mk_{ab}+C_7^2\mk_a\mk_b \\\\
\end{matrix}\right)\,,
\end{align}
with 
\begin{align}
C_1^2&=\frac{2\tr^2}{3}\left(1+B^2\right),\nonumber	&	C_2^2&=\tr^2\left(1+2B^2+4C^2\right),\nonumber\\
C_3&=\frac{2}{3}\tr^2\left(1+2C\right),	&	C_4&=\tr^2\left(1+4C\right),\nonumber\\
C_5^2&=16A^2\tr^2,\nonumber & C_6^2&=\frac{2}{3}\tr^2\left(B^2+4C^2\right),\nonumber\\
C_7^2&=\tr^2\left(B^2+8C^2+144A^2\right);
\end{align}
and that
\begin{align}
\label{gZ1eta}
\left(\p_\alpha\vec{\g_r}^t\right) \hat{\bf Z}_1\vec{\eta_\beta}&=e^K\left(\begin{matrix}
0&0&0&0\\
0 &  \sigma_1\mk_{ab}  &  0  &     -B\sigma_1\mk_{ab}\\
  \\
0 & 0	&	A\sigma_2\mk\left(\p_\alpha K	\p_\beta K-4 K_{\alpha\beta}\right)	&	0\\
  \\
 0 &  -B\sigma_1\mk_{ab}		&	0	& -\left(B\s_3+2C\s_1\right)\mk_{ab}\\
 \end{matrix}\right)\, ,
\end{align}
with $\sigma_1=\frac{4C\mk}{3}\tilde{\rho}^2$,  $\sigma_2=\left(\frac{1}{3}-2A\right)2\mk \tr^2$ and  $\sigma_3=\frac{2B}{3}\mathcal{K}\tilde{\rho}^2$.

Already from this expression one can see that modes of the form
\begin{equation}\label{flat}
    ( \Xi^\mu,\, 0,\,  0,\,  0)  \qquad \text{such that} \qquad \Xi^\mu \partial_\mu K\rvert_{\rm vac} = 0\, ,
\end{equation}
are zero modes of the Hessian. Since in the branch {\bf S1} $\partial_\mu K\rvert_{\rm vac} \propto h_\mu$, such zero modes correspond to axionic modes of the the complex structure moduli that do not appear in the superpotential \eqref{WQ}. In fact, one can easily see that such directions do not appear in \eqref{VF}, and therefore are flat directions of the potential. 

In the following we will analyse further specific properties of {\bf H} for the branches \textbf{A1-S1} and \textbf{A2-S1}. For the former we will compute the mass spectrum for canonically normalised fields, finding that all tachyons satisfy the BF bound. For the latter we will directly show that {\bf H} is positive semidefinite, and therefore it contains no tachyons. Instead of tachyons, we will see that it contains additional zero modes compared to the other branches, in such a way that massless modes arrange into complex scalars. 

\section{Branch \textbf{A2-S1}}
\label{ap:ha2s1}

Let us first consider the Hessian in the branch \textbf{A2-S1} and, as stated above, show that it is positive semidefinite. By Sylverster's law of inertia, for showing that one may consider {\bf H} in any basis, without the need to express it in the basis of canonically normalised fields. Consider the expression \eqref{ap:Hfin}, which in the case at hand reads:
\begin{align}
\label{HA2S1}
{\bf H}\rvert_{\rm A2-S1}=e^K\tr^2\left(\begin{matrix}
 \frac{1}{18}\mk^2\p_\nu K\p_\mu K& -\frac{1}{6}\mk\partial_\nu K\mk_a& 0& 0\\
  \\
-\frac{1}{6}\mk\partial_\nu K\mk_a &-\frac{7}{3}\mk\mk_{ab}+\frac{7}{2}\mk_a\mk_b  &0 & \frac{4B}{3}\mk \mk_{ab}\\
  \\
0 & 0&\frac{1}{18}\mk^2\p_\alpha K	\p_\beta K & \frac{1}{6}\mk\partial_\alpha K\mk_b\\
  \\
 0& \frac{4B}{3}\mk \mk_{ab}&\frac{1}{6}\mk\partial_\alpha K\mk_b &-\frac{4}{3}\mk\mk_{ab}+\frac{7}{2}\mk_a\mk_b  \\
 \end{matrix}\right)\ , 
\end{align}
with $B=\pm1/2$.

Now, any (real) positive semidefinite matrix is a $n\times n$  symmetric matrix $M$ such that, for all non-zero  $x$ in $\mathds{R}^n$ satisfies $x^T Mx \geq0$. If one decomposes it as $M=\sum_i M_i$, and ech of the components satisfy
\begin{align}
x^T M_ix \geq0
\end{align} 
then it is straigtforard to see that
\begin{align}
x^T &M_ix \geq0 \quad \longrightarrow \quad x^TMx=x^T\sum_i M_ix\geq0 \, ,
\end{align}
which proves that $M$ is positive semidefinite. 

In the following we will use this property to show that \eqref{HA2S1} is positive semidefinite. We first decompose \eqref{HA2S1}  as
\begin{align}
    {\bf H}\rvert_{\rm A2-S1}= e^K\tr^2 \left({\bf X} + {\bf Y} + {\bf Z}\right)\,,
\end{align}
where
\begin{align}
{\bf X} & = 
\left(\begin{matrix}
 \frac{1}{18}\mk^2\p_\nu K\p_\mu K& -\frac{1}{6}\mk\partial_\nu K\mk_a& 0& 0\nonumber\\
  \\
-\frac{1}{6}\mk\partial_\nu K\mk_a &\frac{1}{2}\mk_a\mk_b  &0 & 0\\
  \\
0 & 0&0 & 0\\
  \\
0 & 0&0 & 0\\
 \end{matrix}\right)\,,
 {\bf Y}  = 
 \left(\begin{matrix}
0 & 0&0 & 0\\  \\
0 & 0&0 & 0\\
  \\
0 & 0&\frac{1}{18}\mk^2\p_\alpha K	\p_\beta K & \frac{1}{6}\mk\partial_\alpha K\mk_b\\
  \\
 0& 0&\frac{1}{6}\mk\partial_\alpha K\mk_b &\frac{1}{2}\mk_a\mk_b \\
 \end{matrix}\right)\, ,
\end{align}
and
\begin{align}
{\bf Z} & = 
\left(\begin{matrix}
0 & 0&0 & 0\\  \\
0 & -\frac{7}{3}\mk\mk_{ab}+3\mk_a\mk_b &0 & \frac{4B}{3}\mk \mk_{ab}\\
  \\
0 & 0&0 & 0\\
  \\
 0& \frac{4B}{3}\mk \mk_{ab}&0 &-\frac{4}{3}\mk\mk_{ab}+3\mk_a\mk_b  \\
 \end{matrix}\right)\, .\nonumber
\end{align}
We need to prove that each of these three matrices is positive semidefinite. Starting with {\bf X}, one can see that the non-trivial block can be decomposed as the following product 
\begin{align}
&\left(\begin{matrix}
 \frac{1}{18}\mk^2\p_\nu K\p_\mu K& -\frac{1}{6}\mk\partial_\mu K\mk_b\\
  \\
-\frac{1}{6}\mk\partial_\nu K\mk_a &\frac{1}{2}\mk_a\mk_b \\
 \end{matrix}\right)=\left(\begin{matrix}
 \frac{\sqrt{2}}{12}\mk\p_\mu K& 0\\
  \\
-\frac{\sqrt{2}}{4}\mk_a &0\\
 \end{matrix}\right)\left(\begin{matrix}
 4& 0\\
  \\
0 &K^{ab}\\
 \end{matrix}\right)\left(\begin{matrix}
 \frac{\sqrt{2}}{12}\mk\p_\nu K& -\frac{\sqrt{2}}{4}\mk_b\\
  \\
0 &0\\
 \end{matrix}\right)\, .
\end{align}
That is, it can be written as a Gramian matrix, which implies its positive-semidefiniteness. The same statement applies to the non-trivial block of the matrix {\bf Y}, which reads
\begin{align}
&\left(\begin{matrix}
 \frac{1}{18}\mk^2\p_\nu K	\p_\mu K & \frac{1}{6}\mk\partial_\mu K\mk_b\\
  \\
\frac{1}{6}\mk\partial_\nu K\mk_a &\frac{1}{2}\mk_a\mk_b  \\
 \end{matrix}\right)=\left(\begin{matrix}
 \frac{\sqrt{2}}{12}\mk\p_\mu K& 0\\
  \\
\frac{\sqrt{2}}{4}\mk_a &0\\
 \end{matrix}\right)\left(\begin{matrix}
 4& 0\\
  \\
0 &K^{ab}\\
 \end{matrix}\right)\left(\begin{matrix}
 \frac{\sqrt{2}}{12}\mk\p_\nu K& \frac{\sqrt{2}}{4}\mk_b\\
  \\
0 &0\\
 \end{matrix}\right)\, .
\end{align}

Things are slightly more involved for the non-trivial block of the matrix {\bf Z}. This reads
\begin{align}
\left(\begin{matrix}
-\frac{7}{3}\mk\mk_{ab}+3\mk_a\mk_b  & \pm \frac{2}{3}\mk \mk_{ab}\\
  \\
\pm \frac{2}{3}\mk \mk_{ab} &-\frac{4}{3}\mk\mk_{ab}+3\mk_a\mk_b  \\
 \end{matrix}\right)\, ,
\label{Zblock}
\end{align}
where we have considered for both choices of sign in $B=\pm 1/2$. In this case one can rewrite \eqref{Zblock} as:
\begin{align}
&\frac{2}{3} \left(\mk_a\mk_b-\mk \mk_{ab}\right)
\left(\begin{matrix}
1 & \mp 1 \\
  \\
\mp 1 & 1  \\
 \end{matrix}\right) +\left(\begin{matrix}
-\frac{5}{3}\mk\mk_{ab}+\frac{7}{3}\mk_a\mk_b  & \pm \frac{2}{3}\mk_{a}\mk_{b}\\
  \\
\pm \frac{2}{3}\mk_{a}\mk_{b} &-\frac{2}{3}\mk\mk_{ab}+\frac{7}{3}\mk_a\mk_b  \\
 \end{matrix}\right)\, .
\end{align}
The first matrix is a tensor product of two positive semidefinite matrices. The second one satisfies:
\begin{align}
&\left(\begin{matrix}
q^b	&	p^b
\end{matrix}\right)\left(\begin{matrix}
-\frac{5}{3}\mk\mk_{ab}+\frac{7}{3}\mk_a\mk_b  & \pm \frac{2}{3}\mk_{a}\mk_{b}\\
  \\
\pm \frac{2}{3}\mk_{a}\mk_{b} &-\frac{2}{3}\mk\mk_{ab}+\frac{7}{3}\mk_a\mk_b  \\
 \end{matrix}\right)\left(\begin{matrix}
q^a	\\
	p^a
\end{matrix}\right)\nonumber\\&=-\frac{5}{3}\mk\mk_{ab}q^aq^b+\frac{7}{3}\mk_a\mk_bq^aq^b \pm \frac{4}{3}\mk_{a}\mk_{b}q^ap^b-\frac{2}{3}\mk\mk_{ab}p^ap^b+\frac{7}{3}\mk_a\mk_bp^ap^b\nonumber\\ 
&=\frac{2}{3}\mk_a\mk_b \left(q^a \pm p^a\right)\left(q^b\pm p^b\right) +\frac{5}{3}\left(\mk_a\mk_b-\mk\mk_{ab}\right)q^aq^b +\frac{2}{3}\left(\mk_a\mk_b-\mk\mk_{ab}\right)p^ap^b+\mk_a\mk_b p^a p^b \geq 0
\end{align}
where we have used that all the metrics involved are positive semidefinite. Therefore {\bf Z} is also positive semidefinite.

Notice that the Hessian matrix \eqref{HA2S1} has further zero modes beyond the ones corresponding to the flat directions \eqref{flat}. These are of the form
\begin{equation}\label{zm}
    ( 0, \, 0,\,  \Xi^\mu,\,  0)  \qquad \text{such that} \qquad \Xi^\mu \partial_\mu K\rvert_{\rm vac} = 0\, ,
\end{equation}
and are nothing but the complex structure saxions that pair up with the axionic flat directions into complex scalar field. This time, as these fields appear in the potential via \eqref{ZAB}, they will not be flat directions of the potential. One can check that they develop a quartic potential, as discussed in section \ref{ap:complex} below.

\section{Branch \textbf{A1-S1}}\label{ap:ha1s1}

In this branch the Hessian \eqref{ap:Hfin} takes a block-diagonal form, namely
\begin{align}
    {\bf H}\rvert_{\rm A1-S1}= e^K\tr^2 
    \left(\begin{matrix}
    {\bf A} & 0 \\ 0 & {\bf S}\\
     \end{matrix}\right)\,,
\end{align}
where
\begin{align}\label{AxiM}
{\bf A} & = 
\left(\begin{matrix}
 \frac{8}{225}\mk^2\p_\nu K\p_\mu K& \frac{8C}{15}\mk\partial_\nu K\mk_a& \\
  \\
\frac{8C}{15}\mk\partial_\nu K\mk_a &\left(-\frac{4}{3}+\frac{8C}{3}\right)\mk\mk_{ab}+\frac{68}{25}\mk_a\mk_b \\
 \end{matrix}\right)\,,
 \end{align}
 \begin{align}\label{SaxiM}
 {\bf S}  = 
 \left(\begin{matrix}
\mk^2\left(\frac{4}{75}\p_\alpha K	\p_\beta K-\frac{16}{225} K_{\alpha\beta}\right) & \frac{8}{75}\mk\partial_\alpha K\mk_b\\
  \\
\frac{8}{75}\mk\partial_\alpha K\mk_b &-\frac{24}{25}\mk\mk_{ab}+\frac{68}{25}\mk_a\mk_b  \\
 \end{matrix}\right)\, ,
\end{align}
with $C=\pm 3/10$. Therefore, one can analyse the spectrum of axions or saxions separately. 


\subsubsection*{Axionic sector}

Let us first analyse the axionic sector. One can rewrite {\bf A} as:
\begin{align}
{\bf A}&=\left(\begin{matrix}
 \frac{\sqrt{2}}{15}\mk\p_\mu K& 0\\
  \\
C\sqrt{2}\mk_a &0\\
 \end{matrix}\right)\left(\begin{matrix}
 4& 0\\
  \\
0 &K^{ab}\\
 \end{matrix}\right)\left(\begin{matrix}
 \frac{\sqrt{2}}{15}\mk\p_\nu K& C\sqrt{2}\mk_b\\
  \\
0 &0\\
 \end{matrix}\right)+\nonumber
 \\&+\left(\begin{matrix}
 0&0\\
  \\
0 &\left(\frac{4}{3}-\frac{8C}{3}\right)\left(\mk_a\mk_b-\mk\mk_{ab}\right)+\left(\frac{2+8C}{3}\right)\mk_a\mk_b\\
 \end{matrix}\right)\, ,
 \label{tachC}
\end{align}
so for $C=\frac{3}{10}$ (i.e., the supersymmetric branch) the matrix {\bf A} \eqref{tachC} is a sum of positive semidefinite matrices, whereas for $C=-\frac{3}{10}$ the second one is not positive semidefinite. 

In order to compute the physical mass spectrum we need to express the Hessian in a basis of canonically normalised fields. For this, notice that the K\"ahler metrics for the K\"ahler and complex structure fields can be decomposed as:
\begin{align}
&K_{ab}=\frac{3}{2\mathcal{K}}\left(\frac{3\mathcal{K}_a\mathcal{K}_b}{2\mathcal{K}}-\mathcal{K}_{ab}\right)=\frac{3}{4}\frac{\mk_a\mk_b}{\mk^2} + \frac{3}{2\mk}\left(\frac{\mk_a\mk_b}{\mk}-\mk_{ab}\right)=K_{ab}^{\rm NP}+K_{ab}^{\rm P}\, ,\\
&K_{\mu\nu}=\frac{1}{16}\frac{\p_\mu G\p_\nu G}{G^2} + \frac{1}{4}\left(\frac{3}{4}\frac{\p_\mu G\p_\nu G}{G^2}-\frac{\p_\mu\p_\nu G}{G}\right)=K_{\mu\nu}^{\rm NP}+K_{\mu\nu}^{\rm P}\, ,
\end{align}
with $G = {\cal G}_T^2$, as defined below \eqref{KQ}. Here $K_{ab}^{\rm P}$ and $K_{ab}^{\rm NP}$ stand for the primitive and non-primitive factors of the K\"ahler moduli metric, which act on orthogonal subspaces of dimension $h^{1,1}_--1$ and 1. A similar decomposition holds for the metric of the dilaton-complex structure sector, now acting on spaces of dimension $N$ and 1, with $N$ the number of complex structure moduli. In terms of this decomposition, the matrix {\bf A} in the non-SUSY branch $C=-\frac{3}{10}$ reads
\begin{align}\label{AnonSUSY}
{\bf A} &= \left(\begin{matrix}
 \frac{128}{225}\mk^2K_{\mu\nu}^{\rm NP}& -\frac{4}{25}\mk\p_\nu K\mk_a\\
  \\
-\frac{4}{25}\mk\p_\nu K\mk_a& \frac{64}{45}\mk^2 K^{\rm P}_{ab}+\frac{176}{225}\mk^2 K_{ab}^{\rm NP}\\
 \end{matrix}\right)\, .
\end{align}

Now, the effective Lagrangian describing the axion spectrum will be of the form
\begin{equation}
L\supset\left(\p \xi^\mu \ \p b^a\right)\left( \begin{matrix}
K_{\mu\nu}\rvert_{\rm vac} &	0\\
0	&	K_{ab}\rvert_{\rm vac}
\end{matrix}\right) \left(\begin{matrix}\p \xi^\nu \\ \p b^b \end{matrix}\right)
+\frac{1}{2}\left( \xi^\nu \  b^a\right) \left[e^K \tr^2 {\bf A}\right]_{\rm vac} \left(\begin{matrix} \xi^\nu \\  b^b \end{matrix}\right)\, ,
\end{equation}
with {\bf A} given by \eqref{AnonSUSY} in the non-supersymmetric case. One can now define a basis of canonically normalised fields by performing the change of basis
\begin{equation}\label{canon}
    ( \xi^\mu \quad b^a) \quad \longrightarrow \quad  (\hat{\xi}\quad \hat{b} \quad \xi^{\hat{\mu}} \quad b^{\hat{a}} )\, ,
\end{equation}
where $\hat{b}$ is the vector along the subspace corresponding to $K_{ab}^{\rm NP}\rvert_{\rm vac}$, with unit norm, and similarly for $\hat{\xi}$ with $K_{\mu\nu}^{\rm NP}\rvert_{\rm vac}$. Finally, $\xi^{\hat{\mu}}$ with $\hat{\mu} = 1, \dots, N$ and $b^{\hat{a}}$ with $\hat{a} = 1, \dots h_-^{1,1} -1$ correspond to vectors of unit norm with respect to $K_{\mu\nu}^{\rm P}\rvert_{\rm vac}$ and $K_{ab}^{\rm P}\rvert_{\rm vac}$, respectively. One can see that in this new basis {\bf A} has the form
\begin{align}
    \hat{\bf A} & = \frac{16}{5}
    \left( \begin{matrix}
    \frac{8}{45} & \frac{2}{5\sqrt{3}} & & \\
    \frac{2}{5\sqrt{3}} & \frac{11}{45} & & \\
    & & 0 & \\
    & & & \frac{4}{9}
    \end{matrix}\right) \mk^2\, ,
\end{align}
and so the Hessian eigenvalues in the canonically normalised basis are
\begin{align}
   e^K \cK^2 \tr^2 \frac{8}{45} \left\{-\frac{1}{5}\, , \quad  4\, , \quad 0 \, ,  \quad 4\right\}\, .
\end{align}
Finally, one must compare such masses with the BF bound
\begin{equation}
    |m_{\rm BF}|^2 \, =\, - \frac{3}{4} V_{\rm vac} \, =\, e^K \frac{\CK^2\tr^2}{25}\, .
\end{equation}
In term of it one finds that the spectrum reads
\begin{align}
  m^2 =   \left\{-\frac{8}{9}\, ,  \quad  \frac{160}{9}\, , \quad 0  \, , \quad  \frac{160}{9}\right\}  |m_{\rm BF}|^2 \, ,
\end{align}
and so the tachyon in this sector does not induce an instability. 

For completeness, let us finish this section by computing also the spectrum for the SUSY case.  Proceeding exactly as before but taking $C=\frac{3}{10}$ it is straightforward to obtain the following eigenvalues for the canonically normalised Hessian:
\begin{align}
   e^K \cK^2 \tr^2 \frac{8}{45} \left\{\frac{44}{5}\, , \quad   1\, , \quad 0 \, ,  \quad 1\right\}\, ,
\end{align}
or in terms of the BF bound:
\begin{align}
  m^2 =   \left\{\frac{352}{9}\, ,  \quad  \frac{40}{9}\, , \quad 0  \, , \quad  \frac{40}{9}\right\}  |m_{\rm BF}|^2 \, .
\end{align}
\subsubsection*{Saxionic sector}

Let us now analyse the spectrum in the saxionic sector. Notice that this time the matrix \eqref{SaxiM} is independent of the sign of $C$, and so the tachyonic directions that one may find will be common to the supersymmetric and non-supersymmetric branches of the kind {\bf A1-S1}. Since the supersymmetric branch should not contain any classical instability, neither should there be one for its non-supersymmetric counterpart. Let us nevertheless confirm this expectation explicitly. 

As before we first rewrite \eqref{SaxiM} as
\begin{align}
\label{Snew}
{\bf S} &=\left(\begin{matrix}
\frac{176}{225}\mk^2K_{\mu\nu}^{\rm NP}-\frac{16}{225}\mk^2K^{\rm P}_{\mu\nu}	&	\frac{8}{75}\mk\partial_\alpha K\mk_b\\
  \\
\frac{8}{75}\mk\partial_\alpha K\mk_b & \frac{48}{75} \cK^2 K^{\rm P}_{ab} + \frac{176}{75} \cK^2 K^{\rm NP}_{ab}  \\
 \end{matrix}\right)\, .
\end{align}
Then we perform a change of basis for the saxions
\begin{equation}
    ( u^\mu \quad t^a) \quad \longrightarrow \quad  (\hat{u}\quad \hat{t} \quad u^{\hat{\mu}} \quad t^{\hat{a}} )\, ,
\end{equation}
with analogous definitions as in \eqref{canon}. In this basis the matrix {\bf S} reads
\begin{align}
    \hat{\bf S} & = \frac{16}{75}
    \left( \begin{matrix}
    \frac{11}{3} & -\frac{4}{\sqrt{3}} & & \\
    -\frac{4}{\sqrt{3}} & 11 & & \\
    & & - \frac{1}{3} & \\
    & & & 3
    \end{matrix}\right) \mk^2\, ,
\end{align}
and so the Hessian eigenvalues in the canonically normalised basis are
\begin{align}
   e^K \cK^2 \tr^2 \frac{8}{75} \left\{3 \, ,  \quad \frac{35}{3}\, ,\quad  -\frac{1}{3}\, , \quad  3 \right\}\, ,
\end{align}
where now the tachyonic eigenvalue has a degeneracy of $N$, as it corresponds to the `primitive' complex structure saxions $u^{\hat{\mu}}$. Comparing with the BF bound one finds 
\begin{align}
  m^2 =   \left\{8 \, , \quad  \frac{280}{9}\, , \quad  -\frac{8}{9}\, ,  \quad  8 \right\}  |m_{\rm BF}|^2 \, .
\end{align}
As expected, the tachyonic directions in this sector do not induce a classical instability. 
\section{Complex structure saxions}
\label{ap:complex}

In the superpotential \eqref{WQ} only one linear combination of dilaton and complex structure moduli appear. As a direct consequence we have $N$ axionic flat directions of the potential, where $N$ is the number of complex structure moduli. In the following we would like to analyse the potential that it is induced for their saxionic partners. This question is particularly relevant for the branch {\bf A2-S1} of vacua, where such saxionic modes are found to be massless.

Let us consider the linear combinations of complex structure and dilaton moduli $U^i = \xi^i + i u^i$ not appearing in the superpotential \eqref{WQ}. Then, one can check that they satisfy the property
\begin{equation}
\left[\p_{u^i} K\right]_{\rm vac}=0\, .
\end{equation}
Using this it is straightforward to see that at the vacuum
\begin{align}
\p_{u^i} V\rvert_{\rm vac} & = \left[e^K \left(\p_{u^i} K \tilde{V}+ e^K\p_{u^i}\tilde{V}\right)\right]_{\rm vac}=0\, , \\
\p_{u^i}\p_{u^j} V\rvert_{\rm vac} & = \left[e^K\left(\p_{u^i}\p_{u^j} K \tilde{V}+\p_{u^i}\p_{u^j} \tilde{V}\right)\right]_{\rm vac} = {\cal A} \left[e^K\p_{u^i}\p_{u^j} K\right]_{\rm vac}\, ,
\end{align}
where we have defined  $\tilde{V}=e^{-K}V$ and 
\begin{align}
{\cal A} = \left(8A^2-\frac{2B^2}{27}-\frac{16C^2}{27}\right)\, .
\label{combi}
\end{align}
Replacing the values for the constants $A, B, C$ for the different branches in the second equation, one recovers the corresponding sector of the Hessian. In particular, one can check that \eqref{combi} vanishes for the branch {\bf A2-S1}, as expected. 

One may then proceed and compute further derivatives of the potential at the vacuum:
\begin{align}
\p_{u^i}\p_{u^j}\p_{u^l} V\rvert_{\rm vac}&=\left[e^K\left(\p_{u^i}\p_{u^j}\p_{u^l} K \tilde{V}+\p_{u^l}\p_{u^i}\p_{u^j} \tilde{V}\right) \right]_{\rm vac} = {\cal A} \left[e^K\p_{u^i}\p_{u^l}\p_{u^j} K\right]_{\rm vac}\, ,\\
\p_{u^i}\p_{u^j}\p_{u^l}\p_{u^m} V\rvert_{\rm vac}&=128A^2 \left[e^K \left( K_{u^ju^l}K_{u^iu^m}+K_{u^ju^m}K_{u^iu^l}+K_{u^lu^m}K_{u^iu^j} \right)\right]_{\rm vac}+\dots
\end{align}
where the dots stand for terms proportional to ${\cal A}$. As the term in brackets is a product of kinetic terms, in the case ${\cal A} = 0$ we obtain a non-vanishing, positive quartic coupling. This completes the proof that the branch {\bf A2-S1} features a positive semidefinite potential in the vicinity of the vacuum.

\section{Adding mobile D6-branes}
\label{ap:openH}

In the presence of mobile D6-branes and for each extremum found in section \ref{s:D6branes}, one can show that the formalism developed in section \ref{ss:hessian} is still valid. The matrix of second derivatives takes the form: $$\p_\alpha\p_\beta V'\rvert_{\rm vac}=\p_\alpha\p_\beta \left(V_1'+V_2'\right)\rvert_{\rm vac}=2\left(\p_\alpha\vec{\g'}^t\right) \hat{\bf Z'}_{1} \left(\p_\beta\vec{\g'}+\vec{\eta'_\beta}\right)\, ,$$ with the correspondent redefinition of $\{\hat{\bf Z}'_1, \p_\alpha \vec{\gamma}', \eta_{\beta}'\}$ incorporating the open string moduli and $\{V_1'$, $V_2'\}$ introduced in \eqref{openpotential}. The matrix $\hat{\bf Z}_1'$ is defined, analogously to \eqref{Vsplit}, such that $V_1'=\vec{\gamma}^{t'}\hat{\bf Z}'_1\vec{\gamma}'$ is quadratic on quantities that vanish in the vacuum. Looking at \eqref{openpotential} it is straightforward to see that:
\begin{equation}
\hat{\bf Z}_1' = 
\left(
\begin{matrix}
 4 & 0 & 0 & 0 & 0 & 0\\
 \\
 0 & K^{ab} & 0 & 0 & 0 & 0 \\
 \\
 0 & 0 & \frac{4}{9} \mk^2 K_{ab} & 0 & 0 & 0 \\
 \\
  0 & 0 & 0 & K^{\mu \nu } & 0 & 0\\
 \\
  0 & 0 & 0 & 0 &  G^{ij} & 0 \\
  \\
0 & 0 & 0 & 0 & 0 & G^{ij}    \\
\end{matrix}
\right)\, .
\end{equation}
Regarding the new $\p_\alpha \vec{\gamma}'$'s its explicit expression can be computed directly from \eqref{gammaopen}:
\begin{align}
\partial_{\xi^\mu}\vec{\g}^t=&\left(\begin{matrix}
h_\mu ,&0 ,&0, &0,&0 ,&0 
\end{matrix}\right)\, ,\nonumber \\
\partial_{b^c}\vec{\g}^t=&\left(\begin{matrix}
\rho_c-\hth^i\r_{ci}, &\mk_{acd}\tr^d -f^i_a \rho_{c\, i},&\delta^a_c\tr ,&0, & -\rho_{c\, i},&0
\end{matrix}\right)\, ,\nonumber \\
\partial_{\hth^i}\vec{\g}^t=&\left(\begin{matrix}
\r_i, &-\r_{ai},&0 ,&0, & 0,&0
\end{matrix}\right)\, ,\nonumber \\
\partial_{u^\alpha}\vec{\g}^t=&\left(\begin{matrix}
0, &0, &0, &-\tr A\mk\p_\alpha\partial_\nu K-\tr\mk\p_\alpha\tilde{\epsilon}^p_\nu,&0 ,&0 
\end{matrix}\right)\, ,\nonumber \\
\partial_{t^c}\vec{\g}^t=&\left(\begin{matrix}
0, & -2\tr C\mk_{ac}, &-\tr B\delta^a_c,&-3\tr A\mk_c\partial_\nu K -3\tr\mk_c\tilde{\epsilon}^p_\mu,0,0
\end{matrix}\right)+\nonumber \\+&\left(\begin{matrix}
0, & \p_{t^c}\left(f^i_a\r_i-\frac{1}{2}H^\mu_a\hr_\mu\right), &-\p_{t^c}\left(\mk^{ab}\phi^i+\mk^{ad}t^bf^i_d\right)\r_{bi},&-\frac{1}{2}\p_{t^c}g^\mu_i\hr_\mu,&\r_{ci}
\end{matrix}\right)\, ,\nonumber\\
\partial_{\phi^i}\vec{\g}^t=&\left(\begin{matrix}
0,&\p_{\phi^i}\left(f^j_a\r_j-\frac{1}{2}H^\mu_a\hr_\mu\right),& -\mk^{ab}\r_{bi}-\mk^{ad}t^b\p_{\phi^i}f^j_d\r_{bj},&0
\end{matrix}\right)\, .
\end{align}
Finally, the  $\eta_{\alpha}'$'s are obatined by direct computation rewriting the second derivatives of $V'_2$  as:
\begin{equation}
\partial_\alpha \partial_\beta V'_2\rvert_{\rm vac}=2\vec{\eta'_\alpha}^t \hat{\bf Z'}_{1} \p_\beta\vec{\gamma'}= 2\p_\alpha\vec{\g'}^{\, t} \hat{\bf Z'}_{1} \vec{\eta'_\beta}\, .
\end{equation}
The  result is:
\begin{align}\nonumber
\vec{\eta}_{\xi^\mu}{}^t=& \left(\begin{matrix}
0,&0, &0, &0,	&	0,	&	0
\end{matrix}   \right)\, ,
\\
\vec{\eta}_{b^d}{}^t=&\tr\left(\begin{matrix}
0, &0,	&	\frac{3C}{\mk} K^{bc}\mk_{cd},	&	0,	&	0,	&	-\frac{4C}{3}\mk f^k_cG_{kj}
\end{matrix}\right)\, ,\nonumber \\ \nonumber
\vec{\eta}_{\hth^i}{}^t=& \tr\left(\begin{matrix}
0,&0, &0, &0,	&	0,	&	-\frac{4C}{3}\mk\frac{G_{ij}}{\delta^i_i}
\end{matrix}   \right)\, ,
\\
\vec{\eta}_{u^\alpha}{}^t=&\tr\left(\begin{matrix}
0, &C\p_\alpha K\mk_a,	&B\p_\alpha K t^a,	  	&	\left(\frac{2}{3}-4A\right)\mk\left(K_{\alpha\mu}-\frac{1}{4}\p_\mu K \p_\alpha K\right), &0, &0\end{matrix}\right)+ \nonumber
\\&+\tr\left(\begin{matrix}
0, &0,	&0,	  	&	e^{-K}\mk\tr K_{\beta\mu}\p_\alpha \left(e^KK^{\gamma\beta}\te^p_\gamma\right), &0, &0\end{matrix}\right)\nonumber \, ,
 \\
\vec{\eta}_{t^d}{}^t=&\tr\left(\begin{matrix}
0, & \frac{4C\mk}{3} K_{bd},	&	\frac{3B}{2\mk} K^{bc}\mk_{cd}, 	&	\mk\p_{t^d}\te^p_\mu, &-\frac{4C}{3}\mk^2 f^k_c G_{kj}, &\frac{2B}{3}\mk^2 f^k_c G_{kj}
\end{matrix}\right)+\frac{\left(H^\alpha_d-f^i_dg^\alpha_i\right)}{4} \vec{\eta_{u^\alpha}}^t\, ,\nonumber \\
\vec{\eta}_{\phi^i}{}^t=& \tr\left(\begin{matrix}
0,&0, &0, &0,	&\frac{4C}{3}\mk^2 \frac{G_{ij}}{\delta^i_i}, &-\frac{2B}{3}\mk^2 \frac{G_{ij}}{\delta^i_i}
\end{matrix}   \right)+\frac{1}{4}g^\alpha_i{\eta_{u^\alpha}}^t\, .
\end{align}


\chapter{Fluxes and axion polynomials}
\label{ap:convl}

\section{Axionic flux orbits and the $P$-matrices}
\label{ap:convm}
From the superpotential it is easy to read the gauge-invariant flux-axion polynomials \eqref{RRrhos} and \eqref{NSrhos}. Then, as in the Calabi--Yau case \cite{Herraez:2018vae}, one can check that all the remaining entries of $\rho_{\cal A}$ can be generated by taking derivatives of the {\it master polynomial} $\rho_0$. Indeed, in our more general case one finds that
\bea
\label{eq:drho0}
& & \frac{\partial\rho_0}{\partial b^a} = \rho_a\, , \quad \frac{\partial\rho_0}{{\partial b^a}{\partial b^b}} = {\cal K}_{abc} \, \tilde\rho^c\, , \quad \frac{\partial\rho_0}{{\partial b^a}{\partial b^b}{\partial b^c}} = {\cal K}_{abc} \, \tilde\rho\,, \quad \frac{\partial\rho_0}{\partial \xi^K} = \rho_K\,  , \\
& & \frac{\partial\rho_0}{{\partial b^a}{\partial \xi^K}} = \rho_{aK}\, , \quad \frac{\partial\rho_0}{{\partial b^a}{\partial b^b}{\partial \xi^K}} = {\cal K}_{abc} \, \tilde\rho^c{}_K\, , \quad \frac{\partial\rho_0}{{\partial b^a}{\partial b^b}{\partial b^c}{\partial \xi^K}} = {\cal K}_{abc} \, \tilde\rho_K \, ,\nonumber
\eea
while all the other derivatives vanish. Just like in \cite{Herraez:2018vae}, one can understand these relations from the fact that the matrix ${\cal R}$ in relating quantised and gauge invariant fluxes can be written as
\be
{\cal R}\equiv e^{b^a P_a + \xi^K P_K} \,,
\ee
with $P_a$ and $P_K$ nilpotent matrices.  Indeed, given \eqref{eq:invRmat} one can check that
\bea
\label{eq:P-matrices}
& & \hskip-0cm P_a = \begin{bmatrix}
0 & \vec\delta_a^t & 0 & 0 & 0 & 0 & 0 & 0 \\
0 & 0 & {\cal K}_{abc} & 0 & 0 & 0 & 0 & 0 \\
0 & 0 & 0 & \vec\delta_a & 0 & 0 & 0 & 0 \\
0 & 0 & 0 & 0 & 0 & 0 & 0 & 0 \\
0 & 0 & 0 & 0 & 0 & \vec\delta_a^t  \, \delta_K^L & 0 & 0 \\
0 & 0 & 0 & 0 & 0 & 0 & {\cal K}_{abc} \, \delta_K^L & 0 \\
0 & 0 & 0 & 0 & 0 & 0 & 0 & \vec\delta_a \, \delta_K^L \\
0 & 0 & 0 & 0 & 0 & 0 & 0 & 0 \\
\end{bmatrix}\, ,
\eea
and
\bea
& & P_K = \begin{bmatrix}
0 & 0 & 0 & 0 & \vec\delta_K^t & 0 & 0 & 0 \\
0 & 0 & 0 & 0 & 0 &  \vec\delta_a^t  \, \delta_K^L & 0 & 0 \\
0 & 0 & 0 & 0 & 0 & 0 &  \vec\delta_a  \, \delta_K^L & 0 \\
0 & 0 & 0 & 0 & 0 & 0 & 0 &  \vec\delta_K^t \\
0 & 0 & 0 & 0 & 0 & 0 & 0 & 0 \\
0 & 0 & 0 & 0 & 0 & 0 & 0 & 0 \\
0 & 0 & 0 & 0 & 0 & 0 & 0 &  0 \\
0 & 0 & 0 & 0 & 0 & 0 & 0 & 0 \\
\end{bmatrix}\, .
\eea

\section{Curvature and sGoldstino masses} \label{ap:curvature} In this appendix we will show  that the directions \eqref{partialmax} minimise respectively $R_{a\bar c d\bar d}g^ag^bg^cg^d$ and $R_{\mu\hat\nu\rho\hat\sigma}g^{\mu}g^{\hat\rho}g^{\nu}g^{\hat\sigma}$. To do so we will follow closely \cite{Covi:2008ea,Covi:2008zu}.

\subsubsection*{Curvature}
\label{ap:curvature1}
Before talking about the extrema conditions, there are some relations that must be introduced. Consider a K\"ahler  potential depending on some set of complex chiral fields $\phi^A$ obeying a no-scale type condition:
\begin{align}
\label{noscale}
    K^A K_A=p\, ,
\end{align}
where $K_A=\nabla_A K$, $K^A=G^{A\bar B} K_{\bar B}$ and $G_{A\bar B}=\partial_A\partial_{\bar B} K$.
Taking the derivative with respect to $\nabla_B$ in \eqref{noscale} one obtains:
\begin{align}
K_B+K^A\nabla_B K_A=0\, ,
\end{align}
and deriving now with respect to $\nabla_{ C}$ we find:
\begin{align}
2\nabla_CK_B+K^A\nabla_C\nabla_B K_A=0\, . \label{hola0}
\end{align}
Equation \eqref{hola0} can be contracted with $K^CK^{\bar D}$ and $K^{\bar D}$  to obtain respectively
\begin{align}
R_{C\bar{D} M\bar{N}}K^CK^MK^{\bar{N}}K^{\bar D}&=\textcolor{black}{2p}\, ,&		R_{C\bar{D} M\bar{N}}K^MK^{\bar{N}}K^{\bar D}&=\textcolor{black}{2}K_C \label{curva}\, .
\end{align}
We will need these two last relations to study the extrema of $R_{A\bar B C\bar D}g^Ag^{\bar B}g^Cg^{\bar D}$

\subsubsection*{sGoldstino masses}
\label{ap:curvature2}
As discussed in section \ref{ss:fterms},
the relevant parameter to compute the sGoldstino masses is
\be
\hat{\sig} = \frac{2}{3}-R_{A\bar B C \bar D} f^{A} f^{\bar B} f^{C} f^{\bar D}\, ,
\label{app:sigma}
\ee
which we are interested in maximise. In this sense, it was shown in \cite{Covi:2008zu} that the extrema of \eqref{app:sigma} are  given  by the $f_{0A}$ satisfying the implicit relation: 
\begin{align}
    f_{0A}=\frac{R_{A\bar B C\bar D}f^{\bar B }_0f^{C}_0f^{\bar D }_0}{R_{A\bar B C\bar D}f_0^{A }f^{\bar B }_0f^{C}_0f^{\bar D }_0}\, .
    \label{app:ex}
\end{align}
Using the results above it is now straightforward to see that $f_{0A}=e^{i\alpha}\frac{K_A}{\sqrt{p}}$, $\alpha\in \mathds{R}$ are solutions of \eqref{app:ex} and therefore extrema of \eqref{app:sigma}.

\subsubsection*{Type IIA on a CY$_3$} \label{ap:curvature3}
The moduli space metric of IIA on a CY$_3$ orientifold is described  from the K\"ahler  potential:
\begin{align}
    K=K_K+K_Q\, ,
\end{align}
where the subindex $K$ refers to the K\"ahler sector whereas we use $Q$ for the complex sector. All the relations discussed above can be applied independently to $K_K$ with $p=3$ and to $K_Q$ with $p=4$. In particular, this shows that \eqref{partialmax} extremise respectively $R_{a\bar c d\bar d}g^ag^bg^cg^d$ and $R_{\mu\hat\nu\rho\hat\sigma}g^{\mu}g^{\hat\rho}g^{\nu}g^{\hat\sigma}$. Regarding the character of the points one can show that they are minima by doing small perturbations around these directions.

If one just considered the  K\"ahler sector or the complex sector (meaning taking $K_Q=0$ in the first case and $K_T=0$ in the second case) this would be the end of the story. Nevertheless, since in general we want to have both contributions, there appear some subtleties one has to take into account. The point is that now  $R_{A\bar B C\bar D}g^Ag^Bg^Cg^D$ does not have just ``one" contribution but two independent contributions:
\be
R_{A\bar B C\bar D}g^Ag^Bg^Cg^D=R_{a\bar c d\bar d}g^ag^bg^cg^d+R_{\mu\hat\nu\rho\hat\sigma}g^{\mu}g^{\hat\rho}g^{\nu}g^{\hat\sigma}\, ,
\label{app:sigma2}
\ee
and the novelty is that it new extremum appears :
\begin{align}
    f_0^A=\frac{1}{\sqrt{7}}\left\{K_a,e^{i\alpha}K_\mu\right\}
    \label{app:max}
\end{align}
with $\alpha\in\mathds{R}$, which is precisely the one discussed below \eqref{solsfmax}. Doing again a small perturbation around the points, it can be shown that now both $f_0^A=\left\{e^{i\alpha}\frac{K_a}{\sqrt{3}},0\right\}$ and $f_0^A=\left\{0,e^{i\alpha}\frac{K_\mu}{\sqrt{4}}\right\}$ are saddle points of \eqref{app:sigma2} whereas \eqref{app:max} is a  minimum.

\section{Analysis of the Hessian}
\label{ap:Hessian}

In this appendix we will compute the Hessian of the scalar potential and study its properties. We will first focus on the F-term potential, whose complexity will require a detailed analysis and the use of a simplified version of our Ansatz. Once the associated Hessian matrix has been found, we will evaluate the result in both the SUSY and the non-SUSY branches independently, in order to obtain information regarding their stability. Finally, we will briefly discuss the general behaviour of the D-term potential Hessian matrix.

\subsection*{F-term Potential}
\label{ap:Hessianf}
Starting from \eqref{eq:potentialgeom} and evaluating the second derivatives along the vacuum equations we obtain:
\bes
\begin{align}
    e^{-K}\frac{\partial^2 V_F}{\partial \xi^\sigma \partial \xi^\lambda}|_{\text{vac}}=&8\rho_\lambda\rho_\sigma+2g^{ab}\rho_{a\sigma}\rho_{b\lambda}\, ,\\
    e^{-K}\frac{\partial^2 V_F}{\partial\xi^\sigma \partial b^a}|_{\text{vac}}=&8\rho_\sigma \rho_a+8\rho_0\rho_{a\sigma}+2g^{bc}\mathcal{K}_{abd}\rho_{c\sigma}\tilde{\rho}^d\, ,\\
    e^{-K}\frac{\partial^2 V_F}{\partial \xi^\lambda \partial u^\sigma}|_{\text{vac}}=&0\, ,\\
    e^{-K}\frac{\partial^2 V_F}{\partial \xi^\sigma\partial t^a}|_{\text{vac}}=&2\partial_a g^{bc}\rho_{b\sigma}\rho_c\, ,\\
    e^{-K}\frac{\partial^2 V_F}{\partial b^a \partial b^b}|_{\text{vac}}=&8\rho_a\rho_b+8\rho_0\mathcal{K}_{abc}\tilde{\rho}^c+2g^{cd}\mathcal{K}_{ace}\mathcal{K}_{bdf}\tilde{\rho}^e\tilde{\rho}^f+2g^{cd}\mathcal{K}_{abc}\rho_d\tilde{\rho}+\frac{8\mathcal{K}^2}{9}g_{ab}\tilde{\rho}^2\nonumber\\
    &+2c^{\mu\nu}\rho_{a\mu}\rho_{b\nu}\, ,\\
    e^{-K}\frac{\partial^2 V_F}{\partial u^\sigma \partial b^a}|_{\text{vac}}=&2\partial_\sigma c^{\mu\nu}\rho_{a\mu}\rho_\nu\, ,\\
    e^{-K}\frac{\partial^2 V_F}{\partial b^a\partial t^b}|_{\text{vac}}=&2\partial_b g^{cd}\mathcal{K}_{ace}\rho_d\tilde{\rho}^e+\left(\frac{16\mathcal{K}}{3}\mathcal{K}_b g_{ac}+\frac{8\mathcal{K}^2}{9}\partial_b g_{ac}\right)\tilde{\rho}^c\tilde{\rho}\, ,\\
    \frac{\partial^2 V_F}{\partial u^\sigma\partial u^\lambda}|_{\text{vac}}=&V_F\partial_\sigma \partial_\lambda K-V_F\partial_\sigma K\partial_\lambda K\nonumber\\
    &+e^{K}\left[\partial_\sigma \partial_\lambda c^{\mu\nu}\rho_\mu\rho_\nu +t^at^b(\partial_\lambda\partial_\sigma c^{\mu\nu}\rho_{a\mu}\rho_{b\nu}-8\rho_{a\sigma}\rho_{b\lambda})+2g^{ab}\rho_{a\sigma}\rho_{b\lambda}\right]\, ,\\
    \frac{\partial^2 V_F}{\partial t^a\partial u^\sigma}|_{\text{vac}}=&V_F\partial_\sigma \partial_a K-V_F\partial_\sigma K \partial_a K+e^{K}\left[-4\mathcal{K}_a\tilde{\rho}^b\rho_{b\sigma}+4\mathcal{K}_a\tilde{\rho}\rho_\sigma\right.\nonumber\\
    &\left.-8\rho_{a\sigma}\rho_{b\mu}u^\mu t^b-8\rho_{b\sigma}\rho_{a\mu}u^\mu t^b+2\partial_\sigma c^{\mu\nu}\rho_{a\mu}\rho_{b\nu}t^b+2\partial_ag^{bc}\rho_{b\mu} u^\mu\rho_{c\sigma}\right]\, ,\\
     \frac{\partial^2 V_F}{\partial t^a\partial t^b}|_{\text{vac}}=&V_F\partial_a \partial_b K-V_F\partial_a K \partial_b K+e^{K}\left[\partial_a\partial_b g^{cd}\rho_c\rho_d+2\mathcal{K}_a\mathcal{K}_b\tilde{\rho}^2\right.\nonumber\\
     &\left.+\left(8\mathcal{K}_a\mathcal{K}_b g_{cd}+\frac{16\mathcal{K}}{3}\mathcal{K}_{ab}g_{cd}+\frac{8\mathcal{K}}{3}\mathcal{K}_a\partial_{b}g_{cd}+\frac{8\mathcal{K}}{3}\mathcal{K}_b\partial_{a}g_{cd}+\frac{4\mathcal{K}^2}{9}\partial_a\partial_bg_{cd}\right)\tilde{\rho}^c\tilde{\rho}^d\right.\nonumber\\
     &\left.+\frac{4\mathcal{K}}{3}\mathcal{K}_{ab}\tilde{\rho}^2-8\mathcal{K}_{ab}\tilde{\rho}^c\rho_{c\nu}u^\nu+8\mathcal{K}_{ab}\tilde{\rho}\rho_\nu u^\nu+2\tilde{c}^{\mu\nu}\rho_{a\mu}\rho_{b\nu}+\partial_a\partial_b g^{cd}\rho_{c\mu}\rho_{d\nu}u^\mu u^\nu\right]\, .
\end{align}
\ees
If we now introduce the ansatz \eqref{Ansatz} and make use of the decomposition of the metric in its primitive and non primitive parts -see  \eqref{eq: primitive metric}- we are left with:
\bes
\label{eq: partial second derivatives anst general}
\begin{align}
\label{ap1}
    e^{-K}\frac{\partial^2 V_F}{\partial \xi^\sigma \partial \xi^\lambda}|_{\text{vac}}=&(8E^2 +\frac{1}{6}F^2)\mathcal{K}^2\partial_\lambda K\partial_\sigma K +2g_{P}^{ab}\rho_{a\sigma}\rho_{b\lambda}\, ,\\
    e^{-K}\frac{\partial^2 V_F}{\partial \xi^\sigma \partial b^a}|_{\text{vac}}=&(8BE-\frac{4}{3}CF)\mathcal{K}^2\partial_aK\partial_\sigma K +(8A-\frac{4}{3}C)\mathcal{K}\rho_{a\sigma}\, ,\\
    e^{-K}\frac{\partial^2 V_F}{\partial \xi^\lambda \partial u^\sigma}|_{\text{vac}}=&0\, ,\\
    e^{-K}\frac{\partial^2 V_F}{\partial \xi^\sigma \partial t^a}|_{\text{vac}}=&-16B\mathcal{K}\rho_{a\sigma}\, ,\\
     e^{-K}\frac{\partial^2 V_F}{\partial b^a\partial b^b}|_{\text{vac}}=&2c^{\mu\nu}_{P}\rho_{a\mu}\rho_{b\nu}+(8B^2+\frac{4}{9}C^2+\frac{2}{9}D^2+\frac{2}{9}F^2)\mathcal{K}^2\partial_a K\partial_b K\nonumber\\
     &+(8AC-8BD-\frac{4}{3}C^2-\frac{4}{3}D^2)\mathcal{K}\mathcal{K}_{ab}\, ,\\
    e^{-K}\frac{\partial^2 V_F}{\partial b^a\partial u^\sigma}|_{\text{vac}}=&-16E\mathcal{K}\rho_{a\sigma}\, ,\\
     e^{-K}\frac{\partial^2 V_F}{\partial b^a \partial t^b}|_{\text{vac}}=&(-16BC+\frac{8}{3}CD)\mathcal{K}\mathcal{K}_{ab}\, ,\\
     e^{-K}\frac{\partial^2 V_F}{\partial u^\sigma\partial u^\lambda}|_{\text{vac}}=&(8E^2+\frac{F^2}{6})\mathcal{K}^2\partial_\sigma K\partial_\lambda K-\frac{G_{\mu\nu}}{G}(16E^2-\frac{1}{3}F^2-\frac{4}{3}DE+\frac{1}{3}CF)\mathcal{K}^2\nonumber\\
     &+2g^{ab}_{P}\rho_{a\sigma}\rho_{b\lambda} \label{eq: uu}\, ,\\
    e^{-K}\frac{\partial^2 V_F}{\partial u^\sigma\partial t^a}|_{\text{vac}}=&(-8E^2+\frac{1}{6}F^2)\mathcal{K}^2\partial_a K\partial_\sigma K-\frac{4}{3}F\mathcal{K}\rho_{a\sigma}\, ,\\
     e^{-K}\frac{\partial^2 V_F}{\partial t^a \partial t^b}|_{\text{vac}}=&(8B^2+\frac{4}{9}C^2+\frac{2}{9}D^2+\frac{2}{9}F^2)\mathcal{K}^2\partial_aK \partial_b K+ (-96B^2-\frac{8}{3}C^2+\frac{4}{3}F^2)\mathcal{K}\mathcal{K}_{ab}\nonumber\\
     &+2c^{\mu\nu}_P\rho_{a\mu}\rho_{b\nu}\label{apf}\, ;
\end{align}
\ees
where we have used the following relations
\begin{align}
    \partial_b g_{ac}t^c=-2g_{ab}\, ,\\
    \partial_\sigma\partial_\lambda c^{\mu\nu}\partial_\mu K \partial_\nu K=32 c_{\mu\nu}\, ,\\
    \partial_a\partial_b g^{cd}\partial_c K \partial_d K=32 g_{ab}\, ,\\
    \partial_a\partial_b g_{cd}t^ct^d=6g_{ab}\, .
\end{align}
Unfortunately, it is not possible to provide a general description of the stability using the results above. As discussed in section \ref{s:stabalidity}, for an arbitrary $\rho_{a\mu}$ one needs to know explicitly the internal metric. Only if we restrict ourselves to the case in which $\rho_{a\mu}$ has rank one are we able to derive a universal analysis. Therefore, from now on we will set
\begin{align}
\rho_{a\mu}&=-\frac{F}{12}\mathcal{K}\partial_aK_T \partial_\mu K_Q \label{eq: geom r1}\, .
\end{align}
Plugging this expression back into \eqref{eq: partial second derivatives anst general} the on-shell second derivatives of the potential are finally reduced to:
\bes
\begin{align}
    e^{-K}\frac{\partial^2 V_F}{\partial \xi^\sigma\partial \xi^\lambda}|_{\text{vac}}=&(8E^2+\frac{1}{6}F^2)\mathcal{K}^2\partial_\sigma K\partial_\lambda K\, ,\\
    e^{-K}\frac{\partial^2 V_F}{\partial \xi^\sigma\partial b^a}|_{\text{vac}}=&(8EB-\frac{2}{3}AF-\frac{2}{9}CF)\mathcal{K}^2\partial_\sigma K\partial_a K\, ,\\
    e^{-K}\frac{\partial^2 V_F}{\partial \xi^\sigma \partial u^\lambda}|_{\text{vac}}=&0\, ,\\
    e^{-K}\frac{\partial^2 V_F}{\partial \xi^\sigma\partial t^a}|_{\text{vac}}=&\frac{4}{3}BF\mathcal{K}^2\partial_a K\partial_\sigma K\, ,\\
    e^{-K}\frac{\partial^2 V_F}{\partial b^a\partial b^b}|_{\text{vac}}=&(8B^2+\frac{4}{9}C^2+\frac{2}{9}D^2+\frac{2}{9}F^2)\mathcal{K}^2\partial_a K\partial_b K\nonumber\\
    &+(8AC-8BD-\frac{4}{3}C^2-\frac{4}{3}D^2)\mathcal{K}\mathcal{K}_{ab}\, ,\\
    e^{-K}\frac{\partial^2 V_F}{\partial u^\sigma \partial b^a}|_{\text{vac}}=&\frac{4}{3}EF\mathcal{K}^2\partial_aK\partial_\sigma K\, ,\\
    e^{-K}\frac{\partial^2 V_F}{\partial b^a \partial t^b}|_{\text{vac}}=&(-16BC+\frac{8}{3}CD)\mathcal{K}\mathcal{K}_{ab}\, ,\\
    e^{-K}\frac{\partial^2 V_F}{\partial u^\sigma\partial u^\lambda}|_{\text{vac}}=&(8E^2+\frac{1}{6}F^2)\mathcal{K}^2\partial_\sigma K\partial_\lambda K-\frac{G_{\mu\nu}}{G}(16E^2-\frac{1}{3}F^2-\frac{4}{3}DE+\frac{1}{3}CF)\mathcal{K}^2\, ,\\
    e^{-K}\frac{\partial^2 V_F}{\partial u^\sigma\partial t^a}|_{\text{vac}}=&(-8E^2+\frac{5}{18}F^2)\mathcal{K}^2\partial_\sigma K\partial_a K\, ,\\
    e^{-K}\frac{\partial^2 V_F}{\partial t^a\partial t^b}|_{\text{vac}}=&(8A^2+16B^2+\frac{2}{9}C^2+32E^2-\frac{8}{9}F^2)\mathcal{K}^2\partial_aK\partial_b K\nonumber\, ,\\
    &+(-96B^2-\frac{8}{3}C^2+\frac{4}{3}F^2)\mathcal{K}\mathcal{K}_{ab}\, .
\end{align}
\ees
In order to make the computations manageable, we follow the same procedure as in \ref{ap:Hessiannsrr} and consider a basis of canonically normalised fields by performing the following change of basis:
\begin{align}
   \left(\xi^\mu, b^a\right)\rightarrow \left(\hat{\xi}, \hat{b}, \xi^{\hat{\mu}}, b^{\hat{a}}\right)&\, , &     \left( u^\mu, t^a\right)\rightarrow \left(\hat{u}, \hat{t}, u^{\hat{\mu}}, t^{\hat{a}}\right)& \, ,
\end{align}
where $\left\{ \hat{b}, \hat{t}\right\}$ $\left(\left\{ \hat{\xi},\hat{u}\right\}\right)$ are unit vectors along the subspace corresponding to $g_{ab}^{NP}|_{\text{vac}}$ $\left(c_{\mu\nu}^{NP}|_{\text{vac}}\right)$ and  $\left\{b^{\hat{a}}, t^{\hat{a}}\right\}$ $\left(\left\{ \xi^{\hat{\mu}},u^{\hat{\mu}}\right\}\right)$\footnote{Notice that $\hat a=1,\dots,h^{1,1}_--1$; $\hat \mu=1,\dots,h^{2,1}$.} correspond analogously to vectors of unit norm with respect to $g_{ab}^{P}|_{\text{vac}}$ $\left(c_{\mu\nu}^{P}|_{\text{vac}}\right)$. We can then rearrange the Hessian $\hat H$ in a $8\times 8$ matrix with basis $(\hat{\xi}, \hat{b}, \xi^{\hat{\mu}}, b^{\hat{a}},\hat{u}, \hat{t}, u^{\hat{\mu}}, t^{\hat{a}})$ so that it reads
\begin{equation}
    \hat{H}_F=e^K\mathcal{K}^2F^2\begin{pmatrix}\frac{384{E_F}^{2}+8}{3} & H_{12} & 0 & 0 & 0 & \frac{32B}{\sqrt{3}} & 0 & 0\cr 
    H_{12} & H_{22} & 0 & 0 & \frac{32E_F}{\sqrt{3}} & H_{26} & 0 & 0\cr 
    0 & 0 & 0 & 0 & 0 & 0 & 0 & 0\cr 0 & 0 & 0 & H_{44} & 0 & 0 & 0 & H_{48}\cr 
    0 & \frac{32E_F}{\sqrt{3}} & 0 & 0 & H_{55} & H_{56} & 0 & 0\cr 
    \frac{32B_F}{\sqrt{3}} & H_{26} & 0 & 0 & H_{56} & H_{66} & 0 & 0\cr 
    0 & 0 & 0 & 0 & 0 & 0 & H_{77} & 0\cr 
    0 & 0 & 0 & H_{48} & 0 & 0 & 0 & H_{88}\end{pmatrix}\, ,
    \label{eq: hessian}
\end{equation}
where we have defined:
\begin{align}
    H_{22}=&\frac{8{D_F}^{2}-96B_FD_F+32{C_F}^{2}+96A_FC_F+864{B_F}^{2}+24}{9}\, ,\\
    H_{44}=&\frac{8{D_F}^{2}+48B_FD_F+8{C_F}^{2}-48A_FC_F}{9}\, ,\\
    H_{55}=&-\frac{192{E_F}^{2}-48D_FE_F+12C_F-20}{3}\, ,\\
    H_{66}=&\frac{3456{E_F}^{2}-8{C_F}^{2}+576{B_F}^{2}+864{A_F}^{2}-80}{9}\, ,\\
    H_{77}=&\frac{192{E_F}^{2}-16D_FE_F+4C_F-4}{3}\, ,\\
    H_{88}=&\frac{16{C_F}^{2}+576{B_F}^{2}-8}{9}\, ,\\
    H_{12}=&8\sqrt{3}\left(8B_FE_F-\frac{2C_F}{9}-\frac{2A_F}{3}\right)\\
    H_{26}=& \frac{32C_FD_F-192B_FC_F}{9}\, ,\\
    H_{48}=& -\frac{16C_FD_F-96B_FC_F}{9}\, ,\\
    H_{56}=& 8\sqrt{3}\left( \frac{5}{18}-8{E_F}^{2}\right)\, .
\end{align}
Note that \eqref{eq: hessian} defines a symmetric matrix whose components are determined once we chose a vacuum. In other words, given an extremum of the potential, one just needs to plug the correspondent $\left\{A_F,B_C,C_F,D_F\right\}$ into \eqref{eq: hessian} to analyse its perturbative stability. The physical masses of the moduli will be given by $1/2$  of the eigenvalues of the Hessian.

Once the  explicit form of Hessian has been introduced, we are ready to discuss the spectrum of the two branches obtained in the main text. This will be done in detail below.

\subsubsection*{SUSY light spectrum}
\label{sap:SUSYH}
We consider now the Hessian of the F-term potential associated to the supersymemtric branch of solutions. As explained in sections \ref{branchvacu} and \ref{sec:summary} this solution is characterised by
\begin{align}
\label{sussysol}
    A_F&=-3/8\, ,   &    B_F&=-3E_F/2\, ,   &    C_F&=1/4\, ,   &    D_F&=15E_F\, .
\end{align}
Then, one just has to plug \eqref{sussysol} into \eqref{eq: hessian}, diagonalize and divide by $1/2$ to obtain the corresponding mass spectrum. The result is:
\begin{equation}
\label{susymass}
   m^2= F^2e^K\mathcal{K}^2\left\{0,-\frac{1}{2}(1+16E_F^2),-\frac{1}{18}+56E_F^2\pm 
    \frac{1}{3}\sqrt{1+160E_F^2+2304E_F^4},\lambda_5,\lambda_6,\lambda_7,\lambda_8\right\}\, ,
\end{equation}
where the  $\lambda_i$ are the four roots of
\begin{align}
   0=&-160380+18662400E_F^2+62547240960E_F^4+2721784135680E_F^6+29797731532800E_F^8\nonumber\\
   &+(-19971-33191568E_F^2-4174924032E_F^4-74992988160E_F^6)18\lambda\nonumber\\
   &+(4483+1392480E_F^2+55800576E_F^4)\left(18\lambda\right)^2+(-133-13392E_F^2)\left(18\lambda\right)^3+\left(18\lambda\right)^4
    \label{eq: implicit}\, .
\end{align}
In order to discuss the stability, we must compare \eqref{susymass} to the BF bound, which for this case takes the value:
\begin{equation}
    m_{BF}^2=\frac{3}{4}V|_{\text{vac}}=-(\frac{9}{16}+9E_F^2)e^K\mathcal{K}^2 F^2\, .
\end{equation}
It is straightforward to see that the first non-zero eigenvalue can be rewritten as:
\begin{align}
  m_2^2= -\frac{1}{2}(1+16E_F^2)=\frac{8}{9}m_{BF}^2\, .
\end{align}
Regarding the other masses, although they can also be written as functions of the $m_{BF}$ their expressions are not that illuminating. In this sense, one can check that the third eigenvalue is always positive, whereas  $m_4^2$ has a  negative region -respecting the the BF bound- for $|E_F|\lesssim 0.1$. Finally, the dependence of the four remaining eigenvalues with $E_F$, conveyed as implicit solutions of \eqref{eq: implicit}, has to be studied numerically. One finds that only one of them enters in a negative region -again above $m_{BF}^2$- for $|E_F|\lesssim 0.04$.

We conclude that the SUSY vacuum may have up to three tachyons, though only one is preserved for $|E_F| \gtrsim 0.1$. None of them violates the BF bound, as it is expected for this class of vacua. To finish this part of the appendix, let us also write the tachyonic directions:
\begin{itemize}
    \item $m_2^2$. Direction: $u^{\hat{\mu}}$.\footnote{For the complex axions, the direction $\xi^{\hat\mu}$ is the one with zero eigenvalue.}
    \item $m_4^2$.  Direction: linear combination of $b^{\hat{a}}$ and $t^{\hat{a}}$.
    \item $m_5^2=F^2e^K\mathcal{K}^2\lambda_5$ (lowest solution of \eqref{eq: implicit}). Direction: combination of all non primitive directions, i.e. $\hat{\xi}$, $\hat{b}$, $\hat{u}$ and $\hat{t}$.
\end{itemize} 

\subsubsection*{Non-SUSY branch}
\label{sap:nonSUSYH}

We end this section of the appendix by analysing the Hessian of the F-term potential associated with the non-SUSY solutions. As it was studied in detail in the main text, this branch has to be defined implicitly in terms of the $A_F$ and $C_F$ solving equation \eqref{eq: megaeqF} (check table \ref{vacuresul} and figure \ref{fig: generalsol} for details). In consequence, trying to explore the stable regions analytically is, in practice, impossible, and things must be computed numerically. What we have done is to extract  the physical $A_F$ and $C_F$ satisfying \eqref{eq: megaeqF}, plug them into \eqref{eq: hessian} -$B_F$, $D_F$ and $E_F$ are determined once $A_F$ and $C_F$ are chosen- and study the mass spectrum. Despite the numerical approach, results can be obtained easily.  

After performing a complete analysis, we conclude that a single mode is responsible for the stability of the solution. In other words, seven out of the eight masses respect the BF bound at every point of the Non-SUSY branch. Therefore, the behaviour of the aforementioned mode is precisely the one which determines the unstable region (red points) in figure \ref{fig: excludsol}. For the sake of completeness, let us write it explicitly:
\begin{align}
\label{tachmass}
  m^2=&-F^2e^K\mathcal{K}^2\left[9(12A_F^2-1)((2A_F+C_F)(6A_F+C_F)-1)\right]^{-1}\left[-9+7776A_F^6+5184A_F^5C_F\right.\nonumber\\
    &+4A_FC_F(2+C_F)(C_F^2-5C_F+9)+1296A_F^4(C_F^2-2)+144A_F^3C_F(C_F^2+C_F-9)\nonumber\\
    &\left.-C_F(C_F-2)(C_F^2+6C_F-1)+6A_F^2(C_F^4+8C_F^3-46C_F^2+4C_F+45)\right]\, .
\end{align}
As it happened in the SUSY case for the mode with mass $\frac{8}{9}m_{BF}^2$, the direction of the mode with mass \eqref{tachmass} is given by  $u^{\hat{\mu}}$. It is worth to point out that we are not saying that the other modes do not yield tachyons, but they are always above the $BF$ bound.  As discussed below figure \ref{fig: excludsol}, these other tachyons are localised close to the regions where $m^2$ defined in  \eqref{tachmass} violates the BF bound.

\subsection*{D-term potential}

We perform a similar analysis with the D-terms. Starting from \eqref{eq: D-potentialgeom} and evaluating the second derivatives along the vacuum equations, we obtain that the only non-vanishing second partial derivatives of the potential $V_D$ are
\begin{align}
     \frac{\partial^2 V_D}{\partial u^\mu \partial u^\nu}=&\frac{3}{\mathcal{K}}\partial_\mu c_{\nu\sigma}\partial_\lambda K \tilde{g}^{\alpha\beta}\hat{\rho}^\sigma_\alpha\hat{\rho}^\lambda_\beta+\frac{12}{\mathcal{K}}c_{\mu\sigma}c_{\nu\lambda} \tilde{g}^{\alpha\beta}\hat{\rho}^\sigma_\alpha\hat{\rho}^\lambda_\beta\, ,\\
   \frac{\partial^2 V_D}{\partial u^\mu \partial t^a}=&\frac{3}{\mathcal{K}}c_{\mu\sigma}\partial_\lambda K\partial_a \tilde{g}^{\alpha\beta}\hat{\rho}^\sigma_\alpha\hat{\rho}^\lambda_\beta\,-\frac{9\mathcal{K}_a}{\mathcal{K}^2}c_{\mu\sigma}\partial_\lambda K \tilde{g}^{\alpha\beta}\hat{\rho}^\sigma_\alpha\hat{\rho}^\lambda_\beta ,\\
    \frac{\partial^2 V_D}{\partial t^a \partial t^b}=&(\partial_\sigma K\partial_\lambda K\hat{\rho}^\sigma_\alpha\hat{\rho}^\lambda_\beta)\left(\frac{3}{8\mathcal{K}} \partial_a\partial_b g^{\alpha\beta}\,-\frac{9\mathcal{K}_a}{8\mathcal{K}^2} \partial_b g^{\alpha\beta}\,\right.\nonumber\\
    &\left.-\frac{9\mathcal{K}_b}{8\mathcal{K}^2} \partial_a g^{\alpha\beta}\,+\frac{27\mathcal{K}_a\mathcal{K}_b}{4\mathcal{K}^3}\partial_\sigma K\partial_\lambda K  g^{\alpha\beta}\,-\frac{9\mathcal{K}_{ab}}{4\mathcal{K}^2}  g^{\alpha\beta}\,\right) .
\end{align}
If we now take into consideration the ansatz \eqref{Ansatz} together with the Bianchi identity $f_{a\mu}\hat{f}_\alpha^\mu=0$, we have that, on-shell, $\partial_\mu K \hat{\rho}^\mu_\alpha=0$. Hence the saxionic sector of the D-term Hessian becomes
\begin{align}
    \p_A\p_B V_D=\left(\begin{matrix}\frac{12}{\mathcal{K}}c_{\mu\sigma}c_{\nu\lambda} \tilde{g}^{\alpha\beta}\hat{\rho}^\sigma_\alpha\hat{\rho}^\lambda_\beta,    &   0\\
    0   &   0
    \end{matrix}\right)\, ,
    \label{dhessian}
\end{align}
which is clearly positive-semidefinite for any choice of the geometric fluxes.

\chapter{Hodge star and 10d   conventions}
\label{ap:10dconv}
The Hodge star operator in $d$ dimensions acting on a $p$-form $\alpha_p, \beta_p$ is defined by
\begin{equation}
\star_{d} \alpha_{p}=\frac{1}{p !(d-p) !} \sqrt{|g|} \epsilon_{i_{1} \ldots i_{d}} \alpha^{i_{d-p+1} \ldots i_{d}} d y^{i_{1}} \wedge \cdots \wedge d y^{i_{d-p}}\, .
\end{equation}
Therefore
\begin{align}
\star_d\alpha\wedge\beta=\left(\alpha\cdot\beta\right)d\text{vol}_d\, ,
\end{align}
where
\begin{align}
\alpha\cdot\beta=\frac{1}{p!}\alpha_{i_i\dots i_p}\beta^{i_i\dots i_p}\, .
\end{align}
The pure spinors $\Psi_1$ and $\Psi_2$ are chosen with the normalisation
\begin{align}
\braket{ \Psi_1|\bar{\Psi}_1}=\left(2i\right)^{d/2}d\text{vol}_d
\end{align}
where the Mukai pairing is given by
\begin{align}
\braket{\xi|\chi}=\xi\wedge  \sigma\left(\chi\right)|_{\text{top}}
\end{align}
$\sigma$ is the operator reversing the order of the indices of a form, and $\{\xi, \chi\}$ are arbitrary (poly)forms.

We choose $J$ and $\Omega$ such that
\begin{align}
d\vol_6=-\frac{1}{6}J\wedge J\wedge J=-\frac{1}{4}\text{Re}\Omega\wedge\text{Im}\Omega\, .
\end{align}
This in turn implies that
\begin{align}
\star_6\Omega&=-i\Omega\, ,	&	\star_6 J&=-\frac{1}{2} J\wedge J\, ,	&	\star_6 W&= J\wedge W
\end{align}
where $W$ is a primitive $\left(1,1\right)$-form.

\chapter{$SU(3)\times SU(3)$-structure compactifications}
\label{ap:SU33}

In this Appendix we will give more details about the pure spinor approach to supersymmetry. We give the general solution for the SUSY equations in manifolds of $SU(3)\times SU(3)$ structure.

\section{$SU(3)\times SU(3)$  $\theta \neq 0$} 
\label{sub:tneq0}

We now show the generic case, $\theta \neq 0$. The local solution in \cite{Saracco:2012wc} belongs to this class; however, we will see in the next section that this is unlikely to be promoted to a global solution.

Again there are first some purely geometric equations:
\begin{subequations}	
\begin{align}
	\label{eq:drho}
    &d(e^{3A-\phi} \cos \psi\,\sin\theta)=0\ , \\ 
	\label{eq:rev}    
	&{\rm Re} v = \frac{e^A}{2 \mu \sin\theta} d \theta \ ,\\
    \label{eq:dJpsi}
    &d\,\left(\frac1{\sin\theta} J_\psi \right) =2\mu e^{-A}{\rm Im} (v\wedge \omega_\psi)
\end{align}
\end{subequations}     
where again $J_\psi$ and $\omega_\psi$ are given by (\ref{eq:Jpsi}).

Then we have the fluxes, which are completely determined:
\begin{equation}
	H = d B   \, ,\qquad \mathbf{G} = e^{B \wedge} \mathbf{F}\,.
\end{equation}
So for example $G_2 = F_2 + B F_0$, but $G_0 = F_0$.  The $F_k$ are given by
\begin{subequations}
\begin{align}
    \label{eq:B}
 	B&= -\cot(\theta) J_\psi + \tan \psi \mathrm{Im}  \omega\ , \\
	F_0 &= -J_\psi\cdot d (e^{-\phi} \cos \psi {\rm Im} v)
	    + 5 \mu  e^{-A-\phi} \cos \psi\cos \theta \label{eq:F0}\ ,\\
	\label{eq:F2}
		F_2 &= F_0 \cot\theta J_\psi -J_\psi\cdot d\, {\rm Re} ( \cos \psi e^{-\phi}  v \wedge \omega_\psi) \\
		\nonumber &+ \mu \cos \psi  e^{-A-\phi}
		\left[(5+2\tan^2\psi) \sin\theta J_\psi +2 \sin\theta {\rm Re} v \wedge {\rm Im} v- 2\cos\theta\tan^2\psi {\rm Im} \omega_\psi\right]\ , \\
	F_4 & = F_0 \frac{J^2_\psi}{2 \sin^2\theta} + 
	d\Big[ \cos \psi \, e^{-\phi} (J_\psi \wedge {\rm Im} v - \cot\theta {\rm Re} (v \wedge\omega_\psi))\Big] \label{eq:F4}\ ,
	\\
	F_6 & = -\frac1{\cos^2\psi}{\rm vol}_6 \left(F_0\frac{\cos\theta}{\sin^3\theta} 
	+ 3 \frac{\mu \cos \psi e^{-\phi}}{\sin\theta}\right)\ .
\end{align}
\end{subequations} 
$F_4$ is automatically closed; this implies the Bianchi identity for $G_4$, which is $d G_4 + H \wedge G_2=0$.

\section{$SU(3)\times SU(3)$  $\theta = 0$} 
\label{sub:teq0}

The case $\theta=0$ has been discussed in subsection \ref{ssub:t0}. Here we will show that for this case, $G_6=\left( e^{b} \bf F\right)|_6= 0$, as claimed in the main text. 

Explicitly, what we have to compute is:
\begin{align}
G_6=\cancelto{0}{F_6}+\cancelto{0}{\frac{1}{3!}F_0b_\phi^3}+\frac{1}{2}F_2\wedge b_\phi\wedge b_\phi+F_4\wedge b_\phi\ ,
\label{g6}
\end{align}
where we are already using \eqref{su(2)prop} and \eqref{f6su3su3}. Taking into account the expression for $F_2$ -eq \eqref{eq:F2t0}-,  $F_4$ -eq.\eqref{eq:F2t0}- and $b=\tan(\psi)\im{\omega}$, \eqref{g6} reads: 
\begin{align}
\label{g6expand}
G_6&=\rho e^{-3A}\cos(\psi)^2\im\omega_\psi\wedge\re v\wedge \im v \wedge d\im v\nonumber \\&-\frac{1}{2}\tan^2(\psi)\im\omega\wedge\im\omega\wedge J_\psi^{-1} \llcorner d\re \left(  \rho e^{-3A} v \wedge \om_\psi\right) \, .
\end{align}
Let us massage the second term:
\begin{align}
&-\frac{1}{2}\tan^2(\psi)\im\omega\wedge\im\omega\wedge J_\psi^{-1} \llcorner d\re \left(  \rho e^{-3A} v \wedge \om_\psi\right)\nonumber\\ &=-\frac{1}{2}\tan^2(\psi)j^2\wedge J_\psi^{-1} \llcorner d\re \left(  \rho e^{-3A} v \wedge \om_\psi\right)\label{hola}\, ,
\end{align}
writing $j^2$ as $$j^2=\cos^2(\psi)J_\psi^2-\frac{2\cos^2(\psi)}{\tan^2(\psi)}j\wedge\re v\wedge\im v=\cos^2(\psi)^2J_\psi^2-\frac{2\cos^2(\psi)}{\tan^2(\psi)}J_\psi\wedge\re v\wedge \im v \, ,$$ to obtain:
\begin{align}
\label{tedioso}
-\frac{\sin^2\left(\psi^2\right)}{2} J_\psi^2\wedge J_\psi^{-1} \llcorner d\re \left(  \rho e^{-3A} v \wedge \om_\psi\right)+\cos^2\left(\psi^2\right)\re v\wedge\im v \wedge J_\psi \wedge J_\psi^{-1} \llcorner d\re \left(  \rho e^{-3A} v \wedge \om_\psi\right)\, .
\end{align}
We can use now -see \cite{Saracco:2012wc}-:
\begin{align}
&\label{comm}\left[J_\psi^{-1}\llcorner, J_\psi\wedge\right]=h\, ,\quad\quad 		h\omega_k\equiv (3-k)\omega_k\, ,
\end{align}
to rewrite \eqref{tedioso} as:
\begin{align}
&-\frac{\sin^2\left(\psi^2\right)}{2}J_\psi\wedge\left(J_\psi^{-1}\llcorner J_\psi\wedge+1\right)d\re \left(  \rho e^{-3A} v \wedge \om_\psi\right)+\nonumber\\&+\cos^2\left(\psi^2\right)\re v\wedge\im v\wedge\left(J_\psi^{-1}\llcorner J_\psi\wedge+1\right)d\re \left(  \rho e^{-3A} v \wedge \om_\psi\right)\, .
\label{aux2}
\end{align}
Finally, taking into account that the supersymmetry equations imply:
\begin{align}
J_\psi\wedge d\left(\rho e^{nA}\re (v\wedge\omega_\psi)\right)&=0\, ,   &  d\re v&=dA\wedge \re v\, ,
\end{align}
the second term of \eqref{g6expand} can be written as:
\begin{align}
 -\rho e^{-3A}\cos^2\left(\psi^2\right)\im\omega_\psi\wedge\re v\wedge\im v\wedge d\im v\, ,   
\end{align}
and therefore:
\begin{align}
    G_6=0\, .
\end{align}

\section{Proof of the source balanced equation}
\label{ap:SBEproof}

Let us show how the source balanced equation \eqref{preGOE} can be derived. First consider the following Mukai pairing
\be
\left<d_{H} \hat{F} , e^A \mathrm{Im\;}\Phi_- \right> = \left< \hat{F} , d_{H} \left(e^A \mathrm{Im\;}\Phi_- \right) \right> + dX_5 \;,
\ee
with $X_5$ defined as in \eqref{X5}. We can evaluate the left-hand side using the Bianchi identity \eqref{IIABI} in the presence of O6-planes and D6-branes, while the right-hand side can be evaluated using the supersymmetry equation \eqref{eq:psp-}. We obtain
\be
-3\mu  \left< \hat{F}, \mathrm{Im\;}\Phi_+ \right> + e^{4A} \left<\hat{F} , \star \lambda \left(\hat{F} \right) \right>  + dX_5 
= \left< \delta^{(3)}_{\mathrm{source}} , e^A  \mathrm{Im\;}\Phi_- \right> \;.
\label{pp1}
\ee
This expression can be rewritten by noting that taking the Mukai pairing of \eqref{eq:psp-} with $\Phi_+$ yields
\be
\mu \left< \mathrm{Re\;}\Phi_+ , \mathrm{Im\;}\Phi_+ \right> = e^{4A} \left<\Phi_+,  \star \lambda\left( \hat{F} \right) \right>\;.
\label{f03eq}
\ee
The existence of the $SU(3)\times SU(3)$-structure implies a generalised Hodge decomposition of the space of polyforms, according to their eigenvalues under two generalised complex structures $\left({\cal J}_+,{\cal J}_-\right)$. Under this decomposition $\Phi_+$ is of type $\left(3,0\right)$. This means that the right-hand side of (\ref{f03eq}) only receives a contribution from the $\left(-3,0\right)$ component of $\star \lambda\left( \hat{F} \right)$, and so we can replace $\star \lambda\left( \hat{F} \right)$ with $-i \hat{F}$ (see, for example \cite{Tomasiello:2007zq}). We can therefore write (\ref{f03eq}) as
\be
\mu \left< \mathrm{Re\;}\Phi_+ , \mathrm{Im\;}\Phi_+ \right> = -i e^{4A} \left< \Phi_+,\hat{F}  \right> \;.
\label{pppf}
\ee
Now using (\ref{pppf}) we have that (\ref{pp1}) reads
\be
3\mu^2 e^{-4A} \left< \mathrm{Re\;}\Phi_+ , \mathrm{Im\;}\Phi_+ \right>  - e^{4A} \sum_k \hat{F}_k \wedge \star \hat{F}_k + dX_5 
=  \left< \delta^{(3)}_{\mathrm{source}} , e^A  \mathrm{Im\;}\Phi_- \right> \, ,
 \label{ap:preGOE}
\ee
proving the desired relation.

It is important to note that integrating (\ref{ap:preGOE}) over the manifold leads to a constraint which does not differentiate between a local and smeared source, and therefore can be solved already for the $SU(3)$-structure case. If $X_5$ was a completely general function, then the solution to the integral of (\ref{preGOE}) would guarantee a local solution for some choice of $X_5$. However, $X_5$ is not an independent function, it is fixed by the fluxes and the polyforms, and therefore such a local solution is not guaranteed. 
\chapter{Tools for the non-SUSY analysis}
\label{ap:end}

\section{10d equations of motion}
\label{ap:10deom}

In this appendix we will discuss how the SUSY \eqref{solutionflux} and the non-SUSY backgrounds \eqref{solutionfluxnnosusy} presented in the main text solve the 10d equations of motion. The Bianchi identities, which also must be satisfied, were discussed in great detail in chapter \ref{ch:uplift10d} for the SUSY case so we will limit ourselves to emphasise the changes the non-SUSY case introduces. The full set of equations of motion were presented in sections \ref{subs:rrnseq} and \ref{subsec:einseq}.

\subsubsection*{RR and NSNS field equations, SUSY case}

In our approximation, the internal part of the first equation of \eqref{eq:rrnsns} is
\begin{align}
\label{ap1eq1}
0&=\mathrm{d}\left(e^{4A}\star_{\rm CY} G_{2}\right)+e^{4A}H_{3} \wedge \star_{\rm CY} G_{4}+\CO\left(g_s\right)=0+\CO\left(g_s\right)\, ,
\end{align}
where we have used that $G_2$ is known up to $\CO\left(g_s\right)$ -see  \eqref{G2sol}. Since the natural scaling of a  $p$-form is $g_s^{-p/3}$, the total error we are making in solving  this equation is  $\CO(g_s^{8/3})$.

The internal part of second equation, at our level of approximation, reads
\begin{align}
0=d\left(e^{4A}\star_{\rm CY} G_4\right)=4g_se^{4A}G_0d\varphi\wedge J_{\text{CY}}+\frac{e^{4A}}{g_s}d\star_{\rm CY} \left(J_{\rm CY}\wedge d\im v\right)+\CO(g_s^2)\, ,
\label{eqg4}
\end{align}
it is more or less straightforward to check that
\begin{align}
\label{auxg4}
\frac{1}{g_s} d\star_{\rm CY} \left(J_{\rm CY}\wedge d\im v\right)=4G_0g_s\star_{\rm CY}\left( J_{\rm CY}\wedge d_c\varphi\right)=-4G_0g_s J_{\rm CY}\wedge d\varphi\, ,
\end{align}
which cancels out the first term of  \eqref{eqg4} and satisfies the equation up to order $ \CO(g_s^{3})$ compared to the natural scaling of a three-form.

The third equation of \eqref{eq:rrnsns} is trivial, since $G_6=0$ and so it remains to check equation \eqref{problem}, which is the most cumbersome. We will go term by term and write just the internal parts to make the computation clearer. At the level of approximation that we are working the second term in the RHS  is 
\begin{align}
e^{4A}\star_6 G_4\wedge G_2=\frac{12}{5}G_0 d\varphi\wedge\im\Omega_{\rm CY}-\frac{3}{5}G_0\star_{\rm CY}G_2+\CO(g_s)\, ,
\end{align}
while the first term reads
\begin{align}\nonumber
d\left(e^{-2\phi+4A}\star _6 H\right)& =d\left[ e^{-2\phi+4A}\star_6\left(\frac{2}{5}G_0g_se^{-A}\re\Omega-\frac{1}{2}d\re\left(\bar{v}\cdot\Omega_{\rm CY}\right)\right)\right]+\CO(g_s)\\
& = \frac{2 }{5}G_0g_sd\left(e^{-2\phi+3A}\im\Omega\right)-\frac{e^{-2\phi+4A}}{2}d\star_6d\re\left(\bar{v}\cdot\Omega_{\rm CY}\right)+\CO(g_s)\, .
\label{ap:divided}
\end{align}
The first contribution to \eqref{ap:divided} can then be rewritten as
\begin{align}
\frac{2 }{5}G_0g_sd\left(e^{-2\phi+3A}\im\Omega\right)=\frac{2G_0}{5}\left(4d\varphi\im\Omega_{\rm CY}-\star_{\rm CY}G_2\right)+\CO(g_s)\, ,
\end{align}
whereas for the second contribution, a long calculation shows that
\begin{align}
-\frac{e^{-2\phi+4A}}{2}d\star_6d\re\left(\bar{v}\cdot\Omega_{\rm CY}\right)=-4G_0 d\varphi\wedge\im\Omega_{\rm CY}+\CO(g_s)\, .
\end{align}
Finally, putting everything together \eqref{problem} reduces to 
\begin{align}
0=\left(\frac{12}{5}+\frac{8}{5}-4\right)e^{4A}G_0d\varphi\wedge\im\Omega_{\rm CY}+\left(-\frac{2}{5}-\frac{3}{5}+1\right)e^{4A}G_0\star_{\rm CY}G_2+\CO(g_s)\, ,
\end{align}
which,  as a $4$-form equation, we are solving it with an error  $\CO(g_s^{7/3})$.

\subsubsection*{Non-supersymmetric case}

In the non-SUSY solution, only the fields $H$ and $G_4$ change, so it is enough to check the equations involving these quantities.

Let us start by the Bianchi identities, which we ignored in the previous section. To start with we can look at
\begin{align}
\label{bianH}
d G_4= G_2\wedge H\, .
\end{align}
The changes in $G_4^{\text{non-SUSY}}$ appear in the harmonic and the closed parts, so the LHS is the same as the  $G_4^{\text{SUSY}}$. The changes in $H^{\text{non-SUSY}}$ are of order $\CO(g_s^{7/3})$, giving a  contribution beyond the order at which \eqref{bianH} is being solved: we can ignore them and recover the RHS of the SUSY solution as well. The other BIs which could be sensitive to the non-SUSY novelties are  $dG_2$ and $dH_3$. For both of them, the changes appear beyond the order of approximation in which they are being solved, so we can just neglect them.

Regarding the equations of motion, for $G_4$  the internal part now reads
\begin{align}
d\left(e^{4A}\star_{\rm CY} G_4\right)=-\frac{24}{5} g_se^{4A}G_0d\varphi\wedge J_{\text{CY}}-\frac{6e^{4A}}{5g_s}d\star_{\rm CY} \left(J_{\rm CY}\wedge d\im v\right)+\CO(g_s^2)=0+\CO(g_s^2)\, ,
\end{align}
where we have used \eqref{auxg4}. As in the SUSY case, it is solved at total order $\CO(g_s^3)$. 

Finally, the equation for $H$ is again the most tedious. Following the reasoning of the previous section, we will directly write each of the contributions to the internal part. On the one side
\begin{align}
e^{4A}\star_6 G_4\wedge G_2=-\frac{12}{5}G_0 d\varphi\wedge\im\Omega_{\rm CY}+\frac{3}{5}G_0\star_{\rm CY}G_2+\CO(g_s)\, ,
\end{align}
on the other side
\begin{align}
d\left(e^{-2\phi+4A}\star _6 H\right)=\frac{12}{5}G_0d\varphi\wedge\im\Omega_{\rm CY}-\frac{8}{5}G_0\star_{\rm CY}G_2+\mathcal{O}(g_s)\, ,
\end{align}
and \eqref{problem} reduces to
\begin{align}
0=\left(-\frac{12}{5}+\frac{12}{5}\right)e^{4A}G_0d\varphi\wedge\im\Omega_{\rm CY}+\left(-\frac{8}{5}+\frac{3}{5}+1\right)e^{4A}G_0\star_{\rm CY}G_2+\CO(g_s)\, ,
\end{align}
which is again solved  at total order $\CO(g_s^{7/3})$.

\subsubsection*{Einstein and dilaton equations}

To show how our expressions satisfy these two last constraints, equations \eqref{eq:einstein1}-\eqref{eq:einstein3}, we will use the results derived in \cite{Junghans:2020acz}, focusing again on the changes introduced by the non-SUSY case. At leading order the equations evaluated for the non-SUSY solution coincide with the equations evaluated in the SUSY background, so they are satisfied in the first case provided they are solved in the second case -as it happens-. When the changes come into play, they do it at least at order $|F_4|^2\sim e^{-2\phi}|H_3|^2\sim \CO(g_s^{4/3})$. Nevertheless, to solve the equations at this order, we need to consider terms in $e^{A}$ and $e^{-\phi}$ which are beyond our approximation. 
  In other words, the modifications introduced in the non-SUSY case are seen by the Einstein and dilaton equations at the next order in the expansion.


\section{DBI computation}
\label{ap:dbi}

The BIonic D8-brane system of section \ref{s:bion} is defined by the profile \eqref{BIonrel} for the transverse D8-brane position. In this appendix we check that this relation fulfils the basic requirement of a BPS condition, in the sense that it linearises the DBI action of the D8-brane, at least at the level of approximation at which we work in the main text. 

The DBI action of a D8-brane wrapping $X_6$ is given by 
\be
 S_{\rm DBI}^{\rm D8} = - \frac{2\pi}{\ell_s^9} \int dt dx^1 dx^2 \int_{X_6} d^6 \xi   e^{3A - \phi} e^{\frac{3Z}{R}} \sqrt{\det \left(g_{ab} + \p_aZ\p_bZ +\cF_{ab} \right) } \, ,
 \label{ap:DBID8}
\ee
where the D8-brane transverse position $Z$ is seen as a function on $X_6$. For BPS configurations the integrand simplifies, in the sense that the square root linearises and corresponds to integrating a six-form over $X_6$. To see how this happens for the BIon configuration, let us use the matrix determinant lemma to rewrite things as
\be
\det \left(g_{ab} + \p_aZ\p_bZ +\cF_{ab} \right)  =  \det g \, \det \left(\II + g^{-1} \cF\right) \left( 1 + \p Z \cdot (g+\cF)^{-1} \cdot \p Z\right)\, .
\label{inidet}
\ee
Then using that $\cF$ is antisymmetric one can deduce that
\begin{align}
 \det \left(\II + g^{-1} \cF\right) = 1 - \frac{t_2}{2} + \frac{t_2^2}{8} -\frac{t_4}{4} + \frac{\det  \cF}{\det g}\, ,
\end{align}
where  $t_k = \Tr\, g^{-1} \cF^k$. Using in addition the Woodbury matrix identity we obtain 
\be
\p Z \cdot (g+\cF)^{-1} \cdot \p Z = \p Z \cdot \sum_{k=0}^\infty \left(g^{-1} \cF\right)^{2k} g^{-1} \cdot \p Z\, .
\ee

One may then combine all these expressions to compute \eqref{inidet}. Recall however that our unsmeared background description is only accurate below $\cO(g_s^2)$ corrections in the $g_s$ expansion. As pointed out in  \cite{Junghans:2020acz,Marchesano:2020qvg} a flux of the form \eqref{cfsol} is suppressed as $\cO(g_s^{3/2})$ compared to a harmonic two-form and, because of \eqref{BIonrel}, the same suppression holds for $\p Z$. This means that we are only interested in terms up to quadratic order in the worldvolume flux or $\p Z$ in the DBI action, or equivalently up to quartic order in \eqref{inidet}. That is, we are interested in computing the following terms
\be
\left(1 - \oh \Tr \tilde\cF^2 \right) \left(1 + (\p Z)^2\right)   + \frac{1}{8} \left( \Tr \tilde\cF^2\right)^2 -\frac{1}{4}  \Tr \tilde\cF^4 - \left( \p Z\cdot  \tilde \cF \right)^2 \, ,
\ee
where $\tilde \cF \equiv g_{\rm CY}^{-1} \cF$, and $(\p Z)^2 = g_{\rm CY}^{ab} \p_a Z \p_b Z$, etc. To proceed we split the worldvolume flux as in section \ref{ss:bionnosusy}
\be
\tilde \cF_1 \equiv g^{-1}  \cF^{(1,1)} \, , \qquad \tilde \cF_2 \equiv g^{-1} \cF^{(2,0)+(0,2)} \, ,
\ee
assuming that $\cF^{(1,1)}$ is primitive, and use the following identity
\be
\Tr \tilde\cF^4 = \frac{1}{4} \left(\Tr \tilde \cF^2 \right)^2 +  \left(\Tr \tilde \cF_1^2 \right)\left(\Tr \tilde \cF_2^2 \right) + 4 \Tr \left([\tilde \cF_1, \tilde \cF_2]^2\right) \, ,
\ee
to arrive to
\be
\left( 1 - \frac{1}{4} \Tr \tilde\cF^2 + \oh (\p Z)^2 \right)^2 -  \frac{1}{4}\left((\p Z)^2  \Tr \tilde\cF^2 +  \Tr \tilde \cF_1^2 \, \Tr \tilde \cF_2^2 +  (\p Z)^4 \right)  -\Tr \left([\tilde \cF_1, \tilde \cF_2]^2\right) - \left( \p Z\cdot  \tilde \cF \right)^2\, .
\ee
Finally, one can see that \eqref{BIonrel} and primitivity imply that
\be
 (\p Z)^2 = -  \Tr \tilde\cF_2^2\, , \qquad \left( \p Z\cdot  \tilde \cF \right)^2 = \left( \p Z\cdot  \tilde \cF_1 \right)^2 = - \Tr \left([\tilde \cF_1, \tilde \cF_2]^2\right)\, ,
\ee
and so we are left with
\be
\left( 1 - \frac{1}{4} \Tr \tilde\cF^2 + \oh (\p Z)^2 \right)^2 = \left( 1 - \frac{1}{4} \Tr \left(\tilde\cF_1^2  - \tilde\cF_2^2\right) + (\p Z)^2 \right)^2\, .
\ee
When plugged into \eqref{ap:DBID8} this translates into
\be
 S_{\rm DBI}^{\rm D8} = - \frac{2\pi}{\ell_s^9} \int dt dx^1 dx^2  g_s^{-1} e^{\frac{3z_0}{R}} \int_{X_6} \left[-\frac{1}{6}J_{\rm CY}^3 + \oh J_{\rm CY} \wedge \cF^2 + *_{\rm CY} dZ \wedge dZ + \cO(g_s^{4/3}) \right]
 \label{ap:DBID8fin}
\ee
where we used that in our approximation $\cF_1 \equiv \cF^{(1,1)}$ is a primitive (1,1)-from, and as a result $-\oh \Tr \tilde \cF_1^2 d{\rm vol}_{X_6} = \star_{\rm CY} \cF_1 \wedge \cF_1 = J_{\rm CY} \wedge \cF_1 \wedge \cF_1$. Finally, we have expanded $e^{3A-\phi} = g_s^{-1} + \cO(g_s)$ and $e^{\frac{3Z}{R}} = e^{\frac{3z_0}{R}}\left(1 - \frac{12 \ell_s}{|m|R}  \varphi \right) + \cO(g_s^{8/3})$, and used that $\int_{X_6} \varphi = 0$.


\section{BIonic strings and SU(4) instantons}
\label{ap:IIBion}

The BIonic solution found in section \ref{s:bion} is not unique of type IIA flux compactifications. It can also be found when one wraps a D7-brane on the whole internal manifold of type IIB warped Calabi--Yau compactifications with background three-form fluxes. The advantage of this type IIB setup compared to the type IIA one considered in the main text is two-fold: {\it i)} we know the exact 10d background and  {\it ii)} we can directly connect it to the Abelian $SU(4)$ instanton solutions that define Donaldson--Thomas theory \cite{Donaldson:1996kp}.

\subsubsection*{IIB BIonic strings}

Let us consider a type IIB warped Calabi--Yau compactification, namely a metric background of the form
\be
ds^2 = e^{2A}ds^2_{\pr^{1,3}} + e^{-2A} ds^2_{X_6}\, ,
\ee
where $X_6$ is endowed with a Calabi--Yau metric. On top of it we can add background fluxes $H_3$ and $F_3$ which are quantised harmonic three-forms of $X_6$ sourcing the warp factor. Let us consider the case in which $\ell_s^{-2} [H]$ is Poincar\'e dual to a three-cycle class with a special Lagrangian representative $\Pi$ calibrated by $\im \Omega_{\rm CY}$. That is:
\be
 \ell_s^{-2} [H] = {\rm P.D.} [\Pi]  = \ell_s^{-3} \d (\Pi)\, ,
\ee
where $\d (\Pi)$ is the bump delta-function of $X_6$ with support in $\Pi$. 

We now wrap a D7-brane on the internal six-dimensional space, as in \cite[section 6]{Evslin:2007ti}, and extended along $(t,x^1,0,0)$. The Freed--Witten anomaly induced by the $H$-flux can be cured by a D5-brane wrapping $-\Pi$, extended along $(t, x^1,0, x^3>0)$ and ending on the D7-brane. This configuration describes a 4d string to which a 4d membrane is attached. Microscopically this is due to the Freed--Witten anomaly. Macroscopically it as a result of of the type IIB axion $C_0$ gaining an F-term axion-monodromy potential generated by the internal $H$-flux \cite{BerasaluceGonzalez:2012zn,Marchesano:2014mla,Blumenhagen:2014gta}. 

The Bianchi identity for the D7-brane worldvolume flux reads
\be
d\cF =  H - \ell_s^{-1}  \d (\Pi)\, ,
\label{ap:BIF}
\ee
and finding its solution works as in \cite[section 5]{Marchesano:2020qvg}, see also \cite[section 3.4]{Hitchin:1999fh}. We have that
\be
\ell_s^{-1}\cF = d^{\dag}_{\rm CY} K = - J_{\rm CY} \cdot d \left( \hat\varphi \im \Omega_{\rm CY} - *_{\rm CY} K\right) \, ,
\ee
up to a harmonic piece. Here the function $\hat{\varphi}$ satisfies $\int_{X_6} \hat{\varphi} =0$ and 
\be
\Delta_{\rm CY}  \hat \varphi = \left(\frac{{\cal V}_{\Pi}}{{\cal V}_{\rm CY}} - \delta^{(3)}_{\Pi}\right) \, , \qquad \delta_{\Pi}^{(3)} = *_{\rm CY} \left[ \im \Omega \wedge \d (\Pi)\right]\, ,
\ee
while the three-form current $K$ is defined as in \eqref{formK} with the replacement $\varphi \to \hat{\varphi}/4$. The main difference with respect to the type IIA solution is that this one is exact. The 10d BPS configurations is therefore described by a BIon solution with profile
\be
*_{\rm CY} dX^3 =  \im \Omega_{\rm CY} \wedge \cF\, ,
\ee
from where we deduce that $X^3 = -  \ell_s \hat{\varphi}$. This would correspond to a DBI action such that
\bea\nonumber
S_\text{DBI}^{\rm D7} &= & - \frac{2\pi}{\ell_s^9} \int dt dx^1 dx^2  g_s^{-1}  \int_{X_6} e^{2A} \sqrt{\det \left(g_{ab} + e^{2A} \p_a X^3 \p_b X^3  + \mathcal{F}_{ab}\right)} \\ 
& =&- \frac{2\pi}{\ell_s^9} \int dt dx^1 dx^2  g_s^{-1} \int_{X_6}  - \frac{e^{-4A}}{6}J^3_{\rm CY} + \frac{1}{2} \cF \wedge \cF \wedge J_{\rm CY}  +  *_{\rm CY} dX^3 \wedge  dX^3\, ,
\label{DBI}
\eea
as would follow from the results of \cite{Evslin:2007ti}. 

Besides being an exact solution, the D7-brane setup has the interesting feature that the transverse space to the D7 is given by $\pr\times S^1$. As a result one is able to relate the D7 BIon system to a gauge configuration that is defined on $\pr\times S^1 \times X_6$. The natural object where such a gauge theory is defined is a  D9-brane dual to the BIonic D7-brane. As we will now discuss, this construction leads us directly to the setup where Donaldson--Thomas theory is defined. 

\subsubsection*{The Donaldson--Thomas setup}

In a Calabi--Yau four-fold $X_8$ we can define a complex star operator $\star$ that maps a $(0,q)$-form $\alpha$ to a $(0,4-q)$-form $\star \a$ such that 
\be
\alpha \wedge \star \alpha =\frac{1}{4} |\alpha|^2 \bar{\Omega}
\ee
where $\Omega$ is the holomorphic four-form of $X_8$, normalised such that $\Omega \wedge \bar{\Omega} = 16 \, d{\rm vol}_{X_8}$. It turns out that $\star$ maps $(0,2)$-forms to $(0,2)$-forms, and that $\star^2 =1$. One can then define two eigenspaces of $(0,2)$-forms such that $\star \alpha_\pm = \pm \a_\pm$. In particular, one may take the $(0,2)$-component of a real non-Abelian gauge flux $F$ on $X_8$ and demand that $\star F^{0,2} = - F^{0,2}$, or in other words that $F^{0,2}_+ =0$. This is one of the conditions of Donaldson--Thomas $SU(4)$ instanton equations  \cite{Donaldson:1996kp}, that read
\begin{subequations}
\label{DT}
\begin{align}
\label{DT1}
       F_+^{0,2}  & = 0\, ,\\
       F \wedge J^3 & = 0 \, .
        \label{DT2}
\end{align}
\end{subequations}

To connect with the more familiar Hodge star operator $*$, one can use that, when acting on $(0,q)$-forms, $\bar{*} = \frac{1}{4}  \Omega \wedge \star$ \cite{Baulieu:1997jx}. Therefore we deduce that
\be
* F^{0,2}_\pm = \pm \frac{1}{4} \bar \Omega \wedge F^{2,0}_\pm\, .
\label{DTH}
\ee
From here we deduce that $ F_\pm^{0,2} =0$ is equivalent to
\begin{subequations}
\label{DTre}
\begin{align}
\label{DTrere}
*  \re F^{0,2} & = \pm\frac{1}{4} \re \Omega \wedge F\, , \\
*  \im F^{0,2} & = \mp \frac{1}{4}\im \Omega \wedge F \ \implies \ F \wedge F \wedge \im \Omega = 0\, .
\label{DTreim}
\end{align}
\end{subequations}
and also implies
\be
\Tr \left(\re F^{0,2} \wedge * \re F^{0,2}\right) = \frac{1}{4} \Tr \left(\re F^{0,2}_+ \wedge \re F^{0,2}_+  - \re F^{0,2}_- \wedge \re F^{0,2}_- \right) \wedge \re \Omega\, .
\label{splitFpm}
\ee

\subsubsection*{The dictionary}

To connect with the D7 BIon configuration, we consider the Donaldson--Thomas equations for an Abelian gauge theory in the following Calabi--Yau background
\be
  \pr  \times S^1 \times X_6\, ,
\label{DTsetup}
\ee
with with complex coordinates $\{ \omega =  x + i\theta, z^1, z^2, z^3\}$ and holomorphic four-form
\be
\Omega_4 = \left(dx + i d\theta\right) \wedge \Omega_3 \, .
\ee
We now consider a gauge field strength of the form 
\be
\cF = \cF_{X_6} + \cF_{\rm Bion}\, ,
\label{fluxDT}
\ee
where $\cF_{X_6}$ is a two-form on $X_6$ and 
\be
\cF_{\rm Bion} = F_{xi} \, dx \wedge dz^i + {\rm c.c.}
\ee
so that there is no component of the flux along $d\theta$, and as a result $\cF^4 = 0$. 

The dictionary with the D7 BIon configuration can then be done by simple dimensional reduction along $\pr \times S^1$. After that, we recover a gauge theory on $X_6$ with gauge field strength $\cF_{X_6}$ and a non-trivial profile for the transverse position field $X$, seen as a function on $X_6$
\be
\p X = - F_{xi} dz^i\, .
\ee
Notice that
\be
\cF_{\rm BIon} =  dZ \wedge dx =  \frac{1}{2}\left( \p X + \bar{\p} X \right) \wedge \left( d\omega + d\bar{\omega}\right)  \implies \cF_{\rm BIon}^{0,2} = - \frac{1}{2} d\bar{\omega} \wedge  \bar{\p} Z\, .
\ee
Therefore to satisfy \eqref{DT1} we need to impose 
\be
d\bar{\omega} \wedge  \bar{\p} X = - \oh *_4 \left(\bar{\Omega}_4 \wedge \cF_{X_6}\right) \implies   \bar{\p} X  = \frac{i}{2}  *_{X_6}  \left( \bar{\Omega}_3 \wedge \cF_{X_6}\right)\, ,
\ee
from where we deduce the following relations
\bea
\label{BIon1}
*_{X_6} dZ & = &   \im \Omega_3 \wedge \cF_{X_6} \, ,\\
*_{X_6} d^c Z  & = &   \re \Omega_3 \wedge \cF_{X_6}\, .
\label{BIon2}
\eea
Eq.\eqref{BIon1} corresponds to the BIon equation of section \ref{s:bion}, while \eqref{BIon2} looks like a new, independent equation. In principle we would expect that it also satisfied by the BIon solution, and so it would be interesting to understand its implications. Notice that we can also translate \eqref{DT1} into the condition 
\be
\cF_{X_6}^{0,2} = \frac{1}{8} *_4 \left( \bar{\Omega}_4 \wedge \p X \wedge d\omega\right) \implies \cF_{X_6}^{0,2} = -\frac{i}{4} *_3  \left(\bar{\Omega}_3 \wedge \p X \right) \, ,
\ee
which in turn implies
\bea
\label{BIon3}
\re \cF_{X_6}^{2,0}  & = & - \frac{1}{4} *_3  \left(dX \wedge\im \Omega_3 \right) = \frac{1}{4} *_3  d\left(\hat{\varphi} \im \Omega_3 \right) \, ,\\
\im \cF_{X_6}^{2,0}  & = & - \frac{1}{4} *_3  \left(dX \wedge\re \Omega_3 \right) = \frac{1}{4} *_3  d \left(\hat{\varphi} \re \Omega_3 \right) \, .
\label{BIon4}
\eea
Eq.\eqref{BIon3} corresponds to \eqref{cF2} adapted to this setup, while \eqref{BIon4} is equivalent to \eqref{BIon2}. Finally, imposing \eqref{DT2} amounts to require that $\cF_{X_6}$ is primitive, as the BIon solution fulfils. 

The relation between the solutions to the Bianchi identity of the form \eqref{ap:BIF} and the Abelian SU(4) instanton equations of \cite{Donaldson:1996kp} was already pointed out in \cite[section 3.4]{Hitchin:1999fh}. We find it quite amusing that a BIonic D7-brane and the corresponding worldvolume flux on a D9-brane give a neat physical realisation of this correspondence. It would be interesting to understand if this description has any implications for the theory of invariants developed in \cite{Donaldson:1996kp}.


\section{A toroidal orbifold example}
\label{ap:torus}

In this appendix we compute the BIon correction to the D8/D6-system tension  $\Delta_{\rm D8}^{\rm Bion}$ defined in \eqref{QTbionnosusyexp}, for the particular geometry $X_6=T^6/(\mathbb{Z}_2\times \mathbb{Z}_2)$ in the orbifold limit. This case was already analysed in \cite{Marchesano:2020qvg} whose notation we follow up to small modifications. The Calabi--Yau structure is defined as
\bea
J_{\rm CY} & = & 4\pi^2 t_i dx^i \wedge dy^i \, \label{eq: torus J},\\
\re \Om_{\rm CY} & = &  \rho \left(\tau_1\tau_2\tau_3 \b^0 - \tau_1 \b^1 - \tau_2 \b^2 - \tau_3 \b^3 \right) \, ,\\
\im \Om_{\rm CY} & = &   \rho \left( \a_0 - \tau_2\tau_3 \a_1 - \tau_1\tau_3 \a_2 - \tau_1\tau_2 \a_3 \right) \, ,
\eea
where
\be
t^i  =  R_{x^i} R_{y^i}\, , \qquad \tau_i = \frac{R_{y^i}}{R_{x^i}}\, , \qquad \rho =  8\pi^3 \sqrt{\frac{t_1 t_2 t_3}{\tau_1\tau_2\tau_3}} =  8\pi^3 R_{x^1}R_{x^2}R_{x^3 }\, ,
\ee
and we have the following basis of bulk three-forms
\bea\nonumber
\a_0 = dx^1 \wedge dx^2 \wedge dx^3\, , & \quad & \b^0 = dy^1 \wedge dy^2 \wedge dy^3 \, ,\\ \nonumber
\a_1 = dx^1 \wedge dy^2 \wedge dy^3\, , & \quad & \b^1 = dy^1 \wedge dx^2 \wedge dx^3 \, ,\\ \nonumber
\a_2 = dy^1 \wedge dx^2 \wedge dy^3\, , & \quad & \b^2 = dx^1 \wedge dy^2 \wedge dx^3 \, ,\\ \nonumber
\a_3 = dy^1 \wedge dy^2 \wedge dx^3\, , & \quad & \b^3 = dx^1 \wedge dx^2 \wedge dy^3 \, .
\eea

The BIon worldvolume flux can be derived from \eqref{cfsol} and the results in  \cite[section 6.2]{Marchesano:2020qvg}, generalised to the case where the torus radii are not equal. We obtain
\begin{align}
\cF =& \, \frac{\ell_s}{m} d^\dag_{\text{CY}} K=h \ell_s \star_{\text{CY}} d \star_{\text{CY}}\left(B_0\beta^0-B_1\beta^1-B_2\beta^2-B_3\beta^3\right)\nonumber\\ =&\, h  \sum_{\eta}\sum_{\vec{0}\neq\vec{n}\in \IZ^3}\frac{1}{4\pi^2}  2\pi i\left[\frac{e^{i 2\pi  \vec{n}\cdot \left[(y_1, y_2,y_3)+\vec{\eta}\right]}}{ |\vec{n}_{y1,y2,y3}|^2} \left(\frac{n_1}{\tau_1 t_1} dy_2\wedge dy_3-\frac{n_2}{\tau_2 t_2}dy_1\wedge dy_3+\frac{n_3}{\tau_3 t_3}dy_1\wedge dy_2\right)\right. \nonumber\\ &\left. \textcolor{black}{-} \frac{ e^{i 2\pi  \vec{n}\cdot \left[(y_1, x_2, x_3)+\vec{\eta}\right]}}{|\vec{n}_{y1,x2,x3}|^2}\left(\frac{n_1}{\tau_1 t_1} dx_2\wedge dx_3-\frac{n_2\tau_2}{t_2} dy_1\wedge dx_3+\frac{n_3\tau_3}{t_3} dy_1\wedge dx_2\right)\right. \nonumber\\ &\left. \textcolor{black}{-}\frac{e^{i 2\pi  \vec{n}\cdot\left[ (x_1, y_2, x_3)+\vec{\eta}\right]}}{|\vec{n}_{x1,y2,x3}|^2}  \left(\frac{n_1 \tau_1}{t_1} dy_2\wedge dx_3 -\frac{n_2}{t_2\tau_2} dx_1\wedge dx_3+\frac{n_3\tau_3}{t_3} dx_1\wedge dy_2\right)\right. \nonumber\\ &\left. \textcolor{black}{-}\frac{ e^{i 2\pi  \vec{n}\cdot \left[(x_1, x_2, y_3)+\vec{\eta}\right]}}{|\vec{n}_{x1,x2,y3}|^2} \left(\frac{n_1\tau_1}{t_1} dx_2\wedge dy_3 -\frac{n_2\tau_2}{t_2} dx_1\wedge dy_3+\frac{n_3}{\tau_3 t_3} dx_1\wedge dx_2\right)\right]\, ,
\end{align}
where the functions $B_i$ are defined as in \cite[eq.(6.18)]{Marchesano:2020qvg}, $\vec{\eta}$ has entries that are either $0$ or $1/2$ and $|\vec{n}_{x1,x2,y3}|^2=\left(n_1/R_{x^1} \right)^2+\left(n_2/R_{x^2}\right)^2+\left(n_3/R_{y^3}\right)^2$ and similar for the other indices $\{x_i, y_j\}$. 

Using the above expression together with \eqref{eq: torus J}, we arrive to
\begin{align}
\mathcal{F}^2\wedge J_\cy&=2 h^2\sum_{\vec\eta,\vec\eta'} \sum_{\vec{0}\neq\vec{n}\in \IZ^3}\sum_{\vec{0}\neq\vec{m}\in \IZ^3} e^{i2\pi\left(\vec{n}\cdot \vec{\eta}+\vec{m}\cdot \vec{\eta'}\right)}\left[\left( \frac{\tau_1^2}{t_1} \frac{e^{i2\pi\vec{n}\cdot\left(x_1,x_2,y_3\right)+i2\pi\vec{m}\cdot\left(x_1,y_2,x_3\right)}}{|\vec{n}|^2_{x1,x2,y3}|\vec{m}|^2_{x1,y2,x3}}\right. \right.\nonumber \\ &  \left. \left.+\frac{1}{t_1\tau_1^2}\frac{e^{i2\pi\vec{m}\cdot\left(y_1,x_2,x_3\right)+i2\pi\vec{n}\cdot\left(y_1,y_2,y_3\right)}}{|\vec{n}|^2_{y1,x2,x3}|\vec{m}|^2_{y1,y2,y3}}\right)m_1n_1+\dots \right]
 dx_1\wedge dx_2\wedge dx_3\wedge dy_1\wedge dy_2\wedge dy_3\, .
\label{gorda}
\end{align}

Now we would like to compute $\int\mathcal{F}^2\wedge J_\cy$ integrating each piece of \eqref{gorda}, but to perform these integrals we first need to regularise them. We do so by smearing the O6-plane over a region of radius $\sim \ell_s$, which is the region of $X_6$ where the supergravity approximation cannot be trusted. 
In practice this corresponds to a truncation of the summation over the Fourier modes labelled by $\vec{n}$ and $\vec{m}$. This allows us to interchange the order between summation and integration. We then take the limit when the cut-off of the sum $N$ diverges, returning to our original system with a localised O6-plane. For the first piece of the integral we obtain 
\begin{align}
   & \lim_{N\to\infty} \sum_{\substack{\{\vec{0}\neq\vec{n}\in \IZ^3\mid |\vec{n}|\leq N\}\\ \{\vec{0}\neq\vec{m}\in \IZ^3||\vec{n}|\leq N\}}}\sum_{\vec\eta,\vec\eta'} e^{i2\pi\left(\vec{n}\cdot \vec{\eta}+\vec{m}\cdot \vec{\eta'}\right)}\int_{T^6}\frac{\tau_1^2}{t_1} \frac{e^{i2\pi\vec{n}\cdot\left(x_1,x_2,y_3\right)+i2\pi\vec{m}\cdot\left(x_1,y_2,x_3\right)}}{|\vec{n}|^2_{x1,x2,y3}|\vec{m}|^2_{x1,y2,x3}}m_1n_1\nonumber\\
    =& \lim_{N\to\infty} \sum_{\substack{\{\vec{0}\neq\vec{n}\in \IZ^3\mid |\vec{n}|\leq N\}\\ \{\vec{0}\neq\vec{m}\in \IZ^3||\vec{n}|\leq N\}}}\sum_{\vec\eta,\vec\eta'} e^{i2\pi\left(\vec{n}\cdot \vec{\eta}+\vec{m}\cdot \vec{\eta'}\right)}\frac{m_1n_1\tau_1^2}{t_1|\vec{n}|^2_{x1,x2,y3}|\vec{m}|^2_{x1,y2,x3}} \delta(n_1+m_1)\delta(n_2)\delta(n_3)\delta(m_2)\delta(m_3)\nonumber\\
    =& -\sum_{\vec{0}\neq n_1\in \IZ}\sum_{\vec\eta,\vec\eta'} e^{i2\pi n_1\left( \eta_1-\eta_1'\right)}\frac{n_1^2 R_{x^1}^4\tau_1^2}{n_1^4 t_1}\,.
\end{align}

Repeating a similar process for all the contributions in \eqref{gorda} and adding them together we conclude
\begin{align}
    \frac{1}{\ell_s^6}\int_{X_6}  \mathcal{F}^2\wedge J_{CY}=&-2 h^2(t_1+t_2+t_2)\sum_{n=1}^{\infty}\frac{1}{n^2}\sum_{\vec{\eta},\vec{\eta}'}e^{i2\pi n(\eta_1-\eta_1')}\nonumber\\
    =&-64 h^2(t_1+t_2+t_3)\sum_{n=1}^\infty\left(\frac{1}{n^2}+\frac{(-1)^n}{n^2}\right)\nonumber\\
    =&-\frac{(8h)^2}{12}\, \pi^2 (t_1+t_2+t_3)   \, ,
\end{align}
where we have taken into account that the integration space of $X_6 = T^6/(\mathbb{Z}_2 \times \mathbb{Z}_2)$ is $1/4$ of that of $T^6$. Notice that the result goes like the square of the number of D6-branes, and therefore as their pairwise intersection. Indeed, in this case the $\Pi_{\rm O6}$ is composed of $4 \times 8$ different three-cycles, with each group of 8 three-cycles wrapping the same class of $T^6$:
\be
\text{P.D.} [\Pi_{\rm O6}] = 8  [\beta^0] - 8[\beta^1] - 8[\beta^2] - 8[\beta^3]\, .
\ee
Two different three-cycles intersect over one one-cycle, so there are 64 intersections arising from each pair of classes. Finally, because there are  $h$ D6-branes wrapping $\Pi_{\rm O6}$, this factor increases to $(8h)^2$. It would be interesting to generalise this result to more involved D6-brane configurations, and in particular those where they do not lie on top of O6-planes.

\bibliographystyle{JHEP}
\inputencoding{latin2}
\bibliography{biblio}
\inputencoding{utf8}

\newpage
\thispagestyle{empty}
\phantom{lala}
\end{document}